\documentclass[twocolumn,astrosymb]{aastex7}
\usepackage{amsmath}
\usepackage{tikz}
\usepackage{layouts}
\usepackage{booktabs}
\usetikzlibrary{fadings}
\usetikzlibrary{patterns}
\usetikzlibrary{shadows}
\usetikzlibrary{shadows.blur}
\usetikzlibrary{shapes}
\usetikzlibrary{fit}
\usetikzlibrary{calc}
\usetikzlibrary{positioning}
\usetikzlibrary{arrows}
\usetikzlibrary{arrows.meta}

\usepackage{multirow}
\usepackage{rotating}  
\usepackage{todonotes}

\usepackage{zref-clever}
\makeatletter
\usepackage{etoolbox}
\patchcmd\H@refstepcounter{\protected@edef}{\protected@xdef}{}{}
\pretocmd\@sect{\def\@currentcounter{#1}}{}{\fail}
\makeatother

\newcommand{\hi}{H\textsc{i}}
\newcommand{\bh}{\mathcal{W}_{\rm BH}}

\def\nick{\hyperlink{cite.HERA22}{H22a}}
\def\theory{\hyperlink{cite.HERA22theory}{H22b}}
\def\josh{\hyperlink{cite.HERA23}{H23}}
\def\rath{\hyperlink{cite.Rath2025}{R\&P25}}

\makeatletter
\newcommand{\showfont}{encoding: \f@encoding{},
  family: \f@family{},
  series: \f@series{},
  shape: \f@shape{},
  size: \f@size{}
}
\newcommand{\iffont}[3]{\ifthenelse{\equal{\f@family}{#1}}{#2}{#3}}
\makeatother

\begin{document}

\title{First Results from HERA Phase II}

\correspondingauthor{Steven G. Murray}


\author{Zuhra  Abdurashidova}
\affiliation{Department of Astronomy, University of California, Berkeley, CA}
\email{zuhraa470@gmail.com}

\author{Tyrone  Adams}
\affiliation{South African Radio Astronomy Observatory, Black River Park, 2 Fir Street, Observatory, Cape Town, 7925, South Africa}
\email{tadams@sarao.ac.za}

\author[0000-0002-4810-666X]{James E. Aguirre}
\affiliation{Department of Physics and Astronomy, University of Pennsylvania, Philadelphia, PA}
\email{jeaguirre1@gmail.com}

\author{Rushelle  Baartman}
\affiliation{South African Radio Astronomy Observatory, Black River Park, 2 Fir Street, Observatory, Cape Town, 7925, South Africa}
\email{rushelle@sarao.ac.za}

\author[0000-0002-1557-693X]{Rennan Barkana}
\affiliation{School of Physics and Astronomy, Tel-Aviv University, Tel-Aviv, 69978, Israel}
\email{}

\author[0000-0002-2293-9639]{Lindsay M. Berkhout}
\affiliation{School of Earth and Space Exploration, Arizona State University, Tempe, AZ}
\affiliation{Department of Physics and Trottier Space Institute, McGill University, 3600 University Street, Montreal, QC H3A 2T8, Canada}
\email{lberkhou@asu.edu}

\author[0000-0002-0916-7443]{Gianni  Bernardi}
\affiliation{INAF-Istituto di Radioastronomia, via Gobetti 101, 40129 Bologna, Italy}
\affiliation{Department of Physics and Electronics, Rhodes University, PO Box 94, Grahamstown, 6140, South Africa}
\affiliation{South African Radio Astronomy Observatory, Black River Park, 2 Fir Street, Observatory, Cape Town, 7925, South Africa}
\email{giannibernardi75@gmail.com}

\author{Tashalee S. Billings}
\affiliation{Department of Physics and Astronomy, University of Pennsylvania, Philadelphia, PA}
\email{tashalee@sas.upenn.edu}

\author[0000-0001-7794-6599]{Bruno B. Bizarria}
\affiliation{Jodrell Bank Centre for Astrophysics, University of Manchester, Manchester, M13 9PL, United Kingdom}
\affiliation{Astrophysics division -- INPE, Instituto Nacional de Pesquisas Espaciais, S\~ao Jos\'e dos Campos -- SP, Brasil}
\email{bruno.bizarria@postgrad.manchester.ac.uk}

\author[0000-0002-8475-2036]{Judd D. Bowman}
\affiliation{School of Earth and Space Exploration, Arizona State University, Tempe, AZ}
\email{judd.bowman@asu.edu}

\author[0000-0002-2349-3341]{Daniela  Breitman}
\affiliation{Scuola Normale Superiore, 56126 Pisa, PI, Italy}
\email{daniela.breitman@sns.it}

\author[0000-0001-5668-3101]{Philip  Bull}
\affiliation{Jodrell Bank Centre for Astrophysics, University of Manchester, Manchester, M13 9PL, United Kingdom}
\affiliation{Department of Physics and Astronomy,  University of Western Cape, Cape Town, 7535, South Africa}
\email{philbull@gmail.com}

\author[0000-0002-8465-9341]{Jacob  Burba}
\affiliation{Jodrell Bank Centre for Astrophysics, University of Manchester, Manchester, M13 9PL, United Kingdom}
\email{jacob.burba@manchester.ac.uk}

\author[0000-0003-4980-2736]{Ruby  Byrne}
\affiliation{Department of Physics, University of Washington, Seattle, WA}
\email{rlbyrne@uw.edu}

\author{Steven  Carey}
\affiliation{Cavendish Astrophysics, University of Cambridge, Cambridge, UK}
\email{shcarey@mrao.cam.ac.uk}

\author[0009-0002-6260-0055]{Rajorshi Sushovan Chandra}
\affiliation{Raman Research Institute, Bangalore}
\email{rajorshi@rrimail.rri.res.in}

\author[0000-0002-3839-0230]{Kai-Feng  Chen}
\affiliation{MIT Kavli Institute, Massachusetts Institute of Technology, Cambridge, MA}
\affiliation{Department of Physics, Massachusetts Institute of Technology, Cambridge, MA}
\email{kfchen@mit.edu}

\author[0000-0002-2338-935X]{Samir  Choudhuri}
\affiliation{Centre for Strings, Gravitation and Cosmology, Department of Physics, Indian Institute of Technology Madras, Chennai 600036, India}
\email{samir@iitm.ac.in}

\author[0009-0008-2574-3878]{Tyler  Cox}
\affiliation{Department of Astronomy, University of California, Berkeley, CA}
\email{tyler.a.cox@berkeley.edu}

\author[0000-0003-3197-2294]{David R. DeBoer}
\affiliation{Radio Astronomy Lab, University of California, Berkeley, CA}
\email{ddeboer@berkeley.edu}

\author[0000-0001-8530-6989]{Eloy  de Lera Acedo}
\affiliation{Cavendish Astrophysics, University of Cambridge, Cambridge, UK}
\affiliation{Kavli Institute for Cosmology, Madingley Road, Cambridge CB30HA, UK}
\email{eloy@mrao.cam.ac.uk}

\author{Matt  Dexter}
\affiliation{Radio Astronomy Lab, University of California, Berkeley, CA}
\email{mdexter@berkeley.edu}

\author[0000-0002-1481-0907]{Jiten Dhandha}
\affiliation{Institute of Astronomy, University of Cambridge, Madingley Road, Cambridge CB30HA, UK}
\affiliation{Kavli Institute for Cosmology, Madingley Road, Cambridge CB30HA, UK}
\email{jvd29@cam.ac.uk}

\author[0000-0003-3336-9958]{Joshua S. Dillon}
\affiliation{Department of Astronomy, University of California, Berkeley, CA}
\affiliation{Radio Astronomy Lab, University of California, Berkeley, CA}
\email{jsdillon@berkeley.edu}

\author{Scott  Dynes}
\affiliation{MIT Kavli Institute, Massachusetts Institute of Technology, Cambridge, MA}
\email{sbcdynes@mit.edu}

\author{Nico  Eksteen}
\affiliation{South African Radio Astronomy Observatory, Black River Park, 2 Fir Street, Observatory, Cape Town, 7925, South Africa}
\email{neksteen@sarao.ac.za}

\author{John  Ely}
\affiliation{Cavendish Astrophysics, University of Cambridge, Cambridge, UK}
\email{j.ely@mrao.cam.ac.uk}

\author[0000-0002-0086-7363]{Aaron  Ewall-Wice}
\affiliation{Department of Astronomy, University of California, Berkeley, CA}
\affiliation{Department of Physics, University of California, Berkeley, CA}
\email{aaronew@berkeley.edu}

\author{Nicolas  Fagnoni}
\affiliation{Cavendish Astrophysics, University of Cambridge, Cambridge, UK}
\email{nf323@mrao.cam.ac.uk}

\author[0000-0002-1369-633X]{Anastasia  Fialkov}
\affiliation{Institute of Astronomy, University of Cambridge, Madingley Road, Cambridge CB30HA, UK}
\affiliation{Kavli Institute for Cosmology, Madingley Road, Cambridge CB30HA, UK}
\email{afialkov@ast.cam.ac.uk}

\author[0000-0002-0658-1243]{Steven R. Furlanetto}
\affiliation{Department of Physics and Astronomy, University of California, Los Angeles, CA}
\email{sfurlane@ucla.edu}

\author{Kingsley  Gale-Sides}
\affiliation{Cavendish Astrophysics, University of Cambridge, Cambridge, UK}
\email{kingsley@gale-sides.net}

\author[0009-0001-3949-9342]{Hugh  Garsden}
\affiliation{Jodrell Bank Centre for Astrophysics, University of Manchester, Manchester, M13 9PL, United Kingdom}
\email{hugh.garsden@manchester.ac.uk}

\author[0000-0002-1712-737X]{Adelie  Gorce}
\affiliation{Institut d’Astrophysique Spatiale, CNRS, Université Paris-Saclay, 91405 Orsay, France}
\email{adelie.gorce@mail.mcgill.ca}

\author[0000-0002-0829-167X]{Deepthi  Gorthi}
\affiliation{Department of Astronomy, University of California, Berkeley, CA}
\email{deepthigorthi@berkeley.edu}

\author{Ziyaad  Halday}
\affiliation{South African Radio Astronomy Observatory, Black River Park, 2 Fir Street, Observatory, Cape Town, 7925, South Africa}
\email{zhalday@sarao.ac.za}

\author[0000-0001-7532-645X]{Bryna J. Hazelton}
\affiliation{Department of Physics, University of Washington, Seattle, WA}
\affiliation{eScience Institute, University of Washington, Seattle, WA}
\email{brynah@uw.edu}

\author[0000-0002-4117-570X]{Jacqueline N. Hewitt}
\affiliation{MIT Kavli Institute, Massachusetts Institute of Technology, Cambridge, MA}
\affiliation{Department of Physics, Massachusetts Institute of Technology, Cambridge, MA}
\email{jhewitt@mit.edu}

\author{Jack  Hickish}
\affiliation{Radio Astronomy Lab, University of California, Berkeley, CA}
\email{jackhickish@gmail.com}

\author{Tian  Huang}
\affiliation{Cavendish Astrophysics, University of Cambridge, Cambridge, UK}
\email{huangtian44@hotmail.com}

\author[0000-0002-0917-2269]{Daniel C. Jacobs}
\affiliation{School of Earth and Space Exploration, Arizona State University, Tempe, AZ}
\email{daniel.c.jacobs@asu.edu}

\author[0000-0002-4118-6695]{Alec  Josaitis}
\affiliation{Cavendish Astrophysics, University of Cambridge, Cambridge, UK}
\email{atj28@cam.ac.uk}

\author[0000-0002-8211-1892]{Nicholas S. Kern}
\affiliation{MIT Kavli Institute, Massachusetts Institute of Technology, Cambridge, MA}
\affiliation{Department of Physics, University of Michigan, Randall Lab, 450 Church St., Ann Arbor, MI 48109}
\email{nkern@umich.edu}

\author[0000-0002-1876-272X]{Joshua  Kerrigan}
\affiliation{Department of Physics, Brown University, Providence, RI}
\email{Joshua_Kerrigan@alumni.brown.edu}

\author[0000-0003-0953-313X]{Piyanat  Kittiwisit}
\affiliation{Department of Physics and Astronomy,  University of Western Cape, Cape Town, 7535, South Africa}
\affiliation{South African Radio Astronomy Observatory, Black River Park, 2 Fir Street, Observatory, Cape Town, 7925, South Africa}
\email{piyanat.kittiwisit@gmail.com}

\author[0000-0002-2950-2974]{Matthew  Kolopanis}
\affiliation{School of Earth and Space Exploration, Arizona State University, Tempe, AZ}
\email{mjkolopa@asu.edu}

\author[0000-0003-2116-3573]{Adam  Lanman}
\affiliation{Department of Physics, Brown University, Providence, RI}
\email{adam_lanman@brown.edu}

\author[0000-0002-4693-0102]{Paul  La Plante}
\affiliation{Department of Computer Science, University of Nevada, Las Vegas, NV 89154}
\affiliation{Nevada Center for Astrophysics, University of Nevada, Las Vegas, NV 89154}
\email{plaplant@berkeley.edu}

\author[0000-0001-6876-0928]{Adrian  Liu}
\affiliation{Department of Astronomy, University of California, Berkeley, CA}
\affiliation{Department of Physics and Trottier Space Institute, McGill University, 3600 University Street, Montreal, QC H3A 2T8, Canada}
\email{adrian.liu2@mcgill.ca}

\author[0000-0001-8108-0986]{Yin-Zhe  Ma}
\affiliation{Department of Physics, Stellenbosch University, Matieland, Western Cape, 7602, South Africa}
\email{mayinzhe@sun.ac.za}

\author[0000-0001-6950-5072]{David H. E. MacMahon}
\affiliation{Radio Astronomy Lab, University of California, Berkeley, CA}
\email{davidm@berkeley.edu}

\author{Lourence  Malan}
\affiliation{South African Radio Astronomy Observatory, Black River Park, 2 Fir Street, Observatory, Cape Town, 7925, South Africa}
\email{lmalan@sarao.ac.za}

\author{Cresshim  Malgas}
\affiliation{South African Radio Astronomy Observatory, Black River Park, 2 Fir Street, Observatory, Cape Town, 7925, South Africa}
\email{cmalgas@sarao.ac.za}

\author{Keith  Malgas}
\affiliation{South African Radio Astronomy Observatory, Black River Park, 2 Fir Street, Observatory, Cape Town, 7925, South Africa}
\email{kmalgas@sarao.ac.za}

\author{Bradley  Marero}
\affiliation{South African Radio Astronomy Observatory, Black River Park, 2 Fir Street, Observatory, Cape Town, 7925, South Africa}
\email{bmarero@sarao.ac.za}

\author{Zachary E. Martinot}
\affiliation{Department of Physics and Astronomy, University of Pennsylvania, Philadelphia, PA}
\email{zmartinot@gmail.com}

\author[0009-0008-7585-9385]{Lisa  McBride}
\affiliation{Institut d’Astrophysique Spatiale, CNRS, Université Paris-Saclay, 91405 Orsay, France}
\affiliation{Department of Physics and Trottier Space Institute, McGill University, 3600 University Street, Montreal, QC H3A 2T8, Canada}
\email{elizabeth.mcbride@mail.mcgill.ca}

\author[0000-0003-3374-1772]{Andrei  Mesinger}
\affiliation{Scuola Normale Superiore, 56126 Pisa, PI, Italy}
\email{andrei.mesinger@sns.it}

\author[0000-0002-8802-5581]{Jordan  Mirocha}
\affiliation{Jet Propulsion Laboratory, California Institute of Technology, 4800 Oak Grove Drive, Pasadena, CA 91109, USA}
\affiliation{California Institute of Technology,  1200 E. California Boulevard, Pasadena, CA 91125, USA}
\email{jordan.mirocha@jpl.nasa.gov}

\author{Nicel  Mohamed-Hinds}
\affiliation{Department of Physics, University of Washington, Seattle, WA}
\email{nicel@uw.edu}

\author{Mathakane  Molewa}
\affiliation{South African Radio Astronomy Observatory, Black River Park, 2 Fir Street, Observatory, Cape Town, 7925, South Africa}
\email{mmolewa@sarao.ac.za}

\author[0000-0001-7694-4030]{Miguel F. Morales}
\affiliation{Department of Physics, University of Washington, Seattle, WA}
\email{miguelfm@uw.edu}

\author[0000-0002-8984-0465]{Julian B. Mu\~noz}
\affiliation{Department of Astronomy, The University of Texas at Austin, 2515 Speedway, Stop C1400, Austin, Texas 78712, USA}
\email{julianbmunoz@utexas.edu}

\author[0000-0003-3059-3823]{Steven G. Murray}
\affiliation{Scuola Normale Superiore, 56126 Pisa, PI, Italy}
\email{steven.murray@sns.it}

\author{Bojan  Nikolic}
\affiliation{Cavendish Astrophysics, University of Cambridge, Cambridge, UK}
\email{b.nikolic@mrao.cam.ac.uk}

\author{Hans  Nuwegeld}
\affiliation{South African Radio Astronomy Observatory, Black River Park, 2 Fir Street, Observatory, Cape Town, 7925, South Africa}
\email{hnuwegeld@sarao.ac.za}

\author[0000-0002-5400-8097]{Aaron R. Parsons}
\affiliation{Department of Astronomy, University of California, Berkeley, CA}
\affiliation{Radio Astronomy Lab, University of California, Berkeley, CA}
\email{aparsons@berkeley.edu}

\author[0000-0003-0073-5528]{Robert  Pascua}
\affiliation{Department of Astronomy, University of California, Berkeley, CA}
\affiliation{Department of Physics and Trottier Space Institute, McGill University, 3600 University Street, Montreal, QC H3A 2T8, Canada}
\email{r.pascua@berkeley.edu}

\author[0000-0002-9457-1941]{Nipanjana  Patra}
\affiliation{Department of Astronomy, University of California, Berkeley, CA}
\affiliation{International Centre for Radio Astronomy Research, Curtin University, Bentley WA 6102, Australia}
\email{nipanjana.patra@uwa.edu.au}

\author[0009-0005-8660-0713]{Simon  Pochinda}
\affiliation{Cavendish Astrophysics, University of Cambridge, Cambridge, UK}
\affiliation{Kavli Institute for Cosmology, Madingley Road, Cambridge CB30HA, UK}
\email{sp2053@cam.ac.uk}

\author[0000-0002-4314-1810]{Yuxiang  Qin}
\affiliation{Research School of Astronomy and Astrophysics, Australian National University, Canberra, ACT, Australia}
\email{yuxiang.l.qin@gmail.com}

\author[0009-0008-7886-2766]{Eleanor  Rath}
\affiliation{MIT Kavli Institute, Massachusetts Institute of Technology, Cambridge, MA}
\affiliation{Department of Physics, Massachusetts Institute of Technology, Cambridge, MA}
\email{elerath@mit.edu}

\author{Nima  Razavi-Ghods}
\affiliation{Cavendish Astrophysics, University of Cambridge, Cambridge, UK}
\email{nima@mrao.cam.ac.uk}

\author{Daniel  Riley}
\affiliation{MIT Kavli Institute, Massachusetts Institute of Technology, Cambridge, MA}
\email{dgriley@mit.edu}

\author{Kathryn  Rosie}
\affiliation{South African Radio Astronomy Observatory, Black River Park, 2 Fir Street, Observatory, Cape Town, 7925, South Africa}
\affiliation{South African Astronomical Observatory, 1 Observatory Rd, Observatory, Cape Town, 7925}
\email{krosie@saao.ac.za}

\author[0000-0003-3892-3073]{Mario G. Santos}
\affiliation{South African Radio Astronomy Observatory, Black River Park, 2 Fir Street, Observatory, Cape Town, 7925, South Africa}
\affiliation{Department of Physics and Astronomy,  University of Western Cape, Cape Town, 7535, South Africa}
\email{mariogrs@gmail.com}

\author[0000-0001-7755-902X]{Saurabh  Singh}
\affiliation{Raman Research Institute, Bangalore}
\email{saurabh.ascan@gmail.com}

\author[0000-0003-4092-0103]{Dara  Storer}
\affiliation{Department of Physics, University of Washington, Seattle, WA}
\email{darajstorer@gmail.com}

\author{Hilton  Swarts}
\affiliation{South African Radio Astronomy Observatory, Black River Park, 2 Fir Street, Observatory, Cape Town, 7925, South Africa}
\email{hswarts@sarao.ac.za}

\author[0000-0001-6161-7037]{Jianrong  Tan}
\affiliation{Department of Physics and Astronomy, University of Pennsylvania, Philadelphia, PA}
\email{jianrong@sas.upenn.edu}

\author[0000-0001-8838-1394]{Emilie Th\'elie}
\affiliation{Department of Astronomy, The University of Texas at Austin, 2515 Speedway, Stop C1400, Austin, Texas 78712, USA}
\email{emilie.thelie@austin.utexas.edu}

\author{Pieter  van Wyngaarden}
\affiliation{South African Radio Astronomy Observatory, Black River Park, 2 Fir Street, Observatory, Cape Town, 7925, South Africa}
\email{pvanwyn@sarao.ac.za}

\author[0000-0001-7716-9312]{Michael J. Wilensky}
\affiliation{Department of Physics and Trottier Space Institute, McGill University, 3600 University Street, Montreal, QC H3A 2T8, Canada}
\affiliation{CITA National Fellow}
\email{michael.wilensky@mcgill.ca}

\author[0000-0003-3734-3587]{Peter K. G. Williams}
\affiliation{Center for Astrophysics, Harvard \& Smithsonian, Cambridge, MA}
\affiliation{American Astronomical Society, Washington, DC}
\email{pwilliams@cfa.harvard.edu}

\author{Haoxuan  Zheng}
\affiliation{Department of Physics, Massachusetts Institute of Technology, Cambridge, MA}
\email{jeff.h.zheng@gmail.com}

\collaboration{all}{The HERA Collaboration}



\begin{abstract}
We report the first upper limits on the power spectrum of 21\,cm fluctuations during the Epoch of Reionization and Cosmic Dawn from Phase II of the Hydrogen Epoch of Reionization Array (HERA) experiment.
HERA Phase II constitutes several significant improvements in the signal chain compared to Phase I, most notably resulting in expanded frequency bandwidth, from 50--250\,MHz. 
In these first upper limits, we investigate a small two-week subset of the available Phase II observations, with a focus on identifying new systematic characteristics of the instrument, and establishing an analysis pipeline to account for them.
We report 2$\sigma$ upper limits in eight spectral bands, spanning $5.6 \leq z \leq 24.4$ that are consistent with thermal noise at the $2\sigma$ level for $k \gtrsim 0.6-0.9\,h{\rm Mpc}^{-1}$ (band dependent).
Our tightest limit during Cosmic Dawn ($z>12$) is $1.13\times 10^6\,{\rm mK}^2$ at ($k=0.55\,h{\rm Mpc}^{-1}, z=16.78$), and during the EoR ($5.5<z<12$) it is $1.78\times 10^3\,{\rm mK}^2$ at ($k=0.70\,h{\rm Mpc}^{-1}, z=7.05$).
We find that mutual coupling has become our dominant systematic, leaking foreground power that strongly contaminates the low-$k$ modes, resulting in the loss of modes from $k=0.35-0.55$ compared to Phase I data. 
\end{abstract}


\section{Introduction}
\label{sec:intro}
\setcounter{footnote}{0}
Observations of the redshifted 21\,cm spectral line of neutral hydrogen (\hi) have great potential to inform us about key physical processes throughout cosmic history: from the Dark Ages \citep{Loeb2006} and Cosmic Dawn \citep{Furlanetto2006}, through the Epoch of Reionization \citep{Mesinger2016} and into the post-reionization Universe \citep{Amiri2023}.
In the pre-reionization Universe, the line is sensitive to both cosmological and astrophysical processes, including the assembly of cosmic structure, the birth of the first stars, and the growth of the first galaxies, through its dependence on the local density and thermal conditions of the inter-galactic medium (IGM) \citep{Furlanetto2006,Mesinger2010}. 
By measuring angular and line-of-sight fluctuations in the 21\,cm brightness temperature with respect to the Cosmic Microwave Background (CMB), we gain access to information about the underlying sources of heating and ionization, and may be able to precisely constrain properties of these sources \citep{Mao2008,Patil2014,Pober2014,Liu2016a,Greig2015,Kern2017,Munoz2018}.
For reviews of the physics of the 21\,cm line and its applications, see e.g. \citet{Ciardi2005,Furlanetto2006,Morales2010,Pritchard2012,Mesinger2016,Liu2020c}.

Detection of the cosmological redshifted 21\,cm line is possible with low-frequency radio telescopes operating at frequencies between $10-1420$\,MHz. 
In particular, the frequency range $47-234$\,MHz corresponds to $z=5.1-29.3$, encompassing Cosmic Dawn and Reionization.
Several experiments---past, present and upcoming---have been built to observe the 21\,cm signal from the Epoch of Reionization and beyond over the past two decades.
These experiments can be categorized into two types.
Interferometers measure the spatial fluctuations of the 21\,cm signal, commonly summarized by the spherically-averaged \textit{power spectrum} $\Delta^2_{21}(k)$. 
Such experiments have placed increasingly stringent upper limits on this quantity.
This includes completed experiments, such as the Giant Meter Wave Radio Telescope \citep[GMRT;][]{Paciga2011} and the Donald C. Backer Precision Array for Probing the EoR \citep[PAPER;][]{Parsons2010,Parsons2014,Cheng2018,Kolopanis2019}.
It also includes several ongoing experiments, such as the Low Frequency Array \citep[LOFAR;][]{vanHaarlem2013,Patil2017a,Gehlot2018,Acharya2024,Mertens2020,Mertens2025}, 
the Murchison Widefield Array \citep[MWA;][]{Tingay2013,Dillon2014a,Dillon2015,Jacobs2016,Ewall-Wice2016a,Beardsley2016,Barry2019,Li2019,Trott2020,Yoshiura2021,Rahimi2021,Kolopanis2023, Wilensky2023,Nunhokee2025},
the Long-Wavelength Array \citep[LWA;][]{Eastwood2019,Garsden2021},
the Upgraded Giant Metrewave Ratio Telescope \citep[uGMRT;][]{Gupta2017,Chatterjee2019},
the Nancay Upgrading LOFAR experiment \citep[NenuFAR;][]{Zarka2012,Munshi2024,Munshi2025a}
and the Hydrogen Epoch of Reionization Array \citep[HERA;][]{DeBoer2017,HERA22,HERA23}.
Future experiments include the Square Kilometer Array \citep[SKA;][]{Dewdney2016} and LOFAR 2.0 \citep{Orru2024}.

In addition to these interferometric experiments, a number of single-antenna experiments that measure the sky-averaged 21\,cm temperature as a function of frequency/redshift (the `global signal')  have been constructed. 
Notably, the Experiment to Detect the Global Eor Signature \citep[EDGES;][]{Rogers2012} has claimed a detection of 21\,cm absorption during Cosmic Dawn \citep{Bowman2018}, though the amplitude and profile of the absorption feature are difficult to account for under standard astrophysical and cosmological scenarios, requiring either cooling of the IGM below the adiabatic limit \citep{Barkana2018,Munoz2018a}, a high-redshift radio background in excess of the CMB \citep{Feng2018,Ewall-Wice2018,Mirocha2019a,Fialkov2019}, or an unmodelled systematic contaminating the data analysis \citep{Hills2018,Bradley2019,Singh2019,Sims2020,Murray2022c, Cang2024}.
Other experiments that have published global signal data include the Large-Aperture Experiment to Detect the Dark Ages \citep[LEDA;][]{Price2018a}, the Broadband Instrument for Global Hydrogen Reionization \citep[BIGHORNS;][]{Sokolowski2015} and the Shaped Antenna measurement of the background RAdio Spectrum \citep[SARAS;][]{Patra2015,Singh2017}. 
Notably, the third-generation SARAS instrument recently published a \textit{non-detection} of the EDGES absorption feature during Cosmic Dawn, with a confidence level of 2$\sigma$ \citep{Singh2022,Bevins2022}.
Upcoming experiments of this variety include the Radio Experiment for the Analysis of Cosmic Hydrogen \citep[REACH;][]{DeLeraAcedo2022a}, the Remote \hi\ eNvironment Observer \citep[RHINO;][]{Bull2024} the Mapper of the IGM Spin Temperature \citep[MIST;][]{Monsalve2024} and Probing Radio Intensity at High-Z from Marion \citep[PRIZM;][]{Philip2019}.

In this paper, we present the first power spectrum upper limits from Phase II of the HERA experiment.
HERA Phase I \citep{DeBoer2017} operated with a subset of the full complement of antennas ($<$$60$), and used repurposed dipoles from the PAPER experiment, sensitive between 100--200\,MHz.
Phase I observations culminated with an observing season in 2017--2018 resulting in two successive power spectrum upper limits; first from a two-week subset of the data (\hyperlink{cite.HERA22}{HERA Collaboration 2022a}; hereafter \nick), and finally from the full 94-night dataset (\hyperlink{cite.HERA23}{HERA Collaboration 2023}; hereafter \josh).
In the meantime, Phase II has been steadily rolling out.
Phase II \citep{Berkhout2024} includes several upgrades: a new signal chain, correlator and upgraded Vivaldi feeds that together extend the bandwidth out to 47--234\,MHz.
The redshifts newly observable in Phase II are of particular interest; they now cover the absorption feature reported by EDGES \citep{Bowman2018} as well the tail-end of reionization ($z=5-6$) which has received considerable interest lately, with several studies based on the Ly$\alpha$ forest of quasi-stellar objects (QSO's) placing the end of reionization at $z\sim$ 5.3--5.5 \citep{Bosman2022,Zhu2022, Qin2024}.
In addition to the upgraded system components, Phase II has added more antennas, increasing the instantaneous sensitivity of the telescope. 
Here, we present an analysis of a two-week subset of data observed in October 2022, during which 180 dual-polarized antennas were online.
For more details on the HERA Phase II system design, see \citet{Berkhout2024}.

The greatest challenge to making a detection of 21\,cm fluctuations in the pre-reionization Universe is the confluence of spectrally-structured systematics and bright foregrounds that outshine the signal by up to four orders of magnitude.
Since the bright foregrounds are most readily distinguished from the 21\,cm signal by their different spectral behaviour---with foregrounds intrinsically spectrally smooth and the 21\,cm signal possessing spectral structure on all scales---it is important to design and calibrate 21\,cm experiments to minimize the structure they imprint on the foregrounds.
There are many potential sources of such structure, for example environmental factors such as ionospheric refraction and radio frequency interference \citep[RFI;][]{Wilensky2020, Wilensky2023, Gehlot2024, Munshi2025}, signal-chain effects such as cable reflections \citep{Beardsley2016} and cross-talk, errors in analysis such as primary-beam non-redundancy causing mis-calibration \citep{Orosz2019,Kim2022,Kim2023} or incomplete sky models affecting absolute calibration \citep{Barry2016, Byrne2019}, and structural concerns such as incomplete $uv$ coverage \citep{Liu2014a,Murray2018c}.
One of the motivations of this paper is to uncover and quantify the systematics present in updated Phase II system (especially those that are new or more prominent in comparison to Phase I), and to demonstrate that they can be appropriately modeled and/or mitigated. 
One particular systematic that has emerged as the most detrimental and difficult to handle in Phase II is that of \textit{mutual coupling}: the reflection or re-emission of sky signal by one antenna into another \citep{Camps1998,Kern2019, Kern2020, Josaitis2021,Gueuning2022,Charles2024,Pascua2024,Rath2025,OHara2025}. 
Mutual coupling is present at higher levels in Phase II than was seen in Phase I data.
This is attributable to the increased sensitivity of the new wideband Vivaldi feeds towards the horizon and their suspension above their dish without the shielding of the surrounding cavity used for the Phase I dipoles, which created unwanted reflections within the dish.
Currently, the feeds are in full view of neighboring antennas---both their feeds and the dishes themselves. 
While we have worked to develop tools to understand and mitigate mutual coupling in our analysis pipeline \citep{Kern2019, Kern2020a, Josaitis2021,Charles2024,Rath2025,Pascua2024}, it remains our dominant systematic in the crucial $k$-modes just beyond those dominated by intrinsic foregrounds. 
We will discuss this systematic at some length throughout this work.

To date, the limits presented in \josh~remain the deepest limits of the 21\,cm power spectrum from any experiment at redshifts 8 and 10.
These limits were made with 94 nights of observing (of 12 hours each) with a maximum of 41 unflagged antennas per night.
The limits, specifically $\Delta^2_{21}(k=0.34 h{\rm Mpc}^{-1}) \leq 457 {\rm mK}^2$ at $z=7.9$ and $\Delta^2_{21}(k=0.36 h{\rm Mpc}^{-1}) \leq 3496 {\rm mK}^2$ at $z=10.4$, are consistent with thermal noise over a wide range of $k$, indicating that further integration should yield corresponding improvement.
Notably, both of the previous HERA upper limits were supported by extensive simulation and statistical validation \citep{Aguirre2022,Tan2021}, which cataloged small sources of signal loss that were corrected in the final limits.
This analysis in this paper extends these supporting validation tests with updated and expanded simulations that cover the larger set of antennas in the Phase II dataset.

Although previous HERA results were only upper limits, a range of inferences with different physical models were unanimous in concluding that these limits rule out the class of `cold reionization' models (\hyperlink{cite.HERA22theory}{HERA Collaboration 2022b}, hereafter \theory; strengthened by \josh{}).

Cold reionization is the regime in which large-scale reionization occurs without significant heating of the IGM (due to, for example, inefficient production of X-rays from high-mass X-ray binaries). 
Cold reionization is the scenario that naturally produces the largest amplitude $\Delta^2_{21}$ after reionization begins ($z\sim6-12$) \citep{Mesinger2014,Pober2015a}.
Prior to being ruled out in \josh, cold reionization scenarios were already disfavored by prior limits set by both LOFAR \citep{Ghara2020a,Greig2021,Ghara2025} and the MWA \citep{Ghara2021,Greig2021a}. 
Beyond ruling out cold reionization, \josh{} constrained the properties of early high-mass X-ray binaries (HMXB's);
under the assumption that these systems dominate the heating of the IGM before and during reionization, their soft-band X-ray luminosity per star-formation rate ($L_{X<2keV}/{\rm SFR}$) was inferred to be higher than that of local HMXB's at more than 3$\sigma$, and instead consistent with a population of low-metallicity HMXB's \citep{Fragos2013,Madau2017,Kaur2022}.
However, this conclusion is conditional on the astrophysical model employed.  For example, \citet{Lazare2024} showed that if star formation is very efficient inside the first, molecularly-cooled galaxies, the constraints on $L_{X<2keV}/{\rm SFR}$ could weaken considerably, though this depends strongly on the unknown properties of such Population III dominated galaxies (e.g. \citealt{Lazare2024}; Breitman et al in prep).

Meanwhile, over the past several years, our understanding of the $z>5$ Universe has continued to improve thanks to measurements beyond the 21\,cm line. For reionization, the detection of high-HI~opacity regions in quasar spectra at $z \la 5$ has provided increasingly strong evidence that the tail end of reionization stretches to between $z\sim5-5.5$ \citep{Becker2021,Choudhury2021,Bosman2022,Gaikwad2023,Qin2024,Zhu2022}. 
But the \textit{James Webb Space Telescope} \citep[JWST;][]{Gardner2006} observations have shown that star formation is substantially more common at $z \ga 10$ than expected from models calibrated to HST observations (e.g., \citealt{Leung2023,Donnan2024,Finkelstein2024}) and that the observed galaxy population is extremely blue \citep{Topping2022,Cullen2023}. These observations suggest that $z \ga 10$ galaxies might be more efficient at emitting ionizing photons than previously expected \citep{Gelli2024,Munoz2024,Nikolic2024}, which could result in an earlier start to reionization.  In such a scenario, recombinations inside IGM clumps would be essential in extending the later stages of the EoR in order to match the Lyman alpha forest observations  \citep{Davies2021,Qin2024}. Meanwhile, Lyman-$\alpha$ line emission from galaxies -- which should be absorbed by the IGM once it is substantially neutral -- has proven to be surprisingly common even at very high redshifts (e.g., \citealt{Bunker2023,Witstok2025}). These varied observations have raised even more questions about the reionization process. 

JWST has also raised new questions about the importance of AGN to the Cosmic Dawn era. Signatures of accreting supermassive black holes have been found in a surprisingly large number of high-$z$ sources, including UV-luminous galaxies \citep{Bunker2023,Goulding2023,Maiolino2024,Castellano2024} and the so-called ``Little Red Dots,’’ compact sources with broad emission lines that may be due to gas surrounding accreting black holes  (e.g. \citealt{Matthee2024,Greene2024}, but see also \citealt{Leung2024} for other explanations). The prevalence of these sources (accounting for up to $\sim 10\%$ of the galaxy population) suggests that accreting black holes may play a larger role in the early Universe than previously expected, though this depends on the (heretofore poorly understood) nature of the central sources (e.g., \citealt{Leung2024,Maiolino2025}). 
An additional population of accreting black holes that are X-ray luminous may affect the early thermal history of the IGM, which will be tested by HERA and other 21-cm experiments.

This paper is organized as follows. 
In \zcref{sec:data} we describe the dataset analysed in this work, including the instrumental configuration and data selection. 
In \zcref{sec:methods} we detail the analysis pipeline we have developed, highlighting key differences compared to our previous analyses of Phase I data (\nick{}, \josh{}). 
In \zcref{sec:validation} we demonstrate the validity of our pipeline by applying it to detailed end-to-end mock simulations, as well as performing statistical tests designed to expose biases in our power spectrum estimates. 
In \zcref{sec:results} we present the main results of the paper: both cylindrically-averaged spectra and spherically-averaged upper limits. 
In \zcref{sec:interpretation} we explore the impact of these new limits on our understanding of astrophysics during Cosmic Dawn, and finally in \zcref{sec:conclusions} we conclude with a summary and prospectus for future HERA analyses, given the large amount of data already taken. 

Throughout this work, we adopt the cosmology of \citet{Planck2015Cosmo}, namely $\Omega_\Lambda = 0.68440$, $\Omega_b=0.04911$, $\Omega_c=0.26442$ and $h=0.6727$.

\section{Observations}
\label{sec:data}

\subsection{The HERA Phase II Telescope}
The HERA telescope is located at the SARAO site in the Karoo desert in South Africa.
It consists of 350 wire-mesh parabolic dishes, each of which has a diameter of $\sim$14\,m and contains a suspended feed \citep{DeBoer2017}.
The dishes are arranged in a tightly-packed hexagon \citep{Dillon2016} spaced 14.6\,m apart (center to center), with the dishes almost touching each other. 320 antennas form a core with a maximum inter-antenna distance of 292\,m, along with two concentric layers of outriggers that increase the maximum baseline length to 876\,m. 

This array layout maximizes sensitivity to a small number of (mostly large-scale) $k$-modes, making HERA an extremely sensitive telescope for these cosmologically-relevant modes, as well as enabling the redundant-calibration approach discussed in \zcref{sec:methods}.
\zcref[S]{fig:array-layout} shows HERA's antenna layout, comparing the antennas used for our previous limits to those used here.

\begin{figure*}
    \centering
    \includegraphics{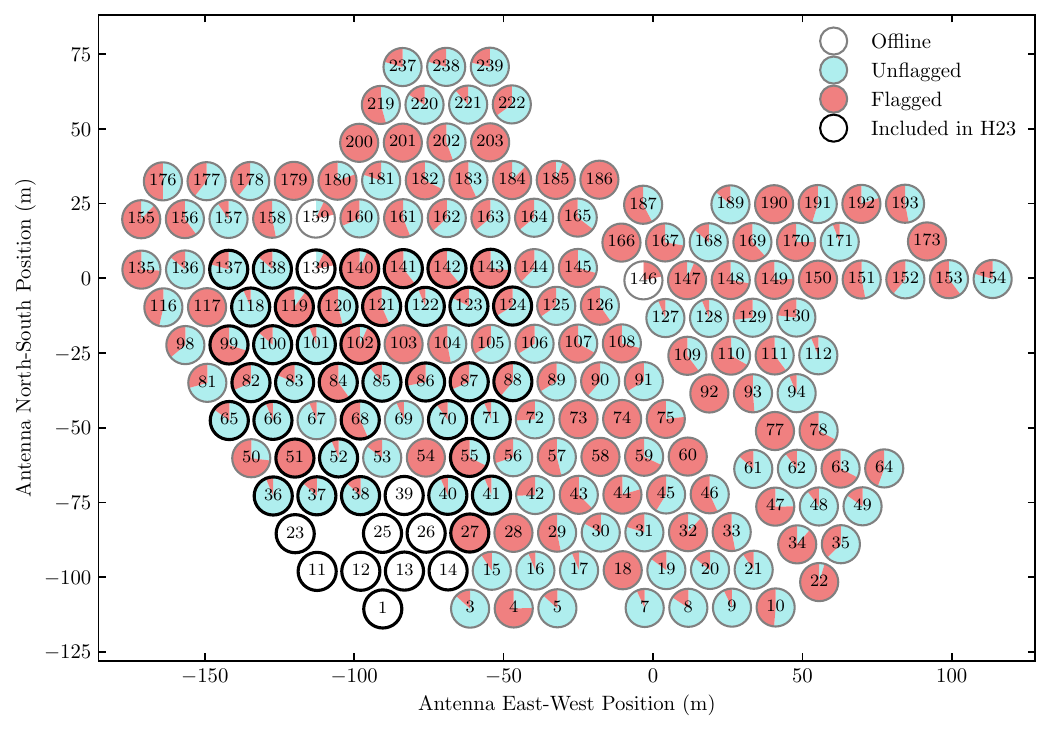}
    \caption{Map of antennas used at least once within this dataset, with inner pie-charts indicating the fraction of time (i) observed without flagging (blue) (ii) flagged (red) and (iii) offline (white).
    Antennas highlighted with a thick black border indicate those used in \josh.
    Two nights completely flagged due to lightning (c.f. \zcref{tab:lightningflags}) are not considered here.
    The indicated flagging amounts include data quality checks (see \zcref{sec:methods:per-file}) and per-antenna flag synthesis (see \zcref{sec:methods:per-night:per-ant}), but not channel-based flags from e.g. RFI. 
    The displayed fractions for each antenna average across both feed polarizations.
    The total time observed across the dataset is 136.3 hours.
    Over 120 antennas observed usable data during this dataset, compared to the 48 in \josh{}, corresponding naively to a $\sim$5$-$10-fold increase in sensitivity. 
    }
    \label{fig:array-layout}
\end{figure*}

Drift scanning from its location in South Africa, HERA observes a stripe across the $-30$\arcdeg\ declination.  This stripe crosses the galactic center, rising to a galactic declination of $-80$\arcdeg (c.f. \zcref{fig:lstcov}). 
Observing constraints include minimizing foreground power from the Galaxy, bright point sources, and the Sun. This results in an optimal season spanning the southern summer, generally from September through April of the following calendar year. 

During a season, the number of antennas that are online is reasonably constant -- generally upgrades and maintenance are most active during the off-season.
In this paper, we report a small subset of a single observing season, 2022-2023.

The most important development between HERA's first two reported upper limits and those presented here is the system change from Phase I to Phase II.
The Phase I system covered 100$-$200\,MHz ($z=6-13$), and reused several components from its predecessor PAPER \citep{Parsons2010}, including the dipole feeds, cables, signal processing boards \citep{Parsons2006}, xGPU correlator \citep{Parsons2008,Clark2013}, and some analog components. 
The new Phase II instrument encompasses several major upgrades aimed at increasing bandwidth and reducing systematics. Bandwidth was increased by changing the feed and upgrading the digitizers.
The  PAPER feeds were replaced with broader-band Vivaldi-style feeds \citep{Fagnoni2021}, extending the usable frequency range on both ends, to 47$-$234\,MHz ($z\sim5-27$).
On the low-band end, this covers the redshift range in which Cosmic Dawn is expected to occur \citep{Pritchard2012,Bowman2018}, while on the upper end this covers the tail-end of reionization, which may extend all the way to $z\sim5.3$ \citep{Bosman2022}.
The analog system, signal processing boards \citep{Hickish2016}, and correlator were also upgraded to match the extended frequency range of the feeds and accommodate a larger number of antennas \citep{Berkhout2024}. 
Finally, the 100m analog cables were replaced with an RF over fiber (RFoF) system. Use of fiber reduced re-radiation from the coaxial cables and allowed the cables to be made 500m long. This increased length moved reflections to delays corresponding to $k$ modes of reduced cosmological significance.
For this reason, cable reflections are no longer fit for as a part of calibration as was necessary in \nick{} and \josh{} (see \citealt{Kern2019,Kern2020a} for the methodology).

\subsection{Data Selection}
\label{sec:data:selection}

HERA records continuously through the observing season from sunset to sunrise for an average of 12 hours per night.
The data used in this paper are a 14-night subset of the 2022-2023 observing season, starting on October $8^{th}-9^{th}$ 2022 (JD 2459861) and ending on October $23^{rd}-24^{th}$ 2022 (JD 2459876).  
Of this contiguous set of nights, two were excluded from the analysis (2459865 and 2459875) due to a high prevalence of strong broadband RFI throughout the nights, which we attribute to lightning storms \cite{Heiligstein2023}. 
Within the remaining 14 nights, several more hours were flagged for the entire array due to similar broadband RFI. 
The quality metrics used for excluding observing times were determined based only on the auto-correlations---a small subset of the full data--- prior to any of the analysis steps described in \zcref{sec:methods}.
\zcref[S]{tab:lightningflags} lists the nights and LSTs that were affected. In total, not including the two full nights that were excluded, $\sim$7.5 observing hours or 5.2\% of the remaining 144 hours  were flagged due to this broadband RFI.

\begin{table}[]
    \centering
    \begin{tabular}{c|l}
         Night (JD) & LSTs Flagged (hours) \\
         \hline
         2459863 & $20^h 46^m$--$23^h 39^m$ \\
         2459865 & ALL \\
         2459869 & $21^h 10^m$ -- $0^h10^m$, $7^h05^m$--$7^h12^m$\\
         2459872 & $21^h22^m$--$22^h20^m$, \\
         2459875 & ALL \\
         2459876 & $21^h38^m$--$21^h52^m$, $7^h26^m$--$7^h39^m$\\
         \hline
    \end{tabular}
    \caption{Nights and LSTS for which the entire observation was flagged due to excess broadband RFI, believed to be attributable to lightning storms. In general, partially flagged nights were only flagged at the beginning or end of the night (or both) so as not to introduce large flagging gaps in an individual night.}
    \label{tab:lightningflags}
\end{table}

\begin{figure*}
    \centering
    \includegraphics{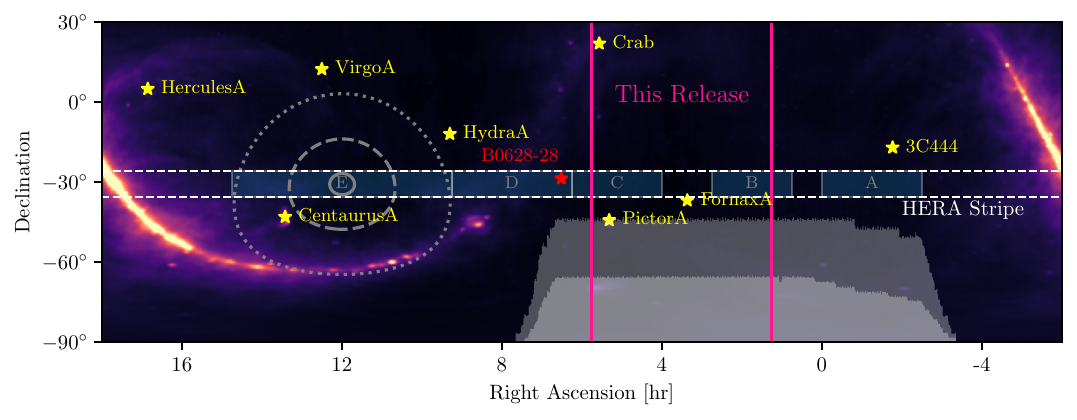}
    \caption{The sky coverage of this dataset, compared to previous HERA limits. 
    The background shows the diffuse sky model of \citet{DeOliveira-Costa2008} computed with \texttt{pygdsm} \citep{Price2016} with an overlay of the ``HERA stripe'' indicating the region of sky HERA can observe. 
    Inset transparent-blue regions indicate the fields included as part of the \josh{} upper limits. In this analysis, we include just one field, from LST 1.25--5.75 hours, indicated by the pink vertical lines. 
    Note that while we denote HERA's field-of-view by the dashed white stripes, corresponding to the full-width at half-maximum of the primary beam (at 150\,MHz), the beam has non-negligible sensitivity out to large zenith-angles, which we illustrate with the gray contours. The dashed (dotted) contour corresponds to the 1\% (0.2\%) sensitivity of the primary beam at 150\,MHz.
    The beam has $\sim0.1\%$ sensitivity down to the horizon.
    The inset gray histograms indicate the LST-coverage of the dataset we describe here. The taller background shaded region has a height proportional to the number of antenna-feed-integrations a particular LST was observed (i.e. the number of integrations observed by any antenna on any feed-polarization at that LST), while the shorter foreground histogram is the resulting coverage after antenna flagging (see \zcref{fig:array-layout}).
    For reference, the maximum of the background histogram is 65,394 antenna-polarization-integrations (10 seconds each) within LST bins of 270.5-seconds (c.f. \zcref{sec:methods:pspec-bl:frf}), while the maximum number of \textit{unflagged} observations is just over half this value.}
    \label{fig:lstcov}
\end{figure*}

~\zcref[S]{fig:lstcov} illustrates the LST coverage of this dataset, comparing it to the Phase I dataset reported in \josh.
The background in this figure represents the foreground intensity, here computed with the \texttt{pygdsm}\footnote{\url{https://github.com/telegraphic/pygdsm}} software.
Since HERA is a fixed, zenith-pointing instrument, its observational footprint is confined to a stripe of constant declination, 
here demarcated by the white dashed lines---limited by the full-width at half-maximum of the primary beam which at the lowest operable frequency is ($\sim$$10^\circ$).  
However, HERA receives emission well beyond this stripe: the primary beam is illustrated in \zcref{fig:lstcov} by the gray contours, which denote the 1\% (dashed) and 0.2\% (dotted) sensitivities respectively, while the sensitivity at the horizon is estimated to be $\sim0.1\%$ \citep{Fagnoni2021}.
Sufficiently bright sources anywhere above the horizon significantly affect measured visibilities.
The inlaid gray histogram indicates the number of independent observations at each LST, where independent observations are drawn from different nights, antennas, polarizations and integrations within an LST bin. 
As can be seen, the dataset we present here covers LSTs between $\sim$21 to 7\,hours, which includes Fields A, B, C and part of D from \josh.
In this work, we do not split the observations into separate fields as in previous works, rather reporting the power spectrum upper limits derived from the full range of 1.25 to 5.75 hours, which have maximum coverage,  have relatively low foreground amplitude, and avoid problematic sources such as the setting of the Galactic centre and the bright, high-rotation-measure pulsar B0628-28.

\zcref[S]{fig:array-layout} shows the antennas that were online during the observations included in this analysis, as well as the relative fraction of time they were unflagged (blue), flagged (red), or offline (white).
In this figure, the flags include the full per-antenna flags, but neglect frequency-dependent flags (e.g. due to RFI, see \zcref{sec:methods:per-night:per-ant}). 
The total for each antenna includes both polarization feeds.
Some antennas (e.g.\ 139, 159) were brought online mid-way through the two-week observing period.
Note that certain antennas (e.g.\ 186), though technically online, were always flagged. \zcref[S]{fig:array-layout} also shows the antennas that were included in \josh{} with darker outlines, highlighting the increased number of antennas in this dataset.
In summary, 180-183 antennas were online for each night in this dataset, of which 138 were unflagged for some portion of the dataset. 
The shortest baselines are 14.6\,m in length, and the longest (unflagged) baseline in the dataset is $\sim$280\,m in the East-West direction. 
We describe our antenna-based flagging metrics in more detail in \zcref{sec:methods:per-night-flag-synth}, and summarize the details of the data selection in \zcref{tab:obschar}.

This dataset represents a small fraction, roughly 10\%, of HERA's 2022--2023 observing season. 
It was not chosen based primarily on its quality, rather it was simply the first set of 14 nights observed that passed basic full-night quality checks (to avoid lightning storms).
During the later part of the season, the addition of new antennas and repairs improved the data quality somewhat, and fall weather brought fewer lightning storms.
Compared to the data set reported here the full season represents a ten-fold increase in the number of observed nights/hours that remain after autocorrelation-based flagging, as well as a modest increase in the average number of antennas that survive quality checks.
In total, the 2022--2023 season has just over 1300 hours of unflagged data, with an average of 140 unflagged antennas, while the dataset analysed here contains 129 hours of data with an average of 131 unflagged antennas.

\begin{table}[]
    \centering
    \begin{tabular}{lcc}
         &\textbf{Here} & \textbf{Full Season}  \\
         \hline\hline
         Array Location & \multicolumn{2}{c}{-30.72$^\circ$S, 21.43$^\circ$E} \\
         Avg.\ Antennas Connected & 180 & 195 \\
         Avg.\ Antennas Unflagged & 131 & 140 \\
         Shortest Baseline & 14.6\,m & 14.6\,m \\
         Longest Avail. Baseline & 280\,m & 755\,m \\
         \hline
         Minimum Frequency & \multicolumn{2}{c}{46.92\,MHz} \\
         Maximum Frequency & \multicolumn{2}{c}{234.30\,MHz} \\
         Channels & \multicolumn{2}{c}{1536} \\
         Channel Width, $\Delta\nu$ & \multicolumn{2}{c}{122\,kHz} \\
         \hline
         Integration Time, $\Delta t$ & \multicolumn{2}{c}{9.6\,sec} \\
         Per-Night Obs. & \multicolumn{2}{c}{12\,hr} \\
         Total Nights Used & 14 & 147 \\
         Unflagged Obs. Hours & 129 & 1312 \\
         \hline
         Raw Data Volume (Night) & 2.7\,TB & 2.7\,TB \\
         Raw Data Volume (Full) & 36\,TB & $\sim$350\,TB \\
         \hline
    \end{tabular}
    \caption{Observation characteristics for HERA's 2022-2023 observing season.}
    \label{tab:obschar}
\end{table}

\section{Analysis Pipeline}
\label{sec:methods}

The data analysis pipeline consists of several stages of data flagging, calibration, filtering and averaging. 
Almost all stages of analysis have been upgraded or modified with respect to \josh.
In this section, we detail the full pipeline, highlighting the main modifications compared to \josh{}.
For the reader already familiar with previous HERA analyses (\nick{}, \josh{}), we provide a compact overview of the main differences introduced in this work following the full description in \zcref{sec:methods:differences}.

\subsection{Definition of terms}

We first define some notation and nomenclature that recurs throughout the upcoming analysis discussion.


The basic measurement of the interferometer is a visibility $V_{ij}^{pq}$ correlating feed $p$ on antenna $i$ with $q$ on $j$.  The correlator integrates the instantaneous visibility over channel width $\Delta \nu$ and over time $\Delta t$ (c.f. \zcref{tab:obschar}). 
Colloquially known as ``integrations'', these visibilities measured between antennas in the HERA hexagonal grid offer repeated or ``redundant'' measurements of the same correlation\footnote{In practice, we define two baselines $\mathbf{b}_{ij}$ and $\mathbf{b}_{kl}$ as redundant if $|b^p_{ij} - b^p_{kl}|<2\,{\rm m}\ \ \forall p\in \{x,y,z\}$.}. 
These redundant measurements of the sky all occupy the same point in $uv$ space and are nominally only different up to thermal noise.  
Often one examines a particular ``redundant group'' $\mathcal{G}_{ij} = \{(i,j), (k,l), \dots \}$ of $|\mathcal{G}_{ij}|$ baselines that all occupy the same $uv$ point or ``unique baseline'' vector.
We uniquely label the baseline group by one of its elements (here $ij$).
The number of unique baselines in the data set is $N_{\rm ubl}$ (this can change over time as antennas are added/removed or flagged). 
Visibilities from redundant baselines are eventually averaged to a single ``redundantly averaged'' visibility. 
Once this averaging is done it is often practical to label the redundantly-averaged visibility according to the `key' baseline, so that $V_{ij}^{pq} \equiv V_{\mathcal{G}_{ij}}^{pq}$. 
The context will make clear when a visibility is a redundant average.

Throughout the pipeline, we often make use of the ``expected" variance of a particular visibility.
Whenever we reference the ``expected variance" of a cross-correlation, we mean that given by the radiometer equation, where the system temperature is estimated as the \textit{calibrated auto-correlation}\footnote{Any time the noise is estimated during the pipeline, the most up-to-date calibration solutions consistent with that level of processing are applied.}:
\begin{equation}
    {\rm Var}(V^{pq}_{ij}) = \frac{|V^{pp}_{ii}||V^{qq}_{jj}|}{\Delta t \Delta \nu N_{\rm samples}}.
    \label{eq:visibility-noise}
\end{equation}
Here $N_{\rm samples}$ is the number of integrations or baselines that have been coherently averaged together into $V_{ij}^{pq}$ (e.g. if a redundantly-averaged visibility from group $\mathcal{G}$ is under consideration, the number of samples includes $|\mathcal{G}|$). 
Very often, since well-calibrated auto-correlations should be nearly equal across antennas, we use the \textit{average} auto-correlation across antennas for a given polarization, $\overline{V^{pp}}$, instead of each $V^{pp}_{ii}$ individually.

\subsection{Overview of Pipeline}
\label{sec:methods:overview}
The analysis approach taken here is to estimate the power spectrum at each $uv$ point and then average these power spectra to get a single result. Averaging in this ``incoherent'' way requires phase agreement between redundant baselines and correctable stability in time and frequency but not between all baselines as would be required in an imaging approach \citep{Morales2019}. 
The remaining challenge, still significant, is to average coherently across redundant baselines, short timescales, and many nights and then average power spectra over sidereal times. 
The coherent average pipeline used here uses filters and statistical tests across time, frequency, and night to solve for antenna gains, mask interference, and check for system failures. The order of operations and choice of algorithms is constrained by the amount of data which can be held in memory, the speed of data access patterns and similar matters. The two week data set processed here, amounting to 36\,TB of data, ran in 3 days on a 16 node cluster.  Here we report the algorithmic approach and give some detail into the also-important technical organization behind this processing. 

\begin{figure*}
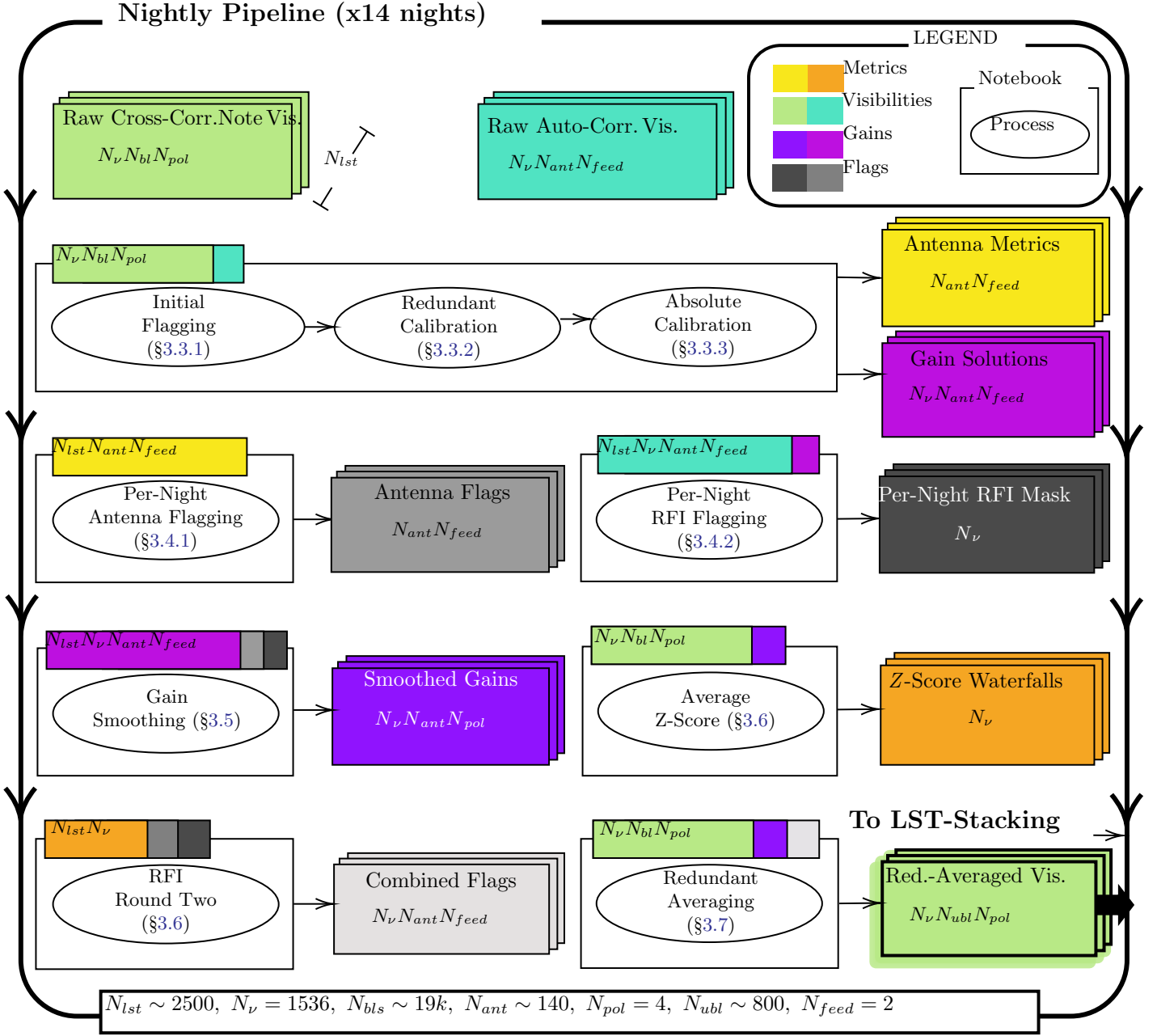

    \centering
    \include{analysis-flowchart}
    \caption{The HERA per-night analysis pipeline (leading to the power spectrum pipeline in \zcref{fig:pspec-flowchart}). 
    As indicated in the title, this pipeline is run separately for each night of observation.
    The general flow of the analysis is from top-to-bottom and left-to-right.
    Large colored rectangles represent data products, with the dimensionality indicated as text in each rectangle, and the stacks representing the fact that the products are stored as multiple files over time (LST).
    Ovals with inset text represent analysis steps (with references to sections of this paper in which they are described), and rectangles surrounding one or more ovals represent single computational ``tasks'' generating Jupyter notebooks for inspection. 
        Small colored labels inset onto the top-left of each task are the input data required for the task, with color specifying the data type (corresponding to a previous data product) and text specifying the cut of data considered simultaneously. 
    }
    \label{fig:analysis-flowchart}
\end{figure*}

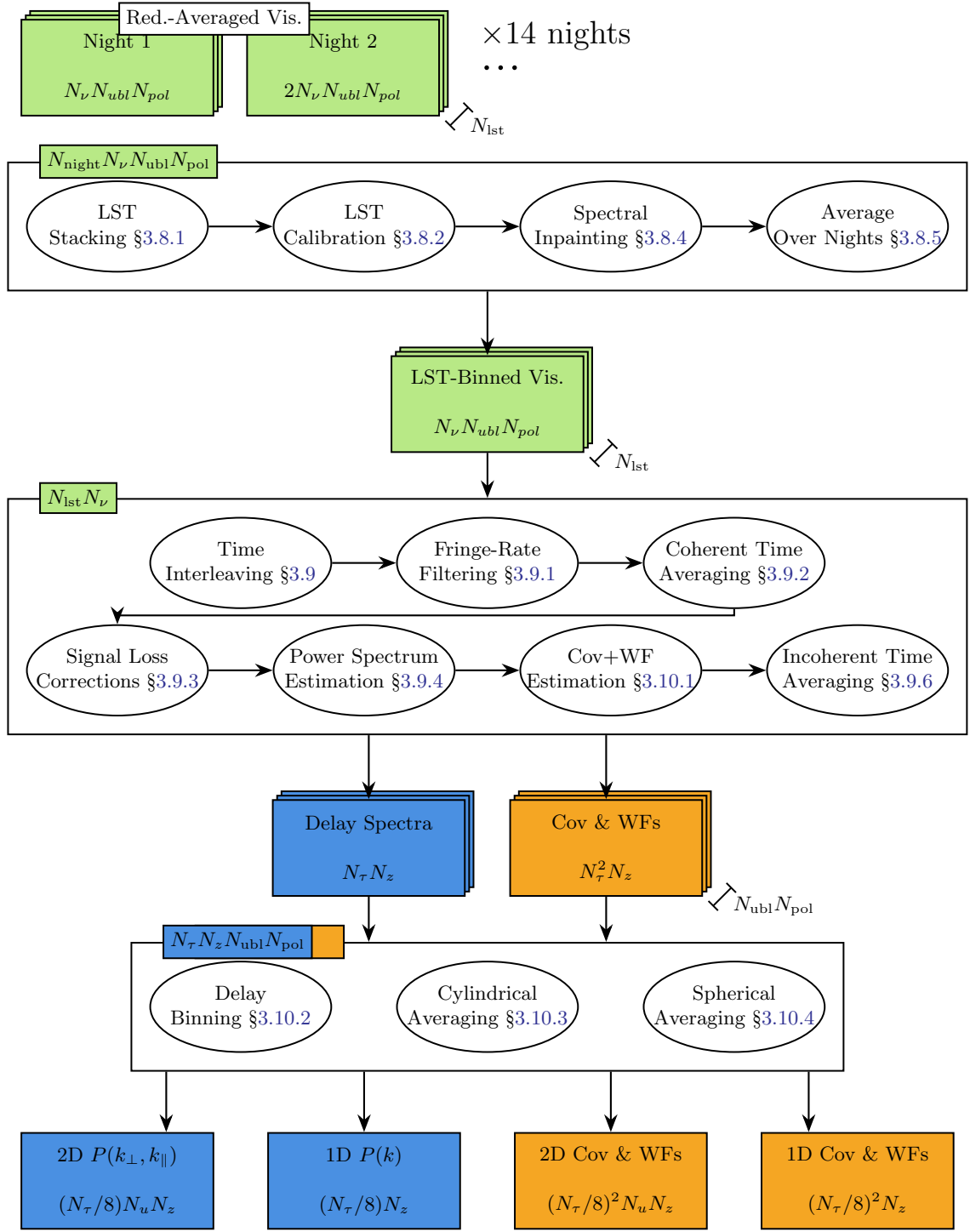
\begin{figure*}
    \centering
    \tikzset{every picture/.style={line width=0.75pt}} 

\definecolor{colorvis}{HTML}{B8E986}
\definecolor{colorspec}{HTML}{4A90E2}
\definecolor{colorcov}{HTML}{F5A623}

\begin{tikzpicture}[
    ell/.style={ellipse, draw=black,minimum height=1.5, minimum width=3, fit={(0,0) (2, 1)}, inner sep=0pt},
    rect/.style={rectangle, draw=black, minimum height=1.5, minimum width=3, inner sep=0pt},
    fls/.style = {double copy shadow, shadow xshift=2pt, shadow yshift=-2pt},
    arrow/.style={-{Stealth[scale=1.2]}, thick}
]

\newcommand{\nfiles}[2]{
    \begin{scope}
    \node[shape=coordinate] at (#1.south east) (#1-arrow) {};
    \draw[|-|]  ($(#1-arrow) + (0.15, -0.15)$) -- ++(0.3, 0.3);
    \node[anchor=west] at ($(#1-arrow) + (0.35, -0.15)$) {#2};
    \end{scope}
}

\node  at (0,0) [rect,fls,fill=colorvis, fit={(0,0) (3, 1.5)}, label={[align=center]center:Night 1\\\\$N_\nu N_{ubl} N_{pol}$}] (night1) {};
\node[
    rect, fls,fill=colorvis, fit={(0,0) (3, 1.5)}, 
    label={[align=center]center:Night 2\\\\$2N_\nu N_{ubl}N_{pol}$},
    right=0.5 of night1
]  (night2) {};

\node[right=0.5] (dots) at (night2.east) {\Huge ...} ;
\node[above=0.5,anchor=west] (n14) at (dots.west) {\Large $\times$14 nights} ;
\nfiles{night2}{$N_{\rm lst}$};
\node[draw, fill=white, anchor=center] at (night1.north east) (redavglab) {Red.-Averaged Vis.};

\node[
    ell, 
    label={[align=center]center:LST\\Stacking \S\ref{sec:methods:lstbin:stacking}}, 
    below=of night1,
] (lststack) {};
\node[
    ell,
    label={[align=center]center:LST\\Calibration \S\ref{sec:methods:lstbin:lstcal}}, 
    right=of lststack,
    node distance=0.25
] (lstcal) {};
\node[
    ell,
    label={[align=center]center:Spectral\\Inpainting \S\ref{sec:methods:lstbin:inpainting}}, 
    right=of lstcal,
    node distance=0.25
] (inpaint) {};
\node[
    ell, 
    label={[align=center]center:Average\\Over Nights \S\ref{sec:methods:lstbin:avg}}, 
    right=of inpaint,
    node distance=0.25
] (nightavg) {};

\draw[arrow] (lststack) -- (lstcal);
\draw[arrow] (lstcal) -- (inpaint);
\draw[arrow] (inpaint) -- (nightavg);

\node[shape=rectangle, draw, fit=(lststack) (nightavg), inner sep=8] (lststacknb) {};

\node[draw, fill=colorvis] at (lststacknb.north west) [right=0.5] (lststack-input) {$N_{\rm night}N_\nu N_{\rm ubl} N_{\rm pol}$};

\node[
    rect, fls, fill=colorvis, fit={(0,0) (3, 1.5)}, 
    label={[align=center]center:LST-Binned Vis.\\\\$N_\nu N_{ubl}N_{pol}$},
    below=of lststacknb, node distance=1
]  (lstavgdata) {};
\nfiles{lstavgdata}{$N_{\rm lst}$};

\draw[arrow] (lststacknb) -- (lstavgdata);


\node[
    ell, 
    label={[align=center]center:Fringe-Rate\\Filtering \S\ref{sec:methods:pspec-bl:frf}}, 
    below=of lstavgdata,
] (frf) {};
\node[
    ell,
    label={[align=center]center:Time\\Interleaving \S\ref{sec:methods:pspec-bl}}, 
    left=of frf,
    node distance=0.25
] (timeint) {};
\node[
    ell,
    label={[align=center]center:Coherent Time\\Averaging \S\ref{sec:methods:pspec-bl:coh-time-avg}}, 
    right=of frf,
    node distance=0.25
] (cohavg) {};

\node[
    ell,
    label={[align=center]center:Signal Loss\\Corrections \S\ref{sec:methods:signal-loss}},
    node distance=0.25,
    below=5.5cm of lststack
] (losscorr) {};

\node[
    ell, 
    label={[align=center]center:Power Spectrum\\Estimation \S\ref{sec:methods:pspec-bl:pspec}}, 
    right=of losscorr,
    node distance=0.25
] (pspecsingle) {};
\node[
    ell, 
    label={[align=center]center:Cov+WF\\Estimation \S\ref{sec:methods:cov_window}}, 
    right=of pspecsingle,
    node distance=0.25
] (covwf) {};

\node[
    ell, 
    label={[align=center]center:Incoherent Time\\Averaging \S\ref{sec:methods:pspec:incoh}}, 
    right=of covwf,
    node distance=0.25
] (incoh) {};

\node[shape=rectangle, draw, fit=(losscorr) (incoh) (cohavg), inner sep=8] (pspecnb) {};
\draw[arrow] (lstavgdata) -- (pspecnb);

\draw[arrow] (timeint) -- (frf);
\draw[arrow] (frf) -- (cohavg);
\draw[arrow] (cohavg.south) |- ($(frf.south) - (0, 0.1)$) -| (losscorr.north); 
\draw[arrow] (losscorr) -- (pspecsingle);
\draw[arrow] (pspecsingle) -- (covwf);
\draw[arrow] (covwf) -- (incoh);


\node[draw, fill=colorvis] at (pspecnb.north west) [right=0.5] (pspecnb-input) {$N_{\rm lst}N_\nu$};

\node[
    rect, fls, fill=colorspec, fit={(0,0) (3, 1.5)}, 
    label={[align=center]center:Delay Spectra\\\\$N_\tau N_z$},
    below=of pspecnb.225, node distance=1
]  (delayspec) {};
\node[
    rect, fls, fill=colorcov, fit={(0,0) (3, 1.5)}, 
    label={[align=center]center:Cov \& WFs\\\\$N^2_\tau N_z$},
    below=of pspecnb.315, node distance=1
]  (covwfsfls) {};
\nfiles{covwfsfls}{$N_{\rm ubl}N_{\rm pol}$};

\draw[arrow] (pspecnb.225) -- (delayspec.north);
\draw[arrow] (pspecnb.315) -- (covwfsfls.north);

\node[
    ell, 
    label={[align=center]center:Delay\\Binning \S\ref{sec:methods:sphavg:delaybin}}, 
    below=5.5cm of timeint,
] (delaybin) {};
\node[
    ell,
    label={[align=center]center:Cylindrical\\Averaging \S\ref{sec:methods:sphavg:cylavg}}, 
    right=of delaybin,
    node distance=0.25
] (cylavg) {};
\node[
    ell,
    label={[align=center]center:Spherical\\Averaging \S\ref{sec:methods:sphavg:sphavg}}, 
    right=of cylavg,
    node distance=0.25
] (sphavg) {};

\node[shape=rectangle, draw, fit=(delaybin) (sphavg), inner sep=8] (sphavgnb) {};
\node[draw, fill=colorcov] at (sphavgnb.north west) [right=1] (sphavg-input-false) {\phantom{$N_\tau N_z N_{\rm ubl} N_{\rm pol}$}};
\node[draw, fill=colorspec] at (sphavgnb.north west) [right=0.5] (sphavg-input) {$N_\tau N_z N_{\rm ubl} N_{\rm pol}$};

\draw[arrow] (delayspec.south) -- (sphavgnb.north -| delayspec.south);
\draw[arrow] (covwfsfls) -- (sphavgnb.north -| covwfsfls.south);

\node[
    rect, fill=colorspec, fit={(0,0) (3, 1.5)}, 
    label={[align=center]center:2D $P(k_\perp, k_\parallel)$\\\\$(N_\tau/8) N_u N_z$},
    below=6.5cm of losscorr
]  (2dpk) {};
\node[
    rect, fill=colorspec, fit={(0,0) (3, 1.5)}, 
    label={[align=center]center:1D $P(k)$\\\\$(N_\tau/8) N_z$},
    below=6.5cm of pspecsingle
]  (1dpk) {};
\node[
    rect, fill=colorcov, fit={(0,0) (3, 1.5)}, 
    label={[align=center]center:2D Cov \& WFs\\\\$(N_\tau/8)^2 N_u N_z$},
    below=6.5cm of covwf
]  (2dcov) {};
\node[
    rect, fill=colorcov, fit={(0,0) (3, 1.5)}, 
    label={[align=center]center:1D Cov \& WFs\\\\$(N_\tau/8)^2 N_z$},
    below=6.5cm of incoh
]  (1dcov) {};

\draw[arrow] (sphavgnb.south -| 2dpk.45) -- (2dpk.45) ;
\draw[arrow] (sphavgnb.south -| 1dpk.north) -- (1dpk.north);
\draw[arrow] (sphavgnb.south -| 2dcov.north) -- (2dcov.north);
\draw[arrow] (sphavgnb.south -| 1dcov.135) -- (1dcov.135);

\end{tikzpicture}
    \caption{The HERA LST-stacking power-spectrum estimation pipeline. The visual language has the same meaning as in \zcref{fig:analysis-flowchart}, with the addition of blue data products representing power spectra, and orange data products representing power spectrum statistics (covariances and window functions). In this pipeline, the size of each stack is indicated at its bottom right corner.
    Additionally, in contrast to \zcref{fig:analysis-flowchart}, this part of the pipeline is run only once, combining all nights in the first processing step.
    }
    \label{fig:pspec-flowchart}
\end{figure*}

The data analysis pipeline is illustrated in \zcref{fig:analysis-flowchart,fig:pspec-flowchart}.
We have broken up the pipeline into two flowcharts, with \zcref{fig:analysis-flowchart} showing the calibration and quality checks performed separately on each nights' data, and \zcref{fig:pspec-flowchart} depicting the combination of the nights and the subsequent power spectrum estimation.
Both charts flow in general from top to bottom, and where applicable from left to right.
In each, data products are represented by colored rectangles, with colors indicating the \textit{type} of data (indicated in the legend), and the data dimensionality indicated with text, as well as the `stacked' representation\footnote{The axis over which the data are `stacked' in the flowcharts corresponds to the axis over which the data is broken into separate files on disk.}. 

Observations are recorded on site and saved to UVH5 files\footnote{\url{https://github.com/RadioAstronomySoftwareGroup/pyuvdata/blob/main/docs/references/uvh5_memo.pdf}} that each include all 1536 frequency channels, 4 linear polarizations, $N_{\rm bl}$ baselines, and two 9.6-s time integrations.
Auto-correlation visibilities are extracted from the datafiles and saved independently, to allow for fast access during the analysis. 
These two raw datasets are represented as colored boxes at the top of \zcref{fig:analysis-flowchart}.

Each major step of the pipeline is implemented as a parameterized Jupyter Notebook \citep{Kluyver2016jupyter}, with formatted tables and figures within the notebooks making a self-documenting visual record of the analysis performed.
All of the notebooks produced by this analysis, as well as the final upper limits, covariances and window functions, are publicly available at \url{https://zenodo.org/records/15799499}.
The underlying algorithms for quality metrics, calibration, averaging, and power spectrum estimation are defined in a suite of libraries maintained by the HERA Collaboration at \url{https://github.com/HERA-Team/}\footnote{Of particular note are the repos \texttt{hera-qm}, \texttt{hera-cal} and \texttt{hera\_pspec}.}.
In the flowcharts, each Jupyter notebook is represented as a rectangle, within which may be several processing steps, represented by ellipses with titles and references to sections in the paper in which we describe the process.
Each notebook takes input data, which are denoted by colored tabs in the top-left corner of each notebook, where the color defines which data product is being input (multiple colors indicates multiple inputs). 
Each notebook also produces a set of outputs which are connected by arrows.

One of the most important aspects of the analysis pipeline to keep in mind is that most of the tasks cannot see all the data at one time, due to its size. 
The HERA analysis pipeline proceeds in stages by slicing data along different axes, aiming to make increasingly sensitive assessments within the bounds of the available computing and memory resources.
We represent these choices in \zcref{fig:analysis-flowchart,fig:pspec-flowchart} by defining the shape of the input data slices in the colored ``input'' tabs for each process, which are in general different to the size/shape of the data product on disk.
Furthermore, by considering condensed data products that are either independent of frequency (e.g. Antenna Metrics) or antenna-dependent instead of baseline-dependent (e.g. Gain Solutions), we are able to simultaneously consider the full set of LSTs for a particular night (e.g. the Smoothed Gains), and then incorporate this full-night information back into the raw data (e.g. Redundant Averaging) without needing to read the full set of raw data simultaneously.


The pipeline can be split into three major conceptual sections spanning the two flowcharts: calibration, redundant averaging and LST averaging, and finally power spectrum estimation. 
Each step includes additional calibration, filtering, flagging and statistical tests tuned to the relevant level of averaging.

The first section---ending after ``RFI Round Two'' near the bottom of \zcref{fig:analysis-flowchart}---is focused on calibrating and flagging the raw data.
The data itself is not modified at all in this section (note the lack of green visibility outputs up to this point); rather, we ultimately derive two new outputs: the frequency- and time-smoothed per-antenna gain solutions (purple; \zcref{sec:methods:abscal,sec:methods:redcal,sec:methods:smoothcal}) and a combined set of flags incorporating both antenna-based (\zcref{sec:methods:per-night:per-ant}) and frequency-dependent (\zcref{sec:methods:per-file:flagging,sec:methods:per-night:rfi,sec:methods:rfi-round-2}) masks (light gray).

The second section---starting with Redundant Averaging and ending after LST-binning---averages over axes in which the data are assumed to be redundant: baseline groups and nights. 
This results in two new intermediate visibility data products: 
(i) calibrated, flagged and redundantly-averaged visibilities, which are compressed by a factor $N_{\rm bl}/N_{\rm ubl}\sim 10$ (c.f. \zcref{sec:methods:redavg}), 
and (ii) LST-binned visibilities, in which the data from each night at the same sidereal time and baseline are averaged together, further compressing the data by a factor of $N_{\rm nights}=14$ (c.f. \zcref{sec:methods:lstbin}).
The LST-binned dataset is approximately 350\,GB.

The third and final section is power spectrum estimation. 
Here, we examine individual baselines independently rather than LSTs, allowing us to consider the full range of LST, polarization and frequency for each baseline simultaneously.
We interleave the dataset in time to produce four noise-independent subsets of data, for each of which we perform systematics mitigation through fringe-rate filtering (\zcref{sec:methods:pspec-bl:frf}), average the visibilities within LST-bins of 270\,sec (\zcref{sec:methods:pspec-bl:coh-time-avg}), compute signal-loss corrections due to these filters (\zcref{sec:methods:signal-loss}), compute cross-correlation delay spectra (\zcref{sec:methods:pspec-bl:pspec}) and their error-bars (\zcref{sec:methods:pspec:error-bars}) between the interleaved subsets,
and incoherently average the spectra and covariances over the LST-bins (\zcref{sec:methods:pspec:incoh}). 
Finally, we combine all the per-baseline delay spectra, average them within larger delay bins to reduce correlations between neighboring delays (\zcref{sec:methods:sphavg:delaybin}),
average over baseline orientations to produce cylindrical power spectra (\zcref{sec:methods:sphavg:cylavg}) and finally average within regular $|k|$-bins to produce the final spherically-averaged power spectrum estimates with their covariances and window functions (\zcref{sec:methods:sphavg:sphavg}).
In the following subsections, we describe these steps in more detail.

\subsection{Calibration and Flagging}
\label{sec:methods:per-file}
The first major step of the analysis pipeline is for each integration---spanning all baselines, channels, and polarizations---to be separately flagged and calibrated. 
Two types of flags are generated: per-antenna flags\footnote{Note that these are \textit{not} per-baseline flags: we find that most unrecoverable systematics affect individual antennas (and all baselines of which they are a part) rather than baselines, and the convenience of carrying around per-antenna flags is worth the small amount of over-flagging arising from baseline-dependent systematics. Also note that these flags are applied independently for the two instrumental polarizations of each antenna.}, which are for all channels, and per-channel flags, which are for all antennas---typically for RFI.
These flags and calibration solutions are later modified (i.e.\ smoothed or otherwise harmonized) in the context of full nights of gains and flags.

First, the data are flagged using the auto-correlations. 
The data that survives flagging is then calibrated; first we perform \textit{redundant} calibration to achieve relative calibration of all antennas, and the remaining degeneracies are then fixed by comparing to a simulated sky model.
We present each of these processes in more detail in the following subsections.

\subsubsection{Initial, Per-Integration Flagging}
\label{sec:methods:per-file:flagging}
Interference and malfunctioning antennas can introduce outliers that do not integrate down. 
For this reason, each calibration or averaging step is preceded by a flagging step in which outliers are identified and masked. In the first of these steps, each time, frequency and antenna are individually evaluated. In later steps averaged products will be used to find smaller outliers. 

Common factors causing outliers visible in individual data points are: RFI (generated by e.g.\ terrestrial FM stations), nearby lightning storms, and a malfunctioning signal chain for a particular antenna or node. 

For each integration\footnote{For practical reasons, we in fact read in files with \textit{two} integrations each, and process these two integrations together. Conceptually, this makes almost no difference to the analysis (each integration is still treated independently), however there is a small caveat that the initial per-antenna flags are generated \textit{per-file} (i.e. for two integrations) rather than per-integration. This has very minimal impact on the results.}, flags are generated on a per-antenna and per-channel basis.
For some metrics, there is an interaction between these modes; for instance, one metric for an antenna's fitness is whether the spectrum of the autocorrelation is similar to the average over all antennas. 
Anomalous spectral structure might indicate any number of underlying issues in the system.
However, such structure might also be a manifestation of RFI, which can show up more strongly on some antennas than others, and should be flagged on a per-channel basis instead
(potentially causing that antenna to pass antenna checks once the problematic channels are flagged).

It should also be noted that unlike the \texttt{auto\_metrics} presented in \citet{Storer2022}, these per-antenna cuts are mostly\footnote{With the exception of the comparison of the auto spectra to the mean just mentioned.} based on absolute metrics, as opposed to their being statistical outliers relative to the rest of the array.

First, we flag entire antennas that are dead (many zero-valued visibilities), have low correlations \citep{Storer2022} indicative of clock distribution issues \citep{Berkhout2024}, are cross-polarized, exhibit digital packet loss, or have anomalous autocorrelation power or slope relative to the rest of the antennas.
These properties are unlikely to be contaminated by per-channel effects. 
We detail the precise metrics used for these cuts in \zcref{app:antenna-metrics}.

Our three remaining metrics for determining per-antenna flags are more likely to interact with the per-channel flags.
To account for this, our strategy is to first determine an initial set of per-channel flags (we refer to this as an ``RFI mask," though it may flag more than just RFI), and then to iteratively refine both this mask and the set of per-antenna flags until these sets converge.
Using this strategy, we flag antennas that we deem to have broken X-engines (each X-engine correlates all antenna pairs within a block of 96 channels), excess RFI compared to other antennas, or an anomalous autocorrelation spectral shape. 
We detail these three metrics, and how they interact with the RFI mask through an iterative refinement, in \zcref{app:rfi-ant-iterative-flags}.
Ultimately, we carry through both these per-antenna flags and the refined RFI mask to the calibration as described in the next two subsections.

\subsubsection{Redundant Calibration}
\label{sec:methods:redcal}

Redundant calibration simultaneously solves for each complex antenna gain, $g_i(\nu)$ as well as a model for each \textit{unique} visibility $\hat{V}_{ij}$, through $chi^2$-minimization of the residual $V^{\rm meas}_{ij} - g_ig_j^\star \hat{V}_{ij}$.
In HERA's hexagonal layout most baselines are sampled many times which makes gain solutions well determined. This is one of the secondary benefits of HERA's layout design \citep{Dillon2016}. However, the method has limitations and challenges. It assumes identical antenna elements situated on a perfect grid. It also cannot solve for a handful of degrees of freedom in calibration, requiring a sky model.

The formalism of solving the non-linear optimization problem of simultaneously finding gains and visibility solutions is described in \citet{Liu2010,Dillon2020a}. The particular approach used for Phase I is described in \nick{} and \josh. Small changes have been made for Phase II to improve efficiency.

In Phase I, redundant calibration proceeded in three steps: (i) \texttt{firstcal}, which found an approximate solution for a per-antenna phase and delay, (ii) \texttt{logcal} in which the full solutions for the gains, including the amplitude, are determined in an approximate but biased way \citep{Liu2010}, and (iii) \texttt{omnical} in which the $\chi^2$ is minimized with fixed-point iteration, starting from the previous solutions \citep{Dillon2020a, Zheng2014}.
In Phase II, we completely eliminate the second step, \texttt{logcal}, instead simply averaging delay-calibrated visibilities from different baselines together to get initial solutions for iterative \texttt{omnical}. This is allowed to iterate for at least 100 cycles. 

However, if any new antennas are rejected in any iteration for having a $\chi^2$ per antenna beyond our threshold of 3 (antennas consistent with noise should have $\chi^2$ per antenna of 1, see \citet{Dillon2020a} for details), we discard the worst antennas and perform another 50 iterations, iterating until all unflagged antennas are below the threshold\footnote{Additionally, outrigger antennas are excluded from redundant calibration, both for computational speed and because we have not yet verified that they can be well-calibrated and brought into the analysis presented here.}.
The $\chi^2$ of each antenna---including those flagged during the calibration process---is propagated to the following steps to be considered for exclusion from the rest of the pipeline.

\subsubsection{Absolute Calibration}
\label{sec:methods:abscal}
``Absolute Calibration'' refers to the process of fixing the average gain and phase-gradient (or `tip-tilt') of the array to the true sky \citep{Kern2020}, after having determined the relative gains via redundant calibration. 

In \josh, the absolute calibration model was based on sky-calibrated HERA data taken from three LST ranges which contained catalog sources and had fields suitable for self-calibration.
Phase II  datasets span a larger range of frequencies, in particular the $\sim$50-100\,MHz low-band. 
With a larger span of data available a more consistent approach was desirable, and to achieve this we use a detailed model of the southern sky. 
This `Southern Sky Model' \citep[SSM;][Chapter 8]{Martinot2022} was propagated through the \texttt{RIMEz} visibility simulator \citep[][Chapter 3]{Martinot2022}\footnote{\url{https://github.com/zacharymartinot/RIMEz}} using the full HERA array layout and antenna beam \citep{Fagnoni2021}.
In particular, this sky model is anchored by the diffuse sky maps of \citet{Remazeilles2015} at 408 MHz and \citet{Guzman2011} at 45 MHz, with a spectral index that varies on scales of 5\arcdeg.  The overall flux scale is calibrated with the EDGES measurements presented in \citet{Monsalve2021}.  Point sources from the GLEAM catalog \citep{Hurley-Walker2016} are merged with the diffuse maps using a spherical harmonic formalism to correctly represent the power already present in the diffuse maps.  

\begin{figure}
    \centering
    \includegraphics[width=\linewidth]{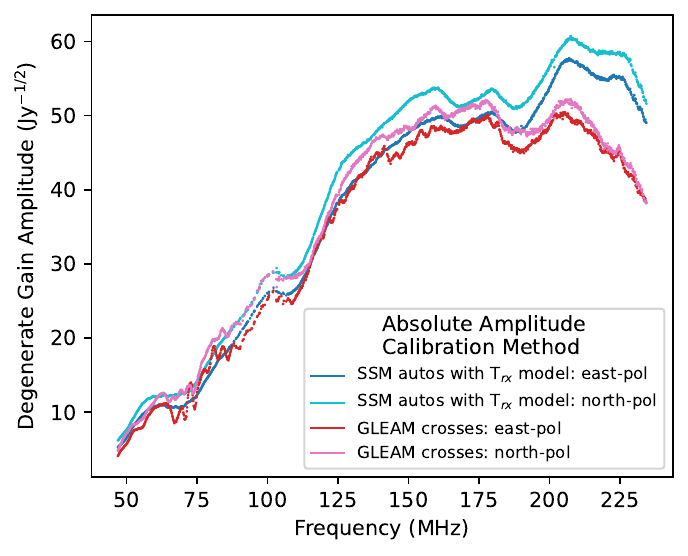}
    \caption{Comparison of the average absolute calibration versus frequency using the SSM from this work and the auto-correlations and using a sky-based calibration from GLEAM and and the cross-correlations.  This comparison is performed for a single integration at $\text{JD}=2459861.4782854$, which corresponds to an LST of 2.0789\,hours. At this LST, the field of view is dominated by two point sources near the center, making it one of HERA's best calibrator fields \citep{Kern2020}. Despite that, diffuse galactic emission is still quite important, and because the simulation using GLEAM lacks a diffuse model, only long baselines (60\,m$< |\mathbf{b}| <$140\,m) are used when performing amplitude calibration with cross-correlations. The agreement is quite good, but the SSM calibration exhibits less spurious spectral structure and is stable across LST, whereas the GLEAM calibration fails in regions with significant diffuse emission or gaps in the catalog.
    }
    \label{fig:ssm-abscal}
\end{figure}

In this analysis, the flux scale is set using the autocorrelations from the SSM simulation, including an estimate of the receiver temperature spectrum, implemented as a cubic spline interpolation of the data from the laboratory measurements performed in Sec.\ III-A of \citet{Fagnoni2021}.
Using autocorrelations instead of cross-correlations also naturally avoids bias in the flux scale due to intrinsic thermal noise \citep[e.g.][]{Aguirre2022}.
The flux scale bias is roughly inversely proportional to the SNR of the data used. 
Cross-correlations have a characteristic SNR of $\sim$10 which produce biases of order 0.1, but auto-correlations achieve an instantaneous SNR of $\sim$1000 so that the order of the thermal noise bias becomes a negligible $10^{-3}$.
The agreement in the absolute flux scale between this model and that achieved by a calibration similar to \cite{Kern2020} is shown in \autoref{fig:ssm-abscal}.

One parameter degenerate in the redundant calibration is tip-tilt phase pointing, which must be fit with the sky model as a spatial gradient in the phase across the array aperture. 
In Phase I, we solved for the phase gradient using a linearized approximation, valid when the model errors are small---as was appropriate for a data-derived model.
However the model used here---which covers a much wider range of frequencies and LSTs---was found to occasionally contain phase errors of more than $\pi$ radians for some baselines. This necessitated a new algorithm for more exact and robust solution of the non-linear least-squares problem to estimate these phase gradient degeneracies. 
Our new method will be explained in detail in an upcoming paper (Z. Martinot \& J. S. Dillon, \textit{in prep}).

While the revised calibration does still have a very small number times and frequencies that end up mis-calibrated due to an insufficiently accurate sky model\footnote{These cases are visually obvious in plots of the phase of the gain solutions as a function of frequency and time, appearing as sharp discontinuities.}, there are significantly fewer than with the previous absolute phase calibration algorithm. These failures are rare enough and sufficiently compact in time and frequency that smoothing of calibration solutions (see \zcref{sec:methods:smoothcal}) provides sufficient mitigation.

\subsection{Per-Night Flagging Synthesis}
\label{sec:methods:per-night-flag-synth}
In the previous step, the flags were determined considering each integration independently. 
Many of the issues triggering flags or poor calibration solutions, such as failed signal chains, are expected to be stable with time and so better identified when more time is considered simultaneously. 

Here, we consider the antenna metrics (upon which the initial per-integration flags were based) for a full night simultaneously.
In doing so, we produce updated sets of flags---both per-antenna/integration flags, and per-channel/integration flags (RFI mask). 
This synthesis uses additional information, such as calibration $\chi^2$, where necessary.

\subsubsection{Per-Antenna Flag Synthesis}
\label{sec:methods:per-night:per-ant}
In this synthesis, a list of antennas to ignore at each integration is generated for each night of data using the auto-correlations, calibration $\chi^2$, time dependent antenna flags, time- and frequency-dependent interference mask. We first re-apply all per-antenna flags previously obtained per time, as well as flagging all antennas at times for which the sun is above the horizon.

Following this, a procedure is needed to combine flags across time with protocols to handle gaps.   
For example, we found integrations that are not flagged, but are surrounded by integrations that are flagged. 
Such cases are suggestive of a low signal-to-noise situation where some integrations may fall just below the threshold for flagging, but due to their proximity to artifacts flagged at higher confidence, they are likely still affected. Consistently harmonizing this information is the aim of the ``smoothed metric flagging" algorithm, described in more detail in \zcref{app:smoothed-metric-flagging}, which smooths antenna metrics and $\chi^2$ from redundant calibration on a 10-minute timescale, and flags the smoothed metrics according to a pre-defined threshold. 
In this way, previously unflagged data can be flagged if in the vicinity of strong outliers. 
Furthermore, to avoid large gaps in the middle of a night (which can negatively affect fringe-rate filtering further on in the pipeline), we also flag all integrations either before or after gaps larger than $\sim$10\,minutes (whichever is smaller) within a night.
Antennas that are flagged for more than 50\% of the integrations on a given night are flagged for the entire night. Antennas flagged in the initial, per-integration flagging are never unflagged as a result of this process.

\subsubsection{Initial Per-Night RFI Mask}
\label{sec:methods:per-night:rfi}
In parallel to the antenna flagging described above, a per-night ``RFI mask'' must also be synthesized. This is a $(N_{\rm integrations}, N_{\rm freq})$-shaped mask that applies to all antennas. 

First a list of good antennas, channels, and times is assembled using the criteria developed in \zcref{sec:methods:per-file:flagging}, as well as requiring the sun to be below the horizon, and omitting the FM band (87.5--108 MHz).
For each antenna, we then perform a 2D high-pass filter on the auto-correlations (as a function of time and frequency) using the \textit{Discrete Prolate Spheroidal Sequence} (DPSS) basis \citep{Slepian1978,Ewall-Wice2021}. 
This basis set and its parameters are presented at length in \citet{Ewall-Wice2021}, with shorter pedagogical guides appearing in \citet{HERAMemo129} and \citet{Chen2025} and applications to data in \citet{Pascua2024} and \citet{Cox2024}.
Here the half-width in the frequency axis is set to 200\,ns (corresponding to a smoothing scale of $\sim$$5\,{\rm MHz}$), and on the time axis is 2.2\,mHz (corresponding to a smoothing scale of $\sim$$450$\,sec).
In this filtering, we consider only the central window for the night outside of which all integrations are flagged (e.g. due to solar elevation).
The weights used for determining the DPSS coefficients are the estimated thermal variance of the autocorrelations, given by the mean autocorrelation (over all candidate antennas).
After filtering, the per-antenna waterfall produced is divided by the expected thermal noise level (c.f.\ \zcref{eq:visibility-noise}) to form a $Z$-score, i.e.\ a metric that should be close to normally distributed. 
Antennas are only used for RFI flagging when their RMS $Z$-score over time and frequency is less than 1.2, or if they are in the best-performing (i.e.\ most stable) quartile of all antennas.

Having obtained a set of `good' antennas, we obtain a new average auto-correlation and associated estimate of thermal noise, as well as the $Z$-scores as computed above, but for the \textit{mean} auto-correlation over the set of ``good" antennas.
With this $Z$-score waterfall in hand, we flag any pixel of the waterfall with $Z>5$, as well as neighboring pixels with $Z>4$.

We then iteratively harmonize the flags over full channels and integrations. 
Here, we compute the mean over \textit{unflagged} $Z$-scores over each axis, and for the axis in which the highest mean exists, we flag all elements whose mean is both higher than the highest mean in the other axis and higher than a threshold of 1.5.
We repeat this process until all flagged means over both axes are below 1.5.
This iterative process allows us to eliminate the most poorly-behaving channels and integrations without counting, for example, an extreme outlier in an otherwise poor channel as evidence of a poor integration (and vice versa).

We then use the flag waterfall we just obtained to determine a new 2D DPSS filter, and repeat the process afresh.
We do this only for two rounds.

\subsection{Per-Night Calibration Smoothing}
\label{sec:methods:smoothcal}
All of the calibration steps discussed so far (redundant and absolute calibration) are performed independently per integration and channel.
This renders the calibration solutions susceptible to both temporal and spectral fluctuations and can result in gain errors, $\hat{g}_i/g_i$, that are spectrally- and temporally-structured. 
While all forms of error in the estimated gains are undesirable, those that are spectrally-structured are particular egregious, causing foreground power to leak from the low-delay wedge into the high-delay 21\,cm window. 
Since apparent spectral structure in gain solutions is likely due to the impact of non-redundancy on redundant calibration \citep{Orosz2019} or absolute calibration \citep{Byrne2019}, we take a first-do-no-harm approach. We thus impose a strong smoothness prior on our calibration solution, relying on the stability and spectral smoothness of the instrument's response.
Thus, any rapid fluctuations in the estimated gains are considered to be spurious.
We thus smooth the gains over both time and frequency.
This process has remained essentially the same as previous HERA limits, but can be summarized as follows.

We first choose a reference antenna (one for each night) -- the antenna that is flagged for the fewest integrations across the night -- and rephase all estimated gains such that the reference antenna has a phase of zero:
\begin{equation}
    g_j(t, \nu) \longrightarrow g_j(t, \nu) \exp\left[-i \phi_{\rm ref}(t, \nu)\right].
\end{equation}
We then smooth each antenna's gain simultaneously over time and frequency using a DPSS model with a half-width in frequency of 100\,ns (corresponding to a smoothing scale of $\sim$$10\,{\rm MHz}$), and in time of  1.65\,$\mu$Hz (corresponding to a smoothing scale of $\sim6\times10^5$\,sec). Though this is longer than a day, the way that DPSS filters are constructed means that the gain solution admits a few temporal modes.

Importantly, it sometimes occurs that during the course of a night, a particular antenna's gain-phase is flipped (rotated by 180$^\circ$) with respect to its phase on the first (unflagged) integration of the night.\footnote{We attribute these flips to the rare accidental triggering of the Walsh switching of HERA feeds---a capacity built into the hardware but not implemented for these observations \citep{Berkhout2024}.} 
In this case, we detect the integrations in which this occurs, and flip all subsequent integrations (until the phase reverts back, if indeed this occurs) before fitting the DPSS model, and flip them back again after smoothing. 
We also flag the integrations in which any flips occur, since the flip might happen at some point during the integration.

\subsection{Deeper RFI flagging}
\label{sec:methods:rfi-round-2}
We find that the detection of RFI in auto-correlations averaged over antennas can miss low-level RFI that is noticeable in cross-correlations after further averaging to increase signal-to-noise.
Hence, we perform a check for this low-level RFI as follows.

We first consider each integration separately, computing a single array-averaged $Z$-score spectrum for each. 
We then combine the $Z$-scores for all integrations on a given night to perform harmonized outlier-detection.

Considering a single integration (all baselines, channels and polarizations), we apply the smoothed calibration solutions and flags, and then redundantly average, as discussed in the next subsection (\zcref{sec:methods:redavg}), yielding $N_{\rm ubl}N_{\rm pol}$ averaged visibility spectra.
We then estimate the thermal noise on each unique baseline group using \zcref{eq:visibility-noise} with $N_{\rm samples} = N^\mathcal{G}_{\rm bl}$, by which we divide each redundantly-averaged visibility yielding an estimated signal-to-noise ratio (SNR).
We then downselect the baseline groups that are carried through to compute the final metric; we include only cross-correlation baselines for which (i) the median $N^\mathcal{G}_{\rm bl}$ (over $t$ and $\nu$) is at least 15\% of the maximum $N^\mathcal{G}_{\rm bl}$ for any baseline group (in practice, the autocorrelations), and (ii) the baseline-delay, $\vec{b_{ij}}/c$, is smaller than the high-pass delay-filter threshold, 750\,ns.
This yields $N_{\rm filt}<N_{\rm ubl}$ baseline groups satisfying the criteria.

To the SNR of each of these baselines, we fit one-dimensional DPSS models over the spectral axis -- one for each unique baseline and integration, fit independently to frequency bands below and above FM -- and subtract the model from the SNR.
As mentioned, this DPSS model has a half width of 750\,ns -- well outside the expected FG-dominated wedge, and including dominant systematics such as mutual coupling. 
If all of the foreground signal is concealed at delays below 750\,ns, the remaining delay-filtered SNRs are expected to be foreground- and systematic-free, and should thus be normally-distributed, i.e. they can be considered $Z$-scores.

There is a small caveat here. 
There is a well-known feature of model fitting in which data towards the edge of the range (or data close to large gaps) are over-fit due to their not being constrained on one side. 
This leads to a systematic \textit{under}-estimate of $|Z|$ close to the edges and large flagging gaps (such as the FM band). 
This can be analytically corrected by dividing the spectrum of $Z_t(\nu_i)$ at each integration $t$ by $L$, where:
\begin{equation}
    L^2_{i} = \frac{\pi}{4} (1- h_{ii}).
\end{equation}
Here, $h_{ii}$ is called the ``leverage'', and is simply $i^{th}$ diagonal element of the ortho-projection (or ``hat") matrix:
\begin{equation}
    H = Z_t (X^T N X)^{-1} X^T N, \\
\end{equation}
where $N$ is the diagonal matrix whose diagonal elements are given by $N^g_{\rm bl}(t)$ and $X$ is the DPSS design matrix.

Following the correction of the set of $Z$ by this leverage-correction factor, we compute a modified mean-$Z$ via the following:
\begin{equation}
    \bar{Z} = (\langle|Z|\rangle - 1)\sqrt{\frac{\pi N_{\rm ubl,filt}}{4 - \pi}},
\end{equation}
where the average over the absolute $Z$-scores is performed over the $N_{\rm filt}$ unique baselines satisfying our selection criteria defined above.
This new, averaged quantity has an expected mean of zero and variance of unity, though it is not normally-distributed.

In a very similar fashion to the flagging performed on the antenna-averaged $Z$-score waterfall in \zcref{sec:methods:per-night:rfi}, we combine the per-integration $Z$-scores over a full night, and perform similar checks but with reduced thresholds (see \zcref{app:rfi-round-2} for details).

\begin{figure*}
    \centering
    \includegraphics[width=\linewidth]{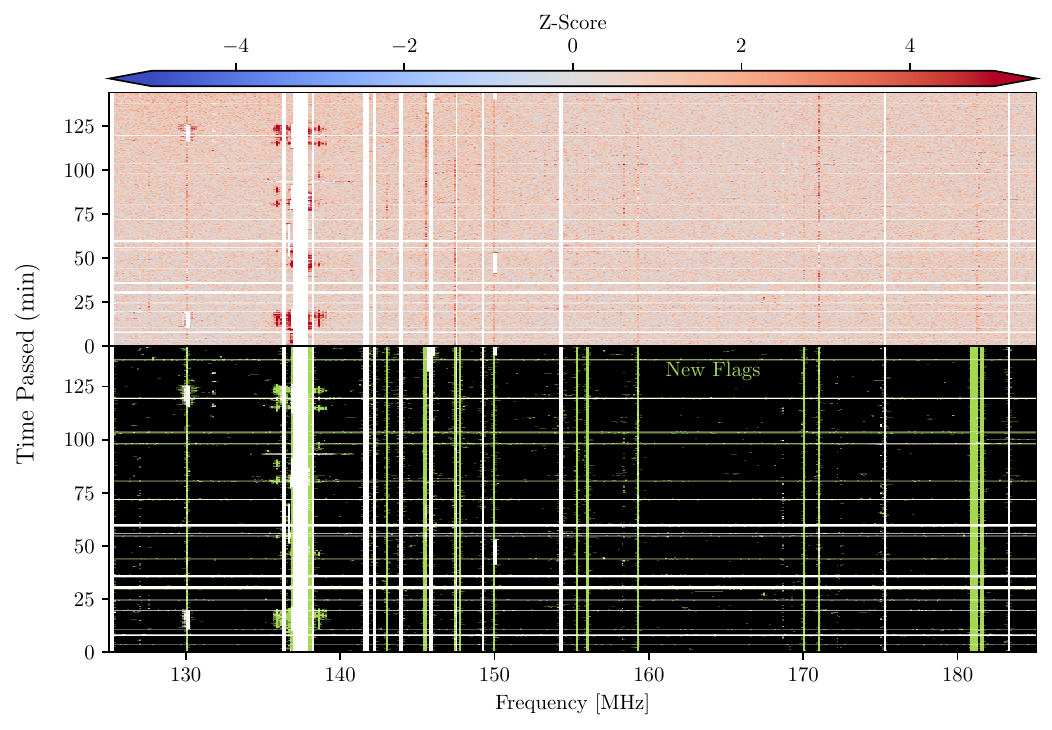}
    \caption{An illustration of per-night RFI flagging, as described in \zcref{sec:methods:per-night:rfi} and \zcref{sec:methods:rfi-round-2}.
    In the upper panel, we show the mean $Z$-score computed over `good' baselines for night 2459861.
    Notice that there are clumps of unflagged high-$Z$, especially surrounding regions flagged for other reasons, e.g.
    at $\sim 137$\,MHz. 
    The synthesis RFI algorithm iteratively flags individual pixels, and either the worst offending entire channels (vertically) or entire integrations (horizontally).
    The culmination of these flags is presented in the lower panel, where new flags from this algorithm are shown in green.
    }    
    \label{fig:rfi-round-2}
\end{figure*}

We show an example of this $Z$-score waterfall in the top panel of \zcref{fig:rfi-round-2}, which highlights a small $\sim40\,$MHz by 2\,hr window. 
Pre-existing flags (see \zcref{sec:methods:per-night:rfi}) are shown in white, while the $Z$-score computed in the manner described above is shown on the color-scale from blue to red, clipped at $5\sigma$. 
Several key features are immediately apparent: there are a number of small `blobs' of high $Z$-score, generally close to channels that were pre-flagged. There are also channels that have intermittent outliers across time. 
The new flags determined by this procedure are depicted in the lower panel of \zcref{fig:rfi-round-2} as the pixels overlaid in orange. 
Note that very often, the resulting flags apply to entire channels for all times, though there are also instances of flags that affect a small region of time and frequency. 

Finally, as an even deeper probe of RFI in particular channels, we take an average of $\bar{Z}$ over integrations, leaving a single spectrum of $Z$-scores, which we further normalize by dividing by the square root of number of unflagged integrations, and subtracting a DPSS model with half-width of 250\,ns, corrected for the leverage as above. Denote this quantity by $Q$.
We then flag any channels where $Q > {\rm max}(4, {\rm max}(Q)/1.5)$, and repeat the averaging and DPSS-fitting until no new flags are found. In this way, large outliers are less likely to cause additional channels to be flagged due to their influence on the filter.

\subsection{Redundant Averaging}
\label{sec:methods:redavg}
A major distinction between this work and previous HERA limits is that here we averaged the visibilities of all redundant baselines prior to forming power spectra.
This is a significant analysis choice, 
making the downstream power spectrum estimation significantly more simplified and efficient.
It also means that the final power spectra include contributions from all baseline pairs within each redundant group, \textit{inluding the auto-pairs} where a baseline is multiplied by itself. 
This is expected to reduce the occurrence of negative-valued systematics\citep[e.g.][]{Kolopanis2023,Morales2023}.
However, it also introduces an increased risk of signal-loss, as imperfections in redundancy between the baselines in a particular group can cause decoherence when averaging their complex visibilities.
We investigate the level of this decoherence in \zcref{sec:sig_loss_coh_incoh_test}, finding that it is of the order a few percent (c.f. \zcref{tab:losses}).

This step is the first in which a new \textit{visibility} data product is created. 
In previous steps, the raw data is read and used to create calibration solutions, and those per-antenna calibration solutions are manipulated in various ways.
In this step, the data are read, the calibration solutions are applied, and a new redundantly-averaged dataset is produced.
This is relatively efficient, as the output product is 
$N_{\rm bl}/N_{\rm ubl}\sim 10$ times smaller than the raw data.

The redundant baseline averaging process itself is quite simple.
For a particular baseline group $\mathcal{G}$ 
we calibrate each visibility at each time and frequency independently\footnote{We omit dependence on $t$ and $\nu$ in the expressions for simplicity.}, and perform a simple masked average:
\begin{equation}
    V^{pq}_\mathcal{G} = n_{\mathcal{G}}^{-1} \sum_{(a, b) \in \mathcal{G}}g^p_a g^{q*}_b \xi_a \xi_b V^{pq}_{ab},
\end{equation}
where $\xi \in \{0, 1\}$ are per-antenna, per-channel, per-time flags, and we propagate the number of accumulated samples as
\begin{equation}
  n_{\mathcal{G}} = \sum_{(a,b) \in \mathcal{G}} \xi_{a} \xi_b \leq |\mathcal{G}|.
\end{equation}

Importantly, note that $n_\mathcal{G}$, which is a per-baseline (time, frequency)-waterfall, is by construction uniform over the frequency axis for any particular baseline group, up to flagging. 
This is because the only per-channel flags we set are in the nightly synthesized RFI mask, which is a single mask \textit{for all antennas} on a particular night. 
Thus, while integrations may be flagged for some antennas and not others, channels are always either flagged or unflagged for all antennas.
Thus, for a particular integration $t$, $n_b$ has a constant value for all channels, except for flagged channels in which it is zero.
This is an important point to which we will return when discussing LST-stacking and inpainting in \zcref{sec:methods:lstbin}.

We note that, beyond the quality metrics and flagging already discussed, no further quality checks are performed at this stage.
We aim to include such checks, which are uniquely possible when the calibrated visibilities from supposedly redundant baselines are considered together, in future analyses.

\subsection{LST-stacking and averaging}
\label{sec:methods:lstbin}
Thus far, we have only considered data from a single night, and have not jointly considered data from multiple nights.
At the same LST each night, we expect to observe the same sky, so that visibilities within the same unique baseline group are random draws from the same distribution.
This affords a few opportunities. 
Firstly, it allows us to jointly compare the calibrated visibilities and adjust the gain solution degeneracies to optimally align them over nights. 
Secondly, it allows us to jointly infer the visibility values that are flagged, with more information at hand than we can use on a single night.
Thirdly, it allows for a deeper level of outlier identification as we compare nights with each other,
and finally, it allows us to average over the nights to increase SNR.

\subsubsection{LST Stacking}
\label{sec:methods:lstbin:stacking}
We refer to the process of identifying and loading the data within an LST-bin as `LST stacking'.
This is primarily a book-keeping problem, which we tackle in the following way.

We first identify all unique baseline groups present on any night in the dataset.
We then generate a grid of LSTs that are evenly-spaced between zero and 24 hours, with a spacing of as close to the integration time of the observations as possible while dividing the total 24 hours evenly (9.6 seconds).
The visibilities for all nights, baselines and polarizations within an LST-bin are gathered in parallel over LST-bins.
Finally, the visibilities are rephased so that phase center is at the Right Ascension and Declination corresponding to zenith at the center of the LST bin:
\begin{equation}
    \label{eq:rephase}
    V^{pq}_{ij} \rightarrow V^{pq}_{ij} \exp(-2\pi i \nu \tau'),
\end{equation}
where
\begin{equation}
    \tau' = (\mathbf{R}\hat{z} - \hat{z})\cdot \mathbf{b}_{ij} / c
\end{equation}
and $\mathbf{R}$ is the rotation matrix that rotates a unit vector towards zenith, $\hat{z}$ at the LST of the observation to a unit vector towards zenith at the central LST of the bin.

\subsubsection{LST Calibration}
\label{sec:methods:lstbin:lstcal}
Having stacked the data within an LST-bin, we implement a final calibration refinement, \texttt{lstcal}, to address systematic variance observed across nights. 
Initial analysis of the dataset \citep{DillonMurray_hera125} revealed that for a particular baseline at a particular LST and channel, the variance measured over nights exceeded the expected variance (computed using \zcref{eq:visibility-noise} where the auto-correlations are averaged over both antennas and nights). We hypothesize that this ``excess variance'' is at least partially a result of day-to-day variations in the per-antenna calibration solutions that manifests as a coherent systematic error in array-wide calibration relative to the average across nights.

The \texttt{lstcal} algorithm addresses this issue by comparing each night's visibilities to the average across all nights within a given LST-bin,
\begin{equation}
V^{\rm avg}_\mathcal{G} = \left(\sum^{N_{\rm nights}}_{k} V^{k}_\mathcal{G} N^{\mathcal{G}, k}_{\rm bl}\right) \left(\sum^{N_{\rm nights}}_{k} N^{\mathcal{G},k}_{\rm bl}\right)^{-1},
\end{equation}
computing a frequency- and polarization-dependent gain that brings each night's visibilities into better agreement with the average. We model the per-night gain correction, $G_k \left(\nu\right)$, as a per-array quantity defined by the absolute calibration degrees of freedom. For each night $k$ and polarization, this correction consists of a per-frequency amplitude, $A_k \left(\nu\right)$, and two phase gradients $\left(\Phi_{x, k}\left(\nu\right), \Phi_{y, k}\left(\nu\right)\right)$. This correction is applied to each redundant group's visibility based on that group's representative baseline vector $\left(b_x, b_y\right)$
\begin{equation}
    G_k(\nu) = A_k(\nu) \exp\left(i \left[ \Phi_{x,k}(\nu) b_x + \Phi_{y,k}(\nu) b_y\right]\right).
\end{equation}
We then solve for the parameters $A_k\left(\nu\right)$, $\Phi_{x, k}\left(\nu\right)$, and $\Phi_{y, k}\left(\nu\right)$ for each night $k$ by minimizing the difference between $V_{\mathcal{G}}^k \left(k\right)$ and $G_k\left(\nu\right)V^{\rm avg}_{\mathcal{G}} \left(k\right)$ across all baseline groups $\mathcal{G}$.

Given that each night's visibility data have already been redundantly averaged prior to LST-stacking, we restrict \texttt{lstcal} to these absolute calibration degrees of freedom, and do not attempt to correct the per-antenna gains. In order to prevent \texttt{lstcal} from introducing spurious spectral structure into the calibrated visibilities, we smooth the gain solutions over frequency using a DPSS model with a smoothing scale of $10\,{\rm MHz}$ before applying them to the data. 

\zcref[S]{fig:lstcal} shows an example of the effects of LST-calibration for two particular baselines at RA=2.908\,hours.
Notice how for both baselines, the spread of the data after LST-calibration (right panels) is reduced compared to the data that has not been LST-calibrated (left panels). To evaluate the stability of the \texttt{lstcal} solutions, we also examined the per-night gain solutions across all nights used in this analysis. The corrections did not show coherent temporal trends and instead appear consistent with random fluctuations expected from nightly calibration variance.

\begin{figure*}
    \centering
    \includegraphics[width=\linewidth]{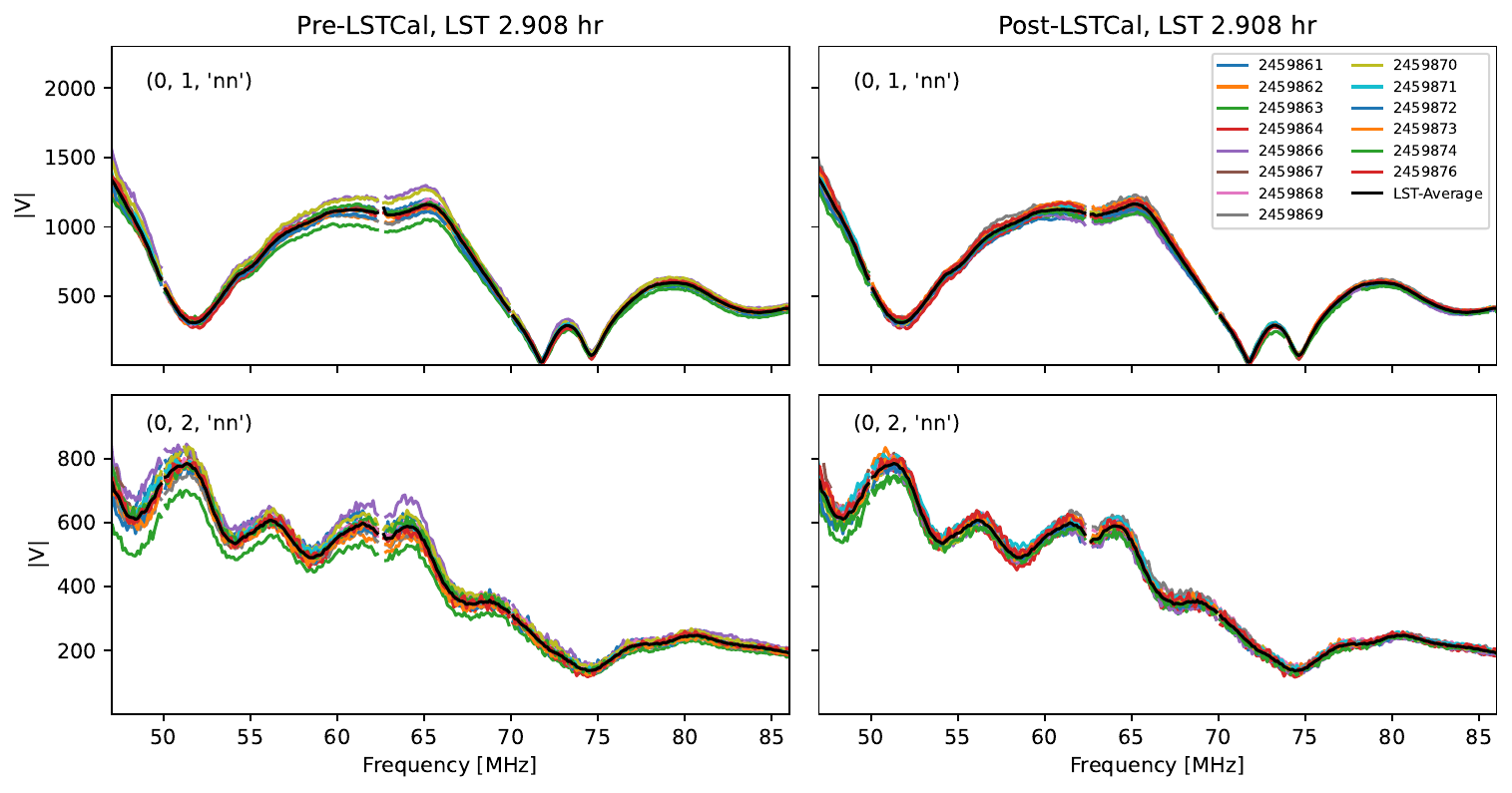}
    \caption{An example of the effects of LST-calibration, described in \zcref{sec:methods:lstbin:lstcal}. The left-hand panels shows the magnitude of the visibilities measured by the 14.6m East-West baseline group (top) and 29.2m East-West group (bottom), on the 14 nights of observation (colored), with their average shown in black. 
    These panels are \textit{before} LST-calibration is applied, while the right hand panels are the same data after LST-calibration is applied.
    Notice how the spread between nights is significantly reduced, while the overall structure remains the same. 
    }
    \label{fig:lstcal}
\end{figure*}

\subsubsection{Choice of Spectral Windows}
\label{sec:methods:lstbin:spws}
Ultimately, our power spectra will be estimated within discrete spectral windows (or bands).
While the spectral windows are primarily intended for power spectrum estimation, it is beneficial to choose them at this stage of processing, to enable more granular flagging after inpainting (see next subsection).

\begin{figure*}
    \centering
    \includegraphics[width=\linewidth]{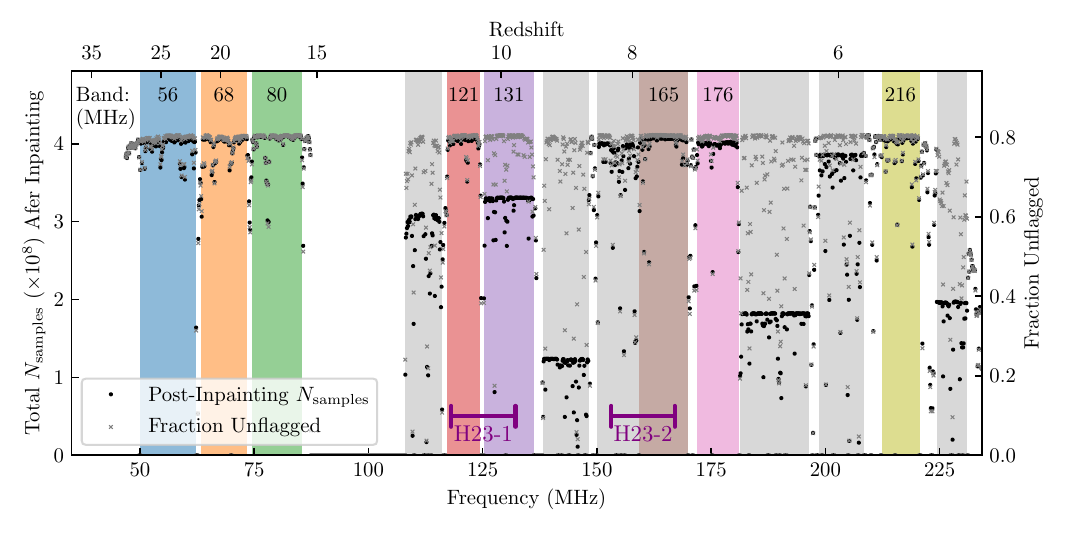}
    \caption{
        Spectral windows defined and used for upper limits in this paper. 
        Gray crosses indicate the (inverse) RFI occupancy in each channel, i.e. the result of integrating the 2D RFI masks over all antennas, feeds and times.
        Black and red dots indicate the total number of unflagged samples (baseline-time-pols) for each frequency channel used in power spectra, i.e. also including flags that omit entire spectral windows if there are intrinsic flag-gaps of sufficient width (c.f. \zcref{sec:methods:lstbin:inpainting}). 
        Colored regions indicate the chosen spectral windows (bands) used for reporting power spectra in this paper, labeled by the nearest integer of the central channel in each band in MHz. 
        Gray bands are bands that were defined for analysis but not ultimately used for reporting power spectra due to their high flagging fraction.
        Red dots indicate channels not included in any band. 
        The two spectral windows used in \josh{} are indicated in purple.
    }
    \label{fig:subbands}
\end{figure*}

We choose 8 spectral windows (in contrast to the single band in \nick{} and the two bands in \josh{}), ranging from 50-231.1 MHz.
Details of the bands are presented in \zcref{tab:subbands}, and their broad characteristics are illustrated in \zcref{fig:subbands}.
The spectral windows were selected using a few criteria; each was required to be less than 15\,MHz in width to alleviate the lightcone effect \citep{Datta2012,Datta2014,Ghara2015,Greig2018,Blamart2025}, and an attempt was made to place wide gaps (more than a few channels) with consistently high flagging fractions between spectral windows. 
These choices were made since strongly flagged channels towards the center of power-spectrum bands have previously been shown to leak foregrounds to high delays \citep{Aguirre2022, Chen2025}, even with spectral in-painting applied.
Further, bands were picked so as to concentrate channels with relatively high levels of flagging into as few bands as possible. 

Note that in \zcref{fig:subbands} two metrics of flagging are included. 
The gray crosses indicate the RFI occupancy of each channel, i.e. the fraction of antennas, times and feeds that are flagged after the ``RFI Round Two'' processing step (\zcref{sec:methods:rfi-round-2}). 
The spectral windows were chosen based on this statistic.
Conversely, the black and red dots indicate the total number of samples (baselines, times and pols) that go into the power spectrum estimates.
These are noticeably disjoint at the edges of the spectral windows, which is an artifact of flagging choices we make after inpainting---with knowledge of the bands---described in \zcref{sec:methods:lstbin:inpainting}. 
Thus, a low overall number of samples within a band (e.g. the band above 131) does \textit{not} necessarily indicate a high flagging fraction due to e.g. RFI, but instead that there were a higher number of large flagging gaps in that window.

After initially specifying 14 spectral windows, we finally chose the eight most well-behaved bands in which we report upper limits. 
The bands that are not ultimately used are marked in \zcref{tab:subbands} and shown as grey in \zcref{fig:subbands}.
Of the six abandoned spectral windows, four of them exhibit very high overall levels of flagging, and the other two (just below bands 165 and 216) have a large variability in flagging fraction within the band.
We found that these bands exhibit weak but noticeable artifacts in histograms of high-delay signal-to-noise for power spectra with low levels of incoherent averaging, most likely arising from inpainting imperfections.


\begin{table*}
    \centering
    \begin{tabular}{llllllll}
    \hline
        \textbf{Channels} & $N_{\rm chans}$ & \textbf{Freq. Range (MHz)} & $\Delta\nu$ (MHz) & \textbf{$z$ range} & \textbf{Center $z$} & \textbf{$\Delta z$} & Used? \\ \hline
        27 — 126 & 99 & 50.2 — 62.2 & 12.1 & 21.82 — 27.3 & 24.6 & 5.5 & \checkmark \\ 
        135 — 218 & 83 & 63.3 — 73.5 & 10.1 & 18.33 — 21.4 & 19.9 & 3.1 & \checkmark \\ 
        227 — 316 & 89 & 74.6 — 85.4 & 10.9 & 15.63 — 18.0 & 16.8 & 2.4 & \checkmark \\ 
        501 — 567 & 66 & 108.0 — 116.1 & 8.1 & 11.24 — 12.1 & 11.7 & 0.9 & $\times$ \\ 
        577 — 635 & 58 & 117.3 — 124.4 & 7.1 & 10.42 — 11.1 & 10.8 & 0.7 & \checkmark \\ 
        643 — 732 & 89 & 125.4 — 136.2 & 10.9 & 9.43 — 10.3 & 9.9 & 0.9 & \checkmark \\ 
        749 — 830 & 81 & 138.3 — 148.2 & 9.9 & 8.59 — 9.3 & 8.9 & 0.7 & $\times$ \\ 
        846 — 920 & 74 & 150.1 — 159.2 & 9.0 & 7.92 — 8.5 & 8.2 & 0.5 & $\times$ \\ 
        921 — 1008 & 87 & 159.3 — 169.9 & 10.6 & 7.36 — 7.9 & 7.6 & 0.6 & \checkmark \\ 
        1024 — 1100 & 76 & 171.9 — 181.1 & 9.3 & 6.84 — 7.3 & 7.1 & 0.4 & \checkmark \\ 
        1102 — 1225 & 123 & 181.4 — 196.4 & 15.0 & 6.23 — 6.8 & 6.5 & 0.6 & $\times$  \\ 
        1242 — 1323 & 81 & 198.5 — 208.4 & 9.9 & 5.82 — 6.2 & 6.0 & 0.3 & $\times$  \\ 
        1355 — 1423 & 68 & 212.3 — 220.6 & 8.3 & 5.44 — 5.7 & 5.6 & 0.3 & \checkmark \\ 
        1454 — 1509 & 55 & 224.3 — 231.1 & 6.7 & 5.15 — 5.3 & 5.2 & 0.2 & $\times$  \\ \hline
    \end{tabular}
    \caption{Spectral window definitions for the power spectra presented in this paper. Channel ranges are non-inclusive on the upper end. While we define 14 bands, we only report upper limits in 8 of the bands, due to high levels of flagging in the other 6 (c.f. \zcref{fig:subbands}).}
    \label{tab:subbands}
\end{table*}

\subsubsection{Inpainting}
\label{sec:methods:lstbin:inpainting}
While flagging data affected by systematics such as RFI is essential, the gaps left in the data by these flags presents its own challenges.
These challenges primarily stem from the fact that if there is systematic variance between the visibilities on $N_{\rm nights}$ different nights (e.g. one night has consistently larger amplitude visibilities for a particular baseline compared to the other nights, while being internally spectrally smooth), then when averaging them together, if the averaging weights are non-uniform across frequency for any particular night (due to flagging), then the resulting average incurs spectral structure.

This effect is well known and studied, with a common solution being to `inpaint' a best guess of the missing data in order to preserve spectral smoothness.
For example, \nick{} used an iterative delay-based convolution filter to in-paint missing data after averaging the raw data over nights, a technique that was numerically validated in \citet{Aguirre2022}.
Other basis sets for inpainting have been considered, including DPSS, Least-Squares Spectral Analysis (LSSA), Gaussian Process Regression (GPR) and Convolutional Neural Networks (CNNs) with varying degrees of success in different regimes (see \citealt{Pagano2023a} for a comparative analysis). 

One problem with the previous applications of inpainting to HERA data is that it was applied \textit{after} averaging of the data over nights; that is, it only filled in channels for which no data was available on \textit{any} night.
While this mitigates the problems encountered when Fourier transforming the data over the frequency axis to produce power spectra, it neglects the problem of induced spectral structure inherent in the averaging of non-stationary data with spectrally-structured weights. 
This problem was explored at length in \citet{Chen2025}.

A simple way to avoid this is to inpaint the calibrated visibilities \textit{before} averaging over nights.
In this work, we adopted the DPSS basis functions to interpolate our data. 
We determine DPSS basis coefficients per-night, per-antenna-pol and per-LST, such that each is interpolating a 1D function of frequency.
An important consideration is the half-width of the basis functions, i.e. the delay-scale out to which to fit spectral structure. 
The aim is to include all relevant foreground and systematic effects, but not to exceed these scales by much, as we do not wish to inpaint the scales at which we hope to measure the cosmic signal.
With this in mind, we choose to use a half-width of 
\begin{equation}
    \tau_{\rm hw} = {\rm max}(500\,{\rm ns}, |\vec{b}|/c),
\end{equation}
which covers all sky-based sources out to the horizon, with a minimum buffer of 500\,ns that covers the bulk of known systematic effects, such as mutual coupling.
The inpainted data takes the value of the true measured data when the data is unflagged, and the value of the smooth inpaint solution when it is flagged.

Unfortunately, inpainting out to such high delays means that the behaviour of the inpaint solutions within moderate-to-large flag gaps can be poorly constrained. 
Naively, we are solving for scales down to $\tau_{\rm hw}^{-1} \approx 2\,{\rm MHz}$, which means that gaps in the spectrum around this size (or larger) can cause poor behaviour of the solutions within the gaps.
While the solutions are constrained to only contain power out to the horizon delay, the final inpainted data product may have sharp transitions between flagged and unflagged regions when the inpaint model is not well-constrained, resulting in power spilling out to much higher delay.
We thus perform some checks to ensure that poorly-behaved inpainted data is flagged before averaging over nights.

These checks centre around identifying regions of the spectrum that are highly flagged over a wide enough region such that the solutions cannot be trusted towards the centre of the region, far away from unflagged data. 
In practice, we define a maximum flagged gap size (in units of channels) as 
\begin{equation}
    s_{\rm max} = \lfloor f_{\rm gap} \tau_{\rm nw}/\Delta_nu \rfloor,
\end{equation}
where $f_{\rm gap}=1$ is a tunable scaling factor.
We then convolve the binary flags $\xi_\nu \in \{0, 1\}$ with a triangular filter of size $2 s_{\rm max} - 1$.
The resulting convolved array, $f_{\rm cnv}$, represents a `flag density' of the surrounding channels, between zero and unity.
We then create a new flag array by thresholding the flag densities at $f_{\rm crit} = 0.4$:
\begin{equation}
    \xi_{\rm inp} = \begin{cases}
    0 & \xi_\nu=0 \ \ {\rm or} \ \ f_{\rm cnv} > 0.4 \\
    1 & {\rm otherwise}.
    \end{cases}
\end{equation}
Any region with more than $s_{\rm max}$ contiguous flags in $\xi_{\rm inp}$ is then identified as needing to be flagged.

Since we require spectrally uniform flags before averaging (which is the entire point of inpainting in the first place), flagging here must be for an entire spectral window per-integration. 
That is, if any such large contiguous gaps exist, any spectral windows overlapping with the gaps must be discarded for that integration.
This is the motivation for choosing the spectral windows prior to inpainting---it allows us to selectively flag per spectral-window rather than discarding the entire spectrum.


We flag any window that overlaps with a contiguously flagged region of sufficient size, as described above. 
However, due to the fact that we use a frequency taper when performing power spectrum estimation (c.f. \zcref{sec:methods:pspec-bl}), we allow flag gaps to overlap with the outer two channels of each band without consequence\footnote{This helps some bands that are adjoined to regions of continuously high flagging fraction to not be flagged every time their first or last channel is flagged.}.

Overall, these choices of spectral windows and our procedure for flagging large gaps results in $\sim10-12\%$ more of the data being flagged, but our tests indicate that these choices are conservative, i.e. they lead to no poorly-constrained solutions inside the flagged gaps.

\subsubsection{Averaging Over Nights}
\label{sec:methods:lstbin:avg}
After inpainting the flagged data and determining the post-inpainting flags, we are ready to average over nights.
This process is reasonably simple. For each unique baseline group, channel and LST-bin, we take the weighted average,
\begin{equation}
    \bar{V}^{pq}_\mathcal{G} = M^{-1} \sum_{k}^{N_{\rm nights}} \xi^\mathcal{G}_{{\rm inp},k} N^\mathcal{G}_{\rm bl} V^{pq}_{\mathcal{G},k},
\end{equation}
with
\begin{equation}
    M = \sum_{k}^{N_{\rm nights}} \xi_{{\rm inp},k} N^{\mathcal{G}}_{{\rm bl}, k}.
\end{equation}
Recall that $N^{\mathcal{G}}_{\rm bl}$ is binary: its value is either some positive constant or zero. 
However, wherever it is zero, we have inputed data via inpainting, and we treat it as being uniform in frequency, taking its non-zero value.
Furthermore, $\xi_{\rm inp}$ is either all zero or all one within a particular spectral window. 
This means that the combined weighting function, $\xi_{\rm inp}N^\mathcal{G}_{\rm bl}$ is uniform within each spectral window. 

However, in computing the effective number of samples that is propagated to power spectrum and covariance estimation, we use
\begin{equation}
    \label{eq:nsamps_before_pspec}
    N = \sum_{k}^{N_{\rm nights}} \xi_k \xi_{\rm inp} N^\mathcal{G}_{\rm bl},
\end{equation}
where here $\xi_k$ are the pre-inpainted flags (i.e. the flags resulting from quality metrics, not the flags resulting from identifying large gaps after inpainting).
Thus, $N$ is not in general spectrally uniform within bands.

\subsection{Per-baseline systematics mitigation and power spectrum estimation}
\label{sec:methods:pspec-bl}
After averaging the data over nights within narrow LST-bins, we begin the process of mitigation of instrumental systematics, and eventually power-spectrum estimation, on a per-baseline basis. 
In practice, we rearrange the data so that we read the visibilities across all LSTs for a particular baseline, independently identifying systematics for each baseline in parallel.

Before any further analysis, we first remove LSTs that have fewer than 20\% integrated samples compared to the maximum samples over all LSTs (for a particular baseline group).
This typically removes the first and last few LSTs in the dataset, which may disproportionately affect any time-based filters used for mitigating systematics.

Our power spectrum estimation method (which we will describe below) does not allow for non-uniform weights over frequency channels (but does allow for non-uniform weights over LSTs). 
To avoid data weights with frequency discontinuities, we use inverse noise variance weights that require the number of samples to be constant \textit{within each spectral window.} To do this, we simply average the number of samples within each window (per-baseline and LST). This means that channels with fewer samples (due to flagging and inpainting) bring down the average across the window for that particular LST.
This does not affect power spectrum estimation, in the sense that our algorithm already ignores relative weightings between channels. 
However, it does impact the estimate of the error bars in a small way. 
We find that this effect is small, with the power spectrum estimates at high delay being consistent with the predicted thermal noise computed with this approximation (c.f. \zcref{fig:upper-limits}) and we leave the handling of non-uniform spectral weights to future work.

Following this, to avoid a noise-power bias in the power spectrum estimate, which arises when cross-multiplying identical (or correlated) visibilities, we split the data into four ``interleaved'' datasets over the LST axis.
That is, we construct four non-overlapping datasets from the data at hand, $V_t$ (where $t$ indexes the LST-bin):
\begin{equation}
    \mathcal{D}_i = \{V_i, V_{i+4}, V_{i + 8}, ... \}, \ \ \ i \in \{0,1,2,3\}.
\end{equation}
Each ``interleave'' spans the full LST range, at a cadence of $4 \Delta t$, where $\Delta t \approx 9.6$\,sec is the LST-bin width. 
The following procedures for mitigation of systematics via fringe-rate filtering are applied independently to each interleave, in order to prevent introduction of correlated noise between the sets.

In previous HERA upper limits, we have avoided this noise bias by cross-correlating different redundant baselines (i.e. cross-multiplying the visibilities $V_{ab}$ and $V_{cd}$, where both $ab$ and $cd$ baselines are in the same redundant group) and ignoring `auto-baseline' pairs (i.e. $V_{ab}V_{ab}^*$). 
While this effectively avoids the noise bias, it allows the introduction of \textit{negative} systematics when different nominally-redundant baselines have systematics whose phases are different, as first observed by \citet{Kolopanis2023} and discussed in \citet{Morales2023}.
Since the probability of phase discontinuities between adjacent times is far smaller than between nominally redundant baselines at the \textit{same} time, this risk is significantly reduced by cross-multiplying adjacent times, and using \textit{all} baseline-pairs within a redundant group (including the auto-baseline pairs).

\subsubsection{Mitigating Instrumental Coupling Systematics via Time Domain Filtering}
\label{sec:methods:pspec-bl:frf}
Two of the major systematics present in HERA's Phase I data were cable reflections and over-the-air coupling \citep[][\nick{}; \josh{}]{Kern2019, Kern2020,Aguirre2022}.
In Phase II, the cable reflections have been mitigated at the instrument level by switching to Radio-Frequency-over-Fiber (RFoF) cables with a sufficient length to push residual reflections outside the delays of cosmological interest ($\gtrsim 2700\,{\rm ns}$). 
However, in Phase II, we have found an increased level of mutual coupling, which we define as the reflection or re-emission of sky signal from one antenna (whether from the feed or glinting off the dish) into surrounding antennas. 
The cause of this increased amplitude of mutual coupling in Phase II is likely the increased sensitivity of the new Vivaldi feeds at low elevation angles (in other words, the Vivaldi feeds have a larger vertical cross-section than their PAPER-style predecessors), in addition to the removal of a cage that surrounded the Phase I feeds, which had the undesirable effect of increased dish-to-feed reflections.

\citet{Kern2019, Kern2020} developed a semi-analytic, re-radiative coupling model that produced a good phenomenological match to the observed Phase I systematics, which only necessitated a two-element coupling model to achieve noise-limited suppression.
Building on this, \citet{Josaitis2021} and \hyperlink{cite.Rath2024}{E. Rath \& R. Pascua et al. (2024)} (hereafter \rath{}) extended the re-radiative mutual coupling model to multi-element terms that are needed to model the more complex Phase II coupling systematics. In fringe rate vs. delay space, where fringe-rate is the Fourier dual of observing time (measured in mHz) and delay is the Fourier-dual of observing frequency (measured in nanoseconds), the effect of mutual coupling can be characterized geometrically.
Since the mutually-coupled signal for baseline $ij$ can be cast as the addition of down-weighted and delayed copies of all other baselines that include $i$ or $j$, we expect that in general, short baselines that probe low fringe-rates will correspondingly exhibit short delays, and vice versa. 
For HERA's array geometry, this results in a characteristic `X' shape of enhanced power when viewed in fringe-rate vs. delay space, with the center of the `X' at a delay of zero, and a fringe-rate corresponding to the East-West projection of the baseline $ij$:
\begin{equation}
    \label{eq:peak-fringe-rate}
    f_{r, {\rm peak}} \approx -0.85 w_\oplus (b_{ij, {\rm EW}}/\lambda),   
\end{equation}
with $w_\oplus$ the angular velocity of Earth's rotation. 

\begin{figure*}
    \centering
    \includegraphics[width=\linewidth]{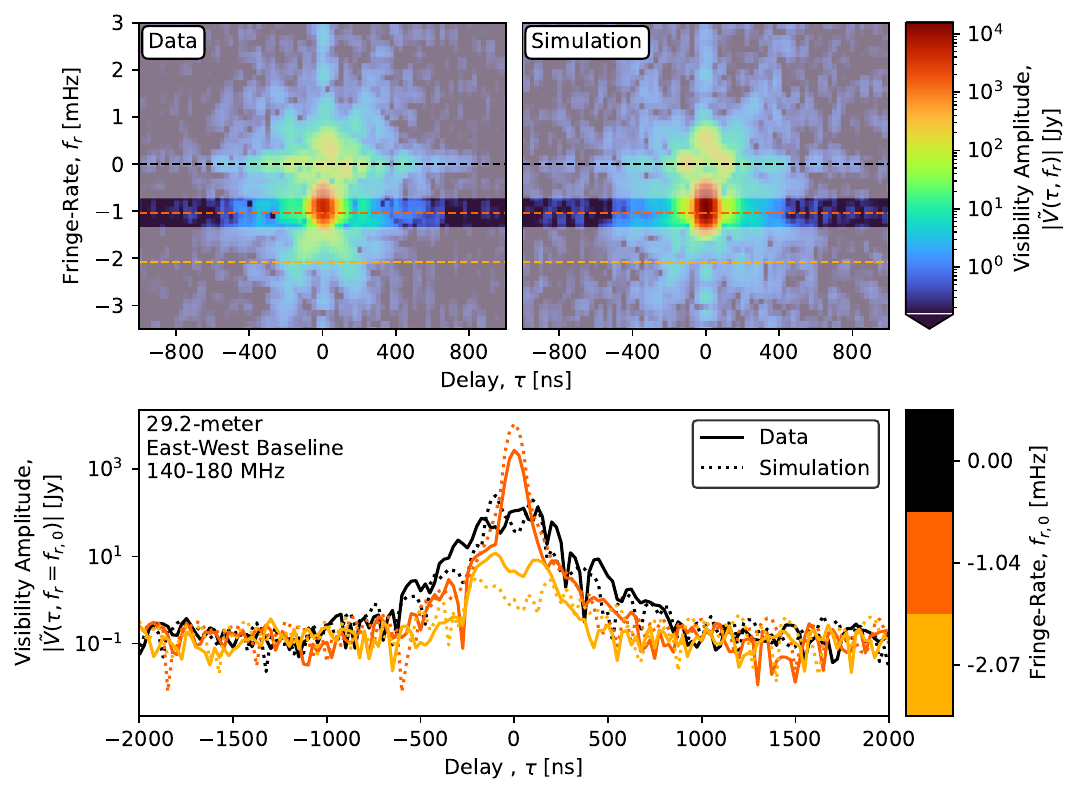}
    \caption{Comparison of real data to validation simulation for a single 29.2\,m East-West baseline. The upper panels display data on the left and simulation on the right. Each panel has delay on the $x$-axis and fringe-rate on the $y$-axis, with colors representing visibility amplitude. Three horizontal dashed lines are shown, at fringe rates of zero (black), $\sim -1\, {\rm mHz}$ (dark orange, representing the peak fringe rate of sky-locked power for this baseline, c.f. \zcref{eq:peak-fringe-rate}) and $\sim -2\,{\rm mHz}$ (yellow, representing non-sky-like power). Semi-opaque white overlay represents fringe-rates that are filtered out using the `main-lobe' filter of \zcref{sec:methods:pspec-bl:frf}.
    Note that the fringe-rates corresponding to the main-lobe (transparent horizontal window in this plot) shift up and down depending on the East-West projected length of the baseline and the frequency of observation.
    The lower panel focuses on cross-sections of the upper panels at the three dashed lines. This illustrates that while there is some disagreement between the mean foreground amplitudes of simulation and data, the overall shape in fringe-rate/delay space is remarkably consistent, and that the noise level (high-$\tau$ amplitude) is well matched.}
    \label{fig:comparison-validation-to-data}
\end{figure*}

This geometric picture, which is laid out in detail in \rath{}, is borne out by this dataset, as seen in \zcref{fig:comparison-validation-to-data}. 
In this figure, the left panel shows the amplitude of the visibilities for the 29.2\,m East-West oriented baseline group, after averaging over both redundant baselines and nights, in the fringe-rate vs. delay space. 
Immediately evident is a bright central core at zero delay and a fringe-rate of -1\,mHz, which represents the bulk of the spectrally smooth foreground emission as observed by this (short) baseline.
At a fringe-rate of zero, there is a horizontal `bar' that captures the non-rotating components of the observed power---in particular the coupling of autocorrelations into cross-correlations \citep{Kern2019}, as well as horizon-based effects such as the `pitchfork' \citep{Thyagarajan2015} and also mutual coupling from North-South oriented baselines.

Conversely, at zero delay there is a tall vertical bar that can be attributed to bright sources crossing the horizon.
Such sources are intrinsically smooth (and therefore appear at low delay), but the sharp transition from below to above the horizon spills power over all fringe-rates. 
This has been confirmed by a lack of such a feature in simulations in which no bright sources cross the horizon.
Finally, there is an `X'-like feature, crossing at the center of the foreground blob, that corresponds to the bulk of mutual coupling, as we have discussed.
This mutual coupling signal appears to be at an amplitude of $\sim$$1\%$ of the foreground amplitude (in visibility units), nominally a factor of $\sim 100\times$ the cosmic signal amplitude. 


Unfortunately, while our semi-analytical models of the mutual coupling confirm our general picture of the origin of this excess power, they are not detailed enough to enable a subtraction of the systematic at the visibility-level. 
While this remains a future goal, in this work we rely on the mitigation techniques laid out in \rath{} and \citet{Pascua2024}.
In particular, we use a pair of filters applied in fringe-rate space to mitigate both the cross-talk and mutual coupling.

Since autocorrelations coupled into cross-correlations present as excess power at zero fringe-rate, we mitigate this by applying a ``notch'' filter that removes power for fringe-rate modes within $\sim \hspace{-1mm}10$ $\mu$Hz (a few bins) of zero.

We additionally apply a ``main-lobe'' filter, as discussed in \rath{}, to mitigate the effects of mutual coupling.
We design the main-lobe filter to nominally retain 90\% of the 21\,cm signal, implemented as a DPSS filter, according to the prescription in \citet{Pascua2024}. In practice, the signal loss is slightly lower than the nominal 10\% target (which we correct for, see \zcref{sec:methods:signal-loss}).
\zcref[S]{fig:comparison-validation-to-data} illustrates this filter by overlaying a transparent screen over the fringe-rates that are filtered out for this particular baseline (29\,m East-West)\footnote{Note that for this baseline, the main-lobe filter overlaps with the notch filter, but this is not true in general, particularly for predominantly North-South baseline orientations.}.
The filter retains the bulk of the sky signal (the bright red regions towards the center) while eliminating much of the characteristic X pattern of the mutual coupling. 

\subsubsection{Coherent Time Averaging}
\label{sec:methods:pspec-bl:coh-time-avg}
Next we coherently average the visibilities from each interleave within larger LST bins of $\sim$270~seconds. 
This range is intended to be as close to 300~seconds as possible while maintaining an integral number of LST-bins in each time-average. 
When performing this average, we first rephase each visibility (see \zcref{eq:rephase}) within each interleave to a common central LST (the mean of the LSTs in \textit{all} interleaves for that bin). 
Then we convert the data in each interleave into pseudo-Stokes representations, from their native instrumental polarizations\footnote{We use the convention in which the total intensity is the sum of polarizations, rather than their average (hence Eq. \ref{eq:pstokes} does not have a factor of 1/2). This is applied consistently during absolutely calibration.}:
\begin{equation}
    \label{eq:pstokes}
    \begin{pmatrix}
    I \\ Q \\ U \\ V
    \end{pmatrix} = \begin{pmatrix}
    V_{xx} + V_{yy} \\
    V_{xx} - V_{yy} \\
    V_{xy} + V_{yx} \\
    i\left[V_{yx} - V_{xy}\right] \\
    \end{pmatrix}.
\end{equation}
Here we note that throughout the analysis we use the `sum' convention for converting instrumental polarization to pseudo-Stokes, as denoted in the above equation, instead of the `average' convention which places a factor of 1/2 on the RHS. Absolute calibration is, by necessity, performed with the same convention.

\subsubsection{Signal Loss Estimates}
\label{sec:methods:signal-loss}

The fringe-rate filters and coherent time average both induce some level of attenuation of the cosmic 21-cm signal \citep{Parsons2016}.
In order to account for this attenuation when reporting our limits on the 21\,cm power spectrum, we compute correction factors based on the signal loss expected for the filters and time averaging applied to the data.
In \nick{}, these were computed via Monte Carlo trials against mock data with a known signal amplitude.
In this work, we follow the procedure laid out in \citet{Pascua2024}, who construct an effective ``filter transfer matrix'' that represents the cumulative effect of fringe-rate filtering and coherently averaging the data, which we then use to compute the expected signal loss for each baseline.

While the main-lobe fringe-rate filter (designed to mitigate mutual coupling) is designed to attenuate less than 10\% of the 21\,cm signal power, this target is computed using a simple top-hat filter in fringe-rate space.
In practice, we use a DPSS filter, which always extends to some extent further than the nominal fringe-rate window, resulting in lower than 10\% loss.
However, for baselines with short East-West projections the main-lobe fringe-rates are close to zero, overlapping with the notch filter.
This increases the combined signal loss from both filters, sometimes exceeding the target 10\%.

The power spectrum estimate for each baseline is corrected for its expected signal loss via
\begin{equation}
    P_{ij} \rightarrow P_{ij}/(1 - L_{ij}),
\end{equation}
where the loss is baseline- and spectral-window-dependent, but LST-independent. 
Ultimately, since the estimates of the signal loss depend on simplified sky and beam models, we conservatively throw away baselines within a spectral window if the estimated signal loss is $>10\%$. 
\zcref[S]{fig:frf-signal-loss} shows the estimate per-baseline signal loss from the two fringe-rate filters,
with baselines outlined in red excluded from the final power spectrum estimates. 
Excluded baselines tend to have short East-West projections, as discussed.

\begin{figure*}
    \centering
    \includegraphics[width=\linewidth]{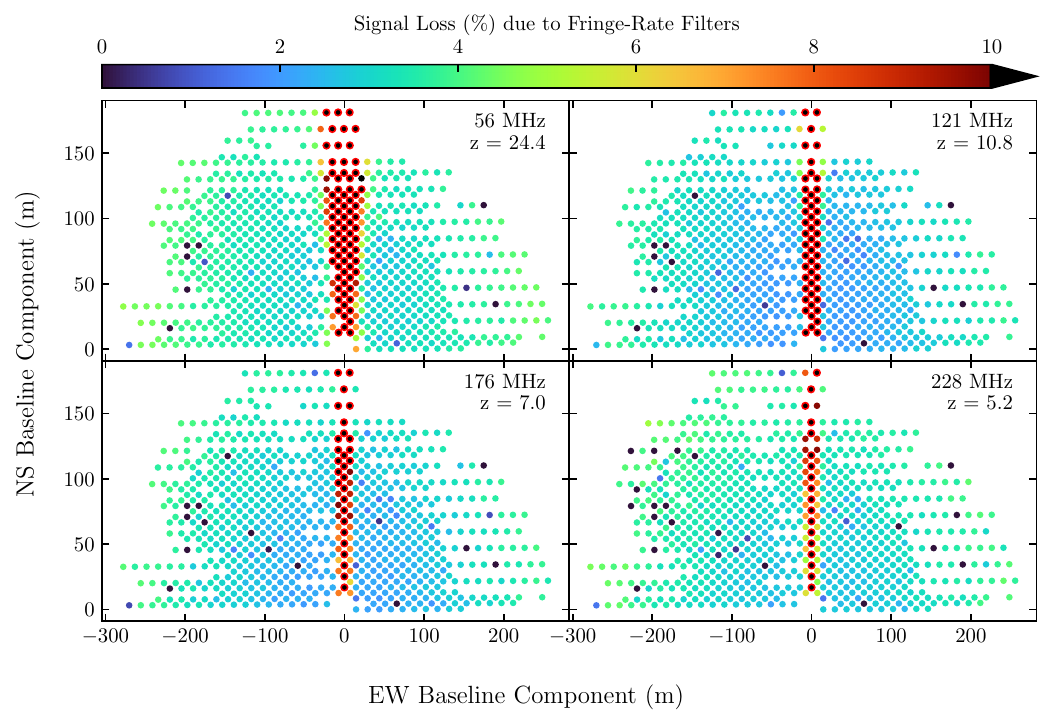}
    \caption{Estimated signal-loss from the combined notch and main-lobe fringe-rate filters described in \zcref{sec:methods:signal-loss}, as a function of baseline vector. Circles with red outlines indicate baselines whose estimate loss exceeds the threshold of 10\% beyond which the baseline is omitted from further analysis. Each panel is a different spectral band, with low frequencies tending to induce more signal loss. Signal-loss tends to be highest for North-South baselines orientations, and also grows modestly with increasing baseline length.}
    \label{fig:frf-signal-loss}
\end{figure*}


\subsubsection{Power Spectrum Estimation}
\label{sec:methods:pspec-bl:pspec}
Our methodology for estimating power spectra from the LST-binned, coherently averaged visibilities we have now obtained (still in separate interleaves) is essentially the same as that used in both \nick{} and \josh{}, and we do not repeat it here.

The only substantive difference in this analysis is that instead of forming cross-spectra between different baselines within the same redundant group, we form them between different interleaves for the same redundantly-averaged unique baseline.
Furthermore, to normalize the power spectrum, we use the updated model of the Phase II Vivaldi primary beam computed in \citet{Fagnoni2021}. 
As in \josh{}, we use a Blackman-Harris spectral taper when Fourier-transforming the visibilities.

We note that, similarly to our previous upper limits, we do not use empirical data covariances (including the number of integrated samples we have thus far propagated) to weight the power spectrum estimate. 
Doing so is subtle and requires great care not to introduce signal loss \citep{Ali2015,Cheng2018,Kolopanis2019}, and we defer this to future work.

\subsubsection{Error Bar Estimation}
\label{sec:methods:pspec:error-bars}
Nevertheless, we do use the number of samples integrated into each averaged datum in order to predict the errors on the estimated power spectra arising from thermal noise.
Here we follow the same formalism presented in \citet{Tan2021} and outlined in \nick{}, with one small correction.
We calculate $P_{N}$ (the expected standard deviation from pure thermal noise) for a given baseline type, interleave pair and spectral window as
\begin{equation}
    \label{eq:pn}
    P_N = \frac{X^2 Y \Omega_{\rm eff} T_{\rm sys}^2}{\chi_{\rm coh} \Delta t N_{\rm cohererent}\sqrt{2 N_{\rm incoherent}}},
\end{equation}
which is the same as Eq. 23 of \nick{}, but with an extra factor $\chi_{\rm coh}$ which we will define shortly.
The system temperature, $T_{\rm sys}$, is estimated using the antenna-averaged autocorrelation, averaged over the band with a Blackman-Harris weighting.
$N_{\rm coherent}$ is the band-averaged number of samples averaged into the data, from different nights and baselines within the redundant group (i.e. $N$ given by \zcref{eq:nsamps_before_pspec}, but averaged within the spectral window). At this stage, $N_{\rm incoherent} = 1$. 

In \zcref{eq:pn} the factor $\chi_{\rm coh}$ is a correction factor that accounts for the fact that neighboring LSTs within a single interleaved stream have become highly correlated by the fringe-rate filter applied for mitigating systematics. One can think of this factor as being, roughly speaking, the ratio of the timescale of the fringe-rate filter to the timescale of coherent averaging applied. More precisely, the factor is computed per spectral window and baseline group as follows. 

First, the visibility noise variance for a particular baseline group can be estimated from the autocorrelations using \zcref{eq:visibility-noise}, where $N_{\rm samples}$ is the sum of the number of unflagged baselines in the baseline group for each night of observation at the LST in question.
%
%
Since the baselines have been redundantly averaged, we construct the noise variance estimates from the LST-stacked, redundantly-averaged autocorrelation measurement. 
We can pack these variances into a diagonal noise covariance matrix for each frequency, baseline group and interleave, $k$, which we denote $\mathbf{N}(\nu, k)$ (omitting baseline subscripts for brevity\footnote{The calculation of $\chi_{\rm coh}$ does not mix frequency channels or interleaves until the very last stage, so we keep these as functional dependencies throughout this derivation. However, baseline groups are \textit{never} mixed and so we omit these for notational clarity.}).

The fringe rate filter, rephasing, and coherent averaging steps are all linear operations on the visibilities. 
This means we can compose the operations into one linear operator, which we denote as $\mathbf{T}(\nu, k)$, which is diagonal in frequency for a given spectral window and operates independently per interleave.
$\mathbf{N}$ is a diagonal square matrix of size $(N_T, N_T)$ for each $\nu$, $k$ and baseline group, where $N_T$ is the number of times in one interleave at the LST-binning resolution. $\mathbf{T}$ is a rectangular matrix of size $(N_t, N_T)$, where $N_t$ is the number of times in an interleave after fringe rate filtering and coherent averaging. Note that $N_t \neq N_{\rm incoherent}$ because some LSTs are not used in the incoherent average.
The noise covariance matrix can be propagated through these processing steps as
\begin{equation}
    \mathbf{N}'(\nu, k) = \mathbf{T}(\nu, k) \mathbf{N}(\nu, k) \mathbf{T}^\dagger(\nu, k).
\end{equation}
From this, we can define a correlation matrix at each frequency and interleave elementwise as
\begin{equation}
    \rho_{tt'}(\nu, k) = \frac{N'_{tt'}(\nu, k)}{\sqrt{N'_{tt}(\nu, k)N'_{t't'}(\nu, k)}}.
\end{equation}
The correction factor at a given frequency and interleave is then defined as
\begin{equation}
    \chi_\mathrm{coh}(\nu, k) = \frac{1}{N_\mathrm{t}}\sum_{tt'} |\rho_{tt'}(\nu, k)|^2.
    \label{eq:chicoh}
\end{equation}

This factor can be thought of as the ratio of total number of coherently averaged times to the ``effective number of independent times'' in the sense that it can be used to preserve the variance of the $\chi^2$ statistic without explicitly modeling the correlations.\footnote{\url{https://reionization.org/manual_uploads/HERA132_chi_square_with_correlated_random_variables.pdf}} It turns out that $\chi_\mathrm{coh}(\nu, k)$ is very smooth as a function of frequency and varies little across different interleaves. So, to save on computational cost, we calculate this for every tenth frequency channel in the spectral window and average over the values computed for each of these frequencies and across interleaves. 
%
%
The correction factor is then applied to the noise variance, $P_N$. However, because these power spectra still have noise that is correlated in time, an additional correction factor will need to be applied after incoherent time averaging, as we discuss below.


\subsubsection{Incoherent Power Spectrum Averaging}
\label{sec:methods:pspec:incoh}
Finally, we average the power spectrum estimates over both LSTs and interleaves for each baseline.

First, we average over LSTs between 1.25--5.75\,hours.
As described in \zcref{sec:data:selection}, we choose this range of LST because it is well-covered over all the nights in our data set and avoids some strong high-rotation-measure sources entering our main beam (e.g. B0628-28, c.f. \zcref{fig:lstcov}). 
LSTs outside this range were found to exhibit spurious high-delay spectral structure, either due to polarization leakage from these sources, or perhaps strong mutual coupling from bright sources in primary-beam side-lobes. 
For these limits, we conservatively choose to ignore these LSTs, though in the future we will seek to mitigate the structure in the extended LST range.

When incoherently averaging (whether over LSTs or other axes) we always weight each sample by its inverse noise variance, $P^{-2}_N$. 
We furthermore update $P_N$ in the resulting average accordingly:
\begin{equation}
    P_N^{\rm avg} = \sqrt{\frac{\chi_{\rm coh}}{\sum P_N^{-2}}}.
\end{equation}
The factor of $\chi_{\rm coh}$ from \zcref{eq:chicoh} accounts for the fact that power spectra still have correlated noise in time (due to the fringe-rate filter) and is only applied to the incoherent average over LSTs. For averages over baseline or interleaved-pairs, it is taken to be unity.

\subsubsection{Refined Error-Bar Estimation}
\label{sec:methods:pspec:psn}

Due to cross-terms between the signal (including foregrounds) and noise, the variance of the measurement includes a term proportional to the signal power, along with the pure-noise power quantified by \zcref{eq:pn}.
As was shown in \citet{Tan2021}, and used in \josh{}, the variance of $\hat{p}$, excluding cosmic variance, can be written
\begin{align}
     {\rm Var}(\hat{p}) \approx \tilde{P}^2_{\rm SN} &= \sqrt{2}P_sP_N + P_N^2 \\
     &\approx \sqrt{2}\hat{P}_sP_N + P_N^2 - P^2_N/\sqrt{\pi},
\end{align}
where $P_s$ is the signal power (including foregrounds), and $\hat{P}_s$ is an estimate of that power from the same noisy data that goes into $\hat{p}$.
\citet{Tan2021} showed that subtraction of $P^2_N/\sqrt{\pi}$ accounts for double-counting of this noise-power, yielding unbiased estimates of the variance for power spectra that have been incoherently averaged.
In this work, to reduce this double-counting, we calculate $P_s$ for interleave $k$ by using \textit{all other} interleaves, i.e.
\begin{equation}
    P_{s, k} = \frac{1}{N_{\rm intpairs} - 1} \sum_{k'\neq k}\hat{p}_{k'},
\end{equation}
where $\hat{p}$ is the LST-averaged power spectrum.
This reduces the double-counting correction by a factor of $N_{\rm intpairs} - 1$, where the number of interleave pairs is $N_{\rm intpairs} = N_{\rm interleaves}(N_{\rm interleaves} -1)/2 = 6$.
To de-bias the estimate of the variance at high delays, where the signal is sub-dominant to the noise, we therefore subtract a modified correction term:
\begin{equation}
    \label{eq:psn}
     {\rm Var}(\hat{p}) \approx \tilde{P}^2_{\rm SN} = \sqrt{2}\hat{P}_sP_N + P_N^2 - \frac{P_N^2}{\sqrt{\pi}(N_{\rm intpairs} - 1)}.
\end{equation}
We use this estimator for our error-bars throughout the rest of the paper.

Finally, we incoherently average over the $N_{\rm intpairs}$ interleave pairs, again weighting by $P^2_N$.
We find that histograms of the power normalized by $P_N$ for $\tau>1000\,{\rm ns}$ on any particular baseline are very well described by a standard normal distribution, as expected.

\subsection{Cylindrical and Spherical Averaging}
\label{sec:methods:sphavg}

Our ultimate products from this analysis are cylindrically- and spherically-averaged power spectra.
Similarly to the analysis of previous limits, we focus on the spherically-averaged spectra and use the cylindrically-averaged spectra predominantly as a diagnostic of systematics. 
However, we note that for data more sensitive than that reported here, it will become more important to consider the cylindrical PS when performing inference due to anisotropies caused by redshift space distortions (Breitman \textit{et al.}, \textit{in prep.}).

\subsubsection{Error Covariance and Window Functions}
\label{sec:methods:cov_window}
Thus far we have used only the variance, $P^2_N$, to weight incoherent averages, treating correlations between the averaged samples in an approximate way\footnote{That is, we ignored correlations between interleaves, and treated correlations between LSTs within a single interleave via the `effective' correction, \zcref{eq:chicoh}.}.
While this has been justifiable for averaging over LSTs and interleaves, we will soon be averaging over delays, which are known to be highly correlated due to the application of a Blackman-Harris frequency taper. 

Here, we describe how we model the covariance of our data, as well as the window functions that relate the true underlying power spectrum to the measurements.
In principle, the data is correlated between LSTs, baselines and delays, resulting in very large and computationally demanding covariance matrices. 
However, we have already treated the correlation between LSTs approximately via the coherent-average correction factor $\chi_{\rm coh}$, and the correlations between baseline types are negligible. 
In that case, the covariance of the LST-averaged data can be defined as an $N_\tau\times N_\tau$-matrix per baseline and spectral window, which greatly reduces computational complexity.

The true data covariance depends in a non-trivial way on the data-inpainting process (c.f. \zcref{sec:methods:lstbin:inpainting}), which fills flagged channels using information from surrounding channels \citep{Chen2025}, correlating neighboring delays in the power spectra.
Nevertheless, we find that the effects of inpainting on the covariance are negligible; not only are our spectral windows specifically chosen to minimize the amount of inpainting required, but the integration times with large gaps are completely flagged. Moreover, the main application of the error covariance in this work is to help us correctly combine neighboring delay bins to reduce residual correlations (c.f. \zcref{sec:methods:sphavg:delaybin}). In this regime, the correlation is dominated by the frequency taper. 

Given these considerations, in this work we choose to model the power spectrum covariance simply using the frequency taper:
\begin{equation}
    \mathbf{C}_{\tau\tau'} = P_{\rm SN}(\tau)P_{\rm SN}(\tau') |\mathcal{F}_\nu\{{\bh^2\}}|^2 (\tau - \tau'),
\end{equation}
where $\mathcal{F}_\nu$ denotes the Fourier transform over frequency.
We found that this approximation is accurate to within a few percent for the four delay-delay bins closest to the diagonal.

To understand the relationship of the true underlying power spectrum to the measurements, we also need to compute the window functions. 
Exact window functions in the absence of data defects such as flags were derived in \citet{Gorce2023}.
The generalization of these window functions to the case in which flagged gaps are in-painted per-night as we have done in this analysis is presented in \citet{Chen2025}.
They found that for the flagging patterns characteristic of this dataset, accounting for per-night inpainting results in differences of $<$1\% in the full-width at half-maximum (FWHM) of the window functions.
We ignore this small effect for the sake of computational performance, using the formalism of \citet{Gorce2023}.

\begin{figure}
    \centering
    \includegraphics[width=\linewidth]{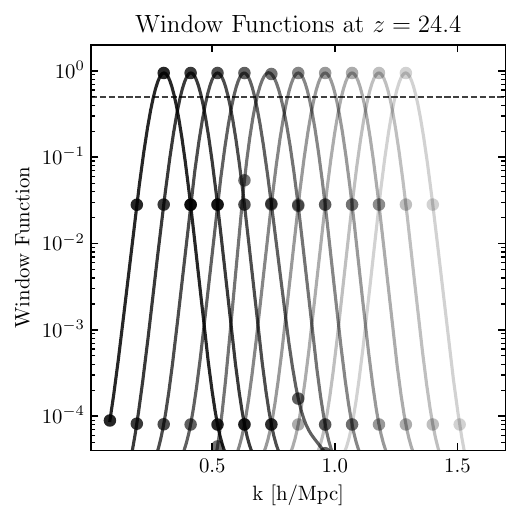}
    \caption{An example set of window functions at $z=24.4$.
    The window functions are computed only sparsely, at the circular markers, and are interpolated at third order in log-space. The window functions are largely symmetrical down to the 0.01\% level, and are generally $\sim3\%$ at the neighbouring $k$-mode.}
    \label{fig:window-functions}
\end{figure}

\zcref[S]{fig:window-functions} shows an example of the window functions of this dataset, for the lowest-frequency band ($z=24.4$), after delay-binning and spherical averaging (which are described in the following subsections).
In \zcref{fig:window-functions} we explicitly show the window function values at the $k$-modes at which they are calculated as circle markers. 
Importantly, since we estimate the window functions at the same $k$-resolution as the power spectra, they are rather sparse, dropping to $10^{-4}$ of the maximum within two $k$-bins. 
When computing the FWHM (which we use only to indicate horizontal error-bars in \zcref{fig:upper-limits}, and for no analysis purposes), we interpolate with a cubic spline in log-space (as visualized in \zcref{fig:window-functions}), which is a good approximation to the analytic form of the window function as presented in \citet{Gorce2023}.

\subsubsection{Delay Binning}
\label{sec:methods:sphavg:delaybin}
Our upper-limit likelihood, as presented in \zcref{sec:interpretation:likelihood}, assumes each $k$-mode to be uncorrelated. 
As described in \zcref{sec:methods:cov_window}, for a particular baseline, the power spectrum estimate between two $k_{||}$ modes is in general correlated, with a correlation length of $\sim$4 modes. 
We reduce correlations between neighboring $k$-modes by averaging within wider $k_{||}$-bins, each containing four modes, using a weighting kernel $\mathcal{K} = \{0.02, 0.48, 0.48, 0.02\}$. This particular kernel was chosen such that the resulting correlations between neighboring (averaged) $k$-bins are less than 1\%\footnote{We find that up to 10\% residual correlation does not affect the posteriors derived from our `upper-limit' likelihood, which assumes zero correlation. We discuss this at more length in \zcref{sec:interpretation:likelihood}.
}.
The $\tau=0$ mode is left un-averaged, so that the first bin corresponds to the first four positive delay modes, and the positive and negative delays are kept symmetrical. 

The delay-delay covariance is also binned appropriately:
\begin{equation}
    \mathbf{C}'_{mn} = \sum_{1\leq i,j \leq 4} \mathcal{K}_i \mathcal{K}_j \mathbf{C}_{4m+i, 4n+j},
\end{equation}
resulting in re-binned error-bars $P_{\rm SN}$ derived from the diagonal of the binned covariance, and re-binned window functions:
\begin{equation}
    W_{k_\perp, k_\parallel,\tau_m, \mathbf{u}} = \sum_{1 \leq i \leq 4} \mathcal{K}_i W_{k_\perp, k_\parallel, \tau}.
\end{equation}
    
\subsubsection{Cylindrical Averaging}
\label{sec:methods:sphavg:cylavg}
Given the per-baseline delay spectra $\hat{p}_{ij}$ the cylindrically-averaged power is given by first averaging together the negative and positive delays, and then averaging together all baselines of the same length (within a 1\,m bin).
Each average is weighted by the noise-variance $P_N^2$ (which itself is averaged in the same way). 
The resulting 2D power spectrum has irregularly-spaced bins of $k_\perp$.

\subsubsection{Spherical Averaging}
\label{sec:methods:sphavg:sphavg}
Finally, we perform spherical averaging into bins of $k \equiv |\vec{k}|$ of width $4 \Delta k_{||}$ (where $\Delta k_{||}$ is the natural bin-width of each delay-bin prior to delay-averaging), starting at $3 \Delta k_{||}$ and ending at $k=2.5h^{-1}$Mpc. 
These bins are designed such that the range of delays covered by each delay-averaged delay-bin across all baselines is predominantly kept within a single spherical bin.

\subsection{Summary of Important Analysis Differences}
\label{sec:methods:differences}
While many of the core techniques have remained the same between HERA's first two limits and this work, a number of techniques and decisions have changed.
Some of these stem from the differing instrumental characteristics of the Phase II instrument, while others reflect improvements in our understanding of systematics, or the implementation of their mitigation. 
Here, we list all major changes and improvements in the pipeline in a compact way, for easy reference.

\paragraph{Absolute Calibration} In previous limits, we performed absolute calibration using pre-calibrated visibilities observed during quiet skies. Here, we use simulated visibilities, based on a tailor-made sky model including both diffuse and compact sources, allowing us to cover more LSTs. Our solver has also received a significant upgrade, making it far more robust to large differences in phase between the data and the model when estimating phase gradient degeneracies.

\paragraph{Deeper RFI Flagging} For these limits our RFI flagging is a deeper, multi-stage process that begins with rough flags on single integrations, and progresses to deeper cuts based on full nights of data, averaged over baselines. This allows us to more precisely flag frequency-dependent artifacts that would otherwise go undetected.

\paragraph{Coherent Redundant Averaging} In previous limits, we conservatively did not average nominally redundant visibilities, instead incoherently averaging the data from these baselines after power spectrum estimation. Here, we perform coherent redundant averaging, resulting in a considerable increase in efficiency of the downstream pipeline.

\paragraph{LST-Calibration} For the first time, in this work we utilize our prior that night-to-night observations at the same LST should be consistent up to thermal noise and calibration errors to perform limited gain corrections informed by the full set of nights. This increases consistency between nightly visibilities, minimizing the spectral structure induced in the average over nights due to frequency- and night-dependent flags.

\paragraph{Per-night inpainting} In previous limits, we performed spectral in-painting only after averaging visibilities over nights. This was done to enable our power spectrum estimator, which is currently unable to handle non-uniform spectral weighting. However, ill effects from non-uniform weighting arise whenever averaging is done, long before power spectrum estimation. In this work, we inpaint each night independently, before averaging them together, to mitigate this spectral leakage.

\paragraph{Fringe-Rate Filtering} In past HERA limits, signal-chain systematics such as cable reflections and over-the-air cross-talk have been mitigated by fitting models in delay space using a CLEAN-like algorithm \citep{Kern2020a,Hogbom1974}. This has the notable drawback that the cleaning is a non-linear transform, so error-bars and window functions are difficult to propagate. Here, we utilize a linear DPSS model in 2D fringe-rate/delay space to mitigate such systematics, including mutual coupling, which has an increased amplitude in this data.

\paragraph{Time-Interleaving} Forming power spectra by cross-multiplying the same visibilities leads to noise bias. In previous upper limits we have avoided this by cross-multiplying different redundant baselines. Here, since we coherently average visibilities within redundant groups, we instead use the approximate redundancy between adjacent 10-second integrations to form our cross-pairs. This also reduces the risk of incurring negative systematics due to decorrelation of systematic phases between nominally redundant baselines, as observed in \citet{Kolopanis2023} and described in \citet{Morales2023}.

\paragraph{Delay binning} For past HERA limits we have formed spherically-averaged power spectra in which the spherical bins had a width of about two native bins of $k_{||}$ under the delay approximation. Since adjacent spherical bins were correlated---a property that our so-called `upper-limit likelihood' cannot handle---we then removed every second $k$-bin. In this dataset, we instead bin delays in the cylindrically-averaged spectra with a weighting kernel specifically designed to suppress correlations between adjacent bins to below 1\%. We then spherically-average the resulting binned spectra.

\section{Validation and Statistical Tests}
\label{sec:validation}

\subsection{Validation of the analysis pipeline}
\label{sec:validation:analysis}

Similarly to previous HERA upper limits, we performed detailed checks of our analysis pipeline for this work via realistic simulations (for similar validation methodologies employed for other 21\,cm experiments see e.g. \citet{Line2024,Line2025}).
Much of the framework used for this updated validation effort remains consistent with our original framework, which was detailed in \citet{Aguirre2022}.
However, there are a number of components that have been updated for this analysis, and we highlight these components before presenting the results of these detailed checks in \zcref{sec:validation:analysis:methods}.

\subsubsection{Pipeline Validation Methodology}
\label{sec:validation:analysis:methods}
To validate our analysis pipeline, we produced a realistic simulation of our data, closely matching the detailed observational characteristics of the entire data set.
In broad terms, this simulation includes full-sky models of unresolved point sources, diffuse Galactic emission and the cosmic 21\,cm signal, passed through an accurate visibility simulator including the full 350-element hexagonal core of the HERA array coupled with an electromagnetic simulation of the Phase II primary beam \citep{Fagnoni2021}.
We produced these ``ideal'' simulations for the exact same channels, polarizations and time stamps as the full dataset, and added instrumental systematics and noise to them: for example a bandpass shape and mutual coupling (see below). 
We then passed this mock data through almost the exact same analysis pipeline as the real data.
Given that the systematic effects included in the simulation conform to the assumptions of the analysis pipeline, the purpose of this effort was not to investigate the optimality of our analysis choices, but rather to check for subtle inconsistencies in the analysis that produce artifacts in the end result. 

In comparison to \citet{Aguirre2022}, there are a couple of novel aspects to the simulations performed for this work.
The first is that the ``ideal'' simulations incorporate all 350 of HERA's antennas, whereas in our previous simulations we produced only the antennas that appeared in the Phase I datasets. 
Simulating all 350 antennas means that our simulations can be adapted for future datasets for which an increasing number of antennas will be online.
However, this came with a significant computational cost, since naively the compute scales as $N_{\rm ant}^2$.
To alleviate this cost, we used the new \texttt{fftvis} simulator \citep{Cox2025} instead of the \texttt{matvis} simulator \citep{Kittiwisit2025} used in previous limits.
This simulator uses fast Non-Uniform FFT algorithms as implemented in the \texttt{finufft}\footnote{\url{https://github.com/flatironinstitute/finufft}} code \citep{Barnett2019} to compute the radio interferometer measurement equation \citep[RIME;][]{Smirnov2011}, leveraging the $N\log N$ scaling of FFT's to achieve high performance with arbitrary precision.
For the simulation configurations we required for this work (350 co-planar antennas, and sky models with a number of sources equivalent to a HEALPix map \citep{Gorski2005} with $N_{\rm side}=1024$) we found that \texttt{fftvis} on CPU completed in a comparable amount of walltime as \texttt{matvis} on GPU\footnote{These simulations and scaling tests were performed on the Bridges-2 Regular Memory HPC at the Pittsburgh Supercomputing Center, via the ACCESS supercomputing scheme \citep{Boerner2023}.}. 
Given that CPUs are vastly more available than GPUs at this time, this represented a significant increase in efficiency, allowing us to produce three full simulations (point sources, diffuse Galactic emission and 21\,cm signal) each with 12.5 million sources, 350 antennas, 1536 frequency channels and 17280 time stamps. 
We refer the reader to \cite{Cox2025} for further details on the \texttt{fftvis} simulator, and its strengths and limitations.

Our sky models consist of three components. 
The diffuse Galactic sky is simulated using the \texttt{pygdsm} software, using the GSM sky model \citep{DeOliveira-Costa2008}.
The native resolution of the GSM sky model is $N_{\rm side}=512$; we smooth this map with a Gaussian filter of size $1^\circ$ to avoid aliasing, before up-sampling to $N_{\rm side}=1024$.
The point-sources have two components: firstly, as in \citet{Aguirre2022}, we include the so-called `A-team' sources---ten of the brightest compact sources on the sky (Centaurus A, Hydra A, Pictor A, Hercules A, Virgo A, Crab, Cygnus A, Cass A Fornax A, and 3C44) with a maximum flux density of 11900 Jy and minimum flux density of 60 Jy at 200\,MHz.
The positions, fluxes and spectral indices of these sources are taken from the GLEAM survey \citep{Hurley-Walker2016}, with the exception of Fornax A, whose position is taken its host galaxy, and whose flux density is taken from \citet{McKinley2015}.
Secondly, we add 12.5 million randomly drawn sources from the flux-density distribution reported in \citet{Franzen2019}, using GLEAM data.
These sources are uniformly placed on the sky. 
This model \textit{statistically} corresponds to the true sky, but is clearly not representative of the \textit{realization} of the true sky. 
This does not affect our fundamental purpose for these simulations: to validate the analysis pipeline for 21\,cm power spectrum estimation.
Finally, our 21\,cm signal model is a Gaussian Random Field following a power-law power spectrum with spectral index of -2.7. 
We compute the redshift-dependent field as HEALPix maps using the \texttt{redshifted\_gaussian\_fields}\footnote{\url{https://github.com/zacharymartinot/redshifted_gaussian_fields}} code \citep[][Chapter 6]{Martinot2022}.
The amplitude of the 21\,cm power spectrum is tuned such that, given the known noise levels of the data, we expect it to be sub-dominant to foregrounds at low-$k$, sub-dominant to thermal noise at high-$k$, but dominant at intermediate $k$ ($\sim 0.5 - 1 h{\rm Mpc}^{-1}$). 
While this is clearly unrealistic, it should not affect the analysis pipeline, and offers the chance to understand how the pipeline behaves in a detection scenario.

After producing the ideal simulations, we interpolated the visibilities to the precise time stamps of the observed dataset (individually for each of the 14 nights). 
Following this, we added thermal noise with an amplitude given by the radiometer equation with $T_{\rm sys} = T_{\rm FG} + T_{\rm rcv}$, where $T_{\rm FG}$ was equivalent to the simulated auto-correlations, and $T_{\rm rcv}$ was set to a frequency- and time-independent value of 100K.

An important new addition to these simulations is mutual coupling.
We used the first-order semi-analytical model of mutual coupling first defined in \citet{Josaitis2021} and extended in \rath{} to inject this systematic.
In this model, the first-order coupled visibilities ${\bf V}^{(1)}$ are related to the uncoupled visibilities ${\bf V}^{(0)}$ through a traceless coupling matrix ${\bf X}$ via
\begin{equation}
    {\bf V}^{(1)} = {\bf V}^{(0)} + {\bf XV}^{(0)} + \Bigl({\bf XV}^{(0)}\Bigr)^\dagger.
\end{equation}
In each term, the rows and columns of the matrix index over antenna-polarization pairs, so this provides a fully polarized description of the coupling to first-order in the coupling coefficients.
Each term in the coupling matrix is computed analytically from a model of the uncoupled beam, a model of the reflection coefficient at the feed-load interface, and information about the antenna positions and sampled frequencies, as described in \rath{}.
When creating mock data for a particular night, we first apply mutual coupling using all of the baselines formed with the antennas that were taking data on that night.
After applying mutual coupling, we then perform a down-select to only keep antennas that were not flagged for the entirety of the night.
The bandpass we applied to the ideal data was derived from a fit to lab measurements of the signal chain bandpass.

With these systematics applied, our mock data are a good approximation of the real data. 
This is illustrated in \zcref{fig:comparison-validation-to-data}, which compares the mock data to real data, represented in fringe-rate vs. delay. 
In the upper panels, both data and simulation are displayed as amplitude waterfalls in fringe-rate/delay, for a 29.2\,m baseline of East-West orientation (2 units East-West in the HERA hexagon), and East-East polarization.

Following \citet{Aguirre2022}, in this work we did not simulate RFI.
RFI is an incredibly complex systematic whose joint statistical distribution over the various axes of the data (frequency, time, baseline, etc.) is very poorly understood. 
Instead of attempting to simulate this systematic, we instead opt to inject the flags obtained via quality metrics on the true dataset into the mock data. 
This allows us to robustly test the impact of flagging gaps on our analysis, but means we cannot explore the impact of residual unflagged RFI in the data.

Ultimately, we passed the mock dataset through \textit{almost} the same analysis pipeline as the true data. 
The only differences in the pipelines centered on the injection of flags that we just described. 
While we ran all the flagging algorithms on the mock data, we did not expect them to detect any significant outliers---and this is indeed what we found.
After the flagging steps, we copied the true data flags into the mock data, and continued the analysis and power spectrum pipeline identically for both true and mock data.

\begin{figure*}
    \centering
    \includegraphics[width=\linewidth]{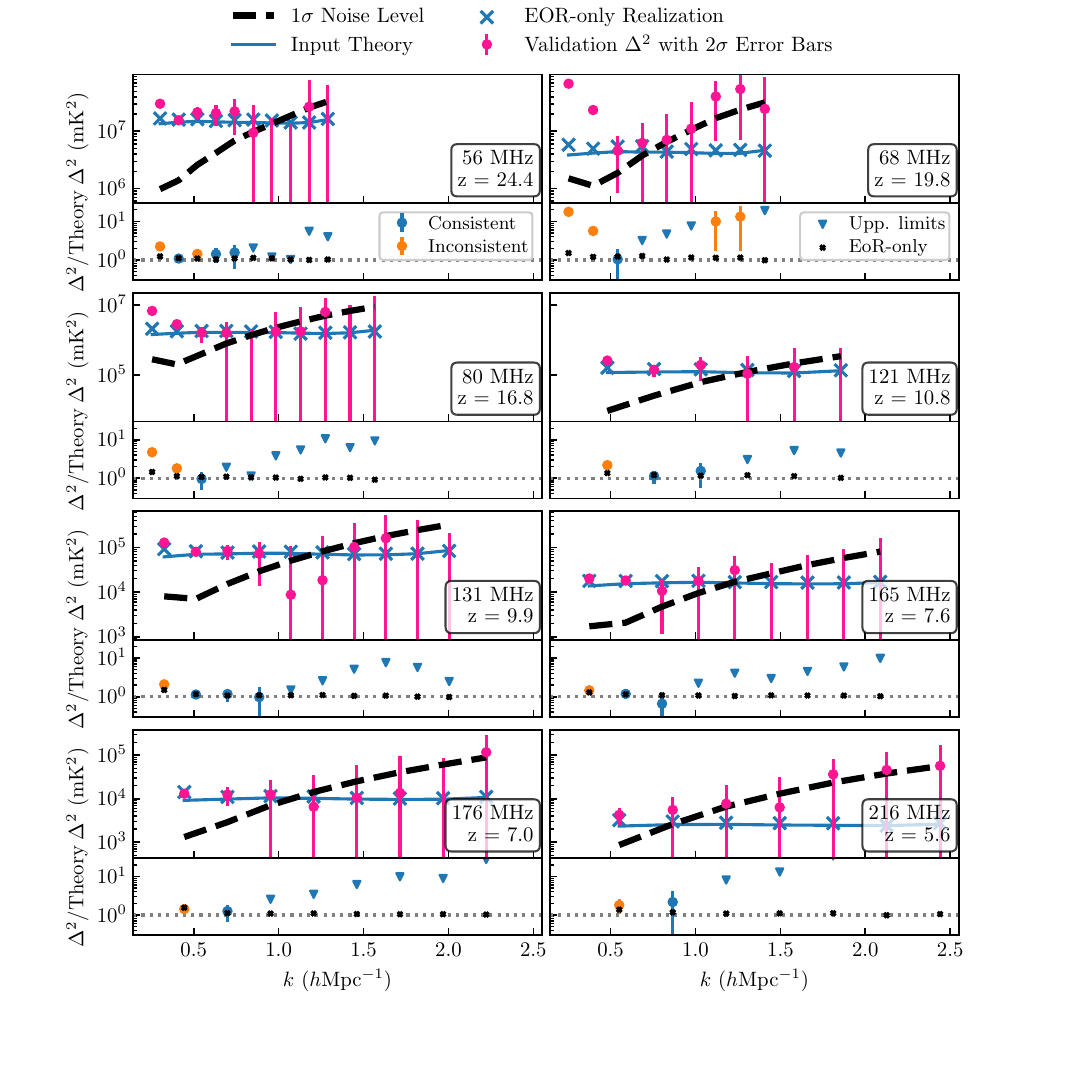}
    \caption{Comparison of estimated power spectra from end-to-end validation simulations (pink circles with error bars) to input analytic 21\,cm power spectrum (blue solid line) and EOR-only realization based on this analytic spectrum (blue crosses). The black dashed line is the estimated 1$\sigma$ thermal noise level, computed using the measured auto-correlations and number of samples integrated together. Each subplot is a different spectral window, representing a different redshift. The lower panel of each subplot shows the ratio of the power spectrum estimate to the input theory. 
    In this panel, blue points indicate estimates that are consistent with the truth to within 2$\sigma$, while orange points are inconsistent at this same significance. Triangles represent 2$\sigma$ upper-limits (i.e. points whose estimate is consistent with zero to within 2$\sigma$) while circles represent the converse (detections of non-zero power at 2$\sigma$). While all orange points indicate the presence of systematics beyond the noise level, and are therefore of some concern, our primary concern here are orange points with a ratio of less than unity, as these indicate significant signal loss. No such points exist in the validation dataset.
    }
    \label{fig:validation-pspec}
\end{figure*}

The resulting spherically-averaged power spectrum estimates for the mock data are shown in \zcref{fig:validation-pspec}, in comparison to both the analytic input power spectrum, and also a power spectrum estimate in which the mock data consisted of purely the 21\,cm signal without noise or systematics (but observed through the full instrumental pipeline).
The 21\,cm realization is generally in very good agreement with the input analytic spectrum, with cosmic variance at the largest scales causing some deviations (see black crosses in lower panel for ratios of 21\,cm-only realization to input theory). 
We recall also that the 21\,cm signal has been boosted to enable testing different regimes: at low-$k$ the foregrounds dominate (resulting in noticeable excess power with respect to the 21\,cm signal), at high-$k$ the noise dominates (indicated by the dashed black line), but at $k\sim 0.5-1\,h{\rm Mpc}^{-1}$ the boosted 21\,cm signal is marginally detectable.
The lower panel of each subplot in \zcref{fig:validation-pspec} shows the ratio of the power spectrum estimates from the full mock data to the analytic form input into the simulations (corrected for expected aliasing and approximate instrumental window functions\footnote{The effect of the window functions is most evident at $k$'s inside the wedge, where the asymmetry in the window functions there (c.f. \autoref{fig:window-functions}) moves power to higher $k$.  There are two (effectively competing) effects at higher $k$: the aliasing of intrinsic power above the Nyquist sampling in frequency to lower $k$ (discussed in \citet{Aguirre2022}, Appendix B), and the behavior of the correlator, which integrates over the power in a frequency channel, rather than sampling it at the midpoint as had been calculated previously.  All these effects are small in the $k$-range  shown here.}).
Here, blue symbols indicate estimates within 2$\sigma$ of the analytic input and orange the converse. Furthermore, triangles indicate points that are consistent with zero at the 2$\sigma$ level (these are marked at the amplitude of the 2$\sigma$ upper-limit) while circles represent the converse, i.e. mock 2$\sigma$ detections. 

We expect that, due to spectral structure from systematics such as mutual coupling, and our imperfect demarcation of foreground-dominated modes when performing spherical power spectrum averaging, low-$k$ modes may have some detections dominated by foregrounds. 
This is indeed what is seen in the lowest five bands (highest redshifts), where the first $1-3$ $k$-modes are orange circles. 
While such biased detections are problematic if interpreted as detections of 21\,cm signal, in this paper---as for previous limits---we adopt a conservative likelihood that essentially treats all estimates as upper limits regardless of their thermal noise. 
In this case, we are most concerned with orange points whose ratio to the 21\,cm-only realization is less than unity---i.e. points which display statistically significant \textit{signal loss}. 
No such points are evident in our validation tests.
This gives us confidence that our analysis techniques are accurate.

\subsection{Mutual Coupling as the Dominant Systematic}
\label{sec:validation:mc}
We have claimed several times that mutual coupling (MC) is the dominant residual systematic in our spherically-averaged power spectra at medium delays of $300\,{\rm ns}< \tau < 600\,{\rm ns}$, and that this effect has been enhanced significantly in Phase II compared to Phase I.

That MC is an important component of the Phase II data is visually evident in \zcref{fig:comparison-validation-to-data}, where a simulation of data with mutual coupling included via the first-order model of \rath{} is compared to real data, and the signature of MC is clearly visible in both.
However, this does not demonstrate that this effect is \textit{new} in Phase II.
Neither does it necessitate that MC is an important effect in the final spherically-averaged power spectrum estimates, since it should be at least partially mitigated by the fringe-rate filter illustrated in that very figure. 

\begin{figure}
    \centering
    \includegraphics[width=\linewidth]{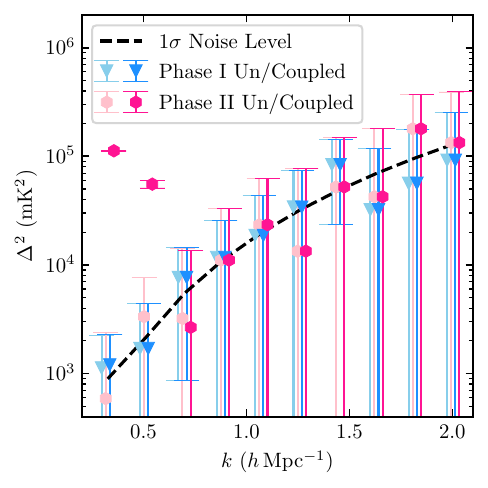}
    \caption{
    Simulated power spectra demonstrating the difference in mutual coupling between Phase I (blue) and Phase II (pink). 
    Data selection parameters have been adjusted to provide datasets with comparable noise levels (dashed black line): details are described in the text. 
    Mutual coupling (dark points) is simulated using the re-radiative model \citep{Josaitis2021} with the amplitude set by the antenna horizon gain, which is much higher for the Phase II Vivaldi antenna. 
    The data is passed through a fringe-rate filter, as described in \zcref{sec:methods:pspec-bl:frf} to mitigate mutual coupling in the spherical power spectrum estimate.
    The larger coupling results in the excess seen at lower $k$ which is morphologically similar to the actual result in Figure  \ref{fig:upper-limits}.
    }
    \label{fig:mutual-coupling-comparison}
\end{figure}

To establish that MC is indeed expected to be the dominant systematic at scales just outside the foreground wedge, we performed like-to-like simulation comparisons between Phase I and Phase II with and without MC.
We show the results in \zcref{fig:mutual-coupling-comparison}.
In this plot, the simulations include only diffuse foregrounds in the sky model as well as thermal noise, and we show just one spectral window, centered at $z=9.9$.
Mutual coupling is modeled using the first-order coupling model of \rath{}, which depends crucially on the sensitivity of the primary beam at the horizon, as well as the layout of the active antennas.
Phase I simulations include the antennas active during Phase I measurements (c.f. \josh{}), and use a model of the Phase I primary beam simulated with CST \citep{Fagnoni2021}.
Phase II simulations include all antennas active in this analysis and the primary beam simulation discussed in \zcref{sec:validation:analysis}.
While the mutual coupling is computed using the set of antennas just described, the power spectra are estimated using a subset of baselines that is the same in all cases, in order to maintain the same signal-to-noise.
The data is passed through a fringe-rate filter, as described in \zcref{sec:methods:pspec-bl:frf} to mitigate mutual coupling in the spherical power spectrum estimate.

Comparison of the dark and light pink hexagons at low $k$ shows that adding MC to the visibilities results in significant extra power.
Since the only difference between these points is that one is uncoupled and the other is coupled, we can reasonably conclude that---despite mitigation by fringe-rate filtering---MC is the dominant cause of excess power just outside the wedge\footnote{This fact is not as readily evident in \zcref{fig:validation-pspec}, since the low-$k$ modes are often dominated by the artificially enhanced 21\,cm power.}.
The similarity between the shape of the excess power at lower $k$ in \zcref{fig:mutual-coupling-comparison} and that seen in the real data (\zcref{fig:upper-limits}) suggests that mutual coupling is a likely culprit for this excess power in the data.

Conversely, comparison of the dark pink hexagons to the dark blue triangles reveals that---at least insofar as the first-order model of MC reflects reality---the magnitude of MC is significantly increased in Phase II with respect to Phase I. 
The reason for this increase, according to the first-order MC model, is the increased sensitivity of the Phase II antennas to the horizon.

\subsection{Statistical Tests of Power Spectra}
\label{sec:validation:stats-tests}
As in previous data releases, we perform a series of tests to check that the statistical behavior of the power spectrum estimates is consistent with the `null hypothesis' interpretation that random noise dominates our measurements far outside the foreground wedge. 
If the null hypothesis was rejected in a significant way by any of our tests, and given that a cosmologically viable 21~cm power spectrum detection is expected to be at least an order of magnitude below our present sensitivity level, we would have cause to suspect that systematic contamination remained in our data. 
Since we marginalize over large positive systematic effects for our cosmological analysis, the presence of systematics in this analysis is less problematic than in a detection scenario, where the precise value of the power spectrum is important and more detailed systematic models would be necessary. In this context, these null tests mainly act to point us to portions of the analysis that may show what these systematic effects are, so that we can better remove or model them in future analyses where it is more critical to do so.

At the same time, we do note that our validation tests (c.f. \zcref{sec:validation:analysis}) indicate that no \textit{modeled} systematics dominate beyond medium $k$ (though several known effects were omitted from the simulations, including low-level residual RFI, sky model errors and non-smooth bandpasses). Positive systematics that may contribute at lower $k$ include mutual coupling, low-level RFI and in-painting residuals.



\subsubsection{Test of Sensitivity to Choice of Spectral Windows}
In \zcref{sec:methods:lstbin:spws}, we present a set of spectral windows chosen to minimize spectral structure introduced by high flagging fractions. To ensure that our results are not finely tuned by these choices, we conduct a statistical test to examine the robustness of our power spectra against variations in the definition of spectral windows. 

This test is carried out as follows. For each spectral window and the visibility data that have gone through all the analysis steps until \zcref{sec:methods:pspec-bl:pspec}, we take two subsets of the frequency channels, one with the upper ten channels removed and the other with the lower ten channels removed. We then form two power spectrum estimators and take the difference. As the foreground emission is chromatic, the difference is not noise-like in the foreground-dominant regime. We therefore examine only the statistics of the power spectrum difference for $|\tau| >1500\,\mathrm{ns}$. To do this, we generate $100$ realizations of noise-only visibility data following \zcref{eq:visibility-noise}, using $N_\mathrm{samples}$ from the LST-binned data (\zcref{sec:methods:lstbin:avg}). These noise simulations undergo the exact same analysis pipeline as our data and are used to estimate the expected behaviors of the difference between the two sub-band power spectrum estimators at high delays. 

Between LSTs of $1.25$ and $5.75$ hours, for the eight spectral windows, we find that $90.90\%$, $93.27\%$, $95.55\%$, $89.69\%$, $95.77\%$, $97.88\%$, $95.47\%$, $94.13\%$ of high-delay bins are within the two-sigma confidence interval given by the noise-only simulations, compared to the expected 95\% if they were purely noise-like. 
For almost all spectral windows, the agreement with the noise is very strong.
We see that the deviation is slightly larger for the two spectral windows at the low frequency end and for the spectral window centered at $z\sim10.8$. The $z\sim10.8$ spectral window is the narrowest among the eight windows and thus is subject to the largest statistical fluctuations because it has the least number of delay bins. Meanwhile, the evolution of foregrounds within the spectral window is larger at lower frequencies. 
Overall, this test indicates that our results are robust against variations in the definition of spectral windows.

\subsubsection{Test of Signal Loss Due to Non-Redundancy}
\label{sec:sig_loss_coh_incoh_test}

Our pipeline makes strong use of the assumed redundancy of different baselines, both for increasing sensitivity and for calibration (c.f. \zcref{sec:methods:redcal}). 
However, several physical factors can lead to baselines that have assumed redundancy to have considerable non-redundancy; for example variations in antenna positioning, pointing offsets, and antenna primary beam variations.
This typically results in baselines within a redundant group containing a distribution of phase offsets from their common mean.
These phase differences lead to decoherence when `redundantly' averaging visibilities (c.f. \zcref{sec:methods:redavg}), resulting in a loss of power and therefore signal.
We estimate the extent of the loss in the same way as \nick{} and \josh{}, by forming the statistic
\begin{equation}
    \hat{L}_{\rm nonred} = 1 - P^{\tau=0}_{\rm coh} / P^{\tau=0}_{\rm incoh}, 
\end{equation}
where $P_{\rm coh}$ is computed by first coherently averaging visibilities within a baseline group, and then forming power spectra (which is the approach used to generate the upper limits we report in this paper), and $P_{\rm incoh}$ is formed by first computing power spectra for each baseline (as an auto-pair with itself), and then averaging over these baselines.

This statistic has some desirable properties under a model in which the data consists purely of an underlying signal $V_{\rm true}$, thermal noise $n$ drawn from a complex Gaussian distribution with a standard deviation of $\sigma=|V_{\rm true}|/{\rm SNR}$, and in which each visibility with a nominally redundant group has an additional phase offset $\phi_{\rm off}$ assumed to derive from its non-redundancy:
\begin{equation}
    V_{\rm model} = V_{\rm true} \exp(-i \phi_{\rm off}) + n.
\end{equation}
In the limit of a wide distribution of $\phi_{\rm off}$, regardless of the SNR, the signal loss tends towards unity (i.e. 100\% loss). 
Furthermore, for high-SNR measurements, the statistic tends towards an unbiased estimate of the true signal loss (defined as $P_{\rm coh}/|V|_{\rm true}^2$). 

Nevertheless, this statistic is not without its problems.
The most obvious source of problems is when the data does not conform to the assumptions of the simple model in which it has the desirable properties;
for example, when there are large phase differences between the visibilities, but they arise primarily due to non-21\,cm sources, such as bright sources in the side-lobes of the primary beam. Unless these phase differences enter via errant gain calibration due to these sources, any signal loss resulting from such an effect would be limited to loss of \textit{foreground} signal, rather than the diffuse all-sky 21\,cm signal. 
In \nick{} and \josh{}, this was overcome by selecting the quietest skies at which to evaluate the statistic (around the Galactic ant-center).
This is not possible in this analysis, since we do not observe this field; we instead simply measure the loss for all LSTs in our full range of $1.25$--$5.75$ \,hours.

Additionally, while the estimator is unbiased under this model for high-SNR data, it is not unbiased for low-SNR data\footnote{This can easily be shown by conducting small monte carlo simulations with the toy model described.}, generally leading to higher predictions for the loss than it should have. 
Low-SNR can occur either when the noise is large, or when the signal is small. 
The former effect was largely overcome in \nick{} and \josh{} by including only highly-redundant baselines, limiting those considered for computing the signal loss to $<$ 60\,m. 
We follow this approach in this analysis.
The latter effect (low SNR due to low signal) can arise at particular LSTs where the signal drops to brief `null'.
This was overcome in \nick{} by smoothing $P_{\rm incoh}$ along the LST-axis with a 1-hour window. 
We do the same in this work.

Finally, there is a curious effect in which, for moderate SNR and a wide range of per-baseline variance (stemming from different per-night flags on each baseline), the statistic can predict signal \textit{gain} (when no such gain is present).
The relative occurrence of this effect (under the toy model) is strongly dependent on the SNR of the data, the distribution of numbers of unflagged nights within a baseline group and the size of the baseline group (the smaller the group, the more likely an estimate of signal gain) and is less sensitive to the distribution of phase offsets within the group. 
We find (relatively rare) occurrences of this signal gain both in simulations of the toy model, and in real data.

After these considerations, we ultimately compute a signal-loss estimate for each baseline group shorter than 60\,m (and which also passes the criterion of having lower than 10\% signal loss from the time-based filtering, c.f. \zcref{sec:methods:pspec-bl:frf}), for each polarization (XX and YY), each LST between $1.25$--$5.75$ hours and each spectral window.
As in \nick{} and \josh{}, our assumption is that the true signal-loss on the \textit{21\,cm component} is very similar between baseline groups, polarizations and LSTs (due to isotropy of the 21\,cm signal, and the likely time-invariance of the sources of non-redundancy).
Thus, we obtain a single loss per spectral window, as an estimate of the true underlying loss that is constistent across baselines, polarizations and LSTs.

\begin{figure}[htbp]
    \centering
    \includegraphics[width=0.5\textwidth]{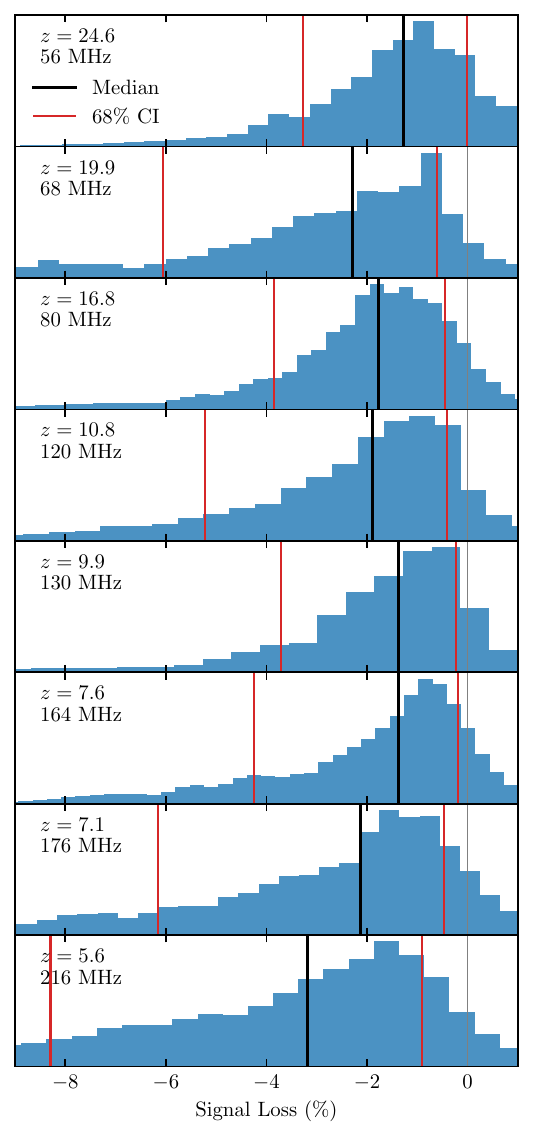} 
    \caption{Histograms of estimates of signal loss incurred by non-redundancy of nominally redundant baselines. Each panel is a spectral window, and samples for each histogram come from different baseline types, LSTs and polarizations. The median, marked as black vertical lines, is taken as the signal loss estimate. The red lines show the 68\% confidence interval. A small fraction of estimates are signal \textit{gains}, consistent with toy models (see text).}
    \label{fig:loss_per_spw}
\end{figure}

\zcref[S]{fig:loss_per_spw} shows histograms of the loss for each spectral window, where the samples for each histogram come from different baseline groups, polarizations and LSTs. 
These histograms exhibit a peaked structure, with long negative tails (out to high signal loss), and short positive tails (including small signal gains). 
We have confirmed that the long negative tail is dominated by the lowest SNR measurements, which calls into question whether they should be given full consideration when deciding a final single loss per spectral window (given the known biases for low-SNR measurements from the toy model). 
On the other hand, high-SNR measurements still have a spread in their estimated signal loss that we have found to be highly correlated in LST (i.e. certain sky distributions tend to produce particular loss estimates that can differ from other skies, even if both are high-SNR). 
The details of the physical source of the fluctuations in the loss estimate as a function of LST and polarization are of great future interest, but in this work we take the same approach as \nick{} and simply use the median of all measurements in the spectral window as our final loss estimate (shown as black vertical lines in \zcref{fig:loss_per_spw}).
The median duly down-weights the long negative tail, which we have reason to be skeptical of, while capturing the main information coming from the measurements.

Regardless of the particular merits of our choice of using the median, we note that the signal loss estimates in each spectral window are quite small: less than 6\% for more than 84\% of the data all cases (leftmost vertical red dashed line in each panel of \zcref{fig:loss_per_spw}).
For the current modality of upper limits, such small losses are not important for astrophysical inference, being dwarfed by other uncertainties, such as theoretical modeling uncertainties. 
Furthermore, the estimates are similar across spectral windows, giving confidence that the estimates are internally consistent and robust. 
Under these considerations, the median in each band is a justifiably sufficient estimate of the signal loss, and we report each such median in \zcref{tab:losses}.

\begin{table}
\centering
\caption{Median \% loss per spectral window.}
\label{tab:losses}
\begin{tabular}{llc}
\hline
$z$ & $\bar \nu$ [MHz] &  Median Loss $[\%]$ \\
\hline
24.4 & 56 & 1.2  \\ 
19.8 & 68 & 2.2 \\
16.8 & 80 & 1.7 \\
10.8 & 121 & 1.8 \\
9.9 & 131 & 1.3 \\
7.6 & 165 & 1.3 \\
7.0 & 177 & 2.1 \\
5.6 & 216 & 3.1 \\
\hline
\end{tabular}
\end{table}


\section{Results}
\label{sec:results}

\subsection{Cylindrical Power Spectra}
\label{sec:results:cylindrical-ps}
\begin{figure*}
    \centering
    \includegraphics[width=\linewidth]{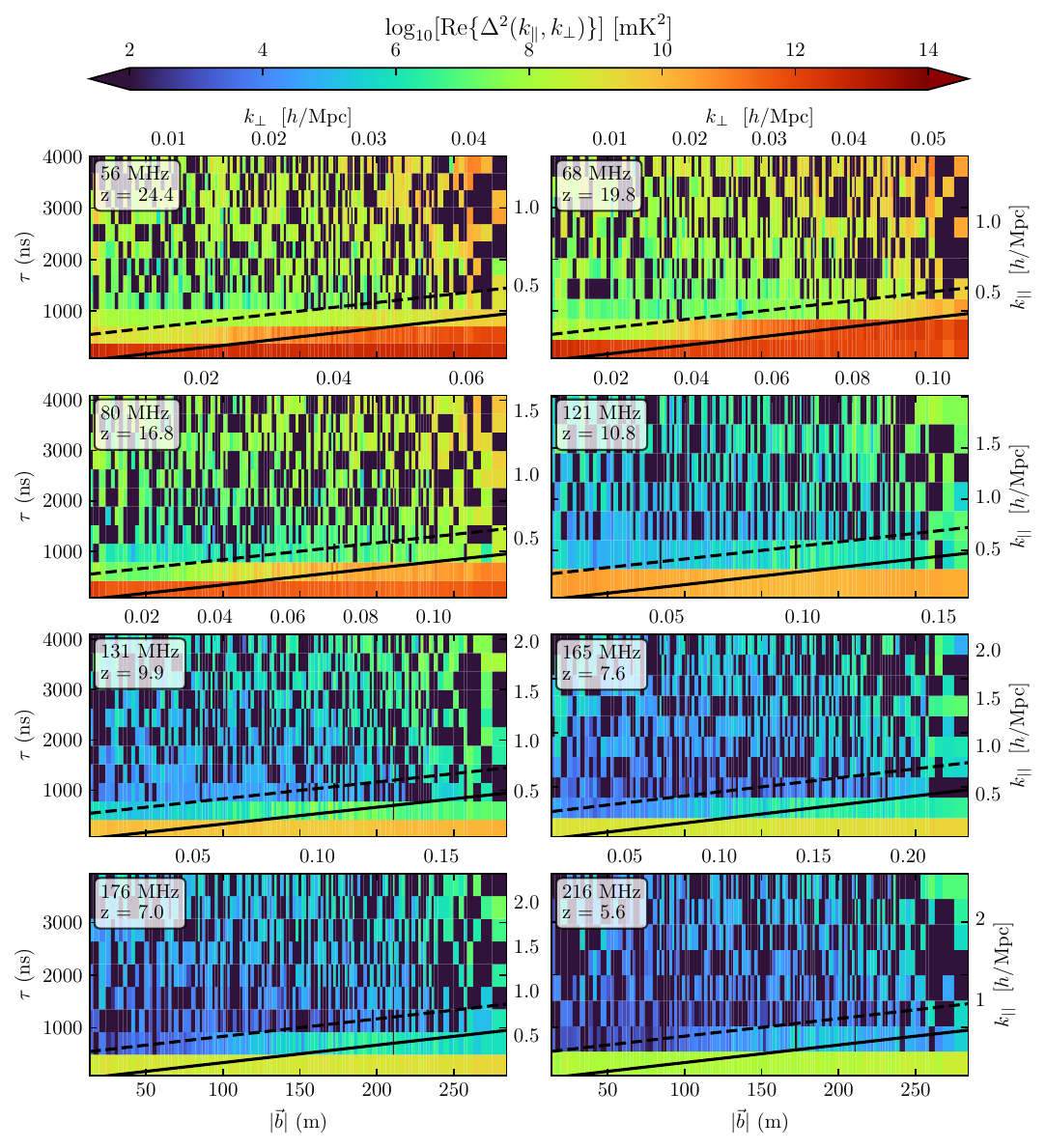}
    \caption{Real parts of the pseudo-Stokes-I cylindrical power spectra estimated from the data in this analysis. Each panel shows a different spectral window (redshift). The $x$-axes are linearly spaced in baseline length, identical for all spectral windows, with equivalent $k_\perp$ for our fiducial cosmology marked on the top of each. The $y$-axes are regularly-spaced delays, with corresponding $k_{||}$ marked on the right axis. The horizon line is marked in solid black, and an additional buffer of 500\,ns is shown as the dashed black line. Values below $10^{-2} {\rm mK}^2$ (including negative values) are shown in navy. Outside the buffer, the estimates are visually noise-like, with values scattered between positive and negative. The gradual rise in power towards larger baseline lengths is indicative of higher $P_N$ due to the decrease in the size of redundant groups at these lengths.}
    \label{fig:cylindrical-power}
\end{figure*}

We show our estimated cylindrically-averaged Stokes-I power spectra for each spectral window in \zcref{fig:cylindrical-power}. 
In this figure, we also mark the delay corresponding to a source on the horizon (solid black), as well as our final foreground cut corresponding to this horizon line plus a buffer of 500\,ns (dashed black line). 
In this plot, each pixel corresponds to the cylindrical average over all baselines within 1\,m-length bins, and over four adjacent delay bins (as described in \zcref{sec:methods:sphavg}). 

In each spectral window, the lowest delay bin is visibly dominated by foregrounds, with a dynamic range of 6--10 orders of magnitude compared to the high-delay modes.
In the highest-redshift bands, the second delay bin (and even the third, for longer baselines) is also foreground dominated. 
As discussed throughout this paper, we attribute the bulk of this power leakage beyond the horizon to mutual coupling.
This is supported by \zcref{fig:cylindrical-power}, which reveals a larger extent of foreground leakage at low frequencies, where mutual coupling is expected to be more severe.
Furthermore, while the ``foreground wedge'' \citep{Parsons2012,Liu2014a,Thyagarajan2015} is certainly present, the dominant foreground structure is rather ``brick''-like; that is, there is a strong contribution to foreground leakage that is roughly baseline-independent. 
This is consistent with predictions from our first-order model of mutual coupling, for which the strength of the coupling power at high delay for a baseline type averaged across the array is primarily determined by its orientation, rather than its length. 

\begin{figure*}
    \centering
    \includegraphics[width=\linewidth]{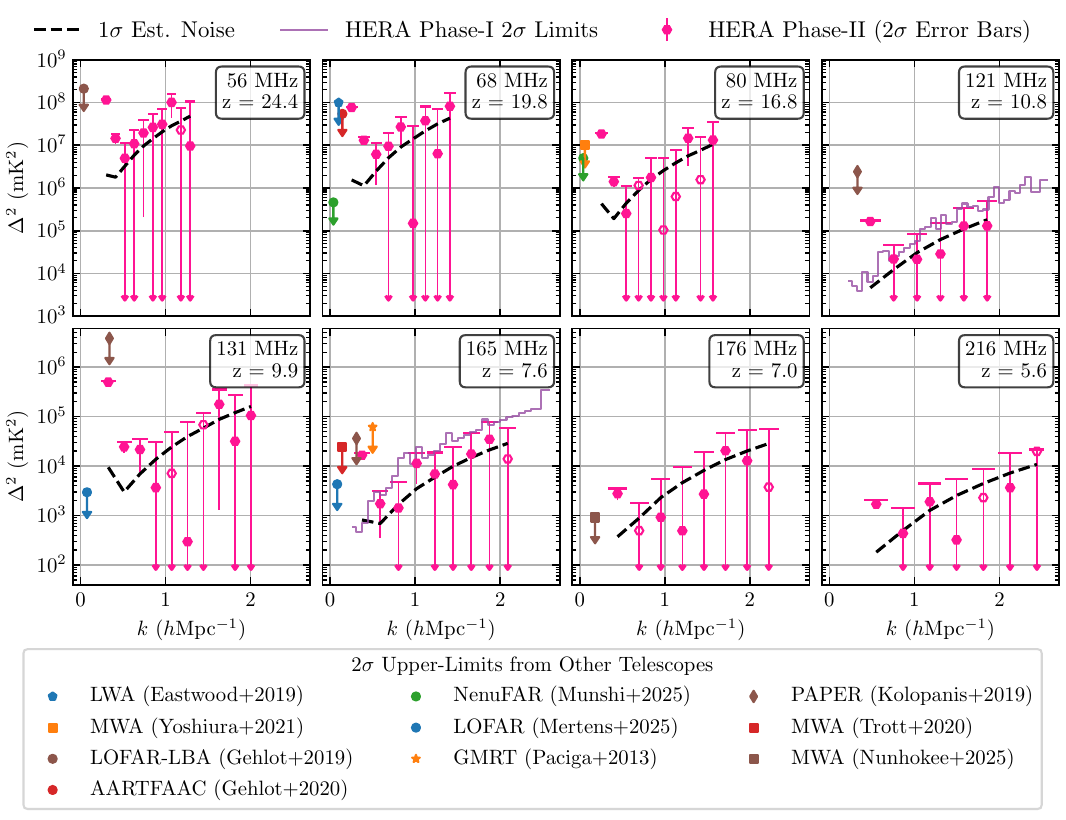}
    \caption{Spherically-averaged power spectrum estimates in each spectral window (pink hexagons with error bars). 
    Vertical error bars are 2$\sigma$ estimates, where the noise is given by $P_{\rm SN}$ (c.f. \zcref{eq:psn}). 
    The width of the top-cap of each error bar represents the full-width at half-maximum of the window function (cf. \zcref{fig:window-functions}).
    Each pink marker represents the power spectrum estimate, with empty markers indicating negative values. 
    Where error bars end with arrows, this indicates that the estimate is consistent with zero at the $2\sigma$-level.
    The thick black dashed line shows the thermal noise level, $P_{\rm SN}$. 
    The purple steps in the panels at $z=10.8$ and $z=7.6$ denote the $2\sigma$ upper-limits reported in \josh{}.
    While sensitivity of this Phase II dataset is similar to the limits of \josh{}, mutual coupling causes spectral leakage of foregrounds out to much higher $k$, causing these limits to be less strong. 
    A selection of the most stringent upper limits across redshift from different experiments are also shown \citep{Eastwood2019,Yoshiura2021,Gehlot2018,Gehlot2020,Munshi2024,Mertens2025,Paciga2013,Kolopanis2019,Wilensky2023,Trott2020,Nunhokee2025}. \josh{} remains the deepest limits to date at any redshift.
    }
    \label{fig:upper-limits}
\end{figure*}

Our choice of foreground cut, illustrated by the dashed black line in each band, was chosen primarily to ensure that leakage from mutual coupling is avoided, based on our observations of delay spectra of much less-averaged data\footnote{For example, we in-painted our data out to 500\,ns to capture foreground structure from mutual coupling in \zcref{sec:methods:lstbin:inpainting}, though this choice is not necessarily tied to our decision on the extent of the foreground cut in the cylindrically-averaged spectra.}.
Visually, \zcref{fig:cylindrical-power} suggests that our choice of foreground cut may be somewhat conservative for our longest baselines, though the fractional sensitivity of these baselines with respect to the shorter baselines is quite low, so this conservatism is not likely to affect our final limits to a large extent. 
On the other hand, we will see evidence in the spherically-averaged spectra (\zcref{fig:upper-limits}) that on short baselines, our foreground cut is not conservative enough, at least in the high-redshift bands ($<100$\,MHz). 
Since in this analysis we only consider upper limits, inclusion of excess foreground power in our estimates is not an issue, however it will become very important to better understand the limits of the foreground leakage in the future when a detection might be claimed. 

\subsection{Upper Limits on Spherical Power Spectra}
\label{sec:results:upper-limits}
\zcref[S]{fig:upper-limits} shows the final spherically-averaged power spectra estimated using the Phase II data analysed in this paper.
In the same figure, we show the predicted noise power (black dashed line), as well as the lowest upper limits from \josh{} at any $k$, shown in the bands closest to those in which they were measured (green triangles).

The lowest $k$-modes at each redshift ($k \lesssim 0.5-0.8\,h{\rm Mpc}^{-1}$ depending on the redshift) are consistently dominated by foregrounds, which have leaked well beyond the horizon (all modes inside the horizon have been cut, as described in the previous subsection). 
Indeed, while the sensitivity of this dataset is comparable to our deepest Phase I upper limit (i.e. the green triangles are roughly consistent with the black lines), our previous limits were significantly lower (approximately an order of magnitude) simply because the foregrounds were not as strong on the lowest $k$-modes we measure here. 
This foreground leakage predominantly comes from mutual coupling, as we established in \zcref{sec:validation:mc}.
This highlights mutual coupling as the greatest cause of sensitivity loss for HERA Phase II, and thus the greatest and most important challenge for future progress.

Nevertheless, while low-$k$ modes are systematics-dominated, for $k \gtrsim 0.5\text{\,}h{\rm Mpc}^{-1}$ our estimates are almost completely consistent with thermal noise in all bands, at the $2\sigma$ level.
This is especially true of the bands above FM frequencies ($>$100\,MHz), which not only have error bars that encompass $P_{\rm SN}$, but also whose estimates are generally scattered evenly about it (estimates not shown with dots are negative, but their error bars extend into the positive far enough to encompass $P_{\rm SN}$).
In the sub-FM bands, there is some evidence of an overall positive bias, since most of the estimated values fall above the noise (even though their error bars generally encompass the noise level). 
Such a positive bias can only affect our inferences in a conservative way in this analysis, but in future work we will investigate these small biases more thoroughly.
In the end, it is not surprising that the largest bias arises at the lowest frequencies, where foregrounds and mutual coupling are the strongest.

We emphasize that on these noise-dominated scales, the sensitivity of this small two-week dataset is similar to that of the full 100-night Phase I limits presented in \josh{}.
The source of this extra sensitivity is the primarily the increase in number of antennas.
The fact that these modes are noise-dominated is promising for future analyses that can utilize a larger portion of the several-hundred nights thus far observed with Phase II.

All estimates and upper limits can be found in \zcref{tab:table-of-limits}\footnote{This data may also be downloaded in a standard YAML format from the \url{https://github.com/EoRImaging/eor_limits} repository.}.

\section{Astrophysical Interpretation}
\label{sec:interpretation}

In this work, we follow the strategy of previous HERA analyses and adopt several theoretical models to infer what our multi-redshift limits tell us about cosmology and astrophysics. 
While our limits at $z<10$ are less constraining than those presented in \josh{} (primarily due to foreground leakage from mutual coupling), we have limits from a much wider range of redshifts in this dataset.
This motivates the adoption of models with mechanisms for producing very strong absorption features during Cosmic Dawn, to see whether such models---motivated by results from the EDGES experiment---are constrained by this multi-redshift data.
Given that the limits at $z<10$ are less constraining than those presented in \josh{}, we do not reconsider any reionization-driven scenarios in this work \citep[e.g.][]{Mirocha2022}. 
Furthermore, we note that the models represented here are not exhaustive, even in the context of semi-numerical/semi-analytical models \citep[e.g.][]{Pagano2020,Ghara2025,Trac2022,Schneider2023,Hirling2024,Cyr2024}, but they do span a range of interesting physical scenarios.

\subsection{Theoretical Models}
\label{sec:interpretation:models}
We adopt five different theoretical models in this work, using three different simulation codes.
The majority of these models include some mechanism for producing values of the volume-averaged ratio of the spin temperature to the background radio temperature that are much lower than standard models allow. 
So-called standard models assume that the minimum spin temperature is achieved via strong coupling to the gas kinetic temperature, whose standard minimum value is given by adiabatic cooling. 
Likewise, the radio background is typically assumed to be dominated by the cosmic microwave background (CMB), which yields a prediction for a \textit{lower limit} on $T_S/T_R$ (see the black dotted line in \zcref{fig:ts_over_tr}). 
However, motivated by results from the EDGES experiment \citep{Bowman2018}, a number of non-standard models have been proposed that can exceed this lower limit, either by reducing $T_S$ via new sources of cooling, or by increasing the radio background via new populations of high-redshift radio sources. 
Given that such models are often focused on achieving such low values of $T_S/T_R$ at redshifts of $z \gtrsim 14$ (where the EDGES experiment made its measurement), we include these models in this work to investigate if their exotic physics might be constrained by the large redshift-range of Phase II measurements.

Specifically, the models we adopt are as follows:

\paragraph{21cmFAST $T_R\equiv T_\gamma$}
This model uses the \texttt{21cmEMUv1} emulator \citep{Breitman2024a}\footnote{\url{https://github.com/21cmFAST/21cmEMU}} to emulate several observational summary statistics computed by \texttt{21cmFASTv3} \citep{Mesinger2010,Murray2020b}\footnote{\url{https://github.com/21cmFAST/21cmFAST}}.
The underlying astrophysical model is defined in \citet{Park2019}, and is defined by nine parameters governing the relationship of UV, X-ray and Lyman-$\alpha$ emissivity to halo mass and redshift. 
This model contains no non-standard contributions to $T_S/T_R$, taking the usual $T_R \equiv T_\gamma$ assumption, and is equivalent to that used to infer astrophysics in \theory{} and \josh{}. We assume the same priors over all parameters as in \theory{} and \citet{Breitman2024a} (see ranges of Figure 6 in both). 

\paragraph{21cmFAST Mini-Halo $T_R$}
This model uses v2 of \texttt{21cmEMU}, trained on  \texttt{21cmFAST} models that include a contribution to the radio background from Population-III stellar remnants hosted in molecularly-cooled galaxies (or mini-halos, c.f \citealt{Qin2020c,Qin2020b}). 
The inhomogeneous radio background sourced by the minihalos is presented in \citet{Cang2024}, and the eventual sterilization of Pop III stars by Lyman-Werner and photo-heating feedback provides a physically-motivated mechanism to avoid violating limits on the present-day radio background \citep[e.g.][]{Fixsen2011}.
Note that while this model contains a superset of the physical parameters defined in \citet{Park2019}, in this work we only vary the five minihalo parameters, setting the parameters governing Population II stars to the fiducial ones of \citet{Munoz2022}. The prior over these five parameters is the same as in \citet{Cang2024} (see $T_{21}$ component in Table 1). 

\paragraph{21cmSPACE Uniform $T_R$}
This model uses an emulator (described below) trained on \texttt{21cmSPACE} simulations \citep[e.g.][]{Visbal2012,Fialkov2012,Fialkov2013,Fialkov2014}\footnote{\url{https://www.cosmicdawnlab.com/21cmSPACE}} that include a spatially uniform synchrotron background which replaces the CMB as $T_R = T_\gamma (1 + z)\left[1 + A_{\rm rad}(\nu_{\rm obs}/78\ {\rm MHz})^{-2.6}\right]$ with the phenomenological parameter $A_{\rm rad}$ \citep{Fialkov2019}. The \texttt{21cmSPACE} simulator includes flexible Pop II and Pop III star-formation in galaxies, contributing to Wouthuysen-Field (WF) coupling, Lyman-Werner feedback, Ly-$\alpha$ heating and ionization, as well as flexible X-ray spectra and efficiencies. The simulations used here are described in more detail in \citet[][see their Table~2 for the nine parameter prior space]{Gessey-Jones2024}.

\paragraph{21cmSPACE Inhomogeneous $T_R$}
This model is identical to the Uniform $T_R$ model, except that the excess radio background is instead sourced by radio luminous galaxies, with an efficiency parameter $f_{\rm rad}$, and is spatially inhomogeneous via the SFR \citep{Reis2020}. For more details on the simulations, see \citet[][and their Section~2.3 for parameter priors]{Pochinda2024}. 
For both the uniform and inhomogeneous $T_R$ models, we employ an updated version of the analysis pipeline from \citet{Pochinda2024}, built using \texttt{PyTorch} and including GPU accelerated distributed training. 

\paragraph{Millicharged Dark Matter}
This model uses the \texttt{Zeus21} analytic code \citep{Munoz2023a}, which includes both atomic- and molecular-cooling galaxies and a flexible kinetic temperature from the Boltzmann solver \texttt{CLASS} \citep{Blas2011a}. 
Here we replace the adiabatic-cooling gas temperature in {\tt Zeus21} with a temperature consistent with cooling from millicharged dark matter \citep{Munoz2018a,Barkana2018,Berlin2018a}, and include the $T_k-T_S$ correction from~\citet{Mittal2021} for very low temperatures\footnote{We conservatively do not include the additional fluctuations from the velocity dependence of the cooling in this analysis~\citep{Munoz2018,Liu2019e}}.
As in the ``Minihalo $T_R$'' model, we keep the Pop II stellar parameters fixed to the fit to the HST and JWST UVLFs from \citet{Munoz2023}, except the X-ray efficiency $L_{40}$ (in units of $10^{40}\,\rm erg/s$ per unit SFR),
and vary the Pop III star-formation efficiency $f_{\star,\rm III}$ (for more details on the code, including implementation of feedback, see \citealt{Cruz2024}). 
In addition, we assume a fraction of 0.5\% for millicharged dark matter and vary over its mass $m_{\rm \chi}$ and charge $Q_{\chi}$ (in units of the electron).
We assume log priors on all parameters over the following ranges: $-3.3<\log_{10}(m_\chi)<-0.7$, $-7<\log_{10}(Q_\chi)<-2$, $-3<\log_{10}(L_{40})<-3$, and $-4<\log_{10}(f_{\star,\rm III})<-0$.

\subsection{Choice of likelihood}
\label{sec:interpretation:likelihood}

Each of the theoretical models employed here compute spherically-averaged power spectra, $\Delta^2_{\rm theory}(\mathbf{k}, \mathbf{z})$ for vectors of $\mathbf{k}$ and $\mathbf{z}$ defined by each theory code. 
These models are interpolated to the scales and redshifts of the data, and processed through the instrumental window function, to obtain theoretical residuals.
We then use Bayesian inference to obtain posterior distributions for the parameters of each model, $\mathbf{\theta}_{\rm mdl}$ as well as functional posteriors for $\Delta^2_{\rm theory}$ and other derived quantities.

Here, as in \theory{} and \josh{}, we use the so-called ``upper-limit likelihood'' \citep{Ghara2020a} when defining our posterior probability (Eq. 11 of \josh{}).
Formally, this likelihood assumes a systematic whose value is uncorrelated between power spectrum bins and may range from 0 to infinity with uniform \textit{a priori} probability. 
It also assumes uncorrelated noise $\sigma_i$ between power spectrum bins, which is not reflective of our actual noise. 
Our power spectrum is weighted such that neighboring bins have 1\% correlation, which is different from previous analyses where we decimated in $k$-space to avoid modeling noise correlations. 

We justify using the same likelihood (i.e. the one corresponding to uncorrelated noise) via a numerical investigation into the marginal distribution of the systematic and noise (jointly between power spectrum bins). 
We detail this investigation in \zcref{app:likelihood-correlations}, but in summary we find that residual correlations between neighboring $k$-bins that are left unmodeled have a negligible effect on the estimated posterior, even when the bins are up to 90\% correlated.
This is mostly due to the systematic model, whose large uncertainty and independence between bins washes out the effect of noise correlations.

In producing the constraints shown in the next subsection, all models were run with two different datasets: one using only the upper limits reported here from Phase II data, and another that also adds the more stringent upper limits at $z=7.9$ and $10.4$ from \josh{}.
In all cases, each upper limit at each scale and redshift is treated independently.

In addition to this likelihood, the models using the \texttt{21cmFAST}/\texttt{21cmEMU} code use a variety of other likelihoods (each treated independently):
(i) \textbf{Thomson scattering optical depth to the CMB}: As in \josh{}, this term is a Gaussian likelihood centred around $\tau_e = 0.0569^{+0.0081}_{-0.0086}$ based on the median and 68\% credible interval (CI) from the posterior obtained in \citealt{Qin2020} from the re-analysis of \citealt{PlanckCollaboration2018} data.  The CMB optical depth is an integral measure of the EoR history, disfavoring early reionization.
(ii) \textbf{Lyman forest dark fraction}: As in \josh{}, this term compares the proposed model's global neutral fraction at $z=5.9$ with the upper bound $\overline{x}_{\textsc{hi}} < 0.06 \pm 0.05$ at 68\% CI obtained with the model-independent QSO dark fraction statistic (introduced in \citealt{Mesinger2010} and applied to observations in \citealt{McGreer2015}). The likelihood function is unity if the proposed global neutral fraction is below the upper bound at $z=5.9$, then it decreases as a one-sided Gaussian for higher values of $\overline{x}_{\textsc{hi}}$. The dark fraction observations disfavor extremely late reionization.
(iii) \textbf{UV luminosity functions} (LFs) which compare the model
with well-established $z = 6, 7, 8, 10$ UVLFs observed with \textit{Hubble} (Bouwens et al. 2015, 2016; Oesch et al. 2018) in the magnitude
range $M_{UV} \in [-20, -10]$. For the UVLFs a Gaussian likelihood is used.  UV LFs constrain how star formation is assigned to dark matter halos, and therefore provide limits on the redshift evolution of the star formation rate density.

The inferences using \texttt{21cmEMUv1} with $T_R \equiv T_\gamma$ include all likelihood terms, while 
inferences using \texttt{21cmEMUv2} involving an excess radio background sourced by minihalos omit (iii), 
since the UV LF data only constrains the more massive galaxies that are held fixed in this model.

\subsection{Inference Results}
\label{sec:interpretation:results}
Since the upper limits at redshifts 8 and 10 from this work are weaker than those reported in \josh{}, we perform inference with all 5 models under different combinations of data: (i) only the data from this upper limit; and (ii) this limit combined with the lowest limits from \josh{} (+Phase I).
Naively, we expect that constraints at $z\leq 10$ would be dominated by the Phase I data, while the limits reported here may have some additional constraining power at higher redshift.
However, given the strong correlations between redshifts for physical models of the power spectrum, strong constraints tied to lower redshifts may dominate even over direct measurements at higher redshifts that are comparatively weaker.

We find that for all models, the constraints on parameters are relatively weak, and very much consistent with the constraints presented in \josh{}. 
That is, parameter constraints are dominated by the more stringent upper limits from Phase I at $z=8, 10$. 
We do not show these constraints directly here, and refer the reader to \josh{} for further discussion of the models ruled out by HERA---typically `cold reionization' models.

\zcref[S]{fig:dsq_constraints} shows the 95\% confidence upper limits on the \textit{inferred} power spectrum at $k=0.5\,h{\rm Mpc}^{-1}$ as a function of redshift.
The first thing to note from this figure is that the low-redshift limits dominate the constraints on each theoretical model. 
Indeed, the $z>15$ limits are not quite within the prior region of any of these models, which is significant since four of these models have ``exotic'' mechanisms for producing significantly boosted power at high redshift (note how far the limits are from the prior of the standard $T_R\equiv T_\gamma$ model---$>3$ orders of magnitude).
Further making this point, the Milli-charged DM model is only constrained by data at $z>10$, as the model does not include reionization processes, and its constraints are by far the weakest.
Since the strongest constraints at lower redshifts come from \josh{}, when they are included they are completely dominant. 
That is, this dataset does not provide any further constraining power---under this set of models---beyond the existing limits.

In \zcref{fig:ts_over_tr} we show the derived posterior limits on the ratio $T_S/T_R$ as a function of redshift, which characterizes the mean thermal evolution of the IGM.
Here it is clear from the prior (illustrated by thin dotted lines) that each model is able to produce very low values of this ratio---well in excess of the adiabatic cooling limit\footnote{In this plot we show only the four models that have mechanisms to produce excess cooling or excess radio background, leaving out the `standard' $T_S=T_R$ model.}.
As a comparison point, we show the value of $T_S/T_R$ implied by the reported global 21\,cm absorption depth reported by EDGES \citep{Bowman2018} as a green marker. 
All models, by construction, include this point in their prior.

\begin{figure*}
    \centering
    \includegraphics[width=\linewidth]{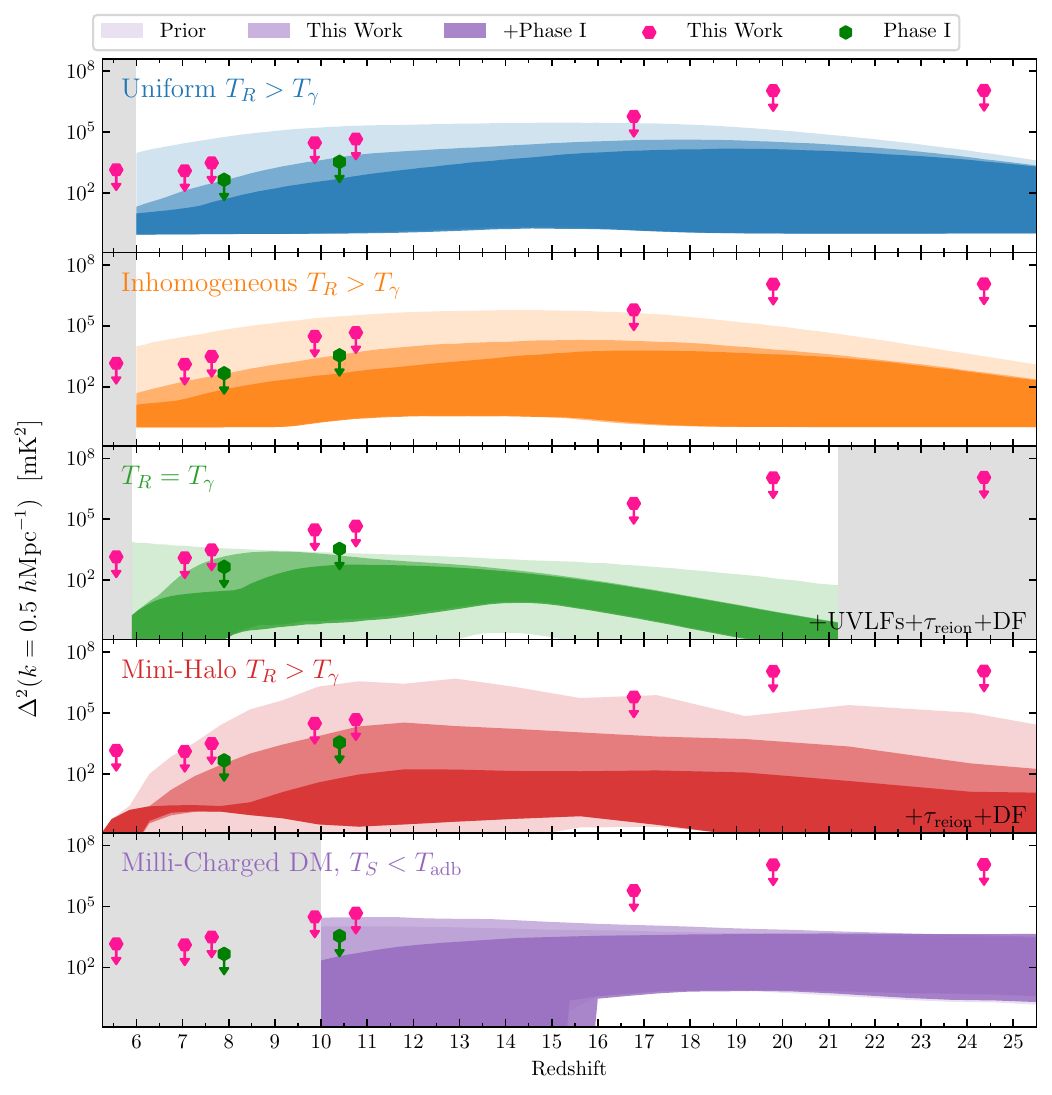}
    \caption{95\% confidence intervals on $\Delta^2(z)$ at $k=0.5\,h{\rm Mpc}^{-1}$. 
    Each panel displays a different theoretical model, and different shading levels represent different combinations of data: the lightest is the model prior, the next layer includes the limits from this work (shown as pink hexagons), and the darkest layer combines these limits with those from \josh{} (shown as green hexagons). 
    The \texttt{21cmFAST}-based model posteriors also include other data, as listed in the lower-right of each panel and described in-text.
    The limits indicated as hexagons are the lowest limit at any $k$ for each particular redshift (generally $k\sim0.5\,h{\rm Mpc}^{-1}$). 
    Not all models are defined at all redshifts where limits were placed: both \texttt{21cmSPACE} models (blue and orange) and the \texttt{21cmFAST} $T_R\equiv T_\gamma$ model only extend down to $z=6$---an artifact of the emulator training for each model. Conversely the \texttt{Zeus21} model (purple) only extends down to $z=10$ as it does not yet include realistic reionization.
    The limits of the predictions of each model are demarcated by gray shaded regions.
    The upper limits, as expected, do not constrain the lower edge of the model priors, although in the case of \texttt{21cmFAST} the lower edges are constrained by the non-21\,cm data. 
    In all cases, the low-redshift limits dominate the constraints at all redshifts, which can be seen most clearly in the \texttt{Zeus21} model, whose constraints are weakest since they do not use any low-$z$ data.
    }
    \label{fig:dsq_constraints}
\end{figure*}

\begin{figure*}
    \centering
    \includegraphics{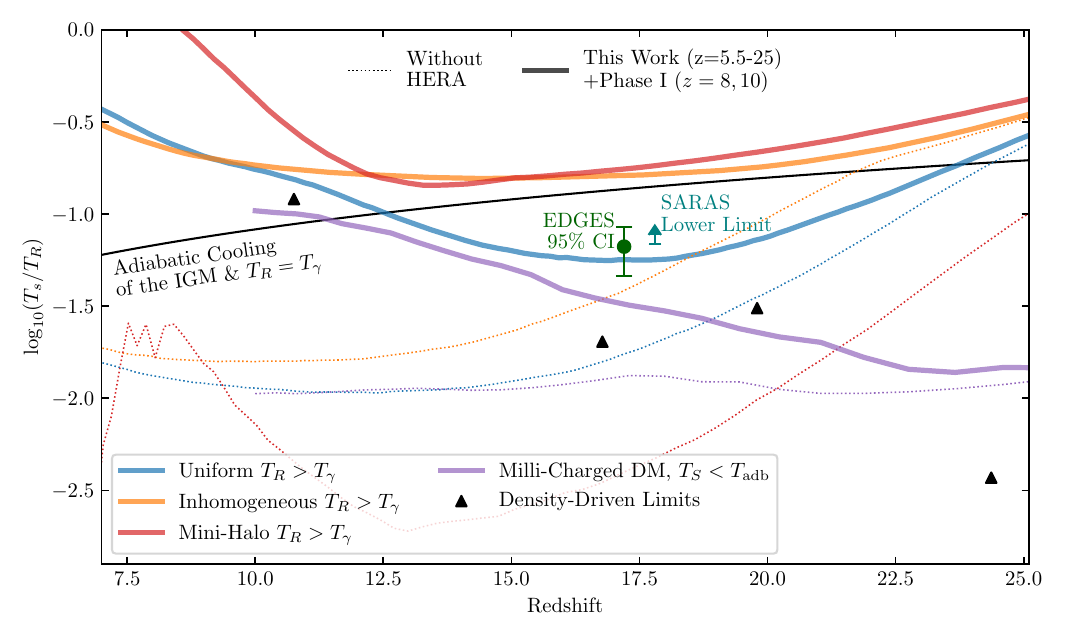}
    \caption{Inferred 95\% lower-limits on the ratio $T_S/T_R$ from several theoretical models. In this figure, different colors represent different theoretical models, all of which include some mechanism for producing $T_S/T_R$ lower than the ``standard'' adiabatic cooling limit with a uniform CMB radio background (black solid line).
    Dotted lines indicate the prior region of each model, while solid lines indicate the constraints from the combination of this data (from $z=5.5-25$) with the limits from \josh{} (at $z=8, 10$).
    Black triangles indicate lower limits derived from a more model-agnostic approach, in which 21\,cm fluctuations before the onset of reionization are considered to be driven by density fluctuations (c.f. Sec. 7.3 of \josh{}). These correspond to 2$\sigma$ lower limits derived directly from the data.
    Also shown is the inferred 95\% confidence interval on $T_s/T_R$ derived from the 21\,cm absorption depth reported in \citet{Bowman2018}, and similarly the lower limit derived in the same way from the limits on the absorption depth presented in \citep{Singh2022}.
    In general, the models that include excess radio backgrounds---whose priors allow for the EDGES-derived ratio---tend to disfavor such small ratios once exposed to the HERA limits. 
    These results are highly model-dependent, since the constraints for the models almost entirely derive from lower-redshift measurements.
    The most model-agnostic limits---those from the density-driven model---are still entirely consistent with excess cooling/radio background, and will require an order of magnitude reduction if they are to begin to rule out the EDGES-derived ratio. 
    }
    \label{fig:ts_over_tr}
\end{figure*}

After constraints from HERA data (both from this work and \josh{}), the excess radio background models (blue, orange and red) all tend to disfavor the EDGES-inspired temperature ratio.
However, we caution against over-interpreting this result, as it is highly model-prior-dependent. 
As mentioned previously in the context of \zcref{fig:dsq_constraints}, almost all the constraining power for the \texttt{21cmFAST}- and \texttt{21cmSPACE}-based models comes from \josh{} at $z=8$ and 10. 
It is the smooth evolution of these models that yields the constraints at higher redshifts. 
Conversely, the \texttt{Zeus21}-based model (purple line) is \textit{only} exposed to data at $z>10$, and is clearly far less constrained. 
The black triangles in \zcref{fig:ts_over_tr} illustrate a less model-dependent constraint.
These are lower-limits obtained directly from the upper-limits on $\Delta_{21}^2$ via a phenomenological argument that relates the 21\,cm power to the underlying density field through a linear bias (c.f. Sec. 7.3 of \josh{}; here we conservatively assume observations at $\mu=1$).
Since each density-driven lower-limit is obtained directly from an upper-limit at that redshift, it provides some measure of how far our upper-limits are from providing direct model-independent constraints on excess cooling/radio background during Cosmic Dawn. 
We find that we require an order-of-magnitude improvement\footnote{Precisely the improvement we might naively expect from processing the full season of data, c.f. \zcref{tab:obschar}.} at $z=17.5$ to yield such a constraint on EDGES-motivated models\footnote{``EDGES-motivated'' here refers to any physical model incorporating excess cooling and/or radio backgrounds during Cosmic Dawn, and does not imply that these models are in fact \textit{favored} by the EDGES data \citep[c.f.][]{Cang2024}, only that their development was motivated by the enhanced absorption reported in \citet{Bowman2018}.}.

\section{Conclusions}
\label{sec:conclusions}
In this work we have presented the first science results from Phase II of the HERA telescope.
Comprising just two weeks of observations, this analysis is aimed at understanding the unique new characteristics of the Phase II system, and developing a robust analysis pipeline that accounts for the spectral and temporal features in the data.
With increased frequency coverage of 47--234\,MHz, this is the largest range of redshifts simultaneously examined in this field, including Cosmic Dawn ($z>20$) and the tail end of reionization ($z<6$).

Our analysis pipeline, composed primarily of reproducible and scalable Jupyter notebooks, first calibrates each integration independently, then performs several quality checks based on metrics compiled over a full night's observations, removing data plagued by RFI, lightning, packet loss, low correlation and other systematics.
The calibration solutions are then smoothed over long temporal and medium spectral scales, taking advantage of our prior that the intrinsic instrument has a smooth response. 
Once the calibration is completed, and the most egregious defects have been flagged, we start averaging the data: first we average visibilities within nominally redundant baseline groups, and then we average visibilities at the same LST on different nights. 
The latter requires us to ``in-paint'' the gaps in the data incurred by frequency-dependent flagging (generally for RFI), which we do on a per-(night, baseline-group, polarization, LST) basis. 
We then attempt to mitigate pervasive systematics such as mutual coupling using filters applied in fringe-rate and delay-space, before computing delay spectra for each baseline independently.
Finally, we average these spectra (incoherently) over LST, orientation (cylindrical average) and baseline group (spherical average) to yield power spectra in a set of eight redshift bands ($z= 24.4$, 19.8, 6.8, 10.8, 9.9, 7.6, 7.0 and 5.6).
For details on these analysis steps, see \zcref{sec:methods}.

Our analysis pipeline was validated using large end-to-end mock simulations, in a similar fashion to \nick{} and \josh, using the framework laid out in \citep{Aguirre2022} and described in \zcref{sec:validation}.
This validation procedure did not uncover any unexpected biases in the estimated 21\,cm power spectrum.
The input spectrum was artificially boosted in the simulations to enable a ``detection'' at scales of $k\sim 0.5-1.0$\,$h{\rm Mpc}^{-1}$, and the mock detection was consistent with the input across all redshifts.

We reported upper limits on the spherically-averaged power spectrum, for which the sensitivity is similar to our deepest Phase I limits (c.f.\ \zcref{fig:upper-limits} and \zcref{tab:table-of-limits}), using only 14 nights of data compared to the 94 used in \josh{}.
This increase in nightly sensitivity is attributable to the increased number of active antennas in this dataset: $\sim$130 vs. $\sim$40 in Phase I.
For $k\gtrsim 0.5$\,$h{\rm Mpc}^{-1}$, our estimated power spectra are consistent with the predicted noise variance, derived from the autocorrelations and accumulated number of samples (c.f. \zcref{eq:psn}), at the $2\sigma$ level.
However, on larger scales, we find that our estimates are foreground/systematics-limited. 
This foreground leakage is more extensive than was found in Phase I, such that even though the overall sensitivity of this data is similar to Phase I, the upper limits are less constraining since the lowest, most constraining, $k$-modes---which were noise-limited in Phase I---are here contaminated by foregrounds.

Nevertheless, these limits are competitive in two primary ways: they are the first to simultaneously cover such a large redshift range, and they are the most constraining at high redshifts ($z>16$) (c.f. \zcref{fig:dsq_constraints}). 
While they do not yet yield further insight into the physics of Cosmic Dawn and the EoR beyond what is known from previous 21\,cm power spectrum upper limits, they confirm the accuracy and efficiency of HERA's analysis pipeline, which is ready to process an order of magnitude more data that has already been taken.

\subsection{Mutual Coupling}
\label{sec:conclusions:mc}
Our primary extant systematic at $k \lesssim 0.7\,h{\rm Mpc}^{-1}$, 
and the cause of the foreground leakage that precludes this dataset from setting limits comparable to \josh{}, is mutual coupling (c.f. \zcref{sec:validation:mc}).
This systematic, which occurs when radiation is reflected or re-emitted from one element of the array before being received into another antenna, is enhanced in Phase II with respect to Phase I due to the larger vertical cross-section of the updated Vivaldi feed, the increased height of the feed within the dish, and the removal of cylindrical mesh guards around the feeds (see e.g. \citealt{Berkhout2024}). 
Taken together, these effects result in an increased interaction between neighbouring antennas and an enhancement of the observed mutual coupling.
Though it is difficult to measure the precise extent to which the effect is enhanced in Phase II, \zcref{fig:mutual-coupling-comparison} suggests via approximate models that it now dominates modes up to $k\sim 0.7\,h{\rm Mpc}^{-1}$ that were previously noise-dominated. 
This is a $\sim1\%$ effect at 500\,ns scales, which is beyond the precision required for 21\,cm science, and should be considered carefully by planned experiments with densely packed antennas. 

The impact of mutual coupling arises from delayed signals coupling back into a baseline's visibility.
Naturally, these delays are super-horizon scale in the cylindrical power spectrum, resulting in foreground leakage into the 21\,cm window.
While a first-order analytic model of this effect (\citealt{Josaitis2021}, \rath{}) has been successful in describing the general characteristics of the systematic, it is not accurate enough to enable us to invert and remove it. 
Instead, we have used the insight gained from our analytic model to partially localize the leaked power in fringe-rate/delay space, and apply a simple filter in this space to remove the bulk of the systematic power. 
This mitigation technique is effective, but is limited in the amount of leaked power it can remove without significant loss of cosmic 21\,cm power, because both mutual coupling and 21\,cm signal exist within the bounds of the filter.

Given the approximate nature of the first-order coupling model, the relative amplitude of mutual coupling to cosmic signal power as a function of scale is highly uncertain.
Figure 10 of \rath{} suggests a smooth decline until $k\sim0.9\,h{\rm Mpc}^{-1}$ before a steeper roll-off, although this roll-off corresponds to the light-crossing time of the array, which is the maximum delay that the \textit{first order} coupling model predicts, and higher-order terms may extend this range.
While we have demonstrated that mutual coupling is dominant below $k\sim0.7\,h{\rm Mpc}^{-1}$, its behavior on smaller scales---how quickly it tapers off with increasing $k$---remains unobserved.
More sensitive observations will be able to reveal this behavior in greater detail.

A key takeaway from this work is that more detailed models of mutual coupling may yield the greatest improvement to HERA's sensitivity, even as more data is averaged. 
The current first-order linear model is clearly insufficient for the purposes of inverting the systematic from the data; 
beyond the possible importance of higher-order terms (i.e. re-reflections and the correlation of two separate reflections),
it is likely that reflections from elements other than feeds may contribute to high delay structure in the visibilities \citep{Fagnoni2021a}. 
While stability of the coupling over nights is broadly expected, it has not been observationally established beyond the resilience of the systematic to data averaging on medium-to-large scales. 
Conversely, differences of the mutual coupling signal between otherwise redundant baselines might be leveraged to help disentangle the effect.
We continue to explore these issues with embedded element antenna simulations, as well as extended analytical modeling.

Happily, our results indicate that the amplitude of foreground leakage from mutual coupling rapidly decays with delay, and so we can expect a clean signal (clean from mutual coupling, at least) at some scale, given enough thermal sensitivity. 
At lower delays, approximate mitigation or inversion techniques will temporarily buy sensitivity for tighter upper limits, but will need to be significantly more accurate to allow detections.
Simulations of ever-increasing realism will play a crucial role in validating these models and algorithms.


\subsection{Other Key Issues for Future Investigation}
\label{sec:conclusions:issues}


Our second most prominent residual systematic is that of night-to-night gain variations that, when combined with non-uniform flagging patterns, result in spectral discontinuities in the average over nights. 
Inconsistency over nights can be measured as a ratio of the variance over nights (for a particular unique baseline group) to the expected variance derived from autocorrelations (c.f. \zcref{eq:visibility-noise}), a metric directly related to the mean $Z^2$-score over the nights \citep{Murray_hera130}.
We generally measure such an ``excess variance'' of $\sim$2, where no excess would yield unity. 
This inconsistency is generally spectrally smooth, evidenced by the fact that the high-delay spectra are consistent with the predicted noise, and so is reasonably associated with errors in the low-delay modes of the gain solutions that fluctuate night-to-night. While such fluctuations are not intrinsically harmful, as long as they remain in foreground-dominated delays, they can be leaked to higher-delay modes when averaging with non-uniform weights. 
We see evidence for this when looking at the more highly-flagged spectral windows that were omitted from this analysis. In these bands, the high-delay spectra consistently have a visibly different distribution than predicted based on the autocorrelations. The correlation of this effect with the higher level of flagging in those bands suggests that the combination of night-to-night inconsistencies and imperfect inpainting is its root cause. We are investigating more sophisticated inpainting techniques to address this in the future.

Another potential issue that is closely related to the inpainting issues we have just described is that while we take great care to maintain spectral smoothness by inpainting over frequency with smooth models, we do not take the same care to maintain temporal smoothness. This is, admittedly, a secondary concern, since spectral structure is the primary distinguishable characteristic of the 21\,cm signal. However, we utilize temporal structures to enable systematics mitigation through the use of fringe-rate filtering, which ultimately feeds back into spectral structure. Future more sophisticated inpainting techniques are likely to improve this issue as well.

While our DPSS-filtering-based technique for identifying low-level RFI in redundantly-averaged cross-correlations is a marked improvement over previous techniques, there are at least two possible avenues for improvement. First, some moderately-broadband RFI (like from TV stations which broadcast in 8\,MHz allocations) may be degenerate with the filter, making it harder to find low-level RFI. Second, the technique does not incorporate any spatial information about the source of the emitter. Instead of an incoherent average over redundant baseline groups, one could use prior information about the locations of known transmitters to coherently combine visibilities for maximum SNR. This can be done both for fixed transmitters like radio and TV stations, as well as for ones that move in predictable ways, like ORBCOMM \citep{Neben2016} or Starlink \citep{Vruno2023,Bassa2024} satellites. 

Another systematic effect that requires further consideration is the leakage of a small number of highly-polarized, high-rotation measure (RM) point sources into our pseudo-Stokes I visibilities. To estimate the unpolarized sky signal, we take a sum of the visibility response from the two orthogonal linear feed polarizations. However, due to asymmetries in the polarized beam responses, this operation introduces leakage from Stokes Q into the pseudo-Stokes I visibilities. In the presence of bright polarized sources with large RMs, this leakage imprints spectral structure that varies approximately as $\exp\left(2 i \rm RM \lambda^2\right)$, contaminating high $k_\parallel$ modes and potentially biasing power spectrum estimates \citep{Moore2017}. Fortunately, this contamination is temporally localized to epochs when such sources enter the primary beam, enabling partial mitigation through careful field selection. However, this imposes constraints on the set of fields that are viable for power spectrum estimation. Catalogs such as those produced by the NRAO VLA Sky Survey (NVSS; \citealt{Taylor2009}) and the POlarised GLEAM Survey (POGS; \citealt{Riseley2020}) have demonstrated that, although most sources are weakly polarized, a non-negligible population exhibits polarization fractions and RMs sufficient to pose a risk for spectral leakage. These catalogs support identification and masking efforts, but complete mitigation will likely require forward modeling of polarized emission, incorporating both the polarized beam response and RM synthesis.
For this work, we have simply avoided observations where known, bright, high-RM pulsars are in the field-of-view (c.f. \zcref{fig:lstcov}). 
A detailed treatment of this effect is left to future work.

\subsection{Future Outlook}

This analysis was aimed at evaluating the updated analysis pipeline as applied to HERA Phase II data. 
As such, we considered only a small subset of the data taken with Phase II, starting in 2022 and still ongoing. 
As described in \zcref{tab:obschar}, the full 2022--2023 observing season alone contains an order of magnitude more data than presented here, and as of July 2025 we have completed an additional two full seasons of observing with even more active antennas.
While our lowest $k$-modes are currently systematics-limited, we expect that the sensitivity of three full years of data (with similar levels of flagging as those encountered in this dataset\footnote{While other instruments have reported an increase in RFI from satellite constellations such as Starlink, it is uncertain whether these will make a significant impact on our future data, as we do not currently localize our sources of RFI.}) at $k\sim0.8$\,$h{\rm Mpc}^{-1}$---which is currently noise-limited in all bands---will be at least 700 times more sensitive (i.e. $\Delta^2_{\rm UL} \approx 2.5\,{\rm mK}^2$ at $z=7$): sufficient to make detections for realistic cosmological scenarios.
Thus, while further mitigation of mutual coupling remains a top priority, it is not necessarily required for an eventual detection, given the enormous raw sensitivity of the HERA array.

In summary, in this work we have demonstrated that 
HERA Phase II is already able to achieve noise-limited limits over multiple $k$-bins (albeit at higher $k$ than in Phase I) and a broad range in redshift. 
With roughly an order of magnitude more data already taken and new techniques for systematics mitigation soon to be validated and implemented, we expect continued progress in the near future towards a first detection of the highly redshifted $21\,\textrm{cm}$ power spectrum.

\begin{acknowledgments}
This material is based upon work supported by the
National Science Foundation under grants \#1636646,
\#1836019, \#1352519 and \#1407804 as well as institutional support from the HERA collaboration partners. 
This research is funded in part by the Gordon and Betty Moore Foundation through Grant GBMF5212 to the Massachusetts Institute of
Technology. 
HERA is hosted by the South African Radio Astronomy Observatory, which is a facility of the National Research Foundation, an agency of the Department of Science and Innovation.

This work used Bridges-2 at Pittsburgh Supercomputing Centre through allocation PHY201142 from the Advanced Cyberinfrastructure Coordination Ecosystem: Services \& Support (ACCESS) program, which is supported by U.S. National Science Foundation grants \#2138259, \#2138286, \#2138307, \#2137603, and \#2138296.
We acknowledge the use of the Ilifu cloud computing facility (www.ilifu.ac.za) and the support from the Inter-University Institute for Data Intensive Astronomy (IDIA; https://www.idia.ac.za).

S. G. Murray has received funding from the European Union’s Horizon 2020 research and innovation programme under the Marie Skłodowska-Curie grant agreement No 101067043.
This result is part of a project that has received funding from the European Research Council (ERC) under the European Union's Horizon 2020 research and innovation programme (Grant agreement No. 948764; PB, JB, MJW).
BBB acknowledges the funding received from FAPESP under process 2024/12902‑3.
NSK acknowledges support from NASA through the NASA Hubble Fellowship grant \# HST-HF2-51533.001-A awarded by the Space Telescope Science Institute, which is operated by the Association of Universities for Research in Astronomy, Incorporated, under NASA contract NAS5-26555.
J.M. was supported by an appointment to the NASA Postdoctoral Program at the Jet Propulsion Laboratory/California Institute of Technology, administered by Oak Ridge Associated Universities under contract with NASA.
A.M. acknowledges support from the
Italian Ministry of Universities and Research (MUR) through the PRIN project
``Optimal inference from radio images of the epoch of reionization'', and the
PNRR project ``Centro Nazionale di Ricerca in High Performance Computing, Big Data e Quantum Computing''.
A.C.L. acknowledges support from the Natural Sciences and Engineering Research Council of Canada through their Discovery Grants Program, as well as the William Dawson Scholar program at McGill University.
\end{acknowledgments}

%

\vspace{5mm}
\facility{HERA}


\software{
\texttt{numpy} \citep{Harris2020},
\texttt{scipy} \citep{Virtanen2020},
\texttt{matplotlib} \citep{Hunter2007},
\texttt{astropy} \citep{AstropyCollaboration2013,AstropyCollaboration2018},  
\texttt{jupyter} \citep{Kluyver2016jupyter},
\texttt{pyuvdata} \citep{Hazelton2017,Keating2025},
\texttt{fftvis} \citep{Cox2025}.
\texttt{hera\_opm} \citep{LaPlante2021},
\texttt{hera\_cal} (\url{https://github.com/hera-team/hera_cal},
\texttt{hera\_pspec} (\url{https://github.com/hera-team/hera_pspec}
          }



\appendix

\section{Antenna Metrics}
\label{app:antenna-metrics}

We flag entire antennas with the following specific characteristics, irrespective of channel-based flags:
\begin{itemize}
\itemsep0em
    \item \textbf{Dead}: Antennas with visibilities whose value is zero more than half of the time are considered ``dead" and flagged.
    \item \textbf{Low Correlation}: Antennas with low correlation coefficients (from 0--0.2) are flagged \citep{Storer2022}. We found that such low correlations are often indicative of clock distribution issues that affect entire nodes \citep{Berkhout2024}.
    \item \textbf{Cross-Polarized}:
    Antennas whose correlation coefficient with other antennas is larger when correlating the other antenna is putatively of the opposite polarization are considered cross-polarized \cite{Storer2022} and flagged (and reported for in-field maintenance).
    \item \textbf{Packet Loss}: While our integration time for snapshot observations is $\sim$10\,s, we also keep ``diff" files containing even-odd differences on a much finer timescale. These can be rearranged into even and odd samples. An excess of zeros in either the evens or the odds is interpreted as a a sign of packet loss. We flag antennas with more than 8 zeros in either the odds or evens\footnote{Note that while packet-loss occurs on a per-baseline basis, we find that it generally affects a few antennas disproportionately because of the ordering of the visibility information. While maintaining per-antenna flags here does result in over-flagging, the utility of carrying forward per-antenna instead of per-baseline flags at this stage is considered worth the small excess data loss.}. 
    \item \textbf{Anomalous Autocorrelation Power}: Antennas with a median amplitude over all channels for each auto-correlation more than 60 times the integration count (i.e. the integration time by the channel width, $\tau_{\rm int} \Delta \nu$), are flagged. These are at risk of a non-linear ADC response. Likewise, antennas with median amplitude below 1 times the integration count are flagged for low sky-response.
    \item \textbf{Anomalous Autocorrelation Slope}: Antennas whose relative absolute slopes (i.e. the absolute value of the ratio of the linear coefficient to the constant coefficient in a least-squares fit to the median-filtered spectrum) are more than 0.6 are flagged. This is generally a sign of low or unusual sky-response.
\end{itemize}

\section{Iterative Algorithm for Antenna and RFI flags}
\label{app:rfi-ant-iterative-flags}
In this appendix we describe in more detail the algorithm used in the per-integration flagging and calibration (c.f. \zcref{sec:methods:per-file}) to iteratively refine the per-antenna and per-channel flags in tandem.

We first determine an initial RFI mask (per-channel flags) as follows.
We subtract neighboring channels of the autocorrelations for each antenna (resulting in spectra close to white-noise, if no RFI were present), and then we identify $20\sigma$ outliers, where $\sigma$ is estimated via \zcref{eq:visibility-noise} (using each auto-correlation to estimate its own noise).
Crucially, the final product is only channel-dependent, not antenna-dependent: channels that are flagged for more than 50\% of antennas are then flagged for \textit{all} antennas, and otherwise the channels are left \textit{unflagged}.

Following this, a linear DPSS model is fit to each autocorrelation, respecting the flags found from the first step. Here, we use the parameters $\tau_c = 0, \Delta \tau = 300\,{\rm ns}, \lambda_{\rm min}=10^{-9}$. After subtraction of this model, the resulting residuals should be white-noise if no RFI were present, and we perform a more aggressive cut at $4\sigma$ (with $\sigma$ computed in the same way as above). 
Again, only channels with a flag fraction of more than 50\% across antennas are ultimately flagged. 

This rough RFI mask is used to identify cross-correlation engines (X-engines) that are systematically aberrant. Each X-engine is responsible for 96 channels. For each set of 96 channels, we take finely time-differenced data, which should be noise-like, and compute a quasi-$Z$-score based on its amplitude: $(|V_{\rm diff}| - \sigma \sqrt(\pi/4))/(\sigma \sqrt{(4 - \pi)/4})$, where $\sigma$ is the predicted standard deviation of the thermal noise of the observation, based on the auto-correlations. 
This quantity is expected to have mean zero and variance unity (though it is not Gaussian distributed),
and we compute the mean of its absolute value within each set of 96 channels controlled by a particular X-engine. 
This mean, $\overline{|Z|}$ should be half-normally distributed, with mean $\sqrt{2}/\pi$ and variance $(1 - 2/\pi)/N_{\rm samples}$, where $N_{\rm samples}\sim96$ is the number of channels in the 96-channel chunk that are \textit{not} flagged by the previously-calculated RFI mask. We re-scale $\overline{|Z|}$ to have mean zero and variance unity before counting the number of times a particular antenna has an X-engine with a rescaled mean $Z$-score above 10. 
Then, starting with the antennas that have the highest count of bad X-engines, we iteratively remove them, until we are left with no baselines with bad X-engines.

Our final two antenna-based cuts are established in conjunction with our final spectral mask---which refines our initial rough RFI mask---in a single iterative loop. 
This loop is as follows, and runs for \textit{at least} three iterations, and until no new channels are flagged and no new antennas are flagged. 
Each operation is perform only on auto-correlations. Initially, the shape-$N_\nu$ vector  of RFI flags $\xi_\nu \in (0,1)$ are those obtained from the `rough' RFI mask described above. 
\begin{enumerate}
    \itemsep0em
    \item Calculate $Z_{ii} = (P_{ii} - M)_{ii}/ ( \sqrt{2} M_{ii} / \sqrt{\tau \Delta \nu})$ for each auto-correlation power $P_{ii}$, where $M_{ii}$ is a DPSS model of the power over frequency, with $\tau_c = 0, \Delta \tau = 300\,{\rm ns}, \lambda_{\rm min}=10^{-9}$.
    \item Compute the RMS, $\sigma_{Z_{ii}} = \sqrt{\sum_\nu \xi_\nu Z^2_{ii}/N_{\xi}}$ for each auto-power, where $N_{\xi}$ is the number of unflagged channels, $\sum_\nu \xi_\nu$ (the same for all antennas). 
    \item Choose which auto-correlations to use for estimating the RFI, starting from a candidate pool of all those that have not yet been flagged (either by previous checks, or previous iterations of this loop), then:
    \begin{itemize}
        \item If this is the \textit{first} iteration: use the half of the candidate pool that has the lowest RMS.
        \item If this is the \textit{second} iteration: use all candidate antennas whose RFI classification is `good' (see below) unless this is less than half of the candidate pool, in which case use the half of the candidate pool that has the lowest RMS.
        \item If this is the \textit{third} or higher iteration: use all candidate antennas whose RFI classification is not `bad' (see below).
    \end{itemize}
    \item New channel-based flags, $\xi_\nu$, are computed exactly in the same way as the `rough' RFI flags above, except that instead of \textit{each} auto-correlation being tested (and then the final flags being combined via thresholding over all antennas), only the mean auto-correlation over the antennas in the set selected above are used.
    \item These flags are used to identify outliers in \textit{spectral shape}. Here, we rescale each RFI-flagged auto-correlation to be mean unity, and then find the mean rescaled auto-correlation across all antennas (per polarization) that have not thus far been flagged. We flag antennas whose RMS difference with respect to the mean is greater than 20\%.
\end{enumerate}

\section{Temporally-Smoothed Metric Flagging}
\label{app:smoothed-metric-flagging}
In \zcref{sec:methods:per-night:per-ant} we outlined an algorithm for temporally smoothing quality metrics, and re-flagging based on these smoothed metrics. Here we describe the algorithm in more detail.
The algorithm is as follows:
\begin{itemize}
    \item Begin with a per-antenna time-series of a particular metric, $m_t$, and associated flags $\xi_t$ when $m_t$ is outside some threshold $(M_{\rm low}, M_{\rm high})$.
    \item Convolve the metric series with a Gaussian of width $\sigma = 60$ integrations (about 10 minutes). 
    \item Iteratively:
    \begin{itemize}
        \item Partition the time-series into contiguous regions that are either completely flagged or completely unflagged, and further categorize the flagged regions as those in which the convolved metric either surpasses (`persistent') the threshold at least once or not (`non-persistent'). 
        \item Re-partition the time-series such that both unflagged and non-persistent flagged regions from the previous partitioning are merged together (type A), and persistent flagged regions remain separate (type B). 
        \item For each region of type A, if the smoothed metric is outside the threshold \textit{everywhere} in the region, flag the entire region. This is meant to flag small unflagged regions embedded in larger flagged regions, as well as regions with a high time-to-time variability of flags.
        \item Finally, if there is any flagged region of size greater than 60 integrations, flag all times before or after the gap, whichever is smaller. 
        \item If there are new flags in this iteration, keep iterating.
    \end{itemize}
\end{itemize}

This procedure is separately performed for each of the autocorrelation power, shape and slope metrics, as well as the RFI RMS (cf. \zcref{sec:methods:per-file:flagging}) and the $\chi^2$ computed from redundant calibration.
For the latter, the initial flags are set to \textit{ignore} times flagged for any other reason (whereas other metrics are considered in isolation). 

After separately performing this procedure for each metric, and obtaining a final set of unified flags (where any of the metrics exceeds the threshold), we once more flag regions either before or after flag gaps of more than 60 integrations (whichever is smaller).
Finally, antennas that are flagged for more than 50\% of the integrations on a given night are flagged entirely.

\section{Iterative Algorithm for Deeper RFI Excision}
\label{app:rfi-round-2}
In \zcref{sec:methods:rfi-round-2} we describe our method for constructing an RFI mask based on baseline-averaged per-night $Z$-scores derived from a high-pass delay filter.
Here we give the details of the precise steps and thresholds of this procedure.
We flag the following (in this order): (1) Iteratively flag the worst offending integrations and channels, in exactly the same way as the iterative procedure outlined in \zcref{sec:methods:per-night:rfi}, but with a threshold of 1.0 instead of 1.5, and using the median over each axis instead of the mean; (2) any channels that are flagged for more than 25\% of integrations; (3) any integrations that are flagged for more than 10\% of channels; (4) any particular integration-channel with $\bar{Z}>4$; (5) any integration-channel neighboring a flagged channel with $\bar{Z}>2$; (6) repeat 1 using the mean instead of median; (7) repeat 2 and 3.

\section{Justification of Residual Bin-to-bin Correlations}
\label{app:likelihood-correlations}
We claimed in \zcref{sec:interpretation:likelihood} that allowing 10\% correlations between neighboring $k$-bins does not significantly affect the inferred posterior.
Here we detail our justification for this claim.

Formally, our prior on the systematic is improper. 
To carry out this investigation, we instead assumed it is uniformly distributed from 0 to 100,000 mK$^2$, far in excess of the noise levels used in this numerical experiment (and therefore well-approximating the improper prior assumed in the likelihood). 
We then simulated samples from the systematic prior and different Gaussian noise distributions for two fictitious ``neighboring'' power spectrum bins with varying correlation coefficient: 0, 0.1, and 0.99, and standard deviation $1000$ mK$^2$ (chosen because they are simple, round numbers at approximately the sensitivity of the $z<8$ limits in \zcref{tab:table-of-limits}). 
In each case, we used the large number of simulated draws to form a kernel density estimate of the marginal density of these contributions to the pair of power spectrum bins. 
Then, for several fictitious measurements and broad uniform priors from 10 to 1000 mk$^2$ on the signal power spectrum in the two bins, we normalized the density estimates so that they were properly normalized posterior probability distributions. 

In summary, we derived marginal posterior distributions for the `21\,cm signal' (parameterized simply as the power spectrum amplitude in two neighboring $k$-bins) for mock data generated with different levels of true correlation between the neighboring $k$-bins, but for which the likelihood model assumed no correlation.
We obtained such posteriors in a range of scenarios: systematics-dominated, noise-dominated and signal-dominated.

We found that, regardless of whether the measurements were noise-limited (e.g. 1000 mK$^2$), or if one (or both) were strongly systematically contaminated (e.g. 10000 mK$^2$), the three resulting posteriors (for the different true correlation levels) were all nearly identical---even for the case where the noise correlation between the two bins was 0.99.
Variations in posterior probability density were at most $\sim2\%$, where the strongest deviations occurred in the noise-limited case with correlation coefficient equal to 0.99. For a correlation coefficient of 0.1, deviations were sub-percent in all cases. This means that posterior inferences would be nearly the same regardless of whether we fully modeled the noise correlations in our likelihood. 
We believe that this occurs because of the strong uncertainty (and lack of correlation) that is reflected by our model for systematic effects, which is not particularly realistic. We suspect that more realistic treatments of the systematic effect (particularly if they do not range by many orders of magnitude and are not totally uncorrelated) may create very different posteriors in the presence of correlated noise, and would therefore demand more careful noise modeling in the astrophysical inference step.

\pagebreak

\section{Table of Power Spectrum Upper Limits}
\label{sec:limit-table}

The following table contains the limits presented in \zcref{fig:upper-limits}.

\begin{longtable}{c|rrrr|rrrr}
\toprule
& $k$ & $\Delta^{2}(k)$ & $1\sigma$ & $\Delta^{2}_{UL}$  & $k$ & $\Delta^{2}(k)$ & $1\sigma$ & $\Delta^{2}_{UL}$  \\
& ($h/$Mpc) & (mK$^2$) & (mK$^2$) & (mK$^2$) & ($h/$Mpc) & (mK$^2$) & (mK$^2$) & (mK$^2$) \\
\midrule
\endfirsthead
\toprule
& $k$ & $\Delta^{2}(k)$ & $1\sigma$ & $\Delta^{2}_{UL}$  & $k$ & $\Delta^{2}(k)$ & $1\sigma$ & $\Delta^{2}_{UL}$  \\
& ($h/$Mpc) & (mK$^2$) & (mK$^2$) & (mK$^2$) & ($h/$Mpc) & (mK$^2$) & (mK$^2$) & (mK$^2$) \\
\midrule
\endhead
\midrule
\multicolumn{9}{r}{Continued on next page} \\
\midrule
\endfoot
\bottomrule
\endlastfoot
\multirow{5}{*}{\rotatebox[origin=c]{90}{$z=24.37$}}& 0.30 & 115,193,441 & 2,024,176 & 119,241,794& 0.85 & 26,288,139 & 14,288,749 & 54,865,638\\
& 0.41 & 14,545,056 & 1,790,638 & 18,126,332& 0.96 & 31,125,314 & 20,641,861 & 72,409,037\\
& 0.52 & 4,975,043 & 3,255,384 & \textbf{11,485,811}& 1.07 & 101,151,265 & 28,620,389 & 158,392,044\\
& 0.63 & 10,890,061 & 5,868,207 & 22,626,477& 1.18 & -22,797,735 & 36,778,435 & 73,556,871\\
& 0.74 & 19,442,615 & 9,617,651 & 38,677,919& 1.29 & 9,617,674 & 48,273,574 & 106,164,823\\
\midrule
\multirow{5}{*}{\rotatebox[origin=c]{90}{$z=19.80$}}& 0.25 & 76,689,186 & 1,543,420 & 79,776,026& 0.98 & 148,711 & 14,129,685 & 28,408,081\\
& 0.40 & 13,100,453 & 1,126,258 & 15,352,970& 1.12 & 37,439,608 & 21,798,781 & 81,037,171\\
& 0.54 & 6,144,314 & 2,491,275 & \textbf{11,126,866}& 1.26 & 6,369,229 & 31,762,166 & 69,893,562\\
& 0.69 & 9,482,752 & 5,083,277 & 19,649,308& 1.41 & 82,269,409 & 43,227,317 & 168,724,043\\
& 0.83 & 26,944,303 & 9,078,540 & 45,101,385& & & & \\
\midrule
\multirow{5}{*}{\rotatebox[origin=c]{90}{$z=16.78$}}& 0.25 & 18,392,322 & 429,193 & 19,250,710& 0.98 & -104,028 & 2,557,758 & 5,115,517\\
& 0.40 & 1,408,317 & 190,982 & 1,790,281& 1.13 & -628,410 & 3,880,214 & 7,760,429\\
& 0.55 & 252,112 & 439,236 & \textbf{1,130,584}& 1.27 & 14,663,521 & 5,668,569 & 26,000,660\\
& 0.69 & -1,146,024 & 875,715 & 1,751,430& 1.42 & -1,565,451 & 7,639,973 & 15,279,947\\
& 0.84 & 1,759,759 & 1,588,757 & 4,937,274& 1.57 & 13,252,746 & 10,600,422 & 34,453,590\\
\midrule
\multirow{3}{*}{\rotatebox[origin=c]{90}{$z=10.76$}}& 0.48 & 164,176 & 4,589 & 173,355& 1.31 & 28,485 & 61,975 & 152,436\\
& 0.76 & 21,703 & 12,291 & \textbf{46,285}& 1.58 & 128,749 & 109,418 & 347,587\\
& 1.03 & 21,326 & 30,541 & 82,409& 1.86 & 129,559 & 180,625 & 490,811\\
\midrule
\multirow{5}{*}{\rotatebox[origin=c]{90}{$z=9.87$}}& 0.33 & 495,706 & 9,428 & 514,562& 1.26 & 296 & 39,414 & 79,125\\
& 0.51 & 24,337 & 2,995 & \textbf{30,328}& 1.44 & -68,169 & 58,410 & 116,821\\
& 0.70 & 21,512 & 6,926 & 35,364& 1.63 & 176,688 & 87,697 & 352,083\\
& 0.89 & 3,656 & 13,695 & 31,048& 1.82 & 31,596 & 119,782 & 271,160\\
& 1.07 & -7,101 & 24,293 & 48,586& 2.00 & 104,834 & 159,778 & 424,392\\
\midrule
\multirow{5}{*}{\rotatebox[origin=c]{90}{$z=7.63$}}& 0.37 & 16,563 & 811 & 18,187& 1.45 & 4,208 & 9,880 & 23,969\\
& 0.59 & 1,734 & 689 & \textbf{3,113}& 1.66 & 17,483 & 14,816 & 47,116\\
& 0.80 & 1,421 & 1,677 & 4,776& 1.87 & 34,627 & 21,188 & 77,004\\
& 1.02 & 11,316 & 3,487 & 18,291& 2.09 & -13,916 & 28,712 & 57,424\\
& 1.23 & 6,891 & 5,942 & 18,777& & & & \\
\midrule
\multirow{4}{*}{\rotatebox[origin=c]{90}{$z=7.05$}}& 0.44 & 2,768 & 372 & 3,513& 1.46 & 2,703 & 8,155 & 19,014\\
& 0.70 & -494 & 887 & \textbf{1,775}& 1.71 & 20,386 & 13,505 & 47,398\\
& 0.95 & 921 & 2,296 & 5,514& 1.97 & 12,768 & 20,132 & 53,032\\
& 1.21 & 490 & 4,604 & 9,699& 2.22 & -3,748 & 28,476 & 56,952\\
\midrule
\multirow{4}{*}{\rotatebox[origin=c]{90}{$z=5.56$}}& 0.55 & 1,695 & 181 & 2,058& 1.81 & -2,295 & 4,434 & 8,869\\
& 0.87 & 435 & 496 & \textbf{1,428}& 2.13 & 3,648 & 7,218 & 18,086\\
& 1.18 & 1,894 & 1,269 & 4,433& 2.44 & -19,753 & 10,793 & 21,586\\
& 1.50 & 322 & 2,536 & 5,394& & & & 
\label{tab:table-of-limits}
\end{longtable}

\bibliography{h6cidr2library}{}

@article{Acharya2024,
  title = {Revised {{LOFAR}} Upper Limits on the 21-Cm Signal Power Spectrum at z {$\approx$} 9.1 Using Machine Learning and Gaussian Process Regression},
  author = {Acharya, Anshuman and Mertens, Florent and Ciardi, Benedetta and Ghara, Raghunath and Koopmans, L{\'e}on V. E. and Zaroubi, Saleem},
  year = 2024,
  month = oct,
  journal = {Monthly Notices of the Royal Astronomical Society},
  volume = {534},
  pages = {L30-L34},
  publisher = {OUP},
  issn = {0035-8711},
  url = {https://ui.adsabs.harvard.edu/abs/2024MNRAS.534L..30A},
  urldate = {2025-04-23}
}

@article{Aguirre2022,
  title = {Validation of the {{HERA Phase I Epoch}} of {{Reionization}} 21 Cm {{Power Spectrum Software Pipeline}}},
  author = {Aguirre, James E. and Murray, Steven G. and Pascua, Robert and Martinot, Zachary E. and Burba, Jacob and Dillon, Joshua S. and Jacobs, Daniel C. and Kern, Nicholas S. and Kittiwisit, Piyanat and Kolopanis, Matthew and Lanman, Adam and Liu, Adrian and Whitler, Lily and Abdurashidova, Zara and Alexander, Paul and Ali, Zaki S. and Balfour, Yanga and Beardsley, Adam P. and Bernardi, Gianni and Billings, Tashalee S. and Bowman, Judd D. and Bradley, Richard F. and Bull, Philip and Carey, Steve and Carilli, Chris L. and Cheng, Carina and DeBoer, David R. and Dexter, Matt and Acedo, Eloy de Lera and Ely, John and {Ewall-Wice}, Aaron and Fagnoni, Nicolas and Fritz, Randall and Furlanetto, Steven R. and {Gale-Sides}, Kingsley and Glendenning, Brian and Gorthi, Deepthi and Greig, Bradley and Grobbelaar, Jasper and Halday, Ziyaad and Hazelton, Bryna J. and Hewitt, Jacqueline N. and Hickish, Jack and Julius, Austin and Kerrigan, Joshua and Kohn, Saul A. and Plante, Paul La and Lekalake, Telalo and Lewis, David and MacMahon, David and Malan, Lourence and Malgas, Cresshim and Maree, Matthys and Matsetela, Eunice and Mesinger, Andrei and Molewa, Mathakane and Morales, Miguel F. and Mosiane, Tshegofalang and Neben, Abraham R. and Nikolic, Bojan and Parsons, Aaron R. and Patra, Nipanjana and Pieterse, Samantha and Pober, Jonathan C. and {Razavi-Ghods}, Nima and Ringuette, Jon and Robnett, James and Rosie, Kathryn and Santos, Mario G. and Sims, Peter and Singh, Saurabh and Smith, Craig and Syce, Angelo and Thyagarajan, Nithyanandan and Williams, Peter K. G. and Zheng, Haoxuan},
  year = 2022,
  month = jan,
  journal = {The Astrophysical Journal},
  volume = {924},
  number = {2},
  pages = {85},
  publisher = {American Astronomical Society},
  issn = {0004-637X},
  url = {https://doi.org/10.3847/1538-4357/ac32cd},
  urldate = {2022-01-14},
  langid = {english}
}

@article{Ali2015,
  title = {{{PAPER-64 Constraints}} on {{Reionization}}: {{The}} 21 Cm {{Power Spectrum}} at z = 8.4},
  shorttitle = {{{PAPER-64 Constraints}} on {{Reionization}}},
  author = {Ali, Zaki S. and Parsons, Aaron R. and Zheng, Haoxuan and Pober, Jonathan C. and Liu, Adrian and Aguirre, James E. and Bradley, Richard F. and Bernardi, Gianni and Carilli, Chris L. and Cheng, Carina and DeBoer, David R. and Dexter, Matthew R. and Grobbelaar, Jasper and Horrell, Jasper and Jacobs, Daniel C. and Klima, Pat and MacMahon, David H. E. and Maree, Matthys and Moore, David F. and Razavi, Nima and Stefan, Irina I. and Walbrugh, William P. and Walker, Andre},
  year = 2015,
  month = aug,
  journal = {The Astrophysical Journal},
  volume = {809},
  pages = {61},
  issn = {0004-637X},
  url = {http://adsabs.harvard.edu/abs/2015ApJ...809...61A},
  urldate = {2017-07-14}
}

@article{Amiri2023,
  title = {Detection of {{Cosmological}} 21 Cm {{Emission}} with the {{Canadian Hydrogen Intensity Mapping Experiment}}},
  author = {Amiri, Mandana and Bandura, Kevin and Chen, Tianyue and Deng, Meiling and Dobbs, Matt and Fandino, Mateus and Foreman, Simon and Halpern, Mark and Hill, Alex S. and Hinshaw, Gary and H{\"o}fer, Carolin and Kania, Joseph and Landecker, T. L. and MacEachern, Joshua and Masui, Kiyoshi and {Mena-Parra}, Juan and Milutinovic, Nikola and Mirhosseini, Arash and Newburgh, Laura and Ordog, Anna and Pen, Ue-Li and {Pinsonneault-Marotte}, Tristan and Polzin, Ava and Reda, Alex and Renard, Andre and Shaw, J. Richard and Siegel, Seth R. and Singh, Saurabh and Vanderlinde, Keith and Wang, Haochen and Wiebe, Donald V. and Wulf, Dallas and {CHIME Collaboration}},
  year = 2023,
  month = apr,
  journal = {The Astrophysical Journal},
  volume = {947},
  pages = {16},
  publisher = {IOP},
  issn = {0004-637X},
  url = {https://ui.adsabs.harvard.edu/abs/2023ApJ...947...16A},
  urldate = {2025-04-16}
}

@article{AstropyCollaboration2013,
  title = {Astropy: {{A}} Community {{Python}} Package for Astronomy},
  shorttitle = {Astropy},
  author = {{Astropy Collaboration} and Robitaille, Thomas P. and Tollerud, Erik J. and Greenfield, Perry and Droettboom, Michael and Bray, Erik and Aldcroft, Tom and Davis, Matt and Ginsburg, Adam and {Price-Whelan}, Adrian M. and Kerzendorf, Wolfgang E. and Conley, Alexander and Crighton, Neil and Barbary, Kyle and Muna, Demitri and Ferguson, Henry and Grollier, Fr{\'e}d{\'e}ric and Parikh, Madhura M. and Nair, Prasanth H. and Unther, Hans M. and Deil, Christoph and Woillez, Julien and Conseil, Simon and Kramer, Roban and Turner, James E. H. and Singer, Leo and Fox, Ryan and Weaver, Benjamin A. and Zabalza, Victor and Edwards, Zachary I. and Azalee Bostroem, K. and Burke, D. J. and Casey, Andrew R. and Crawford, Steven M. and Dencheva, Nadia and Ely, Justin and Jenness, Tim and Labrie, Kathleen and Lim, Pey Lian and Pierfederici, Francesco and Pontzen, Andrew and Ptak, Andy and Refsdal, Brian and Servillat, Mathieu and Streicher, Ole},
  year = 2013,
  month = oct,
  journal = {Astronomy and Astrophysics},
  volume = {558},
  pages = {A33},
  issn = {0004-6361},
  url = {https://ui.adsabs.harvard.edu/abs/2013A&A...558A..33A},
  urldate = {2025-04-18}
}

@article{AstropyCollaboration2018,
  title = {The {{Astropy Project}}: {{Building}} an {{Open-science Project}} and {{Status}} of the v2.0 {{Core Package}}},
  shorttitle = {The {{Astropy Project}}},
  author = {{Astropy Collaboration} and {Price-Whelan}, A. M. and Sip{\H o}cz, B. M. and G{\"u}nther, H. M. and Lim, P. L. and Crawford, S. M. and Conseil, S. and Shupe, D. L. and Craig, M. W. and Dencheva, N. and Ginsburg, A. and VanderPlas, J. T. and Bradley, L. D. and {P{\'e}rez-Su{\'a}rez}, D. and {de Val-Borro}, M. and Aldcroft, T. L. and Cruz, K. L. and Robitaille, T. P. and Tollerud, E. J. and Ardelean, C. and Babej, T. and Bach, Y. P. and Bachetti, M. and Bakanov, A. V. and Bamford, S. P. and Barentsen, G. and Barmby, P. and Baumbach, A. and Berry, K. L. and Biscani, F. and Boquien, M. and Bostroem, K. A. and Bouma, L. G. and Brammer, G. B. and Bray, E. M. and Breytenbach, H. and Buddelmeijer, H. and Burke, D. J. and Calderone, G. and Cano Rodr{\'i}guez, J. L. and Cara, M. and Cardoso, J. V. M. and Cheedella, S. and Copin, Y. and Corrales, L. and Crichton, D. and D'Avella, D. and Deil, C. and Depagne, {\'E}. and Dietrich, J. P. and Donath, A. and Droettboom, M. and Earl, N. and Erben, T. and Fabbro, S. and Ferreira, L. A. and Finethy, T. and Fox, R. T. and Garrison, L. H. and Gibbons, S. L. J. and Goldstein, D. A. and Gommers, R. and Greco, J. P. and Greenfield, P. and Groener, A. M. and Grollier, F. and Hagen, A. and Hirst, P. and Homeier, D. and Horton, A. J. and Hosseinzadeh, G. and Hu, L. and Hunkeler, J. S. and Ivezi{\'c}, {\v Z}. and Jain, A. and Jenness, T. and Kanarek, G. and Kendrew, S. and Kern, N. S. and Kerzendorf, W. E. and Khvalko, A. and King, J. and Kirkby, D. and Kulkarni, A. M. and Kumar, A. and Lee, A. and Lenz, D. and Littlefair, S. P. and Ma, Z. and Macleod, D. M. and Mastropietro, M. and McCully, C. and Montagnac, S. and Morris, B. M. and Mueller, M. and Mumford, S. J. and Muna, D. and Murphy, N. A. and Nelson, S. and Nguyen, G. H. and Ninan, J. P. and N{\"o}the, M. and Ogaz, S. and Oh, S. and Parejko, J. K. and Parley, N. and Pascual, S. and Patil, R. and Patil, A. A. and Plunkett, A. L. and Prochaska, J. X. and Rastogi, T. and Reddy Janga, V. and Sabater, J. and Sakurikar, P. and Seifert, M. and Sherbert, L. E. and {Sherwood-Taylor}, H. and Shih, A. Y. and Sick, J. and Silbiger, M. T. and Singanamalla, S. and Singer, L. P. and Sladen, P. H. and Sooley, K. A. and Sornarajah, S. and Streicher, O. and Teuben, P. and Thomas, S. W. and Tremblay, G. R. and Turner, J. E. H. and Terr{\'o}n, V. and {van Kerkwijk}, M. H. and {de la Vega}, A. and Watkins, L. L. and Weaver, B. A. and Whitmore, J. B. and Woillez, J. and Zabalza, V. and {Astropy Contributors}},
  year = 2018,
  month = sep,
  journal = {The Astronomical Journal},
  volume = {156},
  pages = {123},
  url = {http://adsabs.harvard.edu/abs/2018AJ....156..123A},
  urldate = {2020-09-28}
}

@article{Barkana2018,
  title = {Strong Constraints on Light Dark Matter Interpretation of the {{EDGES}} Signal},
  author = {Barkana, Rennan and Outmezguine, Nadav Joseph and Redigol, Diego and Volansky, Tomer},
  year = 2018,
  month = nov,
  journal = {Physical Review D},
  volume = {98},
  pages = {103005},
  issn = {0556-2821},
  url = {http://adsabs.harvard.edu/abs/2018PhRvD..98j3005B},
  urldate = {2020-11-13}
}

@article{Barnett2019,
  title = {A {{Parallel Nonuniform Fast Fourier Transform Library Based}} on an "{{Exponential}} of {{Semicircle}}" {{Kernel}}},
  author = {Barnett, Alexander H. and Magland, Jeremy and {af Klinteberg}, Ludvig},
  year = 2019,
  month = jan,
  journal = {SIAM Journal on Scientific Computing},
  volume = {41},
  pages = {C479-C504},
  url = {https://ui.adsabs.harvard.edu/abs/2019SJSC...41C.479B},
  urldate = {2025-04-18}
}

@article{Barry2016,
  title = {Calibration Requirements for Detecting the 21 Cm Epoch of Reionization Power Spectrum and Implications for the {{SKA}}},
  author = {Barry, N. and Hazelton, B. and Sullivan, I. and Morales, M. F. and Pober, J. C.},
  year = 2016,
  month = sep,
  journal = {Monthly Notices of the Royal Astronomical Society},
  volume = {461},
  pages = {3135--3144},
  issn = {0035-8711},
  url = {http://adsabs.harvard.edu/abs/2016MNRAS.461.3135B},
  urldate = {2018-04-05}
}

@article{Barry2019,
  ids = {Barry2019b},
  title = {Improving the {{Epoch}} of {{Reionization Power Spectrum Results}} from {{Murchison Widefield Array Season}} 1 {{Observations}}},
  author = {Barry, N. and Wilensky, M. and Trott, C. M. and Pindor, B. and Beardsley, A. P. and Hazelton, B. J. and Sullivan, I. S. and Morales, M. F. and Pober, J. C. and Line, J. and Greig, B. and Byrne, R. and Lanman, A. and Li, W. and Jordan, C. H. and Joseph, R. C. and McKinley, B. and Rahimi, M. and Yoshiura, S. and Bowman, J. D. and Gaensler, B. M. and Hewitt, J. N. and Jacobs, D. C. and Mitchell, D. A. and Udaya Shankar, N. and Sethi, S. K. and Subrahmanyan, R. and Tingay, S. J. and Webster, R. L. and Wyithe, J. S. B.},
  year = 2019,
  month = oct,
  journal = {The Astrophysical Journal},
  volume = {884},
  eprint = {1909.00561},
  pages = {1},
  url = {http://adsabs.harvard.edu/abs/2019ApJ...884....1B},
  urldate = {2020-11-03},
  archiveprefix = {arXiv}
}

@misc{Bassa2024,
  title = {Bright Unintended Electromagnetic Radiation from Second-Generation {{Starlink}} Satellites},
  author = {Bassa, C. G. and Vruno, F. Di and Winkel, B. and Jozsa, G. I. G. and Brentjens, M. A. and Zhang, X.},
  year = 2024,
  month = sep,
  number = {arXiv:2409.11767},
  eprint = {2409.11767},
  primaryclass = {astro-ph},
  publisher = {arXiv},
  url = {http://arxiv.org/abs/2409.11767},
  urldate = {2025-05-20},
  archiveprefix = {arXiv}
}

@article{Beardsley2016,
  title = {First {{Season MWA EoR Power}} Spectrum {{Results}} at {{Redshift}} 7},
  author = {Beardsley, A. P. and Hazelton, B. J. and Sullivan, I. S. and Carroll, P. and Barry, N. and Rahimi, M. and Pindor, B. and Trott, C. M. and Line, J. and Jacobs, Daniel C. and Morales, M. F. and Pober, J. C. and Bernardi, G. and Bowman, Judd D. and Busch, M. P. and Briggs, F. and Cappallo, R. J. and Corey, B. E. and {de Oliveira-Costa}, A. and Dillon, Joshua S. and Emrich, D. and {Ewall-Wice}, A. and Feng, L. and Gaensler, B. M. and Goeke, R. and Greenhill, L. J. and Hewitt, J. N. and {Hurley-Walker}, N. and {Johnston-Hollitt}, M. and Kaplan, D. L. and Kasper, J. C. and Kim, H. S. and Kratzenberg, E. and Lenc, E. and Loeb, A. and Lonsdale, C. J. and Lynch, M. J. and McKinley, B. and McWhirter, S. R. and Mitchell, D. A. and Morgan, E. and Neben, A. R. and Thyagarajan, Nithyanandan and Oberoi, D. and Offringa, A. R. and Ord, S. M. and Paul, S. and Prabu, T. and Procopio, P. and Riding, J. and Rogers, A. E. E. and Roshi, A. and Udaya Shankar, N. and Sethi, Shiv K. and Srivani, K. S. and Subrahmanyan, R. and Tegmark, M. and Tingay, S. J. and Waterson, M. and Wayth, R. B. and Webster, R. L. and Whitney, A. R. and Williams, A. and Williams, C. L. and Wu, C. and Wyithe, J. S. B.},
  year = 2016,
  month = dec,
  journal = {The Astrophysical Journal},
  volume = {833},
  pages = {102},
  issn = {0004-637X},
  url = {http://adsabs.harvard.edu/abs/2016ApJ...833..102B},
  urldate = {2018-08-27}
}

@article{Becker2021,
  title = {The Mean Free Path of Ionizing Photons at 5 {$<$} z {$<$} 6: Evidence for Rapid Evolution near Reionization},
  shorttitle = {The Mean Free Path of Ionizing Photons at 5 {$<$} z {$<$} 6},
  author = {Becker, George D. and D'Aloisio, Anson and Christenson, Holly M. and Zhu, Yongda and Worseck, G{\'a}bor and Bolton, James S.},
  year = 2021,
  month = dec,
  journal = {Monthly Notices of the Royal Astronomical Society},
  volume = {508},
  pages = {1853--1869},
  publisher = {OUP},
  issn = {0035-8711},
  url = {https://ui.adsabs.harvard.edu/abs/2021MNRAS.508.1853B},
  urldate = {2025-04-23}
}

@article{Berkhout2024,
  title = {Hydrogen {{Epoch}} of {{Reionization Array}} ({{HERA}}) {{Phase II Deployment}} and {{Commissioning}}},
  author = {Berkhout, Lindsay M. and Jacobs, Daniel C. and Abdurashidova, Zuhra and Adams, Tyrone and Aguirre, James E. and Alexander, Paul and Ali, Zaki S. and Baartman, Rushelle and Balfour, Yanga and Beardsley, Adam P. and Bernardi, Gianni and Billings, Tashalee S. and Bowman, Judd D. and Bradley, Richard F. and Bull, Philip and Burba, Jacob and Byrne, Ruby and Carey, Steven and Carilli, Chris L. and Chen, Kai-Feng and Cheng, Carina and Choudhuri, Samir and DeBoer, David R. and {de Lera Acedo}, Eloy and Dexter, Matt and Dillon, Joshua S. and Dynes, Scott and Eksteen, Nico and Ely, John and {Ewall-Wice}, Aaron and Fagnoni, Nicolas and Fritz, Randall and Furlanetto, Steven R. and {Gale-Sides}, Kingsley and Garsden, Hugh and Gehlot, Bharat Kumar and Ghosh, Abhik and Glendenning, Brian and Gorce, Adelie and Gorthi, Deepthi and Greig, Bradley and Grobbelaar, Jasper and Halday, Ziyaad and Hazelton, Bryna J. and Hewitt, Jacqueline N. and Hickish, Jack and Huang, Tian and Josaitis, Alec and Julius, Austin and Kariseb, MacCalvin and Kern, Nicholas S. and Kerrigan, Joshua and Kim, Honggeun and Kittiwisit, Piyanat and Kohn, Saul A. and Kolopanis, Matthew and Lanman, Adam and La Plante, Paul and Liu, Adrian and Loots, Anita and Ma, Yin-Zhe and Edward MacMahon, David Harold and Malan, Lourence and Malgas, Cresshim and Malgas, Keith and Marero, Bradley and Martinot, Zachary E. and Mesinger, Andrei and Molewa, Mathakane and Morales, Miguel F. and Mosiane, Tshegofalang and Murray, Steven G. and Neben, Abraham R. and Nikolic, Bojan and Nunhokee, Chuneeta Devi and Nuwegeld, Hans and Parsons, Aaron R. and Pascua, Robert and Patra, Nipanjana and Pieterse, Samantha and Qin, Yuxiang and Rath, Eleanor and {Razavi-Ghods}, Nima and Riley, Daniel and Robnett, James and Rosie, Kathryn and Santos, Mario G. and Sims, Peter and Singh, Saurabh and Storer, Dara and Swarts, Hilton and Tan, Jianrong and Thyagarajan, Nithyanandan and {van Wyngaarden}, Pieter and Williams, Peter K. G. and Zheng, Haoxuan and Xu, Zhilei},
  year = 2024,
  month = apr,
  journal = {Publications of the Astronomical Society of the Pacific},
  volume = {136},
  pages = {045002},
  publisher = {IOP},
  issn = {0004-6280},
  url = {https://ui.adsabs.harvard.edu/abs/2024PASP..136d5002B},
  urldate = {2025-04-17}
}

@misc{Berlin2018a,
  title = {Severely {{Constraining Dark Matter Interpretations}} of the 21-Cm {{Anomaly}}},
  author = {Berlin, Asher and Hooper, Dan and Krnjaic, Gordan and McDermott, Samuel D.},
  year = 2018,
  month = mar,
  eprint = {1803.02804},
  primaryclass = {hep-ph},
  url = {http://arxiv.org/abs/1803.02804},
  urldate = {2025-05-07},
  archiveprefix = {arXiv}
}

@article{Bevins2022,
  title = {A Comprehensive {{Bayesian}} Reanalysis of the {{SARAS2}} Data from the Epoch of Reionization},
  author = {Bevins, H. T. J. and {de Lera Acedo}, E. and Fialkov, A. and Handley, W. J. and Singh, S. and Subrahmanyan, R. and Barkana, R.},
  year = 2022,
  month = jul,
  journal = {Monthly Notices of the Royal Astronomical Society},
  volume = {513},
  pages = {4507--4526},
  issn = {0035-8711},
  url = {https://ui.adsabs.harvard.edu/abs/2022MNRAS.513.4507B},
  urldate = {2022-06-02}
}

@misc{Blamart2025,
  title = {Beyond the {{Power Spectrum}}: {{A New Framework}} for {{Non-Stationary Fields}} with {{Applications}} to {{Light-Cone Effects}} in {{Line Intensity Mapping}}},
  shorttitle = {Beyond the {{Power Spectrum}}},
  author = {Blamart, Matt{\'e}o and Liu, Adrian},
  year = 2025,
  month = may,
  publisher = {arXiv},
  url = {https://ui.adsabs.harvard.edu/abs/2025arXiv250509674B},
  urldate = {2025-07-14}
}

@article{Blas2011a,
  title = {The {{Cosmic Linear Anisotropy Solving System}} ({{CLASS}}) {{II}}: {{Approximation}} Schemes},
  shorttitle = {The {{Cosmic Linear Anisotropy Solving System}} ({{CLASS}}) {{II}}},
  author = {Blas, Diego and Lesgourgues, Julien and Tram, Thomas},
  year = 2011,
  month = jul,
  journal = {Journal of Cosmology and Astroparticle Physics},
  volume = {2011},
  number = {07},
  eprint = {1104.2933},
  primaryclass = {astro-ph},
  pages = {034--034},
  issn = {1475-7516},
  url = {http://arxiv.org/abs/1104.2933},
  urldate = {2025-05-07},
  archiveprefix = {arXiv}
}

@inproceedings{Boerner2023,
  title = {{{ACCESS}}: {{Advancing Innovation}}: {{NSF}}'s {{Advanced Cyberinfrastructure Coordination Ecosystem}}: {{Services}} \& {{Support}}},
  shorttitle = {{{ACCESS}}},
  booktitle = {Practice and {{Experience}} in {{Advanced Research Computing}} 2023: {{Computing}} for the {{Common Good}}},
  author = {Boerner, Timothy J. and Deems, Stephen and Furlani, Thomas R. and Knuth, Shelley L. and Towns, John},
  year = 2023,
  month = sep,
  series = {{{PEARC}} '23},
  pages = {173--176},
  publisher = {Association for Computing Machinery},
  address = {New York, NY, USA},
  url = {https://dl.acm.org/doi/10.1145/3569951.3597559},
  urldate = {2025-07-08},
  isbn = {978-1-4503-9985-2}
}

@article{Bosman2022,
  title = {Hydrogen Reionization Ends by z = 5.3: {{Lyman-$\alpha$}} Optical Depth Measured by the {{XQR-30}} Sample},
  shorttitle = {Hydrogen Reionization Ends by z = 5.3},
  author = {Bosman, Sarah E I and Davies, Frederick B and Becker, George D and Keating, Laura C and Davies, Rebecca L and Zhu, Yongda and Eilers, Anna-Christina and D'Odorico, Valentina and Bian, Fuyan and Bischetti, Manuela and Cristiani, Stefano V and Fan, Xiaohui and Farina, Emanuele P and Haehnelt, Martin G and Hennawi, Joseph F and Kulkarni, Girish and Mesinger, Andrei and Meyer, Romain A and Onoue, Masafusa and Pallottini, Andrea and Qin, Yuxiang and {Ryan-Weber}, Emma and Schindler, Jan-Torge and Walter, Fabian and Wang, Feige and Yang, Jinyi},
  year = 2022,
  month = jul,
  journal = {Monthly Notices of the Royal Astronomical Society},
  volume = {514},
  number = {1},
  pages = {55--76},
  issn = {0035-8711},
  url = {https://doi.org/10.1093/mnras/stac1046},
  urldate = {2025-04-17}
}

@article{Bowman2018,
  title = {An Absorption Profile Centred at 78 Megahertz in the Sky-Averaged Spectrum},
  author = {Bowman, Judd D. and Rogers, Alan E. E. and Monsalve, Raul A. and Mozdzen, Thomas J. and Mahesh, Nivedita},
  year = 2018,
  month = mar,
  journal = {Nature},
  volume = {555},
  number = {7694},
  pages = {67--70},
  issn = {1476-4687},
  url = {https://www.nature.com/articles/nature25792},
  urldate = {2019-03-12},
  copyright = {2018 Nature Publishing Group},
  langid = {english}
}

@article{Bradley2019,
  title = {A {{Ground Plane Artifact}} That {{Induces}} an {{Absorption Profile}} in {{Averaged Spectra}} from {{Global}} 21 Cm {{Measurements}}, with {{Possible Application}} to {{EDGES}}},
  author = {Bradley, Richard F. and Tauscher, Keith and Rapetti, David and Burns, Jack O.},
  year = 2019,
  month = apr,
  journal = {The Astrophysical Journal},
  volume = {874},
  pages = {153},
  issn = {0004-637X},
  url = {https://ui.adsabs.harvard.edu/abs/2019ApJ...874..153B},
  urldate = {2021-10-28}
}

@article{Breitman2024a,
  title = {{{21CMEMU}}: An Emulator of {{21CMFAST}} Summary Observables},
  shorttitle = {{{21CMEMU}}},
  author = {Breitman, Daniela and Mesinger, Andrei and Murray, Steven G. and Prelogovi{\'c}, David and Qin, Yuxiang and Trotta, Roberto},
  year = 2024,
  month = feb,
  journal = {Monthly Notices of the Royal Astronomical Society},
  volume = {527},
  pages = {9833--9852},
  issn = {0035-8711},
  url = {https://ui.adsabs.harvard.edu/abs/2024MNRAS.527.9833B},
  urldate = {2024-01-18}
}

@misc{Bull2024,
  title = {{{RHINO}}: {{A}} Large Horn Antenna for Detecting the 21cm Global Signal},
  shorttitle = {{{RHINO}}},
  author = {Bull, Philip and {El-Makadema}, Ahmed and Garsden, Hugh and Edgley, John and Roddis, Neil and Chluba, Jens and Conselice, Christopher J. and Dutta, Sohini and Glasscock, Katrine A. and Nasirudin, Ainulnabilah and Norris, Jordan and Wilensky, Michael J. and Ye, Isabelle and Zhang, Zheng},
  year = 2024,
  month = sep,
  publisher = {arXiv},
  url = {https://ui.adsabs.harvard.edu/abs/2024arXiv241000076B},
  urldate = {2025-04-17}
}

@article{Bunker2023,
  title = {{{JADES NIRSpec Spectroscopy}} of {{GN-z11}}: {{Lyman-$\alpha$}} Emission and Possible Enhanced Nitrogen Abundance in a z = 10.60 Luminous Galaxy},
  shorttitle = {{{JADES NIRSpec Spectroscopy}} of {{GN-z11}}},
  author = {Bunker, Andrew J. and Saxena, Aayush and Cameron, Alex J. and Willott, Chris J. and {Curtis-Lake}, Emma and Jakobsen, Peter and Carniani, Stefano and Smit, Renske and Maiolino, Roberto and Witstok, Joris and Curti, Mirko and D'Eugenio, Francesco and Jones, Gareth C. and Ferruit, Pierre and Arribas, Santiago and Charlot, Stephane and Chevallard, Jacopo and Giardino, Giovanna and {de Graaff}, Anna and Looser, Tobias J. and L{\"u}tzgendorf, Nora and Maseda, Michael V. and Rawle, Tim and Rix, Hans-Walter and Del Pino, Bruno Rodr{\'i}guez and Alberts, Stacey and Egami, Eiichi and Eisenstein, Daniel J. and Endsley, Ryan and Hainline, Kevin and Hausen, Ryan and Johnson, Benjamin D. and Rieke, George and Rieke, Marcia and Robertson, Brant E. and Shivaei, Irene and Stark, Daniel P. and Sun, Fengwu and Tacchella, Sandro and Tang, Mengtao and Williams, Christina C. and Willmer, Christopher N. A. and Baker, William M. and Baum, Stefi and Bhatawdekar, Rachana and Bowler, Rebecca and Boyett, Kristan and Chen, Zuyi and Circosta, Chiara and Helton, Jakob M. and Ji, Zhiyuan and Kumari, Nimisha and Lyu, Jianwei and Nelson, Erica and Parlanti, Eleonora and Perna, Michele and Sandles, Lester and Scholtz, Jan and Suess, Katherine A. and Topping, Michael W. and {\"U}bler, Hannah and Wallace, Imaan E. B. and Whitler, Lily},
  year = 2023,
  month = sep,
  journal = {Astronomy and Astrophysics},
  volume = {677},
  pages = {A88},
  publisher = {EDP},
  issn = {0004-6361},
  url = {https://ui.adsabs.harvard.edu/abs/2023A&A...677A..88B},
  urldate = {2025-04-17}
}

@article{Byrne2019,
  title = {Fundamental {{Limitations}} on the {{Calibration}} of {{Redundant}} 21 Cm {{Cosmology Instruments}} and {{Implications}} for {{HERA}} and the {{SKA}}},
  author = {Byrne, Ruby and Morales, Miguel F. and Hazelton, Bryna and Li, Wenyang and Barry, Nichole and Beardsley, Adam P. and Joseph, Ronniy and Pober, Jonathan and Sullivan, Ian and Trott, Cathryn},
  year = 2019,
  month = apr,
  journal = {The Astrophysical Journal},
  volume = {875},
  number = {1},
  pages = {70},
  issn = {0004-637X, 1538-4357},
  url = {https://iopscience.iop.org/article/10.3847/1538-4357/ab107d},
  urldate = {2025-04-18},
  langid = {english}
}

@article{Camps1998,
  title = {Mutual Coupling Effects on Antenna Radiation Pattern: {{An}} Experimental Study Applied to Interferometric Radiometers},
  shorttitle = {Mutual Coupling Effects on Antenna Radiation Pattern},
  author = {Camps, A. and Torres, F. and Corbella, I. and Bar{\'a}, J. and {de Paco}, P.},
  year = 1998,
  journal = {Radio Science},
  volume = {33},
  number = {6},
  pages = {1543--1552},
  issn = {1944-799X},
  url = {https://onlinelibrary.wiley.com/doi/abs/10.1029/98RS02813},
  urldate = {2025-04-17},
  langid = {english}
}

@misc{Cang2024,
  title = {The {{EDGES}} Measurement Disfavors an Excess Radio Background during the Cosmic Dawn},
  author = {Cang, Junsong and Mesinger, Andrei and Murray, Steven G. and Breitman, Daniela and Qin, Yuxiang and Trotta, Roberto},
  year = 2024,
  month = nov,
  publisher = {arXiv},
  url = {https://ui.adsabs.harvard.edu/abs/2024arXiv241108134C},
  urldate = {2025-04-18}
}

@article{Castellano2024,
  title = {{{JWST NIRSpec Spectroscopy}} of the {{Remarkable Bright Galaxy GHZ2}}/{{GLASS-z12}} at {{Redshift}} 12.34},
  author = {Castellano, Marco and Napolitano, Lorenzo and Fontana, Adriano and {Roberts-Borsani}, Guido and Treu, Tommaso and Vanzella, Eros and Zavala, Jorge A. and Arrabal Haro, Pablo and Calabr{\`o}, Antonello and Llerena, Mario and Mascia, Sara and Merlin, Emiliano and Paris, Diego and Pentericci, Laura and Santini, Paola and Bakx, Tom J. L. C. and Bergamini, Pietro and Cupani, Guido and Dickinson, Mark and Filippenko, Alexei V. and Glazebrook, Karl and Grillo, Claudio and Kelly, Patrick L. and Malkan, Matthew A. and Mason, Charlotte A. and Morishita, Takahiro and Nanayakkara, Themiya and Rosati, Piero and Sani, Eleonora and Wang, Xin and Yoon, Ilsang},
  year = 2024,
  month = sep,
  journal = {The Astrophysical Journal},
  volume = {972},
  pages = {143},
  publisher = {IOP},
  issn = {0004-637X},
  url = {https://ui.adsabs.harvard.edu/abs/2024ApJ...972..143C},
  urldate = {2025-04-23}
}

@article{Charles2024,
  title = {Mitigating Calibration Errors from Mutual Coupling with Time-Domain Filtering of 21 Cm Cosmological Radio Observations},
  author = {Charles, N. and Kern, N. S. and Pascua, R. and Bernardi, G. and Bester, L. and Smirnov, O. and Acedo, E. D. L. and Abdurashidova, Z. and Adams, T. and Aguirre, J. E. and Baartman, R. and Beardsley, A. P. and Berkhout, L. M. and Billings, T. S. and Bowman, J. D. and Bull, P. and Burba, J. and Byrne, R. and Carey, S. and Chen, K. and Choudhuri, S. and Cox, T. and DeBoer, D. R. and Dexter, M. and Dillon, J. S. and Dynes, S. and Eksteen, N. and Ely, J. and {Ewall-Wice}, A. and Fritz, R. and Furlanetto, S. R. and {Gale-Sides}, K. and Garsden, H. and Gehlot, B. K. and Ghosh, A. and Gorce, A. and Gorthi, D. and Halday, Z. and Hazelton, B. J. and Hewitt, J. N. and Hickish, J. and Huang, T. and Jacobs, D. C. and Josaitis, A. and Kerrigan, J. and Kittiwisit, P. and Kolopanis, M. and Lanman, A. and Liu, A. and Ma, Y. -Z. and MacMahon, D. H. E. and Malan, L. and Malgas, K. and Malgas, C. and Marero, B. and Martinot, Z. E. and McBride, L. and Mesinger, A. and {Mohamed-Hinds}, N. and Molewa, M. and Morales, M. F. and Murray, S. and Nikolic, B. and Nuwegeld, H. and Parsons, A. R. and Patra, N. and Plante, P. L. and Qin, Y. and Rath, E. and {Razavi-Ghods}, N. and Riley, D. and Robnett, J. and Rosie, K. and Santos, M. G. and Sims, P. and Singh, S. and Storer, D. and Swarts, H. and Tan, J. and Wilensky, M. J. and Williams, P. K. G. and v. Wyngaarden, P. and Zheng, H.},
  year = 2024,
  month = nov,
  journal = {Monthly Notices of the Royal Astronomical Society},
  volume = {534},
  pages = {3349--3363},
  publisher = {OUP},
  issn = {0035-8711},
  url = {https://ui.adsabs.harvard.edu/abs/2024MNRAS.534.3349C},
  urldate = {2025-04-17}
}

@article{Chatterjee2019,
  title = {On the Prospects of Measuring the Cosmic Dawn 21-Cm Power Spectrum Using the Upgraded {{Giant Metrewave Radio Telescope}} ({{uGMRT}})},
  author = {Chatterjee, Suman and Bharadwaj, Somnath},
  year = 2019,
  month = feb,
  journal = {Monthly Notices of the Royal Astronomical Society},
  volume = {483},
  pages = {2269--2274},
  publisher = {OUP},
  issn = {0035-8711},
  url = {https://ui.adsabs.harvard.edu/abs/2019MNRAS.483.2269C},
  urldate = {2025-08-05}
}

@article{Chen2025,
  title = {Impacts and {{Statistical Mitigation}} of {{Missing Data}} on the 21 Cm {{Power Spectrum}}: {{A Case Study}} with the {{Hydrogen Epoch}} of {{Reionization Array}}},
  shorttitle = {Impacts and {{Statistical Mitigation}} of {{Missing Data}} on the 21 Cm {{Power Spectrum}}},
  author = {Chen, Kai-Feng and Wilensky, Michael J. and Liu, Adrian and Dillon, Joshua S. and Hewitt, Jacqueline N. and Adams, Tyrone and Aguirre, James E. and Baartman, Rushelle and Beardsley, Adam P. and Berkhout, Lindsay M. and Bernardi, Gianni and Billings, Tashalee S. and Bowman, Judd D. and Bull, Philip and Burba, Jacob and Byrne, Ruby and Carey, Steven and Choudhuri, Samir and Cox, Tyler and DeBoer, {\relax David}. R. and Dexter, Matt and Eksteen, Nico and Ely, John and {Ewall-Wice}, Aaron and Furlanetto, Steven R. and {Gale-Sides}, Kingsley and Garsden, Hugh and Gehlot, Bharat Kumar and Gorce, Ad{\'e}lie and Gorthi, Deepthi and Halday, Ziyaad and Hazelton, Bryna J. and Hickish, Jack and Jacobs, Daniel C. and Josaitis, Alec and Kern, Nicholas S. and Kerrigan, Joshua and Kittiwisit, Piyanat and Kolopanis, Matthew and Plante, Paul La and Lanman, Adam and Ma, Yin-Zhe and MacMahon, David H. E. and Malan, Lourence and Malgas, Cresshim and Malgas, Keith and Marero, Bradley and Martinot, Zachary E. and McBride, Lisa and Mesinger, Andrei and {Mohamed-Hinds}, Nicel and Molewa, Mathakane and Morales, Miguel F. and Murray, Steven G. and Nuwegeld, Hans and Parsons, Aaron R. and Pascua, Robert and Qin, Yuxiang and Rath, Eleanor and {Razavi-Ghods}, Nima and Robnett, James and Santos, Mario G. and Sims, Peter and Singh, Saurabh and Storer, Dara and Swarts, Hilton and Tan, Jianrong and van Wyngaarden, Pieter and Zheng, Haoxuan},
  year = 2025,
  month = jan,
  journal = {The Astrophysical Journal},
  volume = {979},
  number = {2},
  pages = {191},
  publisher = {The American Astronomical Society},
  issn = {0004-637X},
  url = {https://dx.doi.org/10.3847/1538-4357/ad9b91},
  urldate = {2025-05-20},
  langid = {english}
}

@article{Cheng2018,
  title = {Characterizing {{Signal Loss}} in the 21 Cm {{Reionization Power Spectrum}}: {{A Revised Study}} of {{PAPER-64}}},
  shorttitle = {Characterizing {{Signal Loss}} in the 21 Cm {{Reionization Power Spectrum}}},
  author = {Cheng, Carina and Parsons, Aaron R. and Kolopanis, Matthew and Jacobs, Daniel C. and Liu, Adrian and Kohn, Saul A. and Aguirre, James E. and Pober, Jonathan C. and Ali, Zaki S. and Bernardi, Gianni and Bradley, Richard F. and Carilli, Chris L. and DeBoer, David R. and Dexter, Matthew R. and Dillon, Joshua S. and Klima, Pat and MacMahon, David H. E. and Moore, David F. and Nunhokee, Chuneeta D. and Walbrugh, William P. and Walker, Andre},
  year = 2018,
  month = nov,
  journal = {The Astrophysical Journal},
  volume = {868},
  pages = {26},
  issn = {0004-637X},
  url = {https://ui.adsabs.harvard.edu/abs/2018ApJ...868...26C},
  urldate = {2021-11-11}
}

@article{Choudhury2021,
  title = {Studying the {{Lyman}} {$\alpha$} Optical Depth Fluctuations at z   5.5 Using Fast Semi-Numerical Methods},
  author = {Choudhury, T. Roy and Paranjape, Aseem and Bosman, Sarah E. I.},
  year = 2021,
  month = mar,
  journal = {Monthly Notices of the Royal Astronomical Society},
  volume = {501},
  pages = {5782--5796},
  publisher = {OUP},
  issn = {0035-8711},
  url = {https://ui.adsabs.harvard.edu/abs/2021MNRAS.501.5782C},
  urldate = {2025-05-20}
}

@article{Ciardi2005,
  title = {The {{First}} Cosmic Structures and Their Effects},
  author = {Ciardi, Benedetta and Ferrara, Andrea},
  year = 2005,
  journal = {Space Sci. Rev.},
  volume = {116},
  pages = {625--705}
}

@article{Clark2013,
  title = {Accelerating Radio Astronomy Cross-Correlation with Graphics Processing Units},
  author = {Clark, M.A. and Plante, PC La and Greenhill, L.J.},
  year = 2013,
  month = may,
  journal = {The International Journal of High Performance Computing Applications},
  volume = {27},
  number = {2},
  pages = {178--192},
  publisher = {SAGE Publications Ltd STM},
  issn = {1094-3420},
  url = {https://doi.org/10.1177/1094342012444794},
  urldate = {2025-04-17},
  langid = {english}
}

@article{Cox2024,
  title = {Spectral Redundancy for Calibrating Interferometers and Suppressing the Foreground Wedge in 21 Cm Cosmology},
  author = {Cox, Tyler A and Parsons, Aaron R and Dillon, Joshua S and {Ewall-Wice}, Aaron and Pascua, Robert},
  year = 2024,
  month = jul,
  journal = {Monthly Notices of the Royal Astronomical Society},
  volume = {532},
  number = {3},
  pages = {3375--3394},
  issn = {0035-8711, 1365-2966},
  url = {https://academic.oup.com/mnras/article/532/3/3375/7710121},
  urldate = {2025-04-18},
  copyright = {https://creativecommons.org/licenses/by/4.0/},
  langid = {english}
}

@misc{Cox2025,
  title = {Fftvis: {{A Non-Uniform Fast Fourier Transform Based Interferometric Visibility Simulator}}},
  shorttitle = {Fftvis},
  author = {Cox, Tyler A. and Murray, Steven G. and Parsons, Aaron R. and Dillon, Joshua S. and Mandar, Kartik and Martinot, Zachary E. and Pascua, Robert and Kittiwisit, Piyanat and Aguirre, James E.},
  year = 2025,
  month = jun,
  number = {arXiv:2506.02130},
  eprint = {2506.02130},
  publisher = {arXiv},
  url = {http://arxiv.org/abs/2506.02130},
  urldate = {2025-06-04},
  archiveprefix = {arXiv}
}

@misc{Cruz2024,
  title = {The {{First Billion Years}} in {{Seconds}}: {{An Effective Model}} for the 21-Cm {{Signal}} with {{Population III Stars}}},
  shorttitle = {The {{First Billion Years}} in {{Seconds}}},
  author = {Cruz, Hector Afonso G. and Munoz, Julian B. and Sabti, Nashwan and Kamionkowski, Marc},
  year = 2024,
  month = jul,
  number = {arXiv:2407.18294},
  eprint = {2407.18294},
  primaryclass = {astro-ph},
  publisher = {arXiv},
  url = {http://arxiv.org/abs/2407.18294},
  urldate = {2025-05-07},
  archiveprefix = {arXiv}
}

@article{Cullen2023,
  title = {The Ultraviolet Continuum Slopes ({$\beta$}) of Galaxies at z {$\simeq$} 8-16 from {{JWST}} and Ground-Based near-Infrared Imaging},
  author = {Cullen, Fergus and McLure, R. J. and McLeod, D. J. and Dunlop, J. S. and Donnan, C. T. and Carnall, A. C. and Bowler, R. a. A. and Begley, R. and Hamadouche, M. L. and Stanton, T. M.},
  year = 2023,
  month = mar,
  journal = {Monthly Notices of the Royal Astronomical Society, Volume 520, Issue 1, pp.14-23},
  volume = {520},
  number = {1},
  pages = {14},
  issn = {0035-8711},
  url = {https://ui.adsabs.harvard.edu/abs/2023MNRAS.520...14C/abstract},
  urldate = {2025-06-16},
  langid = {english}
}

@article{Cyr2024,
  title = {Soft Photon Heating: A Semi-Analytic Framework and Applications to 21-Cm Cosmology},
  shorttitle = {Soft Photon Heating},
  author = {Cyr, Bryce and Acharya, Sandeep Kumar and Chluba, Jens},
  year = 2024,
  month = oct,
  journal = {Monthly Notices of the Royal Astronomical Society},
  volume = {534},
  pages = {738--757},
  publisher = {OUP},
  issn = {0035-8711},
  url = {https://ui.adsabs.harvard.edu/abs/2024MNRAS.534..738C},
  urldate = {2025-07-26}
}

@article{Datta2012,
  title = {Light-Cone Effect on the Reionization 21-Cm Power Spectrum},
  author = {Datta, Kanan K. and Mellema, Garrelt and Mao, Yi and Iliev, Ilian T. and Shapiro, Paul R. and Ahn, Kyungjin},
  year = 2012,
  month = aug,
  journal = {Monthly Notices of the Royal Astronomical Society},
  volume = {424},
  pages = {1877--1891},
  issn = {0035-8711},
  url = {http://adsabs.harvard.edu/abs/2012MNRAS.424.1877D},
  urldate = {2017-07-06}
}

@article{Datta2014,
  title = {Light Cone Effect on the Reionization 21-Cm Signal - {{II}}. {{Evolution}}, Anisotropies and Observational Implications},
  author = {Datta, Kanan K. and Jensen, Hannes and Majumdar, Suman and Mellema, Garrelt and Iliev, Ilian T. and Mao, Yi and Shapiro, Paul R. and Ahn, Kyungjin},
  year = 2014,
  month = aug,
  journal = {Monthly Notices of the Royal Astronomical Society},
  volume = {442},
  pages = {1491--1506},
  issn = {0035-8711},
  url = {http://adsabs.harvard.edu/abs/2014MNRAS.442.1491D},
  urldate = {2017-07-06}
}

@article{Davies2021,
  title = {The {{Predicament}} of {{Absorption-dominated Reionization}}: {{Increased Demands}} on {{Ionizing Sources}}},
  shorttitle = {The {{Predicament}} of {{Absorption-dominated Reionization}}},
  author = {Davies, Frederick B. and Bosman, Sarah E. I. and Furlanetto, Steven R. and Becker, George D. and D'Aloisio, Anson},
  year = 2021,
  month = sep,
  journal = {The Astrophysical Journal},
  volume = {918},
  pages = {L35},
  publisher = {IOP},
  issn = {0004-637X},
  url = {https://ui.adsabs.harvard.edu/abs/2021ApJ...918L..35D},
  urldate = {2025-04-23}
}

@article{DeBoer2017,
  title = {Hydrogen {{Epoch}} of {{Reionization Array}} ({{HERA}})},
  author = {DeBoer, David R. and Parsons, Aaron R. and Aguirre, James E. and Alexander, Paul and Ali, Zaki S. and Beardsley, Adam P. and Bernardi, Gianni and Bowman, Judd D. and Bradley, Richard F. and Carilli, Chris L. and Cheng, Carina and Acedo, Eloy de Lera and Dillon, Joshua S. and {Ewall-Wice}, Aaron and Fadana, Gcobisa and Fagnoni, Nicolas and Fritz, Randall and Furlanetto, Steve R. and Glendenning, Brian and Greig, Bradley and Grobbelaar, Jasper and Hazelton, Bryna J. and Hewitt, Jacqueline N. and Hickish, Jack and Jacobs, Daniel C. and Julius, Austin and Kariseb, MacCalvin and Kohn, Saul A. and {Telalo Lekalake} and Liu, Adrian and Loots, Anita and MacMahon, David and Malan, Lourence and Malgas, Cresshim and Maree, Matthys and {Zachary Martinot} and Mathison, Nathan and Matsetela, Eunice and Mesinger, Andrei and Morales, Miguel F. and Neben, Abraham R. and Patra, Nipanjana and Pieterse, Samantha and Pober, Jonathan C. and {Razavi-Ghods}, Nima and Ringuette, Jon and {James Robnett} and Rosie, Kathryn and Sell, Raddwine and Smith, Craig and Syce, Angelo and Tegmark, Max and {Nithyanandan Thyagarajan} and Williams, Peter K. G. and Zheng, Haoxuan},
  year = 2017,
  journal = {Publications of the Astronomical Society of the Pacific},
  volume = {129},
  number = {974},
  pages = {045001},
  issn = {1538-3873},
  url = {http://stacks.iop.org/1538-3873/129/i=974/a=045001},
  urldate = {2018-11-20},
  langid = {english}
}

@article{DeLeraAcedo2022a,
  title = {The {{REACH}} Radiometer for Detecting the 21-Cm Hydrogen Signal from Redshift z {$\approx$} 7.5-28},
  author = {{de Lera Acedo}, E. and {de Villiers}, D. I. L. and {Razavi-Ghods}, N. and Handley, W. and Fialkov, A. and Magro, A. and Anstey, D. and Bevins, H. T. J. and Chiello, R. and Cumner, J. and Josaitis, A. T. and Roque, I. L. V. and Sims, P. H. and Scheutwinkel, K. H. and Alexander, P. and Bernardi, G. and Carey, S. and Cavillot, J. and Croukamp, W. and Ely, J. A. and {Gessey-Jones}, T. and Gueuning, Q. and Hills, R. and Kulkarni, G. and Maiolino, R. and Meerburg, P. D. and Mittal, S. and Pritchard, J. R. and Puchwein, E. and Saxena, A. and Shen, E. and Smirnov, O. and Spinelli, M. and {Zarb-Adami}, K.},
  year = 2022,
  month = jul,
  journal = {Nature Astronomy},
  volume = {6},
  pages = {984--998},
  issn = {2397-3366},
  url = {https://ui.adsabs.harvard.edu/abs/2022NatAs...6..984D},
  urldate = {2023-11-06}
}

@article{DeOliveira-Costa2008,
  title = {A Model of Diffuse {{Galactic}} Radio Emission from 10 {{MHz}} to 100 {{GHz}}},
  author = {{De Oliveira-Costa}, Ang{\'e}lica and Tegmark, Max and Gaensler, B. M. and Jonas, Justin and Landecker, T. L. and Reich, Patricia},
  year = 2008,
  journal = {Monthly Notices of the Royal Astronomical Society},
  volume = {388},
  number = {1},
  pages = {247--260},
  issn = {1365-2966},
  url = {https://onlinelibrary.wiley.com/doi/abs/10.1111/j.1365-2966.2008.13376.x},
  urldate = {2021-11-11},
  langid = {english}
}

@article{Dewdney2016,
  title = {{{SKA System Baseline Design}} V2},
  author = {Dewdney, P. and Turner, W. and Braun, R. and {Santander-Vela}, J. and Waterson, M. and Tan, G-H.},
  year = 2016,
  journal = {SKA Document Series}
}

@article{Dillon2014a,
  title = {Overcoming Real-World Obstacles in 21 Cm Power Spectrum Estimation: {{A}} Method Demonstration and Results from Early {{Murchison Widefield Array}} Data},
  shorttitle = {Overcoming Real-World Obstacles in 21 Cm Power Spectrum Estimation},
  author = {Dillon, Joshua S. and Liu, Adrian and Williams, Christopher L. and Hewitt, Jacqueline N. and Tegmark, Max and Morgan, Edward H. and Levine, Alan M. and Morales, Miguel F. and Tingay, Steven J. and Bernardi, Gianni and Bowman, Judd D. and Briggs, Frank H. and Cappallo, Roger C. and Emrich, David and Mitchell, Daniel A. and Oberoi, Divya and Prabu, Thiagaraj and Wayth, Randall and Webster, Rachel L.},
  year = 2014,
  month = jan,
  journal = {Physical Review D},
  volume = {89},
  number = {2},
  eprint = {1304.4229},
  issn = {1550-7998, 1550-2368},
  url = {http://arxiv.org/abs/1304.4229},
  urldate = {2017-08-16},
  archiveprefix = {arXiv}
}

@article{Dillon2015,
  title = {Empirical Covariance Modeling for 21~Cm Power Spectrum Estimation: {{A}} Method Demonstration and New Limits from Early {{Murchison Widefield Array}} 128-Tile Data},
  author = {Dillon, Joshua S. and Neben, Abraham R. and Hewitt, Jacqueline N. and Tegmark, Max and Barry, N. and Beardsley, A. P. and Bowman, J. D. and Briggs, F. and Carroll, P. and {de Oliveira-Costa}, A. and {Ewall-Wice}, A. and Feng, L. and Greenhill, L. J. and Hazelton, B. J. and Hernquist, L. and {Hurley-Walker}, N. and Jacobs, D. C. and Kim, H. S. and Kittiwisit, P. and Lenc, E. and Line, J. and Loeb, A. and McKinley, B. and Mitchell, D. A. and Morales, M. F. and Offringa, A. R. and Paul, S. and Pindor, B. and Pober, J. C. and Procopio, P. and Riding, J. and Sethi, S. and Shankar, N. Udaya and Subrahmanyan, R. and Sullivan, I. and Thyagarajan, Nithyanandan and Tingay, S. J. and Trott, C. and Wayth, R. B. and Webster, R. L. and Wyithe, S. and Bernardi, G. and Cappallo, R. J. and Deshpande, A. A. and {Johnston-Hollitt}, M. and Kaplan, D. L. and Lonsdale, C. J. and McWhirter, S. R. and Morgan, E. and Oberoi, D. and Ord, S. M. and Prabu, T. and Srivani, K. S. and Williams, A. and Williams, C. L.},
  year = 2015,
  journal = {Physical Review D},
  volume = {91},
  number = {12},
  pages = {123011--123011},
  url = {http://adsabs.harvard.edu/abs/2015PhRvD..91l3011D}
}

@article{Dillon2016,
  title = {Redundant {{Array Configurations}} for 21 Cm {{Cosmology}}},
  author = {Dillon, Joshua S. and Parsons, Aaron R.},
  year = 2016,
  month = aug,
  journal = {The Astrophysical Journal},
  volume = {826},
  pages = {181},
  publisher = {IOP},
  issn = {0004-637X},
  url = {https://ui.adsabs.harvard.edu/abs/2016ApJ...826..181D},
  urldate = {2025-04-17}
}

@article{Dillon2020a,
  title = {Redundant-Baseline Calibration of the Hydrogen Epoch of Reionization Array},
  author = {Dillon, Joshua S. and Lee, Max and Ali, Zaki S. and Parsons, Aaron R. and Orosz, Naomi and Nunhokee, Chuneeta Devi and La Plante, Paul and Beardsley, Adam P. and Kern, Nicholas S. and Abdurashidova, Zara and Aguirre, James E. and Alexander, Paul and Balfour, Yanga and Bernardi, Gianni and Billings, Tashalee S. and Bowman, Judd D. and Bradley, Richard F. and Bull, Phil and Burba, Jacob and Carey, Steve and Carilli, Chris L. and Cheng, Carina and DeBoer, David R. and Dexter, Matt and {de Lera Acedo}, Eloy and Ely, John and {Ewall-Wice}, Aaron and Fagnoni, Nicolas and Fritz, Randall and Furlanetto, Steven R. and {Gale-Sides}, Kingsley and Glendenning, Brian and Gorthi, Deepthi and Greig, Bradley and Grobbelaar, Jasper and Halday, Ziyaad and Hazelton, Bryna J. and Hewitt, Jacqueline N. and Hickish, Jack and Jacobs, Daniel C. and Julius, Austin and Kerrigan, Joshua and Kittiwisit, Piyanat and Kohn, Saul A. and Kolopanis, Matthew and Lanman, Adam and Lekalake, Telalo and Lewis, David and Liu, Adrian and Ma, Yin-Zhe and MacMahon, David and Malan, Lourence and Malgas, Cresshim and Maree, Matthys and Martinot, Zachary E. and Matsetela, Eunice and Mesinger, Andrei and Molewa, Mathakane and Morales, Miguel F. and Mosiane, Tshegofalang and Murray, Steven and Neben, Abraham R. and Nikolic, Bojan and Pascua, Robert and Patra, Nipanjana and Pieterse, Samantha and Pober, Jonathan C. and {Razavi-Ghods}, Nima and Ringuette, Jon and Robnett, James and Rosie, Kathryn and Santos, Mario G. and Sims, Peter and Smith, Craig and Syce, Angelo and Tegmark, Max and Thyagarajan, Nithyanandan and Williams, Peter K. G. and Zheng, Haoxuan},
  year = 2020,
  month = dec,
  journal = {Monthly Notices of the Royal Astronomical Society},
  volume = {499},
  pages = {5840--5861},
  publisher = {OUP},
  issn = {0035-8711},
  url = {https://ui.adsabs.harvard.edu/abs/2020MNRAS.499.5840D},
  urldate = {2025-04-17}
}

@techreport{DillonMurray_hera125,
  type = {Memo},
  title = {{{H6C}} Internal Data Release 2.2},
  author = {Dillon, J. S. and Murray, S.},
  year = 2023,
  number = {125},
  address = {reionization.org/science/memos},
  institution = {Department of Astronomy, University of California, Berkeley},
  url = {https://reionization.org/manual_uploads/HERA125_H6C_IDR_2_2_Memo.pdf}
}

@article{Donnan2024,
  title = {{{JWST PRIMER}}: A New Multifield Determination of the Evolving Galaxy {{UV}} Luminosity Function at Redshifts z {$\simeq$} 9 - 15},
  shorttitle = {{{JWST PRIMER}}},
  author = {Donnan, C. T. and McLure, R. J. and Dunlop, J. S. and McLeod, D. J. and Magee, D. and {Arellano-C{\'o}rdova}, K. Z. and Barrufet, L. and Begley, R. and Bowler, R. A. A. and Carnall, A. C. and Cullen, F. and Ellis, R. S. and Fontana, A. and Illingworth, G. D. and Grogin, N. A. and Hamadouche, M. L. and Koekemoer, A. M. and Liu, F. -Y. and Mason, C. and Santini, P. and Stanton, T. M.},
  year = 2024,
  month = sep,
  journal = {Monthly Notices of the Royal Astronomical Society},
  volume = {533},
  pages = {3222--3237},
  publisher = {OUP},
  issn = {0035-8711},
  url = {https://ui.adsabs.harvard.edu/abs/2024MNRAS.533.3222D},
  urldate = {2025-04-23}
}

@article{Eastwood2019,
  title = {The 21 Cm {{Power Spectrum}} from the {{Cosmic Dawn}}: {{First Results}} from the {{OVRO-LWA}}},
  shorttitle = {The 21 Cm {{Power Spectrum}} from the {{Cosmic Dawn}}},
  author = {Eastwood, Michael W. and Anderson, Marin M. and Monroe, Ryan M. and Hallinan, Gregg and Catha, Morgan and Dowell, Jayce and Garsden, Hugh and Greenhill, Lincoln J. and Hicks, Brian C. and Kocz, Jonathon and Price, Danny C. and Schinzel, Frank K. and Vedantham, Harish and Wang, Yuankun},
  year = 2019,
  month = aug,
  journal = {The Astronomical Journal},
  volume = {158},
  pages = {84},
  issn = {0004-6256},
  url = {http://adsabs.harvard.edu/abs/2019AJ....158...84E},
  urldate = {2020-11-13}
}

@article{Ewall-Wice2016a,
  title = {First Limits on the 21 Cm Power Spectrum during the {{Epoch}} of {{X-ray}} Heating},
  author = {{Ewall-Wice}, A. and Dillon, Joshua S. and Hewitt, J. N. and Loeb, A. and Mesinger, A. and Neben, A. R. and Offringa, A. R. and Tegmark, M. and Barry, N. and Beardsley, A. P. and Bernardi, G. and Bowman, Judd D. and Briggs, F. and Cappallo, R. J. and Carroll, P. and Corey, B. E. and {de Oliveira-Costa}, A. and Emrich, D. and Feng, L. and Gaensler, B. M. and Goeke, R. and Greenhill, L. J. and Hazelton, B. J. and {Hurley-Walker}, N. and {Johnston-Hollitt}, M. and Jacobs, Daniel C. and Kaplan, D. L. and Kasper, J. C. and Kim, {\relax HS} and Kratzenberg, E. and Lenc, E. and Line, J. and Lonsdale, C. J. and Lynch, M. J. and McKinley, B. and McWhirter, S. R. and Mitchell, D. A. and Morales, M. F. and Morgan, E. and Thyagarajan, Nithyanandan and Oberoi, D. and Ord, S. M. and Paul, S. and Pindor, B. and Pober, J. C. and Prabu, T. and Procopio, P. and Riding, J. and Rogers, A. E. E. and Roshi, A. and Shankar, N. Udaya and Sethi, Shiv K. and Srivani, K. S. and Subrahmanyan, R. and Sullivan, I. S. and Tingay, S. J. and Trott, C. M. and Waterson, M. and Wayth, R. B. and Webster, R. L. and Whitney, A. R. and Williams, A. and Williams, C. L. and Wu, C. and Wyithe, J. S. B.},
  year = 2016,
  month = aug,
  journal = {Monthly Notices of the Royal Astronomical Society},
  volume = {460},
  pages = {4320--4347},
  issn = {0035-8711},
  url = {http://adsabs.harvard.edu/abs/2016MNRAS.460.4320E},
  urldate = {2020-11-13}
}

@article{Ewall-Wice2018,
  title = {Modeling the {{Radio Background}} from the {{First Black Holes}} at {{Cosmic Dawn}}: {{Implications}} for the 21 Cm {{Absorption Amplitude}}},
  shorttitle = {Modeling the {{Radio Background}} from the {{First Black Holes}} at {{Cosmic Dawn}}},
  author = {{Ewall-Wice}, A. and Chang, T.-C. and Lazio, J. and Dor{\'e}, O. and Seiffert, M. and Monsalve, R. A.},
  year = 2018,
  month = nov,
  journal = {The Astrophysical Journal},
  volume = {868},
  pages = {63},
  url = {http://adsabs.harvard.edu/abs/2018ApJ...868...63E},
  urldate = {2020-08-24}
}

@article{Ewall-Wice2021,
  title = {{{DAYENU}}: A Simple Filter of Smooth Foregrounds for Intensity Mapping Power Spectra},
  shorttitle = {{{DAYENU}}},
  author = {{Ewall-Wice}, Aaron and Kern, Nicholas and Dillon, Joshua S and Liu, Adrian and Parsons, Aaron and Singh, Saurabh and Lanman, Adam and Plante, Paul La and Fagnoni, Nicolas and Acedo, Eloy de Lera and DeBoer, David R and Nunhokee, Chuneeta and Bull, Philip and Chang, Tzu-Ching and Lazio, T Joseph W and Aguirre, James and Weinberg, Sean},
  year = 2021,
  month = feb,
  journal = {Monthly Notices of the Royal Astronomical Society},
  volume = {500},
  number = {4},
  pages = {5195--5213},
  issn = {0035-8711},
  url = {https://doi.org/10.1093/mnras/staa3293},
  urldate = {2025-04-18}
}

@article{Fagnoni2021,
  title = {Design of the {{New Wideband Vivaldi Feed}} for the {{HERA Radio-Telescope Phase II}}},
  author = {Fagnoni, Nicolas and Acedo, Eloy de Lera and Drought, Nick and De Boer, David R. and Riley, Daniel and {Razavi-Ghods}, Nima and Carey, Steven and Parsons, Aaron R.},
  year = 2021,
  journal = {IEEE Transactions on Antennas and Propagation},
  pages = {1--1},
  issn = {1558-2221},
  url = {https://ieeexplore.ieee.org/document/9445678}
}

@article{Fagnoni2021a,
  title = {Understanding the {{HERA Phase I}} Receiver System with Simulations and Its Impact on the Detectability of the {{EoR}} Delay Power Spectrum},
  author = {Fagnoni, Nicolas and {de~Lera~Acedo}, Eloy and DeBoer, David R and Abdurashidova, Zara and Aguirre, James E and Alexander, Paul and Ali, Zaki S and Balfour, Yanga and Beardsley, Adam P and Bernardi, Gianni and Billings, Tashalee S and Bowman, Judd D and Bradley, Richard F and Bull, Phil and Burba, Jacob and Carilli, Chris L and Cheng, Carina and Dexter, Matt and Dillon, Joshua S and {Ewall-Wice}, Aaron and Fritz, Randall and Furlanetto, Steve R and {Gale-Sides}, Kingsley and Glendenning, Brian and Gorthi, Deepthi and Greig, Bradley and Grobbelaar, Jasper and Halday, Ziyaad and Hazelton, Bryna J and Hewitt, Jacqueline N and Hickish, Jack and Jacobs, Daniel C and Josaitis, Alec and Julius, Austin and Kern, Nicholas S and Kerrigan, Joshua and Kim, Honggeun and Kittiwisit, Piyanat and Kohn, Saul A and Kolopanis, Matthew and Lanman, Adam and Plante, Paul La and Lekalake, Telalo and Liu, Adrian and MacMahon, David and Malan, Lourence and Malgas, Cresshim and Maree, Matthys and Martinot, Zachary E and Matsetela, Eunice and Mena~Parra, Juan and Mesinger, Andrei and Molewa, Mathakane and Morales, Miguel F and Mosiane, Tshegofalang and Neben, Abraham R and Nikolic, Bojan and Parsons, Aaron R and Patra, Nipanjana and Pieterse, Samantha and Pober, Jonathan C and {Razavi-Ghods}, Nima and Robnett, James and Rosie, Kathryn and Sims, Peter and Smith, Craig and Syce, Angelo and Thyagarajan, Nithyanandan and Williams, Peter K G and Zheng, Haoxuan},
  year = 2021,
  month = jan,
  journal = {Monthly Notices of the Royal Astronomical Society},
  volume = {500},
  number = {1},
  pages = {1232--1242},
  issn = {0035-8711},
  url = {https://doi.org/10.1093/mnras/staa3268},
  urldate = {2025-11-21}
}

@article{Feng2018,
  title = {Enhanced {{Global Signal}} of {{Neutral Hydrogen Due}} to {{Excess Radiation}} at {{Cosmic Dawn}}},
  author = {Feng, Chang and Holder, Gilbert},
  year = 2018,
  month = may,
  journal = {The Astrophysical Journal},
  volume = {858},
  number = {2},
  pages = {L17},
  url = {https://ui.adsabs.harvard.edu/abs/2018ApJ...858L..17F/abstract},
  urldate = {2019-11-01},
  langid = {english}
}

@article{Fialkov2012,
  title = {Impact of the Relative Motion between the Dark Matter and Baryons on the First Stars: Semi-Analytical Modelling},
  shorttitle = {Impact of the Relative Motion between the Dark Matter and Baryons on the First Stars},
  author = {Fialkov, Anastasia and Barkana, Rennan and Tseliakhovich, Dmitriy and Hirata, Christopher M.},
  year = 2012,
  month = aug,
  journal = {Monthly Notices of the Royal Astronomical Society},
  volume = {424},
  pages = {1335--1345},
  publisher = {OUP},
  issn = {0035-8711},
  url = {https://ui.adsabs.harvard.edu/abs/2012MNRAS.424.1335F},
  urldate = {2025-04-24}
}

@article{Fialkov2013,
  title = {The 21-Cm Signature of the First Stars during the {{Lyman-Werner}} Feedback Era},
  author = {Fialkov, Anastasia and Barkana, Rennan and Visbal, Eli and Tseliakhovich, Dmitriy and Hirata, Christopher M.},
  year = 2013,
  month = jul,
  journal = {Monthly Notices of the Royal Astronomical Society},
  volume = {432},
  pages = {2909--2916},
  publisher = {OUP},
  issn = {0035-8711},
  url = {https://ui.adsabs.harvard.edu/abs/2013MNRAS.432.2909F},
  urldate = {2025-04-24}
}

@article{Fialkov2014,
  title = {The Observable Signature of Late Heating of the {{Universe}} during Cosmic Reionization},
  author = {Fialkov, Anastasia and Barkana, Rennan and Visbal, Eli},
  year = 2014,
  month = feb,
  journal = {Nature},
  volume = {506},
  pages = {197--199},
  issn = {0028-0836},
  url = {https://ui.adsabs.harvard.edu/abs/2014Natur.506..197F},
  urldate = {2025-04-24}
}

@article{Fialkov2019,
  title = {Signature of {{Excess Radio Background}} in the 21-Cm {{Global Signal}} and {{Power Spectrum}}},
  author = {Fialkov, Anastasia and Barkana, Rennan},
  year = 2019,
  month = feb,
  journal = {arXiv:1902.02438 [astro-ph]},
  eprint = {1902.02438},
  primaryclass = {astro-ph},
  url = {http://arxiv.org/abs/1902.02438},
  urldate = {2019-02-11},
  archiveprefix = {arXiv}
}

@article{Finkelstein2024,
  title = {The {{Complete CEERS Early Universe Galaxy Sample}}: {{A Surprisingly Slow Evolution}} of the {{Space Density}} of {{Bright Galaxies}} at z {$\sim$} 8.5--14.5},
  shorttitle = {The {{Complete CEERS Early Universe Galaxy Sample}}},
  author = {Finkelstein, Steven L. and Leung, Gene C. K. and Bagley, Micaela B. and Dickinson, Mark and Ferguson, Henry C. and Papovich, Casey and Akins, Hollis B. and Arrabal Haro, Pablo and Dav{\'e}, Romeel and Dekel, Avishai and Kartaltepe, Jeyhan S. and Kocevski, Dale D. and Koekemoer, Anton M. and Pirzkal, Nor and Somerville, Rachel S. and Yung, L. Y. Aaron and Amor{\'i}n, Ricardo O. and Backhaus, Bren E. and Behroozi, Peter and Bisigello, Laura and Bromm, Volker and Casey, Caitlin M. and Ch{\'a}vez Ortiz, {\'O}scar A. and Cheng, Yingjie and Chworowsky, Katherine and Cleri, Nikko J. and Cooper, M. C. and Davis, Kelcey and {de la Vega}, Alexander and Elbaz, David and Franco, Maximilien and Fontana, Adriano and Fujimoto, Seiji and Giavalisco, Mauro and Grogin, Norman A. and Holwerda, Benne W. and {Huertas-Company}, Marc and Hirschmann, Michaela and Iyer, Kartheik G. and Jogee, Shardha and Jung, Intae and Larson, Rebecca L. and Lucas, Ray A. and Mobasher, Bahram and Morales, Alexa M. and Morley, Caroline V. and Mukherjee, Sagnick and {P{\'e}rez-Gonz{\'a}lez}, Pablo G. and Ravindranath, Swara and Rodighiero, Giulia and Rowland, Melanie J. and Tacchella, Sandro and Taylor, Anthony J. and Trump, Jonathan R. and Wilkins, Stephen M.},
  year = 2024,
  month = jul,
  journal = {The Astrophysical Journal},
  volume = {969},
  pages = {L2},
  publisher = {IOP},
  issn = {0004-637X},
  url = {https://ui.adsabs.harvard.edu/abs/2024ApJ...969L...2F},
  urldate = {2025-04-23}
}

@article{Fixsen2011,
  title = {{{ARCADE}} 2 {{Measurement}} of the {{Absolute Sky Brightness}} at 3-90 {{GHz}}},
  author = {Fixsen, D. J. and Kogut, A. and Levin, S. and Limon, M. and Lubin, P. and Mirel, P. and Seiffert, M. and Singal, J. and Wollack, E. and Villela, T. and Wuensche, C. A.},
  year = 2011,
  month = jun,
  journal = {The Astrophysical Journal},
  volume = {734},
  pages = {5},
  publisher = {IOP},
  issn = {0004-637X},
  url = {https://ui.adsabs.harvard.edu/abs/2011ApJ...734....5F},
  urldate = {2025-05-06}
}

@article{Fragos2013,
  title = {X-{{Ray Binary Evolution Across Cosmic Time}}},
  author = {Fragos, T. and Lehmer, B. and Tremmel, M. and Tzanavaris, P. and {Basu-Zych}, A. and Belczynski, K. and Hornschemeier, A. and Jenkins, L. and Kalogera, V. and Ptak, A. and Zezas, A.},
  year = 2013,
  month = feb,
  journal = {The Astrophysical Journal},
  volume = {764},
  pages = {41},
  publisher = {IOP},
  issn = {0004-637X},
  url = {https://ui.adsabs.harvard.edu/abs/2013ApJ...764...41F},
  urldate = {2025-04-17}
}

@article{Franzen2019,
  title = {Source Counts and Confusion at 72-231 {{MHz}} in the {{MWA GLEAM}} Survey},
  author = {Franzen, T. M. O. and Vernstrom, T. and Jackson, C. A. and {Hurley-Walker}, N. and Ekers, R. D. and Heald, G. and Seymour, N. and White, S. V.},
  year = 2019,
  month = feb,
  journal = {Publications of the Astronomical Society of Australia},
  volume = {36},
  pages = {e004},
  issn = {1323-3580},
  url = {https://ui.adsabs.harvard.edu/abs/2019PASA...36....4F},
  urldate = {2025-04-18}
}

@article{Furlanetto2006,
  title = {Cosmology at Low Frequencies: {{The}} 21 Cm Transition and the High-Redshift {{Universe}}},
  author = {Furlanetto, Steven R. and Peng Oh, S. and Briggs, Frank H.},
  year = 2006,
  journal = {Physics Reports},
  volume = {433},
  number = {4-6},
  pages = {181--301},
  issn = {0370-1573},
  url = {http://arxiv.org/pdf/astro-ph/0608032v2.pdf}
}

@article{Gaikwad2023,
  title = {Measuring the Photoionization Rate, Neutral Fraction, and Mean Free Path of {{H I}} Ionizing Photons at 4.9 {$\leq$} z {$\leq$} 6.0 from a Large Sample of {{XShooter}} and {{ESI}} Spectra},
  author = {Gaikwad, Prakash and Haehnelt, Martin G. and Davies, Fredrick B. and Bosman, Sarah E. I. and Molaro, Margherita and Kulkarni, Girish and D'Odorico, Valentina and Becker, George D. and Davies, Rebecca L. and Nasir, Fahad and Bolton, James S. and Keating, Laura C. and Ir{\v s}i{\v c}, Vid and Puchwein, Ewald and Zhu, Yongda and Asthana, Shikhar and Yang, Jinyi and Lai, Samuel and Eilers, Anna-Christina},
  year = 2023,
  month = nov,
  journal = {Monthly Notices of the Royal Astronomical Society},
  volume = {525},
  pages = {4093--4120},
  publisher = {OUP},
  issn = {0035-8711},
  url = {https://ui.adsabs.harvard.edu/abs/2023MNRAS.525.4093G},
  urldate = {2025-04-23}
}

@misc{Gardner2006,
  title = {The {{James Webb Space Telescope}}},
  author = {Gardner, Jonathan P. and Mather, John C. and Clampin, Mark and Doyon, Rene and Greenhouse, Matthew A. and Hammel, Heidi B. and Hutchings, John B. and Jakobsen, Peter and Lilly, Simon J. and Long, Knox S. and Lunine, Jonathan I. and McCaughrean, Mark J. and Mountain, Matt and Nella, John and Rieke, George H. and Rieke, Marcia J. and Rix, Hans-Walter and Smith, Eric P. and Sonneborn, George and Stiavelli, Massimo and Stockman, H. S. and Windhorst, Rogier A. and Wright, Gillian S.},
  year = 2006,
  month = jun,
  eprint = {astro-ph/0606175},
  url = {http://arxiv.org/abs/astro-ph/0606175},
  urldate = {2025-06-18},
  archiveprefix = {arXiv}
}

@article{Garsden2021,
  title = {A 21-Cm Power Spectrum at 48 {{MHz}}, Using the {{Owens Valley Long Wavelength Array}}},
  author = {Garsden, H. and Greenhill, L. and Bernardi, G. and Fialkov, A. and Price, D. C. and Mitchell, D. and Dowell, J. and Spinelli, M. and Schinzel, F. K.},
  year = 2021,
  month = oct,
  journal = {Monthly Notices of the Royal Astronomical Society},
  volume = {506},
  pages = {5802--5817},
  publisher = {OUP},
  issn = {0035-8711},
  url = {https://ui.adsabs.harvard.edu/abs/2021MNRAS.506.5802G},
  urldate = {2025-04-17}
}

@article{Gehlot2018,
  title = {Wide-Field {{LOFAR-LBA}} Power-Spectra Analyses: {{Impact}} of Calibration, Polarization Leakage and Ionosphere},
  shorttitle = {Wide-Field {{LOFAR-LBA}} Power-Spectra Analyses},
  author = {Gehlot, B. K. and Koopmans, L. V. E. and {de Bruyn}, A. G. and Zaroubi, S. and Brentjens, M. A. and Asad, K. M. B. and Hatef, M. and Jelic, V. and Mevius, M. and Offringa, A. R. and Pandey, V. N. and Yatawatta, S.},
  year = 2018,
  month = aug,
  journal = {Monthly Notices of the Royal Astronomical Society},
  volume = {478},
  number = {2},
  eprint = {1709.07727},
  pages = {1484--1501},
  issn = {0035-8711, 1365-2966},
  url = {http://arxiv.org/abs/1709.07727},
  urldate = {2018-06-20},
  archiveprefix = {arXiv}
}

@article{Gehlot2020,
  title = {The {{AARTFAAC Cosmic Explorer}}: Observations of the 21-Cm Power Spectrum in the {{EDGES}} Absorption Trough},
  shorttitle = {The {{AARTFAAC Cosmic Explorer}}},
  author = {Gehlot, B. K. and Mertens, F. G. and Koopmans, L. V. E. and Offringa, A. R. and Shulevski, A. and Mevius, M. and Brentjens, M. A. and Kuiack, M. and Pandey, V. N. and Rowlinson, A. and Sardarabadi, A. M. and Vedantham, H. K. and Wijers, R. A. M. J. and Yatawatta, S. and Zaroubi, S.},
  year = 2020,
  month = oct,
  journal = {Monthly Notices of the Royal Astronomical Society},
  volume = {499},
  pages = {4158--4173},
  issn = {0035-8711},
  url = {http://adsabs.harvard.edu/abs/2020MNRAS.499.4158G},
  urldate = {2020-11-11}
}

@article{Gehlot2024,
  title = {Transient {{RFI}} Environment of {{LOFAR-LBA}} at 72--75 {{MHz}}: {{Impact}} on Ultra-Widefield {{AARTFAAC Cosmic Explorer}} Observations of the Redshifted 21-Cm Signal},
  shorttitle = {Transient {{RFI}} Environment of {{LOFAR-LBA}} at 72--75 {{MHz}}},
  author = {Gehlot, B. K. and Koopmans, L. V. E. and Brackenhoff, S. A. and Ceccotti, E. and Ghosh, S. and H{\"o}fer, C. and Mertens, F. G. and Mevius, M. and Munshi, S. and Offringa, A. R. and Pandey, V. N. and Rowlinson, A. and Shulevski, A. and Wijers, R. A. M. J. and Yatawatta, S. and Zaroubi, S.},
  year = 2024,
  month = jan,
  journal = {Astronomy \& Astrophysics},
  volume = {681},
  pages = {A71},
  issn = {0004-6361, 1432-0746},
  url = {https://www.aanda.org/10.1051/0004-6361/202346376},
  urldate = {2025-05-12},
  copyright = {https://creativecommons.org/licenses/by/4.0}
}

@article{Gelli2024,
  title = {The {{Impact}} of {{Mass-dependent Stochasticity}} at {{Cosmic Dawn}}},
  author = {Gelli, Viola and Mason, Charlotte and Hayward, Christopher C.},
  year = 2024,
  month = nov,
  journal = {The Astrophysical Journal},
  volume = {975},
  pages = {192},
  publisher = {IOP},
  issn = {0004-637X},
  url = {https://ui.adsabs.harvard.edu/abs/2024ApJ...975..192G},
  urldate = {2025-04-23}
}

@article{Gessey-Jones2024,
  title = {On the Constraints on Superconducting Cosmic Strings from 21-Cm Cosmology},
  author = {{Gessey-Jones}, T. and Pochinda, S. and Bevins, H. T. J. and Fialkov, A. and Handley, W. J. and {de Lera Acedo}, E. and Singh, S. and Barkana, R.},
  year = 2024,
  month = mar,
  journal = {Monthly Notices of the Royal Astronomical Society},
  volume = {529},
  pages = {519--536},
  publisher = {OUP},
  issn = {0035-8711},
  url = {https://ui.adsabs.harvard.edu/abs/2024MNRAS.529..519G},
  urldate = {2025-04-24}
}

@article{Ghara2015,
  title = {21 Cm Signal from Cosmic Dawn - {{II}}. {{Imprints}} of the Light-Cone Effects},
  author = {Ghara, Raghunath and Datta, Kanan K. and Choudhury, T. Roy},
  year = 2015,
  month = nov,
  journal = {Monthly Notices of the Royal Astronomical Society},
  volume = {453},
  pages = {3143--3156},
  issn = {0035-8711},
  url = {http://adsabs.harvard.edu/abs/2015MNRAS.453.3143G},
  urldate = {2017-07-06}
}

@article{Ghara2020a,
  title = {Constraining the Intergalactic Medium at z {$\approx$} 9.1 Using {{LOFAR Epoch}} of {{Reionization}} Observations},
  author = {Ghara, R. and Giri, S. K. and Mellema, G. and Ciardi, B. and Zaroubi, S. and Iliev, I. T. and Koopmans, L. V. E. and Chapman, E. and Gazagnes, S. and Gehlot, B. K. and Ghosh, A. and Jeli{\'c}, V. and Mertens, F. G. and Mondal, R. and Schaye, J. and Silva, M. B. and Asad, K. M. B. and Kooistra, R. and Mevius, M. and Offringa, A. R. and Pandey, V. N. and Yatawatta, S.},
  year = 2020,
  month = feb,
  journal = {Monthly Notices of the Royal Astronomical Society},
  volume = {493},
  pages = {4728--4747},
  publisher = {OUP},
  issn = {0035-8711},
  url = {https://ui.adsabs.harvard.edu/abs/2020MNRAS.493.4728G},
  urldate = {2025-04-23}
}

@article{Ghara2021,
  title = {Constraining the State of the Intergalactic Medium during the {{Epoch}} of {{Reionization}} Using {{MWA}} 21-Cm Signal Observations},
  author = {Ghara, Raghunath and Giri, Sambit K. and Ciardi, Benedetta and Mellema, Garrelt and Zaroubi, Saleem},
  year = 2021,
  month = may,
  journal = {Monthly Notices of the Royal Astronomical Society},
  volume = {503},
  pages = {4551--4562},
  publisher = {OUP},
  issn = {0035-8711},
  url = {https://ui.adsabs.harvard.edu/abs/2021MNRAS.503.4551G},
  urldate = {2025-04-23}
}

@misc{Ghara2025,
  title = {Constraints on the State of the {{IGM}} at \$z\textbackslash sim 8-10\$ Using Redshifted 21-Cm Observations with {{LOFAR}}},
  author = {Ghara, R. and Zaroubi, S. and Ciardi, B. and Mellema, G. and Giri, S. K. and Mertens, F. G. and Mevius, M. and Koopmans, L. V. E. and Iliev, I. T. and Acharya, A. and Brackenhoff, S. A. and Ceccotti, E. and Chege, K. and Georgiev, I. and Ghosh, S. and Hothi, I. and H{\"o}fer, C. and Ma, Q. and Munshi, S. and Offringa, A. R. and Shaw, A. K. and Pandey, V. N. and Yatawatta, S. and Choudhury, M.},
  year = 2025,
  month = may,
  number = {arXiv:2505.00373},
  eprint = {2505.00373},
  primaryclass = {astro-ph},
  publisher = {arXiv},
  url = {http://arxiv.org/abs/2505.00373},
  urldate = {2025-05-05},
  archiveprefix = {arXiv}
}

@article{Gorce2023,
  title = {Impact of Instrument and Data Characteristics in the Interferometric Reconstruction of the 21 Cm Power Spectrum},
  author = {Gorce, Ad{\'e}lie and Ganjam, Samskruthi and Liu, Adrian and Murray, Steven G. and Abdurashidova, Zara and Adams, Tyrone and Aguirre, James E. and Alexander, Paul and Ali, Zaki S. and Baartman, Rushelle and Balfour, Yanga and Beardsley, Adam P. and Bernardi, Gianni and Billings, Tashalee S. and Bowman, Judd D. and Bradley, Richard F. and Bull, Philip and Burba, Jacob and Carey, Steven and Carilli, Chris L. and Cheng, Carina and DeBoer, David R. and Acedo, Eloy de Lera and Dexter, Matt and Dillon, Joshua S. and Eksteen, Nico and Ely, John and {Ewall-Wice}, Aaron and Fagnoni, Nicolas and Fritz, Randall and Furlanetto, Steven R. and {Gale-Sides}, Kingsley and Glendenning, Brian and Gorthi, Deepthi and Greig, Bradley and Grobbelaar, Jasper and Halday, Ziyaad and Hazelton, Bryna J. and Hewitt, Jacqueline N. and Hickish, Jack and Jacobs, Daniel C. and Julius, Austin and Kariseb, MacCalvin and Kern, Nicholas S. and Kerrigan, Joshua and Kittiwisit, Piyanat and Kohn, Saul A. and Kolopanis, Matthew and Lanman, Adam and La Plante, Paul and Loots, Anita and MacMahon, David Harold Edward and Malan, Lourence and Malgas, Cresshim and Malgas, Keith and Marero, Bradley and Martinot, Zachary E. and Mesinger, Andrei and Molewa, Mathakane and Morales, Miguel F. and Mosiane, Tshegofalang and Neben, Abraham R. and Nikolic, Bojan and Nuwegeld, Hans and Parsons, Aaron R. and Patra, Nipanjana and Pieterse, Samantha and Pober, Jonathan C. and {Razavi-Ghods}, Nima and Robnett, James and Rosie, Kathryn and Sims, Peter and Swarts, Hilton and Thyagarajan, Nithyanandan and {van Wyngaarden}, Pieter and Williams, Peter K. G. and Zheng, Haoxuan},
  year = 2023,
  month = mar,
  journal = {Monthly Notices of the Royal Astronomical Society},
  volume = {520},
  pages = {375--391},
  publisher = {OUP},
  issn = {0035-8711},
  url = {https://ui.adsabs.harvard.edu/abs/2023MNRAS.520..375G},
  urldate = {2025-05-12}
}

@article{Gorski2005,
  title = {{{HEALPix}}: {{A Framework}} for {{High-Resolution Discretization}} and {{Fast Analysis}} of {{Data Distributed}} on the {{Sphere}}},
  shorttitle = {{{HEALPix}}},
  author = {G{\'o}rski, K. M. and Hivon, E. and Banday, A. J. and Wandelt, B. D. and Hansen, F. K. and Reinecke, M. and Bartelmann, M.},
  year = 2005,
  month = apr,
  journal = {The Astrophysical Journal},
  volume = {622},
  pages = {759--771},
  publisher = {IOP},
  issn = {0004-637X},
  url = {https://ui.adsabs.harvard.edu/abs/2005ApJ...622..759G},
  urldate = {2025-04-18}
}

@article{Goulding2023,
  title = {{{UNCOVER}}: {{The Growth}} of the {{First Massive Black Holes}} from {{JWST}}/{{NIRSpec-Spectroscopic Redshift Confirmation}} of an {{X-Ray Luminous AGN}} at z = 10.1},
  shorttitle = {{{UNCOVER}}},
  author = {Goulding, Andy D. and Greene, Jenny E. and Setton, David J. and Labbe, Ivo and Bezanson, Rachel and Miller, Tim B. and Atek, Hakim and Bogd{\'a}n, {\'A}kos and Brammer, Gabriel and Chemerynska, Iryna and Cutler, Sam E. and Dayal, Pratika and Fudamoto, Yoshinobu and Fujimoto, Seiji and Furtak, Lukas J. and Kokorev, Vasily and Khullar, Gourav and Leja, Joel and Marchesini, Danilo and Natarajan, Priyamvada and Nelson, Erica and Oesch, Pascal A. and Pan, Richard and Papovich, Casey and Price, Sedona H. and {van Dokkum}, Pieter and Wang, Bingjie and Weaver, John R. and Whitaker, Katherine E. and Zitrin, Adi},
  year = 2023,
  month = sep,
  journal = {The Astrophysical Journal},
  volume = {955},
  pages = {L24},
  publisher = {IOP},
  issn = {0004-637X},
  url = {https://ui.adsabs.harvard.edu/abs/2023ApJ...955L..24G},
  urldate = {2025-04-23}
}

@article{Greene2024,
  title = {{{UNCOVER Spectroscopy Confirms}} the {{Surprising Ubiquity}} of {{Active Galactic Nuclei}} in {{Red Sources}} at z {$>$} 5},
  author = {Greene, Jenny E. and Labbe, Ivo and Goulding, Andy D. and Furtak, Lukas J. and Chemerynska, Iryna and Kokorev, Vasily and Dayal, Pratika and Volonteri, Marta and Williams, Christina C. and Wang, Bingjie and Setton, David J. and Burgasser, Adam J. and Bezanson, Rachel and Atek, Hakim and Brammer, Gabriel and Cutler, Sam E. and Feldmann, Robert and Fujimoto, Seiji and Glazebrook, Karl and {de Graaff}, Anna and Khullar, Gourav and Leja, Joel and Marchesini, Danilo and Maseda, Michael V. and Matthee, Jorryt and Miller, Tim B. and Naidu, Rohan P. and Nanayakkara, Themiya and Oesch, Pascal A. and Pan, Richard and Papovich, Casey and Price, Sedona H. and {van Dokkum}, Pieter and Weaver, John R. and Whitaker, Katherine E. and Zitrin, Adi},
  year = 2024,
  month = mar,
  journal = {The Astrophysical Journal},
  volume = {964},
  pages = {39},
  publisher = {IOP},
  issn = {0004-637X},
  url = {https://ui.adsabs.harvard.edu/abs/2024ApJ...964...39G},
  urldate = {2025-04-23}
}

@article{Greig2015,
  title = {{{21CMMC}}: An {{MCMC}} Analysis Tool Enabling Astrophysical Parameter Studies of the Cosmic 21 Cm Signal},
  shorttitle = {{{21CMMC}}},
  author = {Greig, Bradley and Mesinger, Andrei},
  year = 2015,
  month = jun,
  journal = {Monthly Notices of the Royal Astronomical Society},
  volume = {449},
  number = {4},
  eprint = {1501.06576},
  pages = {4246--4263},
  issn = {0035-8711, 1365-2966},
  url = {http://arxiv.org/abs/1501.06576},
  urldate = {2018-05-16},
  archiveprefix = {arXiv}
}

@article{Greig2018,
  title = {{{21CMMC}} with a {{3D}} Light-Cone: The Impact of the Co-Evolution Approximation on the Astrophysics of Reionisation and Cosmic Dawn},
  shorttitle = {{{21CMMC}} with a {{3D}} Light-Cone},
  author = {Greig, Bradley and Mesinger, Andrei},
  year = 2018,
  month = jul,
  journal = {Monthly Notices of the Royal Astronomical Society},
  volume = {477},
  number = {3},
  pages = {3217--3229},
  issn = {0035-8711, 1365-2966},
  url = {http://arxiv.org/abs/1801.01592},
  urldate = {2018-08-28}
}

@article{Greig2021,
  ids = {Greig2020a},
  title = {Interpreting {{LOFAR}} 21-Cm Signal Upper Limits at z {$\asymp$} 9.1 in the Context of High-z Galaxy and Reionization Observations},
  author = {Greig, Bradley and Mesinger, Andrei and Koopmans, L{\'e}on V. E. and Ciardi, Benedetta and Mellema, Garrelt and Zaroubi, Saleem and Giri, Sambit K. and Ghara, Raghunath and Ghosh, Abhik and Iliev, Ilian T. and Mertens, Florent G. and Mondal, Rajesh and Offringa, Andr{\'e} R. and Pandey, Vishambhar N.},
  year = 2021,
  month = jan,
  journal = {Monthly Notices of the Royal Astronomical Society},
  volume = {501},
  pages = {1--13},
  issn = {0035-8711},
  url = {https://ui.adsabs.harvard.edu/abs/2021MNRAS.501....1G},
  urldate = {2021-11-10}
}

@article{Greig2021a,
  ids = {Greig2020},
  title = {Exploring Reionization and High-z Galaxy Observables with Recent Multiredshift {{MWA}} Upper Limits on the 21-Cm Signal},
  author = {Greig, Bradley and Trott, Cathryn M. and Barry, Nichole and Mutch, Simon J. and Pindor, Bart and Webster, Rachel L. and Wyithe, J. Stuart B.},
  year = 2021,
  month = jan,
  journal = {Monthly Notices of the Royal Astronomical Society},
  volume = {500},
  pages = {5322--5335},
  issn = {0035-8711},
  url = {https://ui.adsabs.harvard.edu/abs/2021MNRAS.500.5322G},
  urldate = {2021-11-10}
}

@inproceedings{Gueuning2022,
  title = {Full-Wave Analysis of Mutual Coupling in the {{HIRAX}} Radio Telescope},
  booktitle = {2022 {{International Conference}} on {{Electromagnetics}} in {{Advanced Applications}} ({{ICEAA}})},
  author = {Gueuning, Q. and Crichton, D. and Sampath, A. and Brown, A. K. and Acedo, E. de Lera and Moodley, K.},
  year = 2022,
  month = sep,
  pages = {288--288},
  url = {https://ieeexplore.ieee.org/document/9899905},
  urldate = {2025-04-17}
}

@article{Gupta2017,
  title = {The Upgraded {{GMRT}}: Opening New Windows on the Radio {{Universe}}},
  shorttitle = {The Upgraded {{GMRT}}},
  author = {Gupta, Y. and Ajithkumar, B. and Kale, H. S. and Nayak, S. and Sabhapathy, S. and Sureshkumar, S. and Swami, R. V. and Chengalur, J. N. and Ghosh, S. K. and {Ishwara-Chandra}, C. H. and Joshi, B. C. and Kanekar, N. and Lal, D. V. and Roy, S.},
  year = 2017,
  month = aug,
  journal = {Current Science},
  volume = {113},
  pages = {707--714},
  url = {https://ui.adsabs.harvard.edu/abs/2017CSci..113..707G},
  urldate = {2025-08-05}
}

@article{Guzman2011,
  title = {All-Sky {{Galactic}} Radiation at 45 {{MHz}} and Spectral Index between 45 and 408 {{MHz}}},
  author = {Guzm{\'a}n, A. E. and May, J. and Alvarez, H. and Maeda, K.},
  year = 2011,
  month = jan,
  journal = {Astronomy \& Astrophysics},
  volume = {525},
  pages = {A138},
  publisher = {EDP Sciences},
  issn = {0004-6361, 1432-0746},
  url = {https://www.aanda.org/articles/aa/abs/2011/01/aa13628-09/aa13628-09.html},
  urldate = {2025-07-09},
  copyright = {\copyright{} ESO, 2010},
  langid = {english}
}

@article{Harris2020,
  title = {Array Programming with {{NumPy}}},
  author = {Harris, Charles R. and Millman, K. Jarrod and {van der Walt}, St{\'e}fan J. and Gommers, Ralf and Virtanen, Pauli and Cournapeau, David and Wieser, Eric and Taylor, Julian and Berg, Sebastian and Smith, Nathaniel J. and Kern, Robert and Picus, Matti and Hoyer, Stephan and {van Kerkwijk}, Marten H. and Brett, Matthew and Haldane, Allan and {del R{\'i}o}, Jaime Fern{\'a}ndez and Wiebe, Mark and Peterson, Pearu and {G{\'e}rard-Marchant}, Pierre and Sheppard, Kevin and Reddy, Tyler and Weckesser, Warren and Abbasi, Hameer and Gohlke, Christoph and Oliphant, Travis E.},
  year = 2020,
  month = sep,
  journal = {Nature},
  volume = {585},
  number = {7825},
  pages = {357--362},
  publisher = {Nature Publishing Group},
  issn = {1476-4687},
  url = {https://www.nature.com/articles/s41586-020-2649-2},
  urldate = {2022-07-25},
  copyright = {2020 The Author(s)},
  langid = {english}
}

@article{Hazelton2017,
  title = {Pyuvdata: An Interface for Astronomical Interferometeric Datasets in Python},
  shorttitle = {Pyuvdata},
  author = {Hazelton, Bryna J. and Jacobs, Daniel C. and Pober, Jonathan C. and Beardsley, Adam P.},
  year = 2017,
  month = feb,
  journal = {The Journal of Open Source Software},
  volume = {2},
  pages = {140},
  url = {https://ui.adsabs.harvard.edu/2017JOSS....2..140H/abstract},
  urldate = {2019-10-31},
  langid = {english}
}

@misc{Heiligstein2023,
  title = {How Often Is Lightning to Blame for {{RFI}} in {{HERA}} Data?},
  author = {Heiligstein, Wren and Jacobs},
  year = 2023,
  month = feb,
  publisher = {https://reionization.org},
  url = {https://reionization.org/manual_uploads/HERA121_lightning2023.pdf},
  archiveprefix = {https://reionization.org}
}

@article{HERA22,
  title = {First {{Results}} from {{HERA Phase I}}: {{Upper Limits}} on the {{Epoch}} of {{Reionization}} 21 Cm {{Power Spectrum}}},
  shorttitle = {First {{Results}} from {{HERA Phase I}}},
  author = {Abdurashidova, Zara and Aguirre, James E. and Alexander, Paul and Ali, Zaki S. and Balfour, Yanga and Beardsley, Adam P. and Bernardi, Gianni and Billings, Tashalee S. and Bowman, Judd D. and Bradley, Richard F. and Bull, Philip and Burba, Jacob and Carey, Steve and Carilli, Chris L. and Cheng, Carina and DeBoer, David R. and Dexter, Matt and {de Lera Acedo}, Eloy and {Dibblee-Barkman}, Taylor and Dillon, Joshua S. and Ely, John and {Ewall-Wice}, Aaron and Fagnoni, Nicolas and Fritz, Randall and Furlanetto, Steven R. and {Gale-Sides}, Kingsley and Glendenning, Brian and Gorthi, Deepthi and Greig, Bradley and Grobbelaar, Jasper and Halday, Ziyaad and Hazelton, Bryna J. and Hewitt, Jacqueline N. and Hickish, Jack and Jacobs, Daniel C. and Julius, Austin and Kern, Nicholas S. and Kerrigan, Joshua and Kittiwisit, Piyanat and Kohn, Saul A. and Kolopanis, Matthew and Lanman, Adam and La Plante, Paul and Lekalake, Telalo and Lewis, David and Liu, Adrian and MacMahon, David and Malan, Lourence and Malgas, Cresshim and Maree, Matthys and Martinot, Zachary E. and Matsetela, Eunice and Mesinger, Andrei and Molewa, Mathakane and Morales, Miguel F. and Mosiane, Tshegofalang and Murray, Steven G. and Neben, Abraham R. and Nikolic, Bojan and Nunhokee, Chuneeta D. and Parsons, Aaron R. and Patra, Nipanjana and Pascua, Robert and Pieterse, Samantha and Pober, Jonathan C. and {Razavi-Ghods}, Nima and Ringuette, Jon and Robnett, James and Rosie, Kathryn and Sims, Peter and Singh, Saurabh and Smith, Craig and Syce, Angelo and Thyagarajan, Nithyanandan and Williams, Peter K. G. and Zheng, Haoxuan and {HERA Collaboration}},
  year = 2022,
  month = feb,
  journal = {The Astrophysical Journal},
  volume = {925},
  pages = {221},
  issn = {0004-637X},
  url = {https://ui.adsabs.harvard.edu/abs/2022ApJ...925..221A},
  urldate = {2024-01-31}
}

@article{HERA23,
  title = {Improved {{Constraints}} on the 21 Cm {{EoR Power Spectrum}} and the {{X-Ray Heating}} of the {{IGM}} with {{HERA Phase I Observations}}},
  author = {Abdurashidova, The Hera Collaboration: Zara and Adams, Tyrone and Aguirre, James E. and Alexander, Paul and Ali, Zaki S. and Baartman, Rushelle and Balfour, Yanga and Barkana, Rennan and Beardsley, Adam P. and Bernardi, Gianni and Billings, Tashalee S. and Bowman, Judd D. and Bradley, Richard F. and Breitman, Daniela and Bull, Philip and Burba, Jacob and Carey, Steve and Carilli, Chris L. and Cheng, Carina and Choudhuri, Samir and DeBoer, David R. and De Lera Acedo, Eloy and Dexter, Matt and Dillon, Joshua S. and Ely, John and {Ewall-Wice}, Aaron and Fagnoni, Nicolas and Fialkov, Anastasia and Fritz, Randall and Furlanetto, Steven R. and {Gale-Sides}, Kingsley and Garsden, Hugh and Glendenning, Brian and Gorce, Ad{\'e}lie and Gorthi, Deepthi and Greig, Bradley and Grobbelaar, Jasper and Halday, Ziyaad and Hazelton, Bryna J. and Heimersheim, Stefan and Hewitt, Jacqueline N. and Hickish, Jack and Jacobs, Daniel C. and Julius, Austin and Kern, Nicholas S. and Kerrigan, Joshua and Kittiwisit, Piyanat and Kohn, Saul A. and Kolopanis, Matthew and Lanman, Adam and La Plante, Paul and Lewis, David and Liu, Adrian and Loots, Anita and Ma, Yin-Zhe and MacMahon, David H. E. and Malan, Lourence and Malgas, Keith and Malgas, Cresshim and Maree, Matthys and Marero, Bradley and Martinot, Zachary E. and McBride, Lisa and Mesinger, Andrei and Mirocha, Jordan and Molewa, Mathakane and Morales, Miguel F. and Mosiane, Tshegofalang and Mu{\~n}oz, Julian B. and Murray, Steven G. and Nagpal, Vighnesh and Neben, Abraham R. and Nikolic, Bojan and Nunhokee, Chuneeta D. and Nuwegeld, Hans and Parsons, Aaron R. and Pascua, Robert and Patra, Nipanjana and Pieterse, Samantha and Qin, Yuxiang and {Razavi-Ghods}, Nima and Robnett, James and Rosie, Kathryn and Santos, Mario G. and Sims, Peter and Singh, Saurabh and Smith, Craig and Swarts, Hilton and Tan, Jianrong and Thyagarajan, Nithyanandan and Wilensky, Michael J. and Williams, Peter K. G. and Van Wyngaarden, Pieter and Zheng, Haoxuan},
  year = 2023,
  month = mar,
  journal = {The Astrophysical Journal},
  volume = {945},
  number = {2},
  pages = {124},
  issn = {0004-637X, 1538-4357},
  url = {https://iopscience.iop.org/article/10.3847/1538-4357/acaf50},
  urldate = {2025-07-04}
}

@techreport{HERAMemo129,
  type = {Memo},
  title = {A Quick Primer on the {{HERA DPSS}} Fringe-Rate Filter Implementation},
  author = {Bull, Phil},
  year = 2024,
  number = {129},
  address = {reionization.org/science/memos},
  institution = {Jodrell Bank Centre for Astrophysics, University of Manchester},
  url = {http://reionization.org/manual_uploads/HERA129_DPSS_fringe_rate_filter_implementation_v1.pdf}
}

@article{Hickish2016,
  title = {A {{Decade}} of {{Developing Radio-Astronomy Instrumentation}} Using {{CASPER Open-Source Technology}}},
  author = {Hickish, Jack and Abdurashidova, Zuhra and Ali, Zaki and Buch, Kaushal D. and Chaudhari, Sandeep C. and Chen, Hong and Dexter, Matthew and Domagalski, Rachel Simone and Ford, John and Foster, Griffin and George, David and Greenberg, Joe and Greenhill, Lincoln and Isaacson, Adam and Jiang, Homin and Jones, Glenn and Kapp, Francois and Kriel, Henno and Lacasse, Rich and Lutomirski, Andrew and MacMahon, David and Manley, Jason and Martens, Andrew and McCullough, Randy and Muley, Mekhala V. and New, Wesley and Parsons, Aaron and Price, Daniel C. and Primiani, Rurik A. and Ray, Jason and Siemion, Andrew and {van Tonder}, Verees{\'e} and Vertatschitsch, Laura and Wagner, Mark and Weintroub, Jonathan and Werthimer, Dan},
  year = 2016,
  month = dec,
  journal = {Journal of Astronomical Instrumentation},
  volume = {5},
  pages = {1641001--12},
  publisher = {WSPC},
  url = {https://ui.adsabs.harvard.edu/abs/2016JAI.....541001H},
  urldate = {2025-04-18}
}

@article{Hills2018,
  title = {Concerns about {{Modelling}} of {{Foregrounds}} and the 21-Cm {{Signal}} in {{EDGES}} Data},
  author = {Hills, Richard and Kulkarni, Girish and Meerburg, P. Daniel and Puchwein, Ewald},
  year = 2018,
  month = may,
  journal = {arXiv:1805.01421 [astro-ph, physics:hep-ph]},
  eprint = {1805.01421},
  primaryclass = {astro-ph, physics:hep-ph},
  url = {http://arxiv.org/abs/1805.01421},
  urldate = {2019-04-09},
  archiveprefix = {arXiv}
}

@article{Hirling2024,
  title = {{{pyC2Ray}}: {{A}} Flexible and {{GPU-accelerated}} Radiative Transfer Framework for Simulating the Cosmic Epoch of Reionization},
  shorttitle = {{{pyC}} 2{$<$}math{$><$}msup Is="true"{$><$}mrow Is="true"{$><$}/Mrow{$><$}mrow Is="true"{$><$}mn Is="true"{$>$}2{$<$}/Mn{$><$}/Mrow{$><$}/Msup{$><$}/Math{$>$} {{Ray}}},
  author = {Hirling, P. and Bianco, M. and Giri, S. K. and Iliev, I. T. and Mellema, G. and Kneib, J. -P.},
  year = 2024,
  month = jul,
  journal = {Astronomy and Computing},
  volume = {48},
  pages = {100861},
  issn = {2213-1337},
  url = {https://www.sciencedirect.com/science/article/pii/S2213133724000763},
  urldate = {2025-05-15}
}

@article{Hogbom1974,
  title = {Aperture {{Synthesis}} with a {{Non-Regular Distribution}} of {{Interferometer Baselines}}},
  author = {H{\"o}gbom, J. A.},
  year = 1974,
  month = jun,
  journal = {Astronomy and Astrophysics Supplement Series},
  volume = {15},
  pages = {417},
  issn = {0365-01380004-6361},
  url = {https://ui.adsabs.harvard.edu/abs/1974A&AS...15..417H},
  urldate = {2025-04-18}
}

@article{Hunter2007,
  title = {Matplotlib: {{A 2D Graphics Environment}}},
  shorttitle = {Matplotlib},
  author = {Hunter, John D.},
  year = 2007,
  month = may,
  journal = {Computing in Science and Engineering},
  volume = {9},
  pages = {90--95},
  url = {https://ui.adsabs.harvard.edu/abs/2007CSE.....9...90H},
  urldate = {2021-07-26}
}

@article{Hurley-Walker2016,
  title = {{{GaLactic}} and {{Extragalactic All-sky Murchison Widefield Array}} ({{GLEAM}}) Survey {{I}}: {{A}} Low-Frequency Extragalactic Catalogue},
  author = {{Hurley-Walker}, Natasha and Callingham, Joseph R. and Hancock, Paul J. and Franzen, Thomas M. O. and Hindson, Luke and Kapinska, Anna D. and Morgan, John and Offringa, Andre R. and Wayth, Randall B. and Wu, Chen and Zheng, Q. and Murphy, Tara and Bell, Martin E. and Dwarakanath, K. S. and For, Bi-Qing and Gaensler, Bryan M. and {Johnston-Hollitt}, Melanie and Lenc, Emil and Procopio, Pietro and {Staveley-Smith}, Lister and Ekers, Ron and Bowman, Judd D. and Briggs, Frank and Cappallo, R. J. and Deshpande, Avinash A. and Greenhill, Lincoln and Hazelton, Brynah J. and Kaplan, David L. and Lonsdale, Colin J. and McWhirter, S. R. and Mitchell, Daniel A. and Morales, Miguel F. and Morgan, Edward and Oberoi, Divya and Ord, Stephen M. and Prabu, T. and Shankar, N. Udaya and Srivani, K. S. and Subrahmanyan, Ravi and Tingay, Steven J. and Webster, Rachel L. and Williams, Andrew and Williams, Christopher L.},
  year = 2016,
  month = oct,
  journal = {Monthly Notices of the Royal Astronomical Society},
  volume = {464},
  pages = {1146--1167},
  url = {http://arxiv.org/abs/1610.08318}
}

@article{Jacobs2016,
  title = {The {{Murchison Widefield Array}} 21 Cm {{Power Spectrum Analysis Methodology}}},
  author = {Jacobs, Daniel C. and Hazelton, B. J. and Trott, C. M. and Dillon, Joshua S. and Pindor, B. and Sullivan, I. S. and Pober, J. C. and Barry, N. and Beardsley, A. P. and Bernardi, G. and Bowman, Judd D. and Briggs, F. and Cappallo, R. J. and Carroll, P. and Corey, B. E. and {de Oliveira-Costa}, A. and Emrich, D. and {Ewall-Wice}, A. and Feng, L. and Gaensler, B. M. and Goeke, R. and Greenhill, L. J. and Hewitt, J. N. and {Hurley-Walker}, N. and {Johnston-Hollitt}, M. and Kaplan, D. L. and Kasper, J. C. and Kim, H. S. and Kratzenberg, E. and Lenc, E. and Line, J. and Loeb, A. and Lonsdale, C. J. and Lynch, M. J. and McKinley, B. and McWhirter, S. R. and Mitchell, D. A. and Morales, M. F. and Morgan, E. and Neben, A. R. and Thyagarajan, N. and Oberoi, D. and Offringa, A. R. and Ord, S. M. and Paul, S. and Prabu, T. and Procopio, P. and Riding, J. and Rogers, A. E. E. and Roshi, A. and Shankar, N. Udaya and Sethi, Shiv K. and Srivani, K. S. and Subrahmanyan, R. and Tegmark, M. and Tingay, S. J. and Waterson, M. and Wayth, R. B. and Webster, R. L. and Whitney, A. R. and Williams, A. and Williams, C. L. and Wu, C. and Wyithe, J. S. B.},
  year = 2016,
  month = may,
  journal = {The Astrophysical Journal},
  volume = {825},
  url = {http://arxiv.org/abs/1605.06978}
}

@misc{Josaitis2021,
  title = {Array {{Element Coupling}} in {{Radio Interferometry I}}: {{A Semi-Analytic Approach}}},
  shorttitle = {Array {{Element Coupling}} in {{Radio Interferometry I}}},
  author = {Josaitis, Alec T. and {Ewall-Wice}, Aaron and Fagnoni, Nicolas and {de Lera Acedo}, Eloy},
  year = 2021,
  month = oct,
  journal = {arXiv e-prints},
  url = {https://ui.adsabs.harvard.edu/abs/2021arXiv211010879J},
  urldate = {2021-11-11}
}

@article{Kaur2022,
  title = {The 21-Cm Signal from the Cosmic Dawn: Metallicity Dependence of High-Mass {{X-ray}} Binaries},
  shorttitle = {The 21-Cm Signal from the Cosmic Dawn},
  author = {Kaur, Harman Deep and Qin, Yuxiang and Mesinger, Andrei and Pallottini, Andrea and Fragos, Tassos and {Basu-Zych}, Antara},
  year = 2022,
  month = jul,
  journal = {Monthly Notices of the Royal Astronomical Society},
  volume = {513},
  number = {4},
  pages = {5097--5108},
  issn = {0035-8711},
  url = {https://doi.org/10.1093/mnras/stac1226},
  urldate = {2023-11-07}
}

@article{Keating2025,
  title = {Pyuvdata v3: An Interface for Astronomical Interferometric Data Sets in {{Python}}},
  shorttitle = {Pyuvdata V3},
  author = {Keating, Garrett and Hazelton, Bryna and Kolopanis, Matthew and Murray, Steven and Beardsley, Adam and Jacobs, Daniel and Kern, Nicholas and Lanman, Adam and La Plante, Paul and Pober, Jonathan and Star, Pyxie},
  year = 2025,
  month = may,
  journal = {The Journal of Open Source Software},
  volume = {10},
  pages = {7482},
  url = {https://ui.adsabs.harvard.edu/abs/2025JOSS...10.7482K},
  urldate = {2025-07-08}
}

@article{Kern2017,
  title = {Emulating {{Simulations}} of {{Cosmic Dawn}} for 21cm {{Power Spectrum Constraints}} on {{Cosmology}}, {{Reionization}}, and {{X-ray Heating}}},
  author = {Kern, Nicholas S. and Liu, Adrian and Parsons, Aaron R. and Mesinger, Andrei and Greig, Bradley},
  year = 2017,
  month = may,
  journal = {arXiv:1705.04688},
  eprint = {1705.04688},
  url = {http://adsabs.harvard.edu/abs/2017arXiv170504688K},
  urldate = {2017-10-07},
  archiveprefix = {arXiv}
}

@article{Kern2019,
  title = {Mitigating {{Internal Instrument Coupling}} for 21 Cm {{Cosmology}}. {{I}}. {{Temporal}} and {{Spectral Modeling}} in {{Simulations}}},
  author = {Kern, Nicholas S. and Parsons, Aaron R. and Dillon, Joshua S. and Lanman, Adam E. and Fagnoni, Nicolas and {de Lera Acedo}, Eloy},
  year = 2019,
  month = oct,
  journal = {The Astrophysical Journal},
  volume = {884},
  pages = {105},
  publisher = {IOP},
  issn = {0004-637X},
  url = {https://ui.adsabs.harvard.edu/abs/2019ApJ...884..105K},
  urldate = {2025-05-20}
}

@article{Kern2020,
  ids = {Kern2020b},
  title = {Absolute {{Calibration Strategies}} for the {{Hydrogen Epoch}} of {{Reionization Array}} and {{Their Impact}} on the 21 Cm {{Power Spectrum}}},
  author = {Kern, Nicholas S. and Dillon, Joshua S. and Parsons, Aaron R. and Carilli, Christopher L. and Bernardi, Gianni and Abdurashidova, Zara and Aguirre, James E. and Alexander, Paul and Ali, Zaki S. and Balfour, Yanga and Beardsley, Adam P. and Billings, Tashalee S. and Bowman, Judd D. and Bradley, Richard F. and Bull, Philip and Burba, Jacob and Carey, Steven and Cheng, Carina and DeBoer, David R. and Dexter, Matt and {de Lera Acedo}, Eloy and Ely, John and {Ewall-Wice}, Aaron and Fagnoni, Nicolas and Fritz, Randall and Furlanetto, Steve R. and {Gale-Sides}, Kingsley and Glendenning, Brian and Gorthi, Deepthi and Greig, Bradley and Grobbelaar, Jasper and Halday, Ziyaad and Hazelton, Bryna J. and Hewitt, Jacqueline N. and Hickish, Jack and Jacobs, Daniel C. and Julius, Austin and Kerrigan, Joshua and Kittiwisit, Piyanat and Kohn, Saul A. and Kolopanis, Matthew and Lanman, Adam and La Plante, Paul and Lekalake, Telalo and Liu, Adrian and MacMahon, David and Malan, Lourence and Malgas, Cresshim and Maree, Matthys and Martinot, Zachary E. and Matsetela, Eunice and Mesinger, Andrei and Molewa, Mathakane and Morales, Miguel F. and Mosiane, Tshegofalang and Murray, Steven G. and Neben, Abraham R. and Nikolic, Bojan and Nunhokee, Chuneeta D. and Patra, Nipanjana and Pieterse, Samantha and Pober, Jonathan C. and {Razavi-Ghods}, Nima and Ringuette, Jon and Robnett, James and Rosie, Kathryn and Sims, Peter and Smith, Craig and Syce, Angelo and Thyagarajan, Nithyanandan and Williams, Peter K. G. and Zheng, Haoxuan},
  year = 2020,
  month = feb,
  journal = {The Astrophysical Journal},
  volume = {890},
  pages = {122},
  issn = {0004-637X},
  url = {http://adsabs.harvard.edu/abs/2020ApJ...890..122K},
  urldate = {2020-11-16}
}

@article{Kern2020a,
  title = {Mitigating {{Internal Instrument Coupling}} for 21 Cm {{Cosmology}}. {{II}}. {{A Method Demonstration}} with the {{Hydrogen Epoch}} of {{Reionization Array}}},
  author = {Kern, Nicholas S. and Parsons, Aaron R. and Dillon, Joshua S. and Lanman, Adam E. and Liu, Adrian and Bull, Philip and {Ewall-Wice}, Aaron and Abdurashidova, Zara and Aguirre, James E. and Alexander, Paul and Ali, Zaki S. and Balfour, Yanga and Beardsley, Adam P. and Bernardi, Gianni and Bowman, Judd D. and Bradley, Richard F. and Burba, Jacob and Carilli, Chris L. and Cheng, Carina and DeBoer, David R. and Dexter, Matt and {de Lera Acedo}, Eloy and Fagnoni, Nicolas and Fritz, Randall and Furlanetto, Steve R. and Glendenning, Brian and Gorthi, Deepthi and Greig, Bradley and Grobbelaar, Jasper and Halday, Ziyaad and Hazelton, Bryna J. and Hewitt, Jacqueline N. and Hickish, Jack and Jacobs, Daniel C. and Julius, Austin and Kerrigan, Joshua and Kittiwisit, Piyanat and Kohn, Saul A. and Kolopanis, Matthew and La Plante, Paul and Lekalake, Telalo and MacMahon, David and Malan, Lourence and Malgas, Cresshim and Maree, Matthys and Martinot, Zachary E. and Matsetela, Eunice and Mesinger, Andrei and Molewa, Mathakane and Morales, Miguel F. and Mosiane, Tshegofalang and Murray, Steven G. and Neben, Abraham R. and Parsons, Aaron R. and Patra, Nipanjana and Pieterse, Samantha and Pober, Jonathan C. and {Razavi-Ghods}, Nima and Ringuette, Jon and Robnett, James and Rosie, Kathryn and Sims, Peter and Smith, Craig and Syce, Angelo and Thyagarajan, Nithyanandan and Williams, Peter K. G. and Zheng, Haoxuan},
  year = 2020,
  month = jan,
  journal = {The Astrophysical Journal},
  volume = {888},
  pages = {70},
  issn = {0004-637X},
  url = {http://adsabs.harvard.edu/abs/2020ApJ...888...70K},
  urldate = {2020-11-16}
}

@article{Kim2022,
  title = {The {{Impact}} of {{Beam Variations}} on {{Power Spectrum Estimation}} for 21 Cm {{Cosmology}}. {{I}}. {{Simulations}} of {{Foreground Contamination}} for {{HERA}}},
  author = {Kim, Honggeun and Nhan, Bang D. and Hewitt, Jacqueline N. and Kern, Nicholas S. and Dillon, Joshua S. and Acedo, Eloy de Lera and Dynes, Scott B. C. and Mahesh, Nivedita and Fagnoni, Nicolas and DeBoer, David R.},
  year = 2022,
  month = dec,
  journal = {The Astrophysical Journal},
  volume = {941},
  number = {2},
  pages = {207},
  publisher = {The American Astronomical Society},
  issn = {0004-637X},
  url = {https://dx.doi.org/10.3847/1538-4357/ac9eaf},
  urldate = {2023-11-02},
  langid = {english}
}

@article{Kim2023,
  title = {The {{Impact}} of {{Beam Variations}} on {{Power Spectrum Estimation}} for 21 Cm {{Cosmology}}. {{II}}. {{Mitigation}} of {{Foreground Systematics}} for {{HERA}}},
  author = {Kim, Honggeun and Kern, Nicholas S. and Hewitt, Jacqueline N. and Nhan, Bang D. and Dillon, Joshua S. and De Lera Acedo, Eloy and Dynes, Scott B. C. and Mahesh, Nivedita and Fagnoni, Nicolas and DeBoer, David R.},
  year = 2023,
  month = aug,
  journal = {The Astrophysical Journal},
  volume = {953},
  number = {2},
  pages = {136},
  issn = {0004-637X, 1538-4357},
  url = {https://iopscience.iop.org/article/10.3847/1538-4357/ace35e},
  urldate = {2025-07-04}
}

@article{Kittiwisit2025,
  title = {Matvis: A Matrix-Based Visibility Simulator for Fast Forward Modelling of Many-Element 21 Cm Arrays},
  shorttitle = {Matvis},
  author = {Kittiwisit, Piyanat and Murray, Steven G. and Garsden, Hugh and Bull, Philip and Wilensky, Michael J. and Cain, Christopher and Parsons, Aaron R. and Sipple, Jackson and Adams, Tyrone and Aguirre, James E. and Baartman, Rushelle and Beardsley, Adam P. and Berkhout, Lindsay M. and Bernardi, Gianni and Billings, Tashalee S. and Bowman, Judd D. and Bradley, Richard F. and Burba, Jacob and Carey, Steven and Carilli, Chris L. and Chen, Kai-Feng and Choudhuri, Samir and Cox, Tyler and DeBoer, David R. and Acedo, Eloy de Lera and Dexter, Matt and Dillon, Joshua S. and Eksteen, Nico and Ely, John and {Ewall-Wice}, Aaron and Fagnoni, Nicolas and Furlanetto, Steven R. and {Gale-Sides}, Kingsley and Gehlot, Bharat Kumar and Glendenning, Brian and Gorce, Adelie and Gorthi, Deepthi and Greig, Bradley and Grobbelaar, Jasper and Halday, Ziyaad and Hazelton, Bryna J. and Hewitt, Jacqueline N. and Hickish, Jack and Jacobs, Daniel C. and Josaitis, Alec and Kern, Nicholas S. and Kerrigan, Joshua and Kim, Honggeun and Kolopanis, Matthew and Lanman, Adam and Plante, Paul La and Liu, Adrian and Ma, Yin-Zhe and MacMahon, David H. E. and Malan, Lourence and Malgas, Cresshim and Malgas, Keith and Marero, Bradley and Martinot, Zachary E. and McBride, Lisa and Mesinger, Andrei and Molewa, Mathakane and Morales, Miguel F. and Mosiane, Tshegofalang and Nunhokee, Chuneeta Devi and Nuwegeld, Hans and Pascua, Robert and Qin, Yuxiang and Rath, Eleanor and {Razavi-Ghods}, Nima and Robnett, James and Santos, Mario G. and Sims, Peter and Singh, Saurabh and Storer, Dara and Swarts, Hilton and Tan, Jianrong and Thyagarajan, Nithyanandan and {van Wyngaarden}, Pieter and Xu, Zhilei and Zheng, Haoxuan},
  year = 2025,
  month = jan,
  journal = {RAS Techniques and Instruments},
  volume = {4},
  pages = {rzaf001},
  url = {https://ui.adsabs.harvard.edu/abs/2025RASTI...4....1K},
  urldate = {2025-04-18}
}

@inproceedings{Kluyver2016jupyter,
  title = {Jupyter {{Notebooks}} -- a Publishing Format for Reproducible Computational Workflows},
  booktitle = {Positioning and Power in Academic Publishing: {{Players}}, Agents and Agendas},
  author = {Kluyver, Thomas and {Ragan-Kelley}, Benjamin and P{\'e}rez, Fernando and Granger, Brian and Bussonnier, Matthias and Frederic, Jonathan and Kelley, Kyle and Hamrick, Jessica and Grout, Jason and Corlay, Sylvain and Ivanov, Paul and Avila, Dami{\'a}n and Abdalla, Safia and Willing, Carol},
  editor = {Loizides, F. and Schmidt, B.},
  year = 2016,
  pages = {87--90},
  publisher = {IOS Press}
}

@article{Kolopanis2019,
  title = {A {{Simplified}}, {{Lossless Reanalysis}} of {{PAPER-64}}},
  author = {Kolopanis, Matthew and Jacobs, Daniel C. and Cheng, Carina and Parsons, Aaron R. and Kohn, Saul A. and Pober, Jonathan C. and Aguirre, James E. and Ali, Zaki S. and Bernardi, Gianni and Bradley, Richard F. and Carilli, Chris L. and DeBoer, David R. and Dexter, Matthew R. and Dillon, Joshua S. and Kerrigan, Joshua and Klima, Pat and Liu, Adrian and MacMahon, David H. E. and Moore, David F. and Thyagarajan, Nithyanandan and Nunhokee, Chuneeta D. and Walbrugh, William P. and Walker, Andre},
  year = 2019,
  month = oct,
  journal = {The Astrophysical Journal},
  volume = {883},
  pages = {133},
  issn = {0004-637X},
  url = {http://adsabs.harvard.edu/abs/2019ApJ...883..133K},
  urldate = {2020-11-16}
}

@article{Kolopanis2023,
  title = {New {{EoR}} Power Spectrum Limits from {{MWA Phase II}} Using the Delay Spectrum Method and Novel Systematic Rejection},
  author = {Kolopanis, Matthew and Pober, Jonathan C. and Jacobs, Daniel C. and McGraw, Samantha},
  year = 2023,
  month = jun,
  journal = {Monthly Notices of the Royal Astronomical Society},
  volume = {521},
  pages = {5120--5138},
  publisher = {OUP},
  issn = {0035-8711},
  url = {https://ui.adsabs.harvard.edu/abs/2023MNRAS.521.5120K},
  urldate = {2025-04-17}
}

@article{LaPlante2021,
  title = {A {{Real Time Processing}} System for Big Data in Astronomy: {{Applications}} to {{HERA}}},
  shorttitle = {A {{Real Time Processing}} System for Big Data in Astronomy},
  author = {La Plante, P. and Williams, P. K. G. and Kolopanis, M. and Dillon, J. S. and Beardsley, A. P. and Kern, N. S. and Wilensky, M. and Ali, Z. S. and Abdurashidova, Z. and Aguirre, J. E. and Alexander, P. and Balfour, Y. and Bernardi, G. and Billings, T. S. and Bowman, J. D. and Bradley, R. F. and Bull, P. and Burba, J. and Carey, S. and Carilli, C. L. and Cheng, C. and DeBoer, D. R. and Dexter, M. and {de Lera Acedo}, E. and Ely, J. and {Ewall-Wice}, A. and Fagnoni, N. and Fritz, R. and Furlanetto, S. R. and {Gale-Sides}, K. and Glendenning, B. and Gorthi, D. and Greig, B. and Grobbelaar, J. and Halday, Z. and Hazelton, B. J. and Hewitt, J. N. and Hickish, J. and Jacobs, D. C. and Julius, A. and Kerrigan, J. and Kittiwisit, P. and Kohn, S. A. and Lanman, A. and Lekalake, T. and Lewis, D. and Liu, A. and MacMahon, D. and Malan, L. and Malgas, C. and Maree, M. and Martinot, Z. E. and Matsetela, E. and Mesinger, A. and Molewa, M. and Morales, M. F. and Mosiane, T. and Murray, S. and Neben, A. R. and Nikolic, B. and Parsons, A. R. and Pascua, R. and Patra, N. and Pieterse, S. and Pober, J. C. and {Razavi-Ghods}, N. and Ringuette, J. and Robnett, J. and Rosie, K. and Santos, M. G. and Sims, P. and Smith, C. and Syce, A. and Thyagarajan, N. and Zheng, H.},
  year = 2021,
  month = jul,
  journal = {Astronomy and Computing},
  volume = {36},
  pages = {100489},
  issn = {2213-1337},
  url = {https://www.sciencedirect.com/science/article/pii/S2213133721000433},
  urldate = {2025-07-08}
}

@article{Lazare2024,
  title = {{{HERA}} Bound on X-Ray Luminosity When Accounting for Population {{III}} Stars},
  author = {Lazare, Hovav and Sarkar, Debanjan and Kovetz, Ely D.},
  year = 2024,
  month = feb,
  journal = {Physical Review D},
  volume = {109},
  pages = {043523},
  publisher = {APS},
  issn = {1550-79980556-2821},
  url = {https://ui.adsabs.harvard.edu/abs/2024PhRvD.109d3523L},
  urldate = {2025-04-17}
}

@article{Leung2023,
  title = {{{NGDEEP Epoch}} 1: {{The Faint End}} of the {{Luminosity Function}} at z   9-12 from {{Ultradeep JWST Imaging}}},
  shorttitle = {{{NGDEEP Epoch}} 1},
  author = {Leung, Gene C. K. and Bagley, Micaela B. and Finkelstein, Steven L. and Ferguson, Henry C. and Koekemoer, Anton M. and {P{\'e}rez-Gonz{\'a}lez}, Pablo G. and Morales, Alexa and Kocevski, Dale D. and Yang, Guang and Somerville, Rachel S. and Wilkins, Stephen M. and Yung, L. Y. Aaron and Fujimoto, Seiji and Larson, Rebecca L. and Papovich, Casey and Pirzkal, Nor and Berg, Danielle A. and Lotz, Jennifer M. and Castellano, Marco and Ch{\'a}vez Ortiz, {\'O}scar A. and Cheng, Yingjie and Dickinson, Mark and Giavalisco, Mauro and Hathi, Nimish P. and Hutchison, Taylor A. and Jung, Intae and Kartaltepe, Jeyhan S. and Natarajan, Priyamvada and Rothberg, Barry},
  year = 2023,
  month = sep,
  journal = {The Astrophysical Journal},
  volume = {954},
  pages = {L46},
  publisher = {IOP},
  issn = {0004-637X},
  url = {https://ui.adsabs.harvard.edu/abs/2023ApJ...954L..46L},
  urldate = {2025-04-23}
}

@article{Leung2024,
  title = {Exploring the {{Nature}} of {{Little Red Dots}}: {{Constraints}} on {{AGN}} and {{Stellar Contributions}} from {{PRIMER MIRI Imaging}}},
  shorttitle = {Exploring the {{Nature}} of {{Little Red Dots}}},
  author = {Leung, Gene C. K. and Finkelstein, Steven L. and {P{\'e}rez-Gonz{\'a}lez}, Pablo G. and Morales, Alexa M. and Taylor, Anthony J. and Barro, Guillermo and Kocevski, Dale D. and Akins, Hollis B. and Carnall, Adam C. and Ch{\'a}vez Ortiz, {\'O}scar A. and Cleri, Nikko J. and Cullen, Fergus and Donnan, Callum T. and Dunlop, James S. and Ellis, Richard S. and Grogin, Norman A. and Hirschmann, Michaela and Koekemoer, Anton M. and Kokorev, Vasily and Lucas, Ray A. and McLeod, Derek J. and Papovich, Casey and Yung, L. Y. Aaron},
  year = 2024,
  month = nov,
  journal = {arXiv e-prints},
  pages = {arXiv:2411.12005},
  url = {https://ui.adsabs.harvard.edu/abs/2024arXiv241112005L/abstract},
  urldate = {2025-04-23},
  langid = {english}
}

@article{Li2019,
  title = {First {{Season MWA Phase II Epoch}} of {{Reionization Power Spectrum Results}} at {{Redshift}} 7},
  author = {Li, W. and Pober, J. C. and Barry, N. and Hazelton, B. J. and Morales, M. F. and Trott, C. M. and Lanman, A. and Wilensky, M. and Sullivan, I. and Beardsley, A. P. and Booler, T. and Bowman, J. D. and Byrne, R. and Crosse, B. and Emrich, D. and Franzen, T. M. O. and Hasegawa, K. and Horsley, L. and {Johnston-Hollitt}, M. and Jacobs, D. C. and Jordan, C. H. and Joseph, R. C. and Kaneuji, T. and Kaplan, D. L. and Kenney, D. and Kubota, K. and Line, J. and Lynch, C. and McKinley, B. and Mitchell, D. A. and Murray, S. and Pallot, D. and Pindor, B. and Rahimi, M. and Riding, J. and Sleap, G. and Steele, K. and Takahashi, K. and Tingay, S. J. and Walker, M. and Wayth, R. B. and Webster, R. L. and Williams, A. and Wu, C. and Wyithe, J. S. B. and Yoshiura, S. and Zheng, Q.},
  year = 2019,
  month = dec,
  journal = {The Astrophysical Journal},
  volume = {887},
  pages = {141},
  url = {http://adsabs.harvard.edu/abs/2019ApJ...887..141L},
  urldate = {2020-11-03}
}

@article{Line2024,
  title = {Verifying the {{Australian MWA EoR}} Pipeline {{I}}: 21-Cm Sky Model and Correlated Measurement Density},
  shorttitle = {Verifying the {{Australian MWA EoR}} Pipeline {{I}}},
  author = {Line, J. L. B. and Trott, C. M. and Cook, J. H. and Greig, B. and Barry, N. and Jordan, C. H.},
  year = 2024,
  month = apr,
  journal = {Publications of the Astronomical Society of Australia},
  volume = {41},
  pages = {e067},
  issn = {1323-3580},
  url = {https://ui.adsabs.harvard.edu/abs/2024PASA...41...67L},
  urldate = {2025-04-18}
}

@article{Line2025,
  title = {Verifying the {{Australian MWA EoR}} Pipeline {{II}}: {{Fundamental}} Limits of the {{AusEoRPipe}} and the Impact of Instrumental Effects},
  shorttitle = {Verifying the {{Australian MWA EoR}} Pipeline {{II}}},
  author = {Line, Jack Laurence Bramble and Trott, Cathryn and Barry, Nichole and Null, Dev and Jordan, Christopher H.},
  year = 2025,
  month = feb,
  journal = {Publications of the Astronomical Society of Australia},
  volume = {42},
  pages = {e024},
  issn = {1323-3580},
  url = {https://ui.adsabs.harvard.edu/abs/2025PASA...42...24L},
  urldate = {2025-04-18}
}

@article{Liu2010,
  title = {Precision Calibration of Radio Interferometers Using Redundant Baselines},
  author = {Liu, Adrian and Tegmark, Max and Morrison, Scott and Lutomirski, Andrew and Zaldarriaga, Matias},
  year = 2010,
  month = oct,
  journal = {Monthly Notices of the Royal Astronomical Society},
  volume = {408},
  pages = {1029--1050},
  publisher = {OUP},
  issn = {0035-8711},
  url = {https://ui.adsabs.harvard.edu/abs/2010MNRAS.408.1029L},
  urldate = {2025-04-17}
}

@article{Liu2014a,
  title = {Epoch of Reionization Window. {{I}}. {{Mathematical}} Formalism},
  author = {Liu, Adrian and Parsons, Aaron R. and Trott, Cathryn M.},
  year = 2014,
  month = jul,
  journal = {Physical Review D},
  volume = {90},
  pages = {023018},
  publisher = {APS},
  issn = {1550-79980556-2821},
  url = {https://ui.adsabs.harvard.edu/abs/2014PhRvD..90b3018L},
  urldate = {2025-04-18}
}

@article{Liu2016a,
  title = {Constraining Cosmology and Ionization History with Combined 21\,Cm Power Spectrum and Global Signal Measurements},
  author = {Liu, Adrian and Parsons, Aaron R.},
  year = 2016,
  month = apr,
  journal = {Monthly Notices of the Royal Astronomical Society},
  volume = {457},
  number = {2},
  pages = {1864--1877},
  issn = {0035-8711},
  url = {https://doi.org/10.1093/mnras/stw071},
  urldate = {2021-11-11}
}

@article{Liu2019e,
  title = {Reviving {{Millicharged Dark Matter}} for 21-Cm {{Cosmology}}},
  author = {Liu, Hongwan and Outmezguine, Nadav Joseph and Redigolo, Diego and Volansky, Tomer},
  year = 2019,
  month = dec,
  journal = {Physical Review D},
  volume = {100},
  number = {12},
  eprint = {1908.06986},
  primaryclass = {hep-ph},
  pages = {123011},
  issn = {2470-0010, 2470-0029},
  url = {http://arxiv.org/abs/1908.06986},
  urldate = {2025-05-07},
  archiveprefix = {arXiv}
}

@article{Liu2020c,
  ids = {Liu2019},
  title = {Data {{Analysis}} for {{Precision}} 21 Cm {{Cosmology}}},
  author = {Liu, Adrian and Shaw, J. Richard},
  year = 2020,
  month = jun,
  journal = {Publications of the Astronomical Society of the Pacific},
  volume = {132},
  number = {1012},
  eprint = {1907.08211},
  pages = {062001},
  issn = {0004-6280, 1538-3873},
  url = {http://arxiv.org/abs/1907.08211},
  urldate = {2020-07-10},
  archiveprefix = {arXiv}
}

@article{Loeb2006,
  title = {The Dark Ages of the {{Universe}}.},
  author = {Loeb, A.},
  year = 2006,
  journal = {Scientific American},
  volume = {295},
  number = {5},
  pages = {5.46},
  url = {https://ui.adsabs.harvard.edu/abs/2006SciAm.295e..46L/abstract},
  urldate = {2019-11-04},
  langid = {english}
}

@article{Madau2017,
  title = {Radiation {{Backgrounds}} at {{Cosmic Dawn}}: {{X-Rays}} from {{Compact Binaries}}},
  shorttitle = {Radiation {{Backgrounds}} at {{Cosmic Dawn}}},
  author = {Madau, Piero and Fragos, Tassos},
  year = 2017,
  month = may,
  journal = {The Astrophysical Journal},
  volume = {840},
  pages = {39},
  publisher = {IOP},
  issn = {0004-637X},
  url = {https://ui.adsabs.harvard.edu/abs/2017ApJ...840...39M},
  urldate = {2025-04-23}
}

@article{Maiolino2024,
  title = {{{JADES}}: {{The}} Diverse Population of Infant Black Holes at 4 {$<$} z {$<$} 11: {{Merging}}, Tiny, Poor, but Mighty},
  shorttitle = {{{JADES}}},
  author = {Maiolino, Roberto and Scholtz, Jan and {Curtis-Lake}, Emma and Carniani, Stefano and Baker, William and {de Graaff}, Anna and Tacchella, Sandro and {\"U}bler, Hannah and D'Eugenio, Francesco and Witstok, Joris and Curti, Mirko and Arribas, Santiago and Bunker, Andrew J. and Charlot, St{\'e}phane and Chevallard, Jacopo and Eisenstein, Daniel J. and Egami, Eiichi and Ji, Zhiyuan and Jones, Gareth C. and Lyu, Jianwei and Rawle, Tim and Robertson, Brant and Rujopakarn, Wiphu and Perna, Michele and Sun, Fengwu and Venturi, Giacomo and Williams, Christina C. and Willott, Chris},
  year = 2024,
  month = nov,
  journal = {Astronomy and Astrophysics},
  volume = {691},
  pages = {A145},
  publisher = {EDP},
  issn = {0004-6361},
  url = {https://ui.adsabs.harvard.edu/abs/2024A&A...691A.145M},
  urldate = {2025-05-20}
}

@article{Maiolino2025,
  title = {{{JWST}} Meets {{Chandra}}: A Large Population of {{Compton}} Thick, Feedback-Free, and Intrinsically {{X-ray}} Weak {{AGN}}, with a Sprinkle of {{SNe}}},
  shorttitle = {{{JWST}} Meets {{Chandra}}},
  author = {Maiolino, Roberto and Risaliti, Guido and Signorini, Matilde and Trefoloni, Bartolomeo and Juod{\v z}balis, Ignas and Scholtz, Jan and {\"U}bler, Hannah and D'Eugenio, Francesco and Carniani, Stefano and Fabian, Andy and Ji, Xihan and Mazzolari, Giovanni and Bertola, Elena and Brusa, Marcella and Bunker, Andrew J. and Charlot, Stephane and Comastri, Andrea and Cresci, Giovanni and DeCoursey, Christa Noel and Egami, Eiichi and Fiore, Fabrizio and Gilli, Roberto and Perna, Michele and Tacchella, Sandro and Venturi, Giacomo},
  year = 2025,
  month = apr,
  journal = {Monthly Notices of the Royal Astronomical Society},
  volume = {538},
  pages = {1921--1943},
  publisher = {OUP},
  issn = {0035-8711},
  url = {https://ui.adsabs.harvard.edu/abs/2025MNRAS.538.1921M},
  urldate = {2025-04-23}
}

@article{Mao2008,
  title = {How Accurately Can 21cm Tomography Constrain Cosmology?},
  author = {Mao, Yi and Tegmark, Max and McQuinn, Matthew and Zaldarriaga, Matias and Zahn, Oliver},
  year = 2008,
  month = jul,
  journal = {Physical Review D},
  volume = {78},
  pages = {023529},
  publisher = {APS},
  issn = {1550-79980556-2821},
  url = {https://ui.adsabs.harvard.edu/abs/2008PhRvD..78b3529M},
  urldate = {2025-04-16}
}

@phdthesis{Martinot2022,
  title = {Improvements in Interferometric Data Modeling for the New Era of Radio Cosmology},
  author = {Martinot, Zachary Egan},
  year = 2022,
  school = {University of Pennsylvania}
}

@article{Matthee2024,
  title = {Little {{Red Dots}}: {{An Abundant Population}} of {{Faint Active Galactic Nuclei}} at z {$\sim$} 5 {{Revealed}} by the {{EIGER}} and {{FRESCO JWST Surveys}}},
  shorttitle = {Little {{Red Dots}}},
  author = {Matthee, Jorryt and Naidu, Rohan P. and Brammer, Gabriel and Chisholm, John and Eilers, Anna-Christina and Goulding, Andy and Greene, Jenny and Kashino, Daichi and Labbe, Ivo and Lilly, Simon J. and Mackenzie, Ruari and Oesch, Pascal A. and Weibel, Andrea and Wuyts, Stijn and Xiao, Mengyuan and Bordoloi, Rongmon and Bouwens, Rychard and {van Dokkum}, Pieter and Illingworth, Garth and Kramarenko, Ivan and Maseda, Michael V. and Mason, Charlotte and Meyer, Romain A. and Nelson, Erica J. and Reddy, Naveen A. and Shivaei, Irene and Simcoe, Robert A. and Yue, Minghao},
  year = 2024,
  month = mar,
  journal = {The Astrophysical Journal},
  volume = {963},
  pages = {129},
  publisher = {IOP},
  issn = {0004-637X},
  url = {https://ui.adsabs.harvard.edu/abs/2024ApJ...963..129M},
  urldate = {2025-04-23}
}

@article{McGreer2015,
  title = {Model-Independent Evidence in Favour of an End to Reionization by z~{$\approx~$}6},
  author = {McGreer, Ian D. and Mesinger, Andrei and D'Odorico, Valentina},
  year = 2015,
  month = feb,
  journal = {Monthly Notices of the Royal Astronomical Society},
  volume = {447},
  number = {1},
  pages = {499--505},
  issn = {0035-8711},
  url = {https://doi.org/10.1093/mnras/stu2449},
  urldate = {2025-05-20}
}

@article{McKinley2015,
  title = {Modelling of the Spectral Energy Distribution of {{Fornax A}}: Leptonic and Hadronic Production of High-Energy Emission from the Radio Lobes},
  shorttitle = {Modelling of the Spectral Energy Distribution of {{Fornax A}}},
  author = {McKinley, B. and Yang, R. and {L{\'o}pez-Caniego}, M. and Briggs, F. and {Hurley-Walker}, N. and Wayth, R. B. and Offringa, A. R. and Crocker, R. and Bernardi, G. and Procopio, P. and Gaensler, B. M. and Tingay, S. J. and {Johnston-Hollitt}, M. and McDonald, M. and Bell, M. and Bhat, N. D. R. and Bowman, J. D. and Cappallo, R. J. and Corey, B. E. and Deshpande, A. A. and Emrich, D. and {Ewall-Wice}, A. and Feng, L. and Goeke, R. and Greenhill, L. J. and Hazelton, B. J. and Hewitt, J. N. and Hindson, L. and Jacobs, D. and Kaplan, D. L. and Kasper, J. C. and Kratzenberg, E. and Kudryavtseva, N. and Lenc, E. and Lonsdale, C. J. and Lynch, M. J. and McWhirter, S. R. and Mitchell, D. A. and Morales, M. F. and Morgan, E. and Oberoi, D. and Ord, S. M. and Pindor, B. and Prabu, T. and Riding, J. and Rogers, A. E. E. and Roshi, D. A. and Udaya Shankar, N. and Srivani, K. S. and Subrahmanyan, R. and Waterson, M. and Webster, R. L. and Whitney, A. R. and Williams, A. and Williams, C. L.},
  year = 2015,
  month = feb,
  journal = {Monthly Notices of the Royal Astronomical Society},
  volume = {446},
  number = {4},
  pages = {3478--3491},
  issn = {0035-8711},
  url = {https://doi.org/10.1093/mnras/stu2310},
  urldate = {2025-04-18}
}

@article{Mertens2020,
  title = {Improved Upper Limits on the 21 Cm Signal Power Spectrum of Neutral Hydrogen at z {$\approx$} 9.1 from {{LOFAR}}},
  author = {Mertens, F. G. and Mevius, M. and Koopmans, L. V. E. and Offringa, A. R. and Mellema, G. and Zaroubi, S. and Brentjens, M. A. and Gan, H. and Gehlot, B. K. and Pandey, V. N. and Sardarabadi, A. M. and Vedantham, H. K. and Yatawatta, S. and Asad, K. M. B. and Ciardi, B. and Chapman, E. and Gazagnes, S. and Ghara, R. and Ghosh, A. and Giri, S. K. and Iliev, I. T. and Jeli{\'c}, V. and Kooistra, R. and Mondal, R. and Schaye, J. and Silva, M. B.},
  year = 2020,
  month = apr,
  journal = {Monthly Notices of the Royal Astronomical Society},
  volume = {493},
  pages = {1662--1685},
  publisher = {OUP},
  issn = {0035-8711},
  url = {https://ui.adsabs.harvard.edu/abs/2020MNRAS.493.1662M},
  urldate = {2025-04-17}
}

@misc{Mertens2025,
  title = {Deeper Multi-Redshift Upper Limits on the {{Epoch}} of {{Reionization}} 21-Cm Signal Power Spectrum from {{LOFAR}} between Z=8.3 and Z=10.1},
  author = {Mertens, F. G. and Mevius, M. and Koopmans, L. V. E. and Offringa, A. R. and Zaroubi, S. and Acharya, A. and Brackenhoff, S. A. and Ceccotti, E. and Chapman, E. and Chege, K. and Ciardi, B. and Ghara, R. and Ghosh, S. and Giri, S. K. and Hothi, I. and H{\"o}fer, C. and Iliev, I. T. and Jeli{\'c}, V. and Ma, Q. and Mellema, G. and Munshi, S. and Pandey, V. N. and Yatawatta, S.},
  year = 2025,
  month = mar,
  publisher = {arXiv},
  url = {https://ui.adsabs.harvard.edu/abs/2025arXiv250305576M},
  urldate = {2025-04-17}
}

@article{Mesinger2010,
  title = {{{21cmFAST}}: {{A Fast}}, {{Semi-Numerical Simulation}} of the {{High-Redshift}} 21-Cm {{Signal}}},
  author = {Mesinger, Andrei and Furlanetto, Steven and Cen, Renyue},
  year = 2010,
  month = mar,
  journal = {Monthly Notices of the Royal Astronomical Society},
  volume = {411},
  pages = {955--972},
  url = {http://arxiv.org/abs/1003.3878}
}

@article{Mesinger2014,
  title = {Reionization and beyond: Detecting the Peaks of the Cosmological 21 Cm Signal},
  shorttitle = {Reionization and Beyond},
  author = {Mesinger, Andrei and {Ewall-Wice}, Aaron and Hewitt, Jacqueline},
  year = 2014,
  month = apr,
  journal = {Monthly Notices of the Royal Astronomical Society},
  volume = {439},
  pages = {3262--3274},
  publisher = {OUP},
  issn = {0035-8711},
  url = {https://ui.adsabs.harvard.edu/abs/2014MNRAS.439.3262M},
  urldate = {2025-04-23}
}

@book{Mesinger2016,
  title = {Understanding the {{Epoch}} of {{Cosmic Reionization}}: {{Challenges}} and {{Progress}}},
  shorttitle = {Understanding the {{Epoch}} of {{Cosmic Reionization}}},
  editor = {Mesinger, Andrei},
  year = 2016,
  series = {Astrophysics and {{Space Science Library}}},
  publisher = {Springer International Publishing},
  url = {//www.springer.com/us/book/9783319219561},
  urldate = {2018-10-06},
  isbn = {978-3-319-21956-1},
  langid = {english}
}

@article{Mirocha2019a,
  ids = {Mirocha2018},
  title = {What Does the First Highly Redshifted 21-Cm Detection Tell Us about Early Galaxies?},
  author = {Mirocha, Jordan and Furlanetto, Steven R.},
  year = 2019,
  month = feb,
  journal = {Monthly Notices of the Royal Astronomical Society},
  volume = {483},
  number = {2},
  pages = {1980},
  url = {https://ui.adsabs.harvard.edu/abs/2019MNRAS.483.1980M/abstract},
  urldate = {2019-11-01},
  langid = {english}
}

@article{Mirocha2022,
  title = {A Galaxy-Free Phenomenological Model for the 21-Cm Power Spectrum during Reionization},
  author = {Mirocha, Jordan and Mu{\~n}oz, Julian B. and Furlanetto, Steven R. and Liu, Adrian and Mesinger, Andrei},
  year = 2022,
  month = aug,
  journal = {Monthly Notices of the Royal Astronomical Society},
  volume = {514},
  pages = {2010--2030},
  publisher = {OUP},
  issn = {0035-8711},
  url = {https://ui.adsabs.harvard.edu/abs/2022MNRAS.514.2010M},
  urldate = {2025-05-20}
}

@article{Mittal2021,
  title = {Lyman-\${$\alpha\$$} Coupling and Heating at {{Cosmic Dawn}}},
  author = {Mittal, Shikhar and Kulkarni, Girish},
  year = 2021,
  month = apr,
  journal = {Monthly Notices of the Royal Astronomical Society},
  volume = {503},
  number = {3},
  eprint = {2009.10746},
  primaryclass = {astro-ph},
  pages = {4264--4275},
  issn = {0035-8711, 1365-2966},
  url = {http://arxiv.org/abs/2009.10746},
  urldate = {2025-05-07},
  archiveprefix = {arXiv}
}

@article{Monsalve2021,
  title = {Absolute {{Calibration}} of {{Diffuse Radio Surveys}} at 45 and 150 {{MHz}}},
  author = {Monsalve, Raul A. and Rogers, Alan E. E. and Bowman, Judd D. and Mahesh, Nivedita and Murray, Steven G. and Mozdzen, Thomas J. and Johnson, Leroy and Barrett, John and Samson, Titu and Lewis, David},
  year = 2021,
  month = feb,
  journal = {The Astrophysical Journal},
  volume = {908},
  pages = {145},
  publisher = {IOP},
  issn = {0004-637X},
  url = {https://ui.adsabs.harvard.edu/abs/2021ApJ...908..145M},
  urldate = {2025-07-09}
}

@article{Monsalve2024,
  title = {Mapper of the {{IGM}} Spin Temperature: Instrument Overview},
  shorttitle = {Mapper of the {{IGM}} Spin Temperature},
  author = {Monsalve, R. A. and Altamirano, C. and Bidula, V. and Bustos, R. and Bye, C. H. and Chiang, H. C. and D{\'i}az, M. and Fern{\'a}ndez, B. and Guo, X. and Hendricksen, I. and Hornecker, E. and Lucero, F. and Mani, H. and McGee, F. and Mena, F. P. and Pess{\^o}a, M. and Prabhakar, G. and Restrepo, O. and Sievers, J. L. and Thyagarajan, N.},
  year = 2024,
  month = jun,
  journal = {Monthly Notices of the Royal Astronomical Society},
  volume = {530},
  pages = {4125--4147},
  publisher = {OUP},
  issn = {0035-8711},
  url = {https://ui.adsabs.harvard.edu/abs/2024MNRAS.530.4125M},
  urldate = {2025-04-17}
}

@article{Moore2017,
  title = {Limits on {{Polarized Leakage}} for the {{PAPER Epoch}} of {{Reionization Measurements}} at 126 and 164 {{MHz}}},
  author = {Moore, David F. and Aguirre, James E. and Kohn, Saul A. and Parsons, Aaron R. and Ali, Zaki S. and Bradley, Richard F. and Carilli, Chris L. and DeBoer, David R. and Dexter, Matthew R. and Gugliucci, Nicole E. and Jacobs, Daniel C. and Klima, Pat and Liu, Adrian and MacMahon, David H. E. and Manley, Jason R. and Pober, Jonathan C. and Stefan, Irina I. and Walbrugh, William P.},
  year = 2017,
  month = feb,
  journal = {The Astrophysical Journal},
  volume = {836},
  number = {2},
  pages = {154},
  publisher = {The American Astronomical Society},
  issn = {0004-637X},
  url = {https://dx.doi.org/10.3847/1538-4357/836/2/154},
  urldate = {2025-04-18},
  langid = {english}
}

@article{Morales2010,
  title = {Reionization and {{Cosmology}} with 21-Cm {{Fluctuations}}},
  author = {Morales, Miguel F. and Wyithe, J. Stuart B.},
  year = 2010,
  month = sep,
  journal = {Annual Review of Astronomy and Astrophysics},
  volume = {48},
  pages = {127--171},
  issn = {0066-4146},
  url = {http://adsabs.harvard.edu/abs/2010ARA%26A..48..127M},
  urldate = {2018-08-27}
}

@article{Morales2019,
  title = {Understanding the Diversity of 21 Cm Cosmology Analyses},
  author = {Morales, Miguel F and Beardsley, Adam and Pober, Jonathan and Barry, Nichole and Hazelton, Bryna and Jacobs, Daniel and Sullivan, Ian},
  year = 2019,
  month = feb,
  journal = {Monthly Notices of the Royal Astronomical Society},
  volume = {483},
  number = {2},
  pages = {2207--2216},
  issn = {0035-8711, 1365-2966},
  url = {https://academic.oup.com/mnras/article/483/2/2207/5144780},
  urldate = {2025-05-20},
  copyright = {https://academic.oup.com/journals/pages/open\_access/funder\_policies/chorus/standard\_publication\_model},
  langid = {english}
}

@article{Morales2023,
  title = {The Statistics of Negative Power Spectrum Systematics in Some 21 Cm Analyses},
  author = {Morales, Miguel F. and Pober, Jonathan and Hazelton, Bryna J.},
  year = 2023,
  month = oct,
  journal = {Monthly Notices of the Royal Astronomical Society},
  volume = {525},
  pages = {2834--2838},
  publisher = {OUP},
  issn = {0035-8711},
  url = {https://ui.adsabs.harvard.edu/abs/2023MNRAS.525.2834M},
  urldate = {2025-07-07}
}

@article{Munoz2018,
  ids = {Munoz2018b},
  title = {21-Cm {{Fluctuations}} from {{Charged Dark Matter}}},
  author = {Mu{\~n}oz, Julian B. and Dvorkin, Cora and Loeb, Abraham},
  year = 2018,
  month = sep,
  journal = {Physical Review Letters},
  volume = {121},
  number = {12},
  pages = {121301},
  url = {https://ui.adsabs.harvard.edu/abs/2018PhRvL.121l1301M/abstract},
  urldate = {2019-11-01},
  langid = {english}
}

@article{Munoz2018a,
  title = {A Small Amount of Mini-Charged Dark Matter Could Cool the Baryons in the Early {{Universe}}},
  author = {Mu{\~n}oz, Julian B. and Loeb, Abraham},
  year = 2018,
  month = may,
  journal = {Nature},
  volume = {557},
  pages = {684--686},
  issn = {0028-0836},
  url = {http://adsabs.harvard.edu/abs/2018Natur.557..684M},
  urldate = {2020-11-13}
}

@article{Munoz2022,
  title = {The Impact of the First Galaxies on Cosmic Dawn and Reionization},
  author = {Mu{\~n}oz, Julian B. and Qin, Yuxiang and Mesinger, Andrei and Murray, Steven G. and Greig, Bradley and Mason, Charlotte},
  year = 2022,
  month = apr,
  journal = {Monthly Notices of the Royal Astronomical Society},
  volume = {511},
  pages = {3657--3681},
  issn = {0035-8711},
  url = {https://ui.adsabs.harvard.edu/abs/2022MNRAS.511.3657M},
  urldate = {2024-01-31}
}

@article{Munoz2023,
  title = {Breaking Degeneracies in the First Galaxies with Clustering},
  author = {Mu{\~n}oz, Julian B. and Mirocha, Jordan and Furlanetto, Steven and Sabti, Nashwan},
  year = 2023,
  month = aug,
  journal = {Monthly Notices of the Royal Astronomical Society: Letters},
  volume = {526},
  number = {1},
  eprint = {2306.09403},
  primaryclass = {astro-ph},
  pages = {L47-L55},
  issn = {1745-3925, 1745-3933},
  url = {http://arxiv.org/abs/2306.09403},
  urldate = {2025-05-07},
  archiveprefix = {arXiv}
}

@article{Munoz2023a,
  title = {An {{Effective Model}} for the {{Cosmic-Dawn}} 21-Cm {{Signal}}},
  author = {Mu{\~n}oz, Julian B.},
  year = 2023,
  month = may,
  journal = {Monthly Notices of the Royal Astronomical Society},
  volume = {523},
  number = {2},
  eprint = {2302.08506},
  primaryclass = {astro-ph},
  pages = {2587--2607},
  issn = {0035-8711, 1365-2966},
  url = {http://arxiv.org/abs/2302.08506},
  urldate = {2025-05-07},
  archiveprefix = {arXiv}
}

@article{Munoz2024,
  title = {Reionization after {{JWST}}: A Photon Budget Crisis?},
  shorttitle = {Reionization after {{JWST}}},
  author = {Mu{\~n}oz, Julian B. and Mirocha, Jordan and Chisholm, John and Furlanetto, Steven R. and Mason, Charlotte},
  year = 2024,
  month = nov,
  journal = {Monthly Notices of the Royal Astronomical Society},
  volume = {535},
  pages = {L37-L43},
  publisher = {OUP},
  issn = {0035-8711},
  url = {https://ui.adsabs.harvard.edu/abs/2024MNRAS.535L..37M},
  urldate = {2025-04-23}
}

@article{Munshi2024,
  title = {First Upper Limits on the 21 Cm Signal Power Spectrum from Cosmic Dawn from One Night of Observations with {{NenuFAR}}},
  author = {Munshi, S. and Mertens, F. G. and Koopmans, L. V. E. and Offringa, A. R. and Semelin, B. and Aubert, D. and Barkana, R. and Bracco, A. and Brackenhoff, S. A. and Cecconi, B. and Ceccotti, E. and Corbel, S. and Fialkov, A. and Gehlot, B. K. and Ghara, R. and Girard, J. N. and Grie{\ss}meier, J. M. and H{\"o}fer, C. and Hothi, I. and M{\'e}riot, R. and Mevius, M. and Ocvirk, P. and Shaw, A. K. and Theureau, G. and Yatawatta, S. and Zarka, P. and Zaroubi, S.},
  year = 2024,
  month = jan,
  journal = {Astronomy \& Astrophysics},
  volume = {681},
  eprint = {2311.05364},
  primaryclass = {astro-ph},
  pages = {A62},
  issn = {0004-6361, 1432-0746},
  url = {http://arxiv.org/abs/2311.05364},
  urldate = {2025-05-15},
  archiveprefix = {arXiv}
}

@misc{Munshi2025,
  title = {Near-Field Imaging of Local Interference in Radio Interferometric Data: {{Impact}} on the Redshifted 21 Cm Power Spectrum},
  shorttitle = {Near-Field Imaging of Local Interference in Radio Interferometric Data},
  author = {Munshi, S. and Mertens, F. G. and Koopmans, L. V. E. and Mevius, M. and Offringa, A. R. and Semelin, B. and Viou, C. and Bracco, A. and Brackenhoff, S. A. and Ceccotti, E. and Chege, J. K. and Fialkov, A. and Gao, L. Y. and Ghara, R. and Ghosh, S. and Shaw, A. K. and Zarka, P. and Zaroubi, S. and Cecconi, B. and Corbel, S. and Girard, J. N. and Griessmeier, J. M. and Konovalenko, O. and Loh, A. and Tokarsky, P. and Ulyanov, O. and Zakharenko, V.},
  year = 2025,
  publisher = {arXiv},
  url = {https://arxiv.org/abs/2503.21728},
  urldate = {2025-05-12},
  copyright = {arXiv.org perpetual, non-exclusive license}
}

@misc{Munshi2025a,
  title = {Improved Upper Limits on the 21-Cm Signal Power Spectrum at \$z=17.0\$ and \$z=20.3\$ from an Optimal Field Observed with {{NenuFAR}}},
  author = {Munshi, S. and Mertens, F. G. and Chege, J. K. and Koopmans, L. V. E. and Offringa, A. R. and Semelin, B. and Barkana, R. and Dhandha, J. and Fialkov, A. and M{\textbackslash}{\'e}riot, R. and Sikder, S. and Bracco, A. and Brackenhoff, S. A. and Ceccotti, E. and Ghara, R. and Ghosh, S. and Hothi, I. and Mevius, M. and Ocvirk, P. and Shaw, A. K. and Yatawatta, S. and Zarka, P.},
  year = 2025,
  month = jul,
  publisher = {arXiv},
  url = {https://ui.adsabs.harvard.edu/abs/2025arXiv250710533M},
  urldate = {2025-08-05}
}

@techreport{Murray_hera130,
  type = {Memo},
  title = {Excess Variance of Visibilities and the Z-Squared Statistic},
  author = {Murray, S. G.},
  year = 2024,
  number = {130},
  address = {reionization.org/science/memos},
  institution = {Scuola Normale Superiore, Pisa, Italy},
  url = {http://reionization.org/manual_uploads/HERA130_statistics_of_visibilities.pdf}
}

@article{Murray2018c,
  title = {The {{Effect}} of {{Baseline Layouts}} on the {{Epoch}} of {{Reionization Foreground Wedge}}: {{A Semianalytical Approach}}},
  shorttitle = {The {{Effect}} of {{Baseline Layouts}} on the {{Epoch}} of {{Reionization Foreground Wedge}}},
  author = {Murray, Steven G. and Trott, C. M.},
  year = 2018,
  month = dec,
  journal = {The Astrophysical Journal},
  volume = {869},
  number = {1},
  pages = {25},
  url = {https://ui.adsabs.harvard.edu/abs/2018ApJ...869...25M/abstract},
  urldate = {2019-10-31},
  langid = {english}
}

@article{Murray2020b,
  title = {{{21cmFAST}} v3: {{A Python-integrated C}} Code for Generating {{3D}} Realizations of the Cosmic 21cm Signal.},
  shorttitle = {{{21cmFAST}} V3},
  author = {Murray, Steven and Greig, Bradley and Mesinger, Andrei and Mu{\~n}oz, Julian and Qin, Yuxiang and Park, Jaehong and Watkinson, Catherine},
  year = 2020,
  month = oct,
  journal = {The Journal of Open Source Software},
  volume = {5},
  pages = {2582},
  url = {http://adsabs.harvard.edu/abs/2020JOSS....5.2582M},
  urldate = {2020-11-16}
}

@article{Murray2022c,
  title = {A {{Bayesian}} Calibration Framework for {{EDGES}}},
  author = {Murray, Steven G. and Bowman, Judd D. and Sims, Peter H. and Mahesh, Nivedita and Rogers, Alan E. E. and Monsalve, Raul A. and Samson, Titu and Vydula, Akshatha Konakondula},
  year = 2022,
  month = dec,
  journal = {Monthly Notices of the Royal Astronomical Society},
  volume = {517},
  pages = {2264--2284},
  issn = {0035-8711},
  url = {https://ui.adsabs.harvard.edu/abs/2022MNRAS.517.2264M},
  urldate = {2023-11-02}
}

@article{Neben2016,
  title = {The {{Hydrogen Epoch}} of {{Reionization Array Dish I}}: {{Beam Pattern Measurements}} and {{Science Implications}}},
  shorttitle = {The {{Hydrogen Epoch}} of {{Reionization Array Dish I}}},
  author = {Neben, Abraham R. and Bradley, Richard F. and Hewitt, Jacqueline N. and DeBoer, David R. and Parsons, Aaron R. and Aguirre, James E. and Ali, Zaki S. and Cheng, Carina and {Ewall-Wice}, Aaron and Patra, Nipanjana and Thyagarajan, Nithyanandan and Bowman, Judd and Dickenson, Roger and Dillon, Joshua S. and Doolittle, Phillip and Egan, Dennis and Hedrick, Mike and Jacobs, Daniel C. and Kohn, Saul A. and Klima, Patricia J. and Moodley, Kavilan and Saliwanchik, Benjamin R. B. and Schaffner, Patrick and Shelton, John and Taylor, H. A. and Taylor, Rusty and Tegmark, Max and Wirt, Butch and Zheng, Haoxuan},
  year = 2016,
  month = aug,
  journal = {The Astrophysical Journal},
  volume = {826},
  number = {2},
  eprint = {1602.03887},
  primaryclass = {astro-ph},
  pages = {199},
  issn = {0004-637X, 1538-4357},
  url = {http://arxiv.org/abs/1602.03887},
  urldate = {2025-05-20},
  archiveprefix = {arXiv}
}

@article{Nikolic2024,
  title = {The Importance of Stochasticity in Determining Galaxy Emissivities and {{UV LFs}} during Cosmic Dawn and Reionization},
  author = {Nikoli{\'c}, Ivan and Mesinger, Andrei and Davies, James E. and Prelogovi{\'c}, David},
  year = 2024,
  month = dec,
  journal = {Astronomy and Astrophysics},
  volume = {692},
  pages = {A142},
  publisher = {EDP},
  issn = {0004-6361},
  url = {https://ui.adsabs.harvard.edu/abs/2024A&A...692A.142N},
  urldate = {2025-04-23}
}

@misc{Nunhokee2025,
  title = {Limits on the 21 Cm Power Spectrum at Z=6.5-7.0 from {{MWA}} Observations},
  author = {Nunhokee, C. D. and Null, D. and Trott, C. M. and Barry, N. and Qin, Y. and Wayth, R. B. and Line, J. L. B. and Jordan, C. H. and Pindor, B. and Cook, J. H. and Bowman, J. and Chokshi, A. and Ducharme, J. and Elder, K. and Guo, Q. and Hazelton, B. and Hidayat, W. and Ito, T. and Jacobs, D. and Jong, E. and Kolopanis, M. and Kunicki, T. and Lilleskov, E. and Morales, M. F. and Pober, J. C. and Selvaraj, A. and Shi, R. and Takahashi, K. and Tingay, S. J. and Webster, R. L. and Yoshiura, S. and Zheng, Q.},
  year = 2025,
  month = may,
  number = {arXiv:2505.09097},
  eprint = {2505.09097},
  primaryclass = {astro-ph},
  publisher = {arXiv},
  url = {http://arxiv.org/abs/2505.09097},
  urldate = {2025-05-15},
  archiveprefix = {arXiv}
}

@article{OHara2025,
  title = {Uncovering the Effects of Array Mutual Coupling in 21-Cm Experiments with the {{SKA-Low}} Radio Telescope},
  author = {O'Hara, Oscar S D and Gueuning, Quentin and {de~Lera~Acedo}, Eloy and Dulwich, Fred and Cumner, John and Anstey, Dominic and Brown, Anthony and Fialkov, Anastasia and Dhandha, Jiten and Faulkner, Andrew and Liu, Yuchen},
  year = 2025,
  month = mar,
  journal = {Monthly Notices of the Royal Astronomical Society},
  volume = {538},
  number = {1},
  pages = {31--48},
  issn = {0035-8711},
  url = {https://doi.org/10.1093/mnras/staf264},
  urldate = {2025-04-17}
}

@article{Orosz2019,
  title = {Mitigating the {{Effects}} of {{Antenna-to-Antenna Variation}} on {{Redundant-Baseline Calibration}} for 21 Cm {{Cosmology}}},
  author = {Orosz, Naomi and Dillon, Joshua S. and {Ewall-Wice}, Aaron and Parsons, Aaron R. and Thyagarajan, Nithyanandan},
  year = 2019,
  month = jul,
  journal = {Monthly Notices of the Royal Astronomical Society},
  volume = {487},
  number = {1},
  eprint = {1809.09728},
  primaryclass = {astro-ph},
  pages = {537--549},
  issn = {0035-8711, 1365-2966},
  url = {http://arxiv.org/abs/1809.09728},
  urldate = {2025-06-18},
  archiveprefix = {arXiv}
}

@inproceedings{Orru2024,
  title = {{{LOFAR}} vs {{LOFAR2}}.0 Operations: New Challenges},
  shorttitle = {{{LOFAR}} vs {{LOFAR2}}.0 Operations},
  booktitle = {Observatory {{Operations}}: {{Strategies}}, {{Processes}}, and {{Systems X}}},
  author = {Orr{\'u}, E. and Norden, M. J. and Iacobelli, M. and ter Veen, S. and Ahmadi, A.},
  year = 2024,
  month = jul,
  volume = {13098},
  pages = {238--248},
  publisher = {SPIE},
  url = {https://www.spiedigitallibrary.org/conference-proceedings-of-spie/13098/130980R/LOFAR-vs-LOFAR20-operations-new-challenges/10.1117/12.3019032.full},
  urldate = {2025-04-17}
}

@article{Paciga2011,
  title = {The {{GMRT Epoch}} of {{Reionization}} Experiment: A New Upper Limit on the Neutral Hydrogen Power Spectrum at Z{$\approx$} 8.6},
  shorttitle = {The {{GMRT Epoch}} of {{Reionization}} Experiment},
  author = {Paciga, Gregory and Chang, Tzu-Ching and Gupta, Yashwant and Nityanada, Rajaram and Odegova, Julia and Pen, Ue-Li and Peterson, Jeffrey B. and Roy, Jayanta and Sigurdson, Kris},
  year = 2011,
  month = may,
  journal = {Monthly Notices of the Royal Astronomical Society},
  volume = {413},
  number = {2},
  pages = {1174},
  url = {https://ui.adsabs.harvard.edu/abs/2011MNRAS.413.1174P/abstract},
  urldate = {2019-11-01},
  langid = {english}
}

@article{Paciga2013,
  title = {A Simulation-Calibrated Limit on the {{H}}\,i Power Spectrum from the {{GMRT Epoch}} of {{Reionization}} Experiment},
  author = {Paciga, Gregory and Albert, Joshua G. and Bandura, Kevin and Chang, Tzu-Ching and Gupta, Yashwant and Hirata, Christopher and Odegova, Julia and Pen, Ue-Li and Peterson, Jeffrey B. and Roy, Jayanta and Shaw, J. Richard and Sigurdson, Kris and Voytek, Tabitha},
  year = 2013,
  month = jul,
  journal = {Monthly Notices of the Royal Astronomical Society},
  volume = {433},
  number = {1},
  pages = {639--647},
  issn = {0035-8711},
  url = {https://doi.org/10.1093/mnras/stt753},
  urldate = {2025-05-15}
}

@article{Pagano2020,
  title = {Quantifying {{Density-Ionization Correlations}} with the 21cm {{Power Spectrum}}},
  author = {Pagano, Michael and Liu, Adrian},
  year = 2020,
  month = oct,
  journal = {Monthly Notices of the Royal Astronomical Society},
  volume = {498},
  number = {1},
  eprint = {2005.11326},
  primaryclass = {astro-ph},
  pages = {373--384},
  issn = {0035-8711, 1365-2966},
  url = {http://arxiv.org/abs/2005.11326},
  urldate = {2025-05-15},
  archiveprefix = {arXiv}
}

@article{Pagano2023a,
  title = {Characterization of Inpaint Residuals in Interferometric Measurements of the Epoch of Reionization},
  author = {Pagano, Michael and Liu, Jing and Liu, Adrian and Kern, Nicholas S and {Ewall-Wice}, Aaron and Bull, Philip and Pascua, Robert and Ravanbakhsh, Siamak and Abdurashidova, Zara and Adams, Tyrone and Aguirre, James E and Alexander, Paul and Ali, Zaki S and Baartman, Rushelle and Balfour, Yanga and Beardsley, Adam P and Bernardi, Gianni and Billings, Tashalee S and Bowman, Judd D and Bradley, Richard F and Burba, Jacob and Carey, Steven and Carilli, Chris L and Cheng, Carina and DeBoer, David R and {de~Lera~Acedo}, Eloy and Dexter, Matt and Dillon, Joshua S and Eksteen, Nico and Ely, John and Fagnoni, Nicolas and Fritz, Randall and Furlanetto, Steven R and {Gale-Sides}, Kingsley and Glendenning, Brian and Gorthi, Deepthi and Greig, Bradley and Grobbelaar, Jasper and Halday, Ziyaad and Hazelton, Bryna J and Hewitt, Jacqueline N and Hickish, Jack and Jacobs, Daniel C and Julius, Austin and Kariseb, MacCalvin and Kerrigan, Joshua and Kittiwisit, Piyanat and Kohn, Saul A and Kolopanis, Matthew and Lanman, Adam and La~Plante, Paul and Loots, Anita and MacMahon, David Harold Edward and Malan, Lourence and Malgas, Cresshim and Malgas, Keith and Marero, Bradley and Martinot, Zachary E and Mesinger, Andrei and Molewa, Mathakane and Morales, Miguel F and Mosiane, Tshegofalang and Neben, Abraham R and Nikolic, Bojan and Nuwegeld, Hans and Parsons, Aaron R and Patra, Nipanjana and Pieterse, Samantha and {Razavi-Ghods}, Nima and Robnett, James and Rosie, Kathryn and Sims, Peter and Smith, Craig and Swarts, Hilton and Thyagarajan, Nithyanandan and {van~Wyngaarden}, Pieter and Williams, Peter K G and Zheng, Haoxuan},
  year = 2023,
  month = apr,
  journal = {Monthly Notices of the Royal Astronomical Society},
  volume = {520},
  number = {4},
  pages = {5552--5572},
  issn = {0035-8711},
  url = {https://doi.org/10.1093/mnras/stad441},
  urldate = {2023-11-06}
}

@article{Park2019,
  title = {Inferring the Astrophysics of Reionization and Cosmic Dawn from Galaxy Luminosity Functions and the 21-Cm Signal},
  author = {Park, Jaehong and Mesinger, Andrei and Greig, Bradley and Gillet, Nicolas},
  year = 2019,
  month = mar,
  journal = {Monthly Notices of the Royal Astronomical Society},
  volume = {484},
  pages = {933--949},
  issn = {0035-8711},
  url = {https://ui.adsabs.harvard.edu/abs/2019MNRAS.484..933P},
  urldate = {2021-11-01}
}

@inproceedings{Parsons2006,
  title = {{{PetaOp}}/{{Second FPGA Signal Processing}} for {{SETI}} and {{Radio Astronomy}}},
  booktitle = {2006 {{Fortieth Asilomar Conference}} on {{Signals}}, {{Systems}} and {{Computers}}},
  author = {Parsons, Aaron and Backer, Donald and Chang, Chen and Chapman, Daniel and Chen, Henry and Crescini, Patrick and {de Jesus}, Christina and Dick, Chris and Droz, Pierre and MacMahon, David and Meder, Kirsten and Mock, Jeff and Nagpal, Vinayak and Nikolic, Borivoje and Parsa, Arash and Richards, Brian and Siemion, Andrew and Wawrzynek, John and Werthimer, Dan and Wright, Melvyn},
  year = 2006,
  month = oct,
  pages = {2031--2035},
  issn = {1058-6393},
  url = {https://ieeexplore.ieee.org/document/4176933},
  urldate = {2025-04-23}
}

@article{Parsons2008,
  title = {A {{Scalable Correlator Architecture Based}} on {{Modular FPGA Hardware}}, {{Reuseable Gateware}}, and {{Data Packetization}}},
  author = {Parsons, Aaron and Backer, Donald and Siemion, Andrew and Chen, Henry and Werthimer, Dan and Droz, Pierre and Filiba, Terry and Manley, Jason and McMahon, Peter and Parsa, Arash and MacMahon, David and Wright, Melvyn},
  year = 2008,
  month = nov,
  journal = {Publications of the Astronomical Society of the Pacific},
  volume = {120},
  pages = {1207},
  publisher = {IOP},
  issn = {0004-6280},
  url = {https://ui.adsabs.harvard.edu/abs/2008PASP..120.1207P},
  urldate = {2025-04-17}
}

@article{Parsons2010,
  title = {The {{Precision Array}} for {{Probing}} the {{Epoch}} of {{Re-ionization}}: {{Eight Station Results}}},
  shorttitle = {The {{Precision Array}} for {{Probing}} the {{Epoch}} of {{Re-ionization}}},
  author = {Parsons, Aaron R. and Backer, Donald C. and Foster, Griffin S. and Wright, Melvyn C. H. and Bradley, Richard F. and Gugliucci, Nicole E. and Parashare, Chaitali R. and Benoit, Erin E. and Aguirre, James E. and Jacobs, Daniel C. and Carilli, Chris L. and Herne, David and Lynch, Mervyn J. and Manley, Jason R. and Werthimer, Daniel J.},
  year = 2010,
  month = apr,
  journal = {The Astronomical Journal},
  volume = {139},
  pages = {1468--1480},
  publisher = {IOP},
  issn = {0004-6256},
  url = {https://ui.adsabs.harvard.edu/abs/2010AJ....139.1468P},
  urldate = {2025-04-17}
}

@article{Parsons2012,
  title = {A {{Per-baseline}}, {{Delay-spectrum Technique}} for {{Accessing}} the 21 Cm {{Cosmic Reionization Signature}}},
  author = {Parsons, Aaron R. and Pober, Jonathan C. and Aguirre, James E. and Carilli, Christopher L. and Jacobs, Daniel C. and Moore, David F.},
  year = 2012,
  journal = {The Astrophysical Journal},
  volume = {756},
  number = {2},
  pages = {165},
  issn = {0004-637X},
  url = {http://stacks.iop.org/0004-637X/756/i=2/a=165},
  urldate = {2018-03-28},
  langid = {english}
}

@article{Parsons2014,
  title = {New {{Limits}} on 21 Cm {{Epoch}} of {{Reionization}} from {{PAPER-32 Consistent}} with an {{X-Ray Heated Intergalactic Medium}} at z = 7.7},
  author = {Parsons, Aaron R. and Liu, Adrian and Aguirre, James E. and Ali, Zaki S. and Bradley, Richard F. and Carilli, Chris L. and DeBoer, David R. and Dexter, Matthew R. and Gugliucci, Nicole E. and Jacobs, Daniel C. and Klima, Pat and MacMahon, David H. E. and Manley, Jason R. and Moore, David F. and Pober, Jonathan C. and Stefan, Irina I. and Walbrugh, William P.},
  year = 2014,
  month = jun,
  journal = {The Astrophysical Journal},
  volume = {788},
  pages = {106},
  issn = {0004-637X},
  url = {http://adsabs.harvard.edu/abs/2014ApJ...788..106P},
  urldate = {2018-08-28}
}

@article{Parsons2016,
  title = {Optimized Beam Sculting with Generalized Fringe-Rate Filters},
  author = {Parsons, Aaron R. and Liu, Adrian and Ali, Zaki S. and Cheng, Carina},
  year = 2016,
  month = mar,
  journal = {The Astrophysical Journal},
  volume = {820},
  number = {1},
  pages = {51},
  publisher = {The American Astronomical Society},
  issn = {0004-637X},
  url = {https://dx.doi.org/10.3847/0004-637X/820/1/51},
  urldate = {2023-11-03},
  langid = {english}
}

@misc{Pascua2024,
  title = {A {{Generalized Method}} for {{Characterizing}} 21-Cm {{Power Spectrum Signal Loss}} from {{Temporal Filtering}} of {{Drift-scanning Visibilities}}},
  author = {Pascua, Robert and Martinot, Zachary E. and Liu, Adrian and Aguirre, James E. and Kern, Nicholas S. and Dillon, Joshua S. and Wilensky, Michael J. and Fagnoni, Nicolas and {de Lera Acedo}, Eloy and DeBoer, David},
  year = 2024,
  month = oct,
  publisher = {arXiv},
  url = {https://ui.adsabs.harvard.edu/abs/2024arXiv241001872P},
  urldate = {2025-04-17}
}

@article{Patil2014,
  title = {Constraining the Epoch of Reionization with the Variance Statistic: Simulations of the {{LOFAR}} Case},
  shorttitle = {Constraining the Epoch of Reionization with the Variance Statistic},
  author = {Patil, Ajinkya H. and Zaroubi, Saleem and Chapman, Emma and Jeli{\'c}, Vibor and Harker, Geraint and Abdalla, Filipe B. and Asad, Khan M. B. and Bernardi, Gianni and Brentjens, Michiel A. and {de Bruyn}, A. G. and Bus, Sander and Ciardi, Benedetta and Daiboo, Soobash and Fernandez, Elizabeth R. and Ghosh, Abhik and Jensen, Hannes and Kazemi, Sanaz and Koopmans, L{\'e}on V. E. and Labropoulos, Panagiotis and Mevius, Maaijke and Martinez, Oscar and Mellema, Garrelt and Offringa, Andre R. and Pandey, Vishhambhar N. and Schaye, Joop and Thomas, Rajat M. and Vedantham, Harish K. and Veligatla, Vamsikrishna and Wijnholds, Stefan J. and Yatawatta, Sarod},
  year = 2014,
  month = sep,
  journal = {Monthly Notices of the Royal Astronomical Society},
  volume = {443},
  pages = {1113--1124},
  publisher = {OUP},
  issn = {0035-8711},
  url = {https://ui.adsabs.harvard.edu/abs/2014MNRAS.443.1113P},
  urldate = {2025-04-17}
}

@article{Patil2017a,
  title = {Upper {{Limits}} on the 21 Cm {{Epoch}} of {{Reionization Power Spectrum}} from {{One Night}} with {{LOFAR}}},
  author = {Patil, A. H. and Yatawatta, S. and Koopmans, L. V. E. and {de Bruyn}, A. G. and Brentjens, M. A. and Zaroubi, S. and Asad, K. M. B. and Hatef, M. and Jeli{\'c}, V. and Mevius, M. and Offringa, A. R. and Pandey, V. N. and Vedantham, H. and Abdalla, F. B. and Brouw, W. N. and Chapman, E. and Ciardi, B. and Gehlot, B. K. and Ghosh, A. and Harker, G. and Iliev, I. T. and Kakiichi, K. and Majumdar, S. and Mellema, G. and Silva, M. B. and Schaye, J. and Vrbanec, D. and Wijnholds, S. J.},
  year = 2017,
  month = mar,
  journal = {The Astrophysical Journal},
  volume = {838},
  pages = {65},
  issn = {0004-637X},
  url = {http://adsabs.harvard.edu/abs/2017ApJ...838...65P},
  urldate = {2017-07-14}
}

@article{Patra2015,
  title = {{{SARAS MEASUREMENT OF THE RADIO BACKGROUND AT LONG WAVELENGTHS}}},
  author = {Patra, Nipanjana and Subrahmanyan, Ravi and Sethi, Shiv and Shankar, N. Udaya and Raghunathan, A.},
  year = 2015,
  month = mar,
  journal = {The Astrophysical Journal},
  volume = {801},
  number = {2},
  pages = {138},
  publisher = {The American Astronomical Society},
  issn = {0004-637X},
  url = {https://dx.doi.org/10.1088/0004-637X/801/2/138},
  urldate = {2025-04-17},
  langid = {english}
}

@article{Philip2019,
  title = {Probing {{Radio Intensity}} at {{High-Z}} from {{Marion}}: 2017 {{Instrument}}},
  shorttitle = {Probing {{Radio Intensity}} at {{High-Z}} from {{Marion}}},
  author = {Philip, L. and Abdurashidova, Z. and Chiang, H. C. and Ghazi, N. and Gumba, A. and Heilgendorff, H. M. and {J{\'a}uregui-Garc{\'i}a}, J. M. and Malepe, K. and Nunhokee, C. D. and Peterson, J. and Sievers, J. L. and Simes, V. and Spann, R.},
  year = 2019,
  month = jan,
  journal = {Journal of Astronomical Instrumentation},
  volume = {8},
  pages = {1950004},
  url = {https://ui.adsabs.harvard.edu/abs/2019JAI.....850004P},
  urldate = {2025-07-14}
}

@article{Planck2015Cosmo,
  title = {Planck 2015 Results. {{XIII}}. {{Cosmological}} Parameters},
  author = {Ade, P. a. R. and Aghanim, N. and Arnaud, M. and Ashdown, M. and Aumont, J. and Baccigalupi, C. and Banday, A. J. and Barreiro, R. B. and Bartlett, J. G. and Bartolo, N. and Battaner, E. and Battye, R. and Benabed, K. and Beno{\^i}t, A. and {Benoit-L{\'e}vy}, A. and Bernard, J.-P. and Bersanelli, M. and Bielewicz, P. and Bock, J. J. and Bonaldi, A. and Bonavera, L. and Bond, J. R. and Borrill, J. and Bouchet, F. R. and Boulanger, F. and Bucher, M. and Burigana, C. and Butler, R. C. and Calabrese, E. and Cardoso, J.-F. and Catalano, A. and Challinor, A. and Chamballu, A. and Chary, R.-R. and Chiang, H. C. and Chluba, J. and Christensen, P. R. and Church, S. and Clements, D. L. and Colombi, S. and Colombo, L. P. L. and Combet, C. and Coulais, A. and Crill, B. P. and Curto, A. and Cuttaia, F. and Danese, L. and Davies, R. D. and Davis, R. J. and {de Bernardis}, P. and {de Rosa}, A. and {de Zotti}, G. and Delabrouille, J. and D{\'e}sert, F.-X. and Di Valentino, E. and Dickinson, C. and Diego, J. M. and Dolag, K. and Dole, H. and Donzelli, S. and Dor{\'e}, O. and Douspis, M. and Ducout, A. and Dunkley, J. and Dupac, X. and Efstathiou, G. and Elsner, F. and En{\ss}lin, T. A. and Eriksen, H. K. and Farhang, M. and Fergusson, J. and Finelli, F. and Forni, O. and Frailis, M. and Fraisse, A. A. and Franceschi, E. and Frejsel, A. and Galeotta, S. and Galli, S. and Ganga, K. and Gauthier, C. and Gerbino, M. and Ghosh, T. and Giard, M. and {Giraud-H{\'e}raud}, Y. and Giusarma, E. and Gjerl{\o}w, E. and {Gonz{\'a}lez-Nuevo}, J. and G{\'o}rski, K. M. and Gratton, S. and Gregorio, A. and Gruppuso, A. and Gudmundsson, J. E. and Hamann, J. and Hansen, F. K. and Hanson, D. and Harrison, D. L. and Helou, G. and {Henrot-Versill{\'e}}, S. and {Hern{\'a}ndez-Monteagudo}, C. and Herranz, D. and Hildebrandt, S. R. and Hivon, E. and Hobson, M. and Holmes, W. A. and Hornstrup, A. and Hovest, W. and Huang, Z. and Huffenberger, K. M. and Hurier, G. and Jaffe, A. H. and Jaffe, T. R. and Jones, W. C. and Juvela, M. and Keih{\"a}nen, E. and Keskitalo, R. and Kisner, T. S. and Kneissl, R. and Knoche, J. and Knox, L. and Kunz, M. and {Kurki-Suonio}, H. and Lagache, G. and L{\"a}hteenm{\"a}ki, A. and Lamarre, J.-M. and Lasenby, A. and Lattanzi, M. and Lawrence, C. R. and Leahy, J. P. and Leonardi, R. and Lesgourgues, J. and Levrier, F. and Lewis, A. and Liguori, M. and Lilje, P. B. and {Linden-V{\o}rnle}, M. and {L{\'o}pez-Caniego}, M. and Lubin, P. M. and {Mac{\'i}as-P{\'e}rez}, J. F. and Maggio, G. and Maino, D. and Mandolesi, N. and Mangilli, A. and Marchini, A. and Maris, M. and Martin, P. G. and Martinelli, M. and {Mart{\'i}nez-Gonz{\'a}lez}, E. and Masi, S. and Matarrese, S. and McGehee, P. and Meinhold, P. R. and Melchiorri, A. and Melin, J.-B. and Mendes, L. and Mennella, A. and Migliaccio, M. and Millea, M. and Mitra, S. and {Miville-Desch{\^e}nes}, M.-A. and Moneti, A. and Montier, L. and Morgante, G. and Mortlock, D. and Moss, A. and Munshi, D. and Murphy, J. A. and Naselsky, P. and Nati, F. and Natoli, P. and Netterfield, C. B. and {N{\o}rgaard-Nielsen}, H. U. and Noviello, F. and Novikov, D. and Novikov, I. and Oxborrow, C. A. and Paci, F. and Pagano, L. and Pajot, F. and Paladini, R. and Paoletti, D. and Partridge, B. and Pasian, F. and Patanchon, G. and Pearson, T. J. and Perdereau, O. and Perotto, L. and Perrotta, F. and Pettorino, V. and Piacentini, F. and Piat, M. and Pierpaoli, E. and Pietrobon, D. and Plaszczynski, S. and Pointecouteau, E. and Polenta, G. and Popa, L. and Pratt, G. W. and Pr{\'e}zeau, G. and Prunet, S. and Puget, J.-L. and Rachen, J. P. and Reach, W. T. and Rebolo, R. and Reinecke, M. and Remazeilles, M. and Renault, C. and Renzi, A. and Ristorcelli, I. and Rocha, G. and Rosset, C. and Rossetti, M. and Roudier, G. and {Rouill{\'e} d'Orfeuil}, B. and {Rowan-Robinson}, M. and {Rubi{\~n}o-Mart{\'i}n}, J. A. and Rusholme, B. and Said, N. and Salvatelli, V. and Salvati, L. and Sandri, M. and Santos, D. and Savelainen, M. and Savini, G. and Scott, D. and Seiffert, M. D. and Serra, P. and Shellard, E. P. S. and Spencer, L. D. and Spinelli, M. and Stolyarov, V. and Stompor, R. and Sudiwala, R. and Sunyaev, R. and Sutton, D. and {Suur-Uski}, A.-S. and Sygnet, J.-F. and Tauber, J. A. and Terenzi, L. and Toffolatti, L. and Tomasi, M. and Tristram, M. and Trombetti, T. and Tucci, M. and Tuovinen, J. and T{\"u}rler, M. and Umana, G. and Valenziano, L. and Valiviita, J. and Van Tent, F. and Vielva, P. and Villa, F. and Wade, L. A. and Wandelt, B. D. and Wehus, I. K. and White, M. and White, S. D. M. and Wilkinson, A. and Yvon, D. and Zacchei, A. and Zonca, A.},
  year = 2016,
  month = sep,
  journal = {Astronomy and Astrophysics},
  volume = {594},
  pages = {A13},
  issn = {0004-6361},
  url = {https://ui.adsabs.harvard.edu/abs/2016A&A...594A..13P/abstract},
  urldate = {2021-07-26},
  langid = {english}
}

@article{PlanckCollaboration2018,
  title = {Planck 2018 Results. {{VI}}. {{Cosmological}} Parameters},
  author = {{Planck Collaboration} and Aghanim, N. and Akrami, Y. and Ashdown, M. and Aumont, J. and Baccigalupi, C. and Ballardini, M. and Banday, A. J. and Barreiro, R. B. and Bartolo, N. and Basak, S. and Battye, R. and Benabed, K. and Bernard, J.-P. and Bersanelli, M. and Bielewicz, P. and Bock, J. J. and Bond, J. R. and Borrill, J. and Bouchet, F. R. and Boulanger, F. and Bucher, M. and Burigana, C. and Butler, R. C. and Calabrese, E. and Cardoso, J.-F. and Carron, J. and Challinor, A. and Chiang, H. C. and Chluba, J. and Colombo, L. P. L. and Combet, C. and Contreras, D. and Crill, B. P. and Cuttaia, F. and {de Bernardis}, P. and {de Zotti}, G. and Delabrouille, J. and Delouis, J.-M. and Di Valentino, E. and Diego, J. M. and Dor{\'e}, O. and Douspis, M. and Ducout, A. and Dupac, X. and Dusini, S. and Efstathiou, G. and Elsner, F. and En{\ss}lin, T. A. and Eriksen, H. K. and Fantaye, Y. and Farhang, M. and Fergusson, J. and {Fernandez-Cobos}, R. and Finelli, F. and Forastieri, F. and Frailis, M. and Fraisse, A. A. and Franceschi, E. and Frolov, A. and Galeotta, S. and Galli, S. and Ganga, K. and {G{\'e}nova-Santos}, R. T. and Gerbino, M. and Ghosh, T. and {Gonz{\'a}lez-Nuevo}, J. and G{\'o}rski, K. M. and Gratton, S. and Gruppuso, A. and Gudmundsson, J. E. and Hamann, J. and Handley, W. and Hansen, F. K. and Herranz, D. and Hildebrandt, S. R. and Hivon, E. and Huang, Z. and Jaffe, A. H. and Jones, W. C. and Karakci, A. and Keih{\"a}nen, E. and Keskitalo, R. and Kiiveri, K. and Kim, J. and Kisner, T. S. and Knox, L. and Krachmalnicoff, N. and Kunz, M. and {Kurki-Suonio}, H. and Lagache, G. and Lamarre, J.-M. and Lasenby, A. and Lattanzi, M. and Lawrence, C. R. and Le Jeune, M. and Lemos, P. and Lesgourgues, J. and Levrier, F. and Lewis, A. and Liguori, M. and Lilje, P. B. and Lilley, M. and Lindholm, V. and {L{\'o}pez-Caniego}, M. and Lubin, P. M. and Ma, Y.-Z. and {Mac{\'i}as-P{\'e}rez}, J. F. and Maggio, G. and Maino, D. and Mandolesi, N. and Mangilli, A. and {Marcos-Caballero}, A. and Maris, M. and Martin, P. G. and Martinelli, M. and {Mart{\'i}nez-Gonz{\'a}lez}, E. and Matarrese, S. and Mauri, N. and McEwen, J. D. and Meinhold, P. R. and Melchiorri, A. and Mennella, A. and Migliaccio, M. and Millea, M. and Mitra, S. and {Miville-Desch{\^e}nes}, M.-A. and Molinari, D. and Montier, L. and Morgante, G. and Moss, A. and Natoli, P. and {N{\o}rgaard-Nielsen}, H. U. and Pagano, L. and Paoletti, D. and Partridge, B. and Patanchon, G. and Peiris, H. V. and Perrotta, F. and Pettorino, V. and Piacentini, F. and Polastri, L. and Polenta, G. and Puget, J.-L. and Rachen, J. P. and Reinecke, M. and Remazeilles, M. and Renzi, A. and Rocha, G. and Rosset, C. and Roudier, G. and {Rubi{\~n}o-Mart{\'i}n}, J. A. and {Ruiz-Granados}, B. and Salvati, L. and Sandri, M. and Savelainen, M. and Scott, D. and Shellard, E. P. S. and Sirignano, C. and Sirri, G. and Spencer, L. D. and Sunyaev, R. and {Suur-Uski}, A.-S. and Tauber, J. A. and Tavagnacco, D. and Tenti, M. and Toffolatti, L. and Tomasi, M. and Trombetti, T. and Valenziano, L. and Valiviita, J. and Van Tent, B. and Vibert, L. and Vielva, P. and Villa, F. and Vittorio, N. and Wandelt, B. D. and Wehus, I. K. and White, M. and White, S. D. M. and Zacchei, A. and Zonca, A.},
  year = 2018,
  month = jul,
  journal = {arXiv e-prints},
  volume = {1807},
  pages = {arXiv:1807.06209},
  url = {http://adsabs.harvard.edu/abs/2018arXiv180706209P},
  urldate = {2020-09-08}
}

@article{Pober2014,
  title = {What {{Next-generation}} 21 Cm {{Power Spectrum Measurements}} Can {{Teach}} Us {{About}} the {{Epoch}} of {{Reionization}}},
  author = {Pober, Jonathan C. and Liu, Adrian and Dillon, Joshua S. and Aguirre, James E. and Bowman, Judd D. and Bradley, Richard F. and Carilli, Chris L. and DeBoer, David R. and Hewitt, Jacqueline N. and Jacobs, Daniel C. and McQuinn, Matthew and Morales, Miguel F. and Parsons, Aaron R. and Tegmark, Max and Werthimer, Dan J.},
  year = 2014,
  month = feb,
  journal = {The Astrophysical Journal},
  volume = {782},
  pages = {66},
  issn = {0004-637X},
  url = {http://adsabs.harvard.edu/abs/2014ApJ...782...66P},
  urldate = {2018-03-28}
}

@article{Pober2015a,
  title = {{{PAPER-64 Constraints On Reionization}}. {{II}}. {{The Temperature}} of the z =8.4 {{Intergalactic Medium}}},
  author = {Pober, Jonathan C. and Ali, Zaki S. and Parsons, Aaron R. and McQuinn, Matthew and Aguirre, James E. and Bernardi, Gianni and Bradley, Richard F. and Carilli, Chris L. and Cheng, Carina and DeBoer, David R. and Dexter, Matthew R. and Furlanetto, Steven R. and Grobbelaar, Jasper and Horrell, Jasper and Jacobs, Daniel C. and Klima, Patricia J. and Kohn, Saul A. and Liu, Adrian and MacMahon, David H. E. and Maree, Matthys and Mesinger, Andrei and Moore, David F. and {Razavi-Ghods}, Nima and Stefan, Irina I. and Walbrugh, William P. and Walker, Andre and Zheng, Haoxuan},
  year = 2015,
  month = aug,
  journal = {The Astrophysical Journal},
  volume = {809},
  pages = {62},
  publisher = {IOP},
  issn = {0004-637X},
  url = {https://ui.adsabs.harvard.edu/abs/2015ApJ...809...62P},
  urldate = {2025-04-23}
}

@article{Pochinda2024,
  title = {Constraining the Properties of {{Population III}} Galaxies with Multiwavelength Observations},
  author = {Pochinda, S. and {Gessey-Jones}, T. and Bevins, H. T. J. and Fialkov, A. and Heimersheim, S. and {Abril-Cabezas}, I. and {de Lera Acedo}, E. and Singh, S. and Sikder, S. and Barkana, R.},
  year = 2024,
  month = jun,
  journal = {Monthly Notices of the Royal Astronomical Society},
  volume = {531},
  pages = {1113--1132},
  publisher = {OUP},
  issn = {0035-8711},
  url = {https://ui.adsabs.harvard.edu/abs/2024MNRAS.531.1113P},
  urldate = {2025-04-24}
}

@article{Price2016,
  title = {{{PyGDSM}}: {{Python}} Interface to {{Global Diffuse Sky Models}}},
  shorttitle = {{{PyGDSM}}},
  author = {Price, Danny C.},
  year = 2016,
  month = mar,
  journal = {Astrophysics Source Code Library},
  pages = {ascl:1603.013},
  url = {https://ui.adsabs.harvard.edu/abs/2016ascl.soft03013P},
  urldate = {2025-04-17}
}

@article{Price2018a,
  title = {Design and Characterization of the {{Large-aperture Experiment}} to {{Detect}} the {{Dark Age}} ({{LEDA}}) Radiometer Systems},
  author = {Price, D. C. and Greenhill, L. J. and Fialkov, A. and Bernardi, G. and Garsden, H. and Barsdell, B. R. and Kocz, J. and Anderson, M. M. and Bourke, S. A. and Craig, J. and Dexter, M. R. and Dowell, J. and Eastwood, M. W. and Eftekhari, T. and Ellingson, S. W. and Hallinan, G. and Hartman, J. M. and Kimberk, R. and Lazio, T. Joseph W. and Leiker, S. and MacMahon, D. and Monroe, R. and Schinzel, F. and Taylor, G. B. and Tong, E. and Werthimer, D. and Woody, D. P.},
  year = 2018,
  month = aug,
  journal = {Monthly Notices of the Royal Astronomical Society},
  volume = {478},
  pages = {4193--4213},
  publisher = {OUP},
  issn = {0035-8711},
  url = {https://ui.adsabs.harvard.edu/abs/2018MNRAS.478.4193P},
  urldate = {2025-04-17}
}

@article{Pritchard2012,
  title = {21 Cm Cosmology in the 21st Century},
  author = {Pritchard, Jonathan R. and Loeb, Abraham},
  year = 2012,
  month = aug,
  journal = {Reports on Progress in Physics},
  volume = {75},
  pages = {086901},
  issn = {0034-4885},
  url = {http://adsabs.harvard.edu/abs/2012RPPh...75h6901P},
  urldate = {2018-08-27}
}

@article{Qin2020,
  title = {Reionization Inference from the {{CMB}} Optical Depth and {{E-mode}} Polarization Power Spectra},
  author = {Qin, Yuxiang and Poulin, Vivian and Mesinger, Andrei and Greig, Bradley and Murray, Steven and Park, Jaehong},
  year = 2020,
  month = jul,
  journal = {Monthly Notices of the Royal Astronomical Society},
  volume = {499},
  pages = {550--558},
  url = {http://adsabs.harvard.edu/abs/2020MNRAS.499..550Q},
  urldate = {2020-11-03}
}

@article{Qin2020b,
  title = {A Tale of Two Sites - {{II}}: {{Inferring}} the Properties of Minihalo-Hosted Galaxies with Upcoming 21-Cm Interferometers},
  shorttitle = {A Tale of Two Sites - {{II}}},
  author = {Qin, Yuxiang and Mesinger, Andrei and Greig, Bradley and Park, Jaehong},
  year = 2020,
  month = nov,
  journal = {Monthly Notices of the Royal Astronomical Society},
  issn = {0035-8711},
  url = {http://adsabs.harvard.edu/abs/2020MNRAS.tmp.3221Q},
  urldate = {2020-11-13}
}

@article{Qin2020c,
  title = {A Tale of Two Sites - {{I}}. {{Inferring}} the Properties of Minihalo-Hosted Galaxies from Current Observations},
  author = {Qin, Yuxiang and Mesinger, Andrei and Park, Jaehong and Greig, Bradley and Mu{\~n}oz, Julian B.},
  year = 2020,
  month = apr,
  journal = {Monthly Notices of the Royal Astronomical Society},
  volume = {495},
  pages = {123--140},
  issn = {0035-8711},
  url = {http://adsabs.harvard.edu/abs/2020MNRAS.495..123Q},
  urldate = {2020-11-13}
}

@misc{Qin2024,
  title = {Percent-Level Timing of Reionization: Self-Consistent, Implicit-Likelihood Inference from {{XQR-30}}+ {{Ly}}\$\textbackslash alpha\$ Forest Data},
  shorttitle = {Percent-Level Timing of Reionization},
  author = {Qin, Yuxiang and Mesinger, Andrei and Prelogovi{\'c}, David and Becker, George and Bischetti, Manuela and Bosman, Sarah E. I. and Davies, Frederick B. and D'Odorico, Valentina and Gaikwad, Prakash and Haehnelt, Martin G. and Keating, Laura and Lai, Samuel and {Ryan-Weber}, Emma and Satyavolu, Sindhu and Walter, Fabian and Zhu, Yongda},
  year = 2024,
  month = dec,
  publisher = {arXiv},
  url = {https://ui.adsabs.harvard.edu/abs/2024arXiv241200799Q},
  urldate = {2025-04-17}
}

@article{Rahimi2021,
  title = {Epoch of Reionization Power Spectrum Limits from {{Murchison Widefield Array}} Data Targeted at {{EoR1}} Field},
  author = {Rahimi, M. and Pindor, B. and Line, J. L. B. and Barry, N. and Trott, C. M. and Webster, R. L. and Jordan, C. H. and Wilensky, M. and Yoshiura, S. and Beardsley, A. and Bowman, J. and Byrne, R. and Chokshi, A. and Hazelton, B. J. and Hasegawa, K. and Howard, E. and Greig, B. and Jacobs, D. and Joseph, R. and Kolopanis, M. and Lynch, C. and McKinley, B. and Mitchell, D. A. and Murray, S. and Morales, M. F. and Pober, J. C. and Takahashi, K. and Tingay, S. J. and Wayth, R. B. and Wyithe, J. S. B. and Zheng, Q.},
  year = 2021,
  month = dec,
  journal = {Monthly Notices of the Royal Astronomical Society},
  volume = {508},
  pages = {5954--5971},
  publisher = {OUP},
  issn = {0035-8711},
  url = {https://ui.adsabs.harvard.edu/abs/2021MNRAS.508.5954R},
  urldate = {2025-04-17}
}

@article{Rath2025,
  title = {Investigating Mutual Coupling in the Hydrogen Epoch of Reionization Array and Mitigating Its Effects on the 21-Cm Power Spectrum},
  author = {Rath, E. and Pascua, R. and Josaitis, A. T. and {Ewall-Wice}, A. and Fagnoni, N. and {de Lera Acedo}, E. and Martinot, Z. E. and Abdurashidova, Z. and Adams, T. and Aguirre, J. E. and Baartman, R. and Beardsley, A. P. and Berkhout, L. M. and Bernardi, G. and Billings, T. S. and Bowman, J. D. and Bull, P. and Burba, J. and Byrne, R. and Carey, S. and Chen, K.-F. and Choudhuri, S. and Cox, T. and DeBoer, D. R. and Dexter, M. and Dillon, J. S. and Dynes, S. and Eksteen, N. and Ely, J. and Fritz, R. and Furlanetto, S. R. and {Gale-Sides}, K. and Garsden, H. and Gehlot, B. K. and Ghosh, A. and Gorce, A. and Gorthi, D. and Halday, Z. and Hazelton, B. J. and Hewitt, J. N. and Hickish, J. and Huang, T. and Jacobs, D. C. and Kern, N. S. and Kerrigan, J. and Kittiwisit, P. and Kolopanis, M. and Lanman, A. and Liu, A. and Ma, Y.-Z. and MacMahon, D. H. E. and Malan, L. and Malgas, C. and Malgas, K. and Marero, B. and McBride, L. and Mesinger, A. and {Mohamed-Hinds}, N. and Molewa, M. and Morales, M. F. and Murray, S. G. and Nikolic, B. and Nuwegeld, H. and Parsons, A. R. and Patra, N. and Plante, P. La and Qin, Y. and {Razavi-Ghods}, N. and Riley, D. and Robnett, J. and Rosie, K. and Santos, M. G. and Sims, P. and Singh, S. and Storer, D. and Swarts, H. and Tan, J. and Wilensky, M. J. and Williams, P. K. G. and {van Wyngaarden}, P. and Zheng, H.},
  year = 2025,
  month = aug,
  journal = {Monthly Notices of the Royal Astronomical Society},
  volume = {541},
  pages = {1125--1144},
  publisher = {OUP},
  issn = {0035-8711},
  url = {https://ui.adsabs.harvard.edu/abs/2025MNRAS.541.1125R},
  urldate = {2025-10-26}
}

@article{Reis2020,
  title = {High-Redshift Radio Galaxies: A Potential New Source of 21-Cm Fluctuations},
  shorttitle = {High-Redshift Radio Galaxies},
  author = {Reis, Itamar and Fialkov, Anastasia and Barkana, Rennan},
  year = 2020,
  month = dec,
  journal = {Monthly Notices of the Royal Astronomical Society},
  volume = {499},
  pages = {5993--6008},
  publisher = {OUP},
  issn = {0035-8711},
  url = {https://ui.adsabs.harvard.edu/abs/2020MNRAS.499.5993R},
  urldate = {2025-04-24}
}

@article{Remazeilles2015,
  title = {An Improved Source-Subtracted and Destriped 408-{{MHz}} All-Sky Map},
  author = {Remazeilles, M. and Dickinson, C. and Banday, A. J. and {Bigot-Sazy}, M.-A. and Ghosh, T.},
  year = 2015,
  month = aug,
  journal = {Monthly Notices of the Royal Astronomical Society},
  volume = {451},
  number = {4},
  pages = {4311--4327},
  issn = {0035-8711},
  url = {https://doi.org/10.1093/mnras/stv1274},
  urldate = {2025-07-09}
}

@article{Riseley2020,
  title = {The {{POlarised GLEAM Survey}} ({{POGS}}) {{II}}: {{Results}} from an All-Sky Rotation Measure Synthesis Survey at Long Wavelengths},
  shorttitle = {The {{POlarised GLEAM Survey}} ({{POGS}}) {{II}}},
  author = {Riseley, C. J. and Galvin, T. J. and Sobey, C. and Vernstrom, T. and White, S. V. and Zhang, X. and Gaensler, B. M. and Heald, G. and Anderson, C. S. and Franzen, T. M. O. and Hancock, P. J. and {Hurley-Walker}, N. and Lenc, E. and Van Eck, C. L.},
  year = 2020,
  month = jul,
  journal = {Publications of the Astronomical Society of Australia},
  volume = {37},
  pages = {e029},
  issn = {1323-3580},
  url = {https://ui.adsabs.harvard.edu/abs/2020PASA...37...29R},
  urldate = {2025-04-18}
}

@article{Rogers2012,
  title = {Absolute Calibration of a Wideband Antenna and Spectrometer for Accurate Sky Noise Temperature Measurements},
  author = {Rogers, Alan E. E. and Bowman, Judd D.},
  year = 2012,
  journal = {Radio Science},
  volume = {47},
  number = {6},
  issn = {1944-799X},
  url = {https://agupubs.onlinelibrary.wiley.com/doi/abs/10.1029/2011RS004962},
  urldate = {2019-08-19},
  copyright = {\copyright 2012. American Geophysical Union. All Rights Reserved.},
  langid = {english}
}

@misc{Schneider2023,
  title = {Cosmological Forecast of the 21-Cm Power Spectrum Using the Halo Model of Reionization},
  author = {Schneider, Aurel and Schaeffer, Timoth{\'e}e and Giri, Sambit K.},
  year = 2023,
  month = feb,
  number = {arXiv:2302.06626},
  eprint = {2302.06626},
  primaryclass = {astro-ph},
  publisher = {arXiv},
  url = {http://arxiv.org/abs/2302.06626},
  urldate = {2025-05-15},
  archiveprefix = {arXiv}
}

@article{Sims2020,
  ids = {Sims2019a},
  title = {Testing for Calibration Systematics in the {{EDGES}} Low-Band Data Using {{Bayesian}} Model Selection},
  author = {Sims, Peter H. and Pober, Jonathan C.},
  year = 2020,
  month = feb,
  journal = {Monthly Notices of the Royal Astronomical Society},
  volume = {492},
  eprint = {1910.03165},
  pages = {22--38},
  url = {http://adsabs.harvard.edu/abs/2020MNRAS.492...22S},
  urldate = {2020-08-24},
  archiveprefix = {arXiv}
}

@article{Singh2017,
  title = {First {{Results}} on the {{Epoch}} of {{Reionization}} from {{First Light}} with {{SARAS}} 2},
  author = {Singh, Saurabh and Subrahmanyan, Ravi and Udaya Shankar, N. and Sathyanarayana Rao, Mayuri and Fialkov, Anastasia and Cohen, Aviad and Barkana, Rennan and Girish, B. S. and Raghunathan, A. and Somashekar, R. and Srivani, K. S.},
  year = 2017,
  month = aug,
  journal = {The Astrophysical Journal Letters},
  volume = {845},
  pages = {L12},
  issn = {0004-637X},
  url = {http://adsabs.harvard.edu/abs/2017ApJ...845L..12S},
  urldate = {2020-11-11}
}

@article{Singh2019,
  title = {The {{Redshifted}} 21 Cm {{Signal}} in the {{EDGES Low-band Spectrum}}},
  author = {Singh, Saurabh and Subrahmanyan, Ravi},
  year = 2019,
  month = jul,
  journal = {The Astrophysical Journal},
  volume = {880},
  pages = {26},
  url = {http://adsabs.harvard.edu/abs/2019ApJ...880...26S},
  urldate = {2020-10-22}
}

@article{Singh2022,
  title = {On the Detection of a Cosmic Dawn Signal in the Radio Background},
  author = {Singh, Saurabh and Nambissan T., Jishnu and Subrahmanyan, Ravi and Udaya Shankar, N. and Girish, B. S. and Raghunathan, A. and Somashekar, R. and Srivani, K. S. and Sathyanarayana Rao, Mayuri},
  year = 2022,
  month = may,
  journal = {Nature Astronomy},
  volume = {6},
  number = {5},
  pages = {607--617},
  publisher = {Nature Publishing Group},
  issn = {2397-3366},
  url = {https://www.nature.com/articles/s41550-022-01610-5},
  urldate = {2022-05-26},
  copyright = {2022 The Author(s), under exclusive licence to Springer Nature Limited},
  langid = {english}
}

@article{Slepian1978,
  title = {Prolate Spheroidal Wave Functions, {{Fourier}} Analysis, and Uncertainty. {{V}} - {{The}} Discrete Case},
  author = {Slepian, D.},
  year = 1978,
  month = jun,
  journal = {AT T Technical Journal},
  volume = {57},
  pages = {1371--1430},
  url = {https://ui.adsabs.harvard.edu/abs/1978ATTTJ..57.1371S},
  urldate = {2025-04-18}
}

@article{Smirnov2011,
  title = {Revisiting the Radio Interferometer Measurement Equation. {{I}}. {{A}} Full-Sky {{Jones}} Formalism},
  author = {Smirnov, O. M.},
  year = 2011,
  month = mar,
  journal = {Astronomy and Astrophysics},
  volume = {527},
  pages = {A106},
  issn = {0004-6361},
  url = {https://ui.adsabs.harvard.edu/abs/2011A&A...527A.106S},
  urldate = {2025-04-18}
}

@article{Sokolowski2015,
  title = {{{BIGHORNS}} - {{Broadband Instrument}} for {{Global HydrOgen ReioNisation Signal}}},
  author = {Sokolowski, Marcin and Tremblay, Steven E. and Wayth, Randall B. and Tingay, Steven J. and Clarke, Nathan and Roberts, Paul and Waterson, Mark and Ekers, Ronald D. and Hall, Peter and Lewis, Morgan and Mossammaparast, Mehran and Padhi, Shantanu and Schlagenhaufer, Franz and Sutinjo, Adrian and Tickner, Jonathan},
  year = 2015,
  month = feb,
  journal = {Publications of the Astronomical Society of Australia},
  volume = {32},
  pages = {e004},
  issn = {1323-3580},
  url = {https://ui.adsabs.harvard.edu/abs/2015PASA...32....4S},
  urldate = {2022-06-02}
}

@article{Storer2022,
  title = {Automated {{Detection}} of {{Antenna Malfunctions}} in {{Large-N Interferometers}}: {{A Case Study With}} the {{Hydrogen Epoch}} of {{Reionization Array}}},
  shorttitle = {Automated {{Detection}} of {{Antenna Malfunctions}} in {{Large-N Interferometers}}},
  author = {Storer, Dara and Dillon, Joshua S. and Jacobs, Daniel C. and Morales, Miguel F. and Hazelton, Bryna J. and {Ewall-Wice}, Aaron and Abdurashidova, Zara and Aguirre, James E. and Alexander, Paul and Ali, Zaki S. and Balfour, Yanga and Beardsley, Adam P. and Bernardi, Gianni and Billings, Tashalee S. and Bowman, Judd D. and Bradley, Richard F. and Bull, Philip and Burba, Jacob and Carey, Steven and Carilli, Chris L. and Cheng, Carina and DeBoer, David R. and {de Lera Acedo}, Eloy and Dexter, Matt and Dynes, Scott and Ely, John and Fagnoni, Nicolas and Fritz, Randall and Furlanetto, Steven R. and {Gale-Sides}, Kingsley and Glendenning, Brian and Gorthi, Deepthi and Greig, Bradley and Grobbelaar, Jasper and Halday, Ziyaad and Hewitt, Jacqueline N. and Hickish, Jack and Huang, Tian and Josaitis, Alec and Julius, Austin and Kariseb, MacCalvin and Kern, Nicholas S. and Kerrigan, Joshua and Kittiwisit, Piyanat and Kohn, Saul A. and Kolopanis, Matthew and Lanman, Adam and La Plante, Paul and Liu, Adrian and Loots, Anita and MacMahon, David and Malan, Lourence and Malgas, Cresshim and Martinot, Zachary E. and Mesinger, Andrei and Molewa, Mathakane and Mosiane, Tshegofalang and Murray, Steven G. and Neben, Abraham R. and Nikolic, Bojan and Nunhokee, Chuneeta Devi and Parsons, Aaron R. and Pascua, Robert and Patra, Nipanjana and Pieterse, Samantha and Pober, Jonathan C. and {Razavi-Ghods}, Nima and Riley, Daniel and Robnett, James and Rosie, Kathryn and Santos, Mario G. and Sims, Peter and Singh, Saurabh and Smith, Craig and Tan, Jianrong and Thyagarajan, Nithyanandan and Williams, Peter K. G. and Zheng, Haoxuan},
  year = 2022,
  journal = {Radio Science},
  volume = {57},
  number = {1},
  pages = {e2021RS007376},
  issn = {1944-799X},
  url = {https://onlinelibrary.wiley.com/doi/abs/10.1029/2021RS007376},
  urldate = {2025-04-17},
  copyright = {\copyright{} 2021. American Geophysical Union. All Rights Reserved.},
  langid = {english}
}

@article{Tan2021,
  title = {Methods of {{Error Estimation}} for {{Delay Power Spectra}} in 21 Cm {{Cosmology}}},
  author = {Tan, Jianrong and Liu, Adrian and Kern, Nicholas S. and Abdurashidova, Zara and Aguirre, James E. and Alexander, Paul and Ali, Zaki S. and Balfour, Yanga and Beardsley, Adam P. and Bernardi, Gianni and Billings, Tashalee S. and Bowman, Judd D. and Bradley, Richard F. and Bull, Philip and Burba, Jacob and Carey, Steven and Carilli, Christopher L. and Cheng, Carina and DeBoer, David R. and Dexter, Matt and {de Lera Acedo}, Eloy and Dillon, Joshua S. and Ely, John and {Ewall-Wice}, Aaron and Fagnoni, Nicolas and Fritz, Randall and Furlanetto, Steve R. and {Gale-Sides}, Kingsley and Glendenning, Brian and Gorthi, Deepthi and Greig, Bradley and Grobbelaar, Jasper and Halday, Ziyaad and Hazelton, Bryna J. and Hewitt, Jacqueline N. and Hickish, Jack and Jacobs, Daniel C. and Julius, Austin and Kerrigan, Joshua and Kittiwisit, Piyanat and Kohn, Saul A. and Kolopanis, Matthew and Lanman, Adam and La Plante, Paul and Lekalake, Telalo and MacMahon, David and Malan, Lourence and Malgas, Cresshim and Maree, Matthys and Martinot, Zachary E. and Matsetela, Eunice and Mesinger, Andrei and Molewa, Mathakane and Morales, Miguel F. and Mosiane, Tshegofalang and Murray, Steven G. and Neben, Abraham R. and Nikolic, Bojan and Nunhokee, Chuneeta D. and Parsons, Aaron R. and Patra, Nipanjana and Pieterse, Samantha and Pober, Jonathan C. and {Razavi-Ghods}, Nima and Ringuette, Jon and Robnett, James and Rosie, Kathryn and Sims, Peter and Singh, Saurabh and Smith, Craig and Syce, Angelo and Thyagarajan, Nithyanandan and Williams, Peter K. G. and Zheng, Haoxuan},
  year = 2021,
  month = aug,
  journal = {The Astrophysical Journal Supplement Series},
  volume = {255},
  pages = {26},
  publisher = {IOP},
  issn = {0067-0049},
  url = {https://ui.adsabs.harvard.edu/abs/2021ApJS..255...26T},
  urldate = {2025-04-17}
}

@article{Taylor2009,
  title = {A {{Rotation Measure Image}} of the {{Sky}}},
  author = {Taylor, A. R. and Stil, J. M. and Sunstrum, C.},
  year = 2009,
  month = sep,
  journal = {The Astrophysical Journal},
  volume = {702},
  pages = {1230--1236},
  publisher = {IOP},
  issn = {0004-637X},
  url = {https://ui.adsabs.harvard.edu/abs/2009ApJ...702.1230T},
  urldate = {2025-04-18}
}

@article{Thyagarajan2015,
  title = {Foregrounds in {{Wide-field Redshifted}} 21 Cm {{Power Spectra}}},
  author = {Thyagarajan, Nithyanandan and Jacobs, Daniel C. and Bowman, Judd D. and Barry, N. and Beardsley, A. P. and Bernardi, G. and Briggs, F. and Cappallo, R. J. and Carroll, P. and Corey, B. E. and {de Oliveira-Costa}, A. and Dillon, Joshua S. and Emrich, D. and {Ewall-Wice}, A. and Feng, L. and Goeke, R. and Greenhill, L. J. and Hazelton, B. J. and Hewitt, J. N. and {Hurley-Walker}, N. and {Johnston-Hollitt}, M. and Kaplan, D. L. and Kasper, J. C. and Kim, Han-Seek and Kittiwisit, P. and Kratzenberg, E. and Lenc, E. and Line, J. and Loeb, A. and Lonsdale, C. J. and Lynch, M. J. and McKinley, B. and McWhirter, S. R. and Mitchell, D. A. and Morales, M. F. and Morgan, E. and Neben, A. R. and Oberoi, D. and Offringa, A. R. and Ord, S. M. and Paul, Sourabh and Pindor, B. and Pober, J. C. and Prabu, T. and Procopio, P. and Riding, J. and Rogers, A. E. E. and Roshi, A. and Udaya Shankar, N. and Sethi, Shiv K. and Srivani, K. S. and Subrahmanyan, R. and Sullivan, I. S. and Tegmark, M. and Tingay, S. J. and Trott, C. M. and Waterson, M. and Wayth, R. B. and Webster, R. L. and Whitney, A. R. and Williams, A. and Williams, C. L. and Wu, C. and Wyithe, J. S. B.},
  year = 2015,
  month = may,
  journal = {The Astrophysical Journal},
  volume = {804},
  pages = {14},
  issn = {0004-637X},
  url = {http://adsabs.harvard.edu/abs/2015ApJ...804...14T},
  urldate = {2018-09-04}
}

@article{Tingay2013,
  ids = {Tingay2012},
  title = {The {{Murchison Widefield Array}}: {{The Square Kilometre Array Precursor}} at {{Low Radio Frequencies}}},
  shorttitle = {The {{Murchison Widefield Array}}},
  author = {Tingay, S. J. and Goeke, R. and Bowman, J. D. and Emrich, D. and Ord, S. M. and Mitchell, D. A. and Morales, M. F. and Booler, T. and Crosse, B. and Wayth, R. B. and Lonsdale, C. J. and Tremblay, S. and Pallot, D. and Colegate, T. and Wicenec, A. and Kudryavtseva, N. and Arcus, W. and Barnes, D. and Bernardi, G. and Briggs, F. and Burns, S. and Bunton, J. D. and Cappallo, R. J. and Corey, B. E. and Deshpande, A. and Desouza, L. and Gaensler, B. M. and Greenhill, L. J. and Hall, P. J. and Hazelton, B. J. and Herne, D. and Hewitt, J. N. and {Johnston-Hollitt}, M. and Kaplan, D. L. and Kasper, J. C. and Kincaid, B. B. and Koenig, R. and Kratzenberg, E. and Lynch, M. J. and Mckinley, B. and Mcwhirter, S. R. and Morgan, E. and Oberoi, D. and Pathikulangara, J. and Prabu, T. and Remillard, R. A. and Rogers, A. E. E. and Roshi, A. and Salah, J. E. and Sault, R. J. and {Udaya-Shankar}, N. and Schlagenhaufer, F. and Srivani, K. S. and Stevens, J. and Subrahmanyan, R. and Waterson, M. and Webster, R. L. and Whitney, A. R. and Williams, A. and Williams, C. L. and Wyithe, J. S. B.},
  year = 2013,
  journal = {Publications of the Astronomical Society of Australia},
  volume = {30},
  publisher = {Cambridge University Press},
  issn = {1323-3580, 1448-6083},
  url = {https://www.cambridge.org/core/journals/publications-of-the-astronomical-society-of-australia/article/murchison-widefield-array-the-square-kilometre-array-precursor-at-low-radio-frequencies/ED20FE56B17C253DAB94836785D887F0},
  urldate = {2020-08-24},
  langid = {english}
}

@article{Topping2022,
  title = {Searching for {{Extremely Blue UV Continuum Slopes}} at z = 7-11 in {{JWST}}/{{NIRCam Imaging}}: {{Implications}} for {{Stellar Metallicity}} and {{Ionizing Photon Escape}} in {{Early Galaxies}}},
  shorttitle = {Searching for {{Extremely Blue UV Continuum Slopes}} at z = 7-11 in {{JWST}}/{{NIRCam Imaging}}},
  author = {Topping, Michael W. and Stark, Daniel P. and Endsley, Ryan and Plat, Adele and Whitler, Lily and Chen, Zuyi and Charlot, St{\'e}phane},
  year = 2022,
  month = dec,
  journal = {The Astrophysical Journal},
  volume = {941},
  pages = {153},
  publisher = {IOP},
  issn = {0004-637X},
  url = {https://ui.adsabs.harvard.edu/abs/2022ApJ...941..153T},
  urldate = {2025-04-23}
}

@article{Trac2022,
  title = {{{AMBER}}: {{A Semi-numerical Abundance Matching Box}} for the {{Epoch}} of {{Reionization}}},
  shorttitle = {{{AMBER}}},
  author = {Trac, Hy and Chen, Nianyi and Holst, Ian and Alvarez, Marcelo A. and Cen, Renyue},
  year = 2022,
  month = mar,
  journal = {The Astrophysical Journal},
  volume = {927},
  number = {2},
  pages = {186},
  publisher = {The American Astronomical Society},
  issn = {0004-637X},
  url = {https://dx.doi.org/10.3847/1538-4357/ac5116},
  urldate = {2025-05-15},
  langid = {english}
}

@article{Trott2020,
  title = {Deep Multiredshift Limits on {{Epoch}} of {{Reionization}} 21 Cm Power Spectra from Four Seasons of {{Murchison Widefield Array}} Observations},
  author = {Trott, Cathryn M. and Jordan, C. H. and Midgley, S. and Barry, N. and Greig, B. and Pindor, B. and Cook, J. H. and Sleap, G. and Tingay, S. J. and Ung, D. and Hancock, P. and Williams, A. and Bowman, J. and Byrne, R. and Chokshi, A. and Hazelton, B. J. and Hasegawa, K. and Jacobs, D. and Joseph, R. C. and Li, W. and Line, J. L. B. and Lynch, C. and McKinley, B. and Mitchell, D. A. and Morales, M. F. and Ouchi, M. and Pober, J. C. and Rahimi, M. and Takahashi, K. and Wayth, R. B. and Webster, R. L. and Wilensky, M. and Wyithe, J. S. B. and Yoshiura, S. and Zhang, Z. and Zheng, Q.},
  year = 2020,
  month = apr,
  journal = {Monthly Notices of the Royal Astronomical Society},
  volume = {493},
  pages = {4711--4727},
  url = {http://adsabs.harvard.edu/abs/2020MNRAS.493.4711T},
  urldate = {2020-11-03}
}

@article{vanHaarlem2013,
  title = {{{LOFAR}}: {{The LOw-Frequency ARray}}},
  author = {{van Haarlem}, M. P. and Wise, M. W. and Gunst, A. W. and Heald, G. and McKean, J. P. and Hessels, J. W. T. and {de Bruyn}, A. G. and Nijboer, R. and Swinbank, J. and Fallows, R. and Brentjens, M. and Nelles, A. and Beck, R. and Falcke, H. and Fender, R. and H{\"o}randel, J. and Koopmans, L. V. E. and Mann, G. and Miley, G. and R{\"o}ttgering, H. and Stappers, B. W. and Wijers, R. A. M. J. and Zaroubi, S. and van den Akker, M. and Alexov, A. and Anderson, J. E. and Anderson, K. and {van Ardenne}, A. and Arts, M. and Asgekar, A. and Avruch, I. M. and Batejat, F. and B{\"a}hren, L. and Bell, M. E. and Bell, M. R. and {van Bemmel}, I. and Bennema, P. and Bentum, M. J. and Bernardi, G. and Best, P. and B{\^i}rzan, L. and Bonafede, A. and Boonstra, A. -J. and Braun, R. and Bregman, J. and Breitling, F. and {van de Brink}, R. H. and Broderick, J. and Broekema, P. C. and Brouw, W. N. and Br{\"u}ggen, M. and Butcher, H. R. and {van Cappellen}, W. and Ciardi, B. and Coenen, T. and Conway, J. and Coolen, A. and Corstanje, A. and Damstra, S. and Davies, O. and Deller, A. T. and Dettmar, R. -J. and {van Diepen}, G. and Dijkstra, K. and Donker, P. and Doorduin, A. and Dromer, J. and Drost, M. and {van Duin}, A. and Eisl{\"o}ffel, J. and {van Enst}, J. and Ferrari, C. and Frieswijk, W. and Gankema, H. and Garrett, M. A. and {de Gasperin}, F. and Gerbers, M. and {de Geus}, E. and Grie{\ss}meier, J. -M. and Grit, T. and Gruppen, P. and Hamaker, J. P. and Hassall, T. and Hoeft, M. and Holties, H. and Horneffer, A. and {van der Horst}, A. and {van Houwelingen}, A. and Huijgen, A. and Iacobelli, M. and Intema, H. and Jackson, N. and Jelic, V. and {de Jong}, A. and Juette, E. and Kant, D. and Karastergiou, A. and Koers, A. and Kollen, H. and Kondratiev, V. I. and Kooistra, E. and Koopman, Y. and Koster, A. and Kuniyoshi, M. and Kramer, M. and Kuper, G. and Lambropoulos, P. and Law, C. and {van Leeuwen}, J. and Lemaitre, J. and Loose, M. and Maat, P. and Macario, G. and Markoff, S. and Masters, J. and {McKay-Bukowski}, D. and Meijering, H. and Meulman, H. and Mevius, M. and Middelberg, E. and Millenaar, R. and {Miller-Jones}, J. C. A. and Mohan, R. N. and Mol, J. D. and Morawietz, J. and Morganti, R. and Mulcahy, D. D. and Mulder, E. and Munk, H. and Nieuwenhuis, L. and {van Nieuwpoort}, R. and Noordam, J. E. and Norden, M. and Noutsos, A. and Offringa, A. R. and Olofsson, H. and Omar, A. and Orr{\'u}, E. and Overeem, R. and Paas, H. and {Pandey-Pommier}, M. and Pandey, V. N. and Pizzo, R. and Polatidis, A. and Rafferty, D. and Rawlings, S. and Reich, W. and {de Reijer}, J. -P. and Reitsma, J. and Renting, A. and Riemers, P. and Rol, E. and Romein, J. W. and Roosjen, J. and Ruiter, M. and Scaife, A. and {van der Schaaf}, K. and Scheers, B. and Schellart, P. and Schoenmakers, A. and Schoonderbeek, G. and Serylak, M. and Shulevski, A. and Sluman, J. and Smirnov, O. and Sobey, C. and Spreeuw, H. and Steinmetz, M. and Sterks, C. G. M. and Stiepel, H. -J. and Stuurwold, K. and Tagger, M. and Tang, Y. and Tasse, C. and Thomas, I. and Thoudam, S. and Toribio, M. C. and {van der Tol}, B. and Usov, O. and {van Veelen}, M. and {van der Veen}, A. -J. and {ter Veen}, S. and Verbiest, J. P. W. and Vermeulen, R. and Vermaas, N. and Vocks, C. and Vogt, C. and {de Vos}, M. and {van der Wal}, E. and {van Weeren}, R. and Weggemans, H. and Weltevrede, P. and White, S. and Wijnholds, S. J. and Wilhelmsson, T. and Wucknitz, O. and Yatawatta, S. and Zarka, P. and Zensus, A. and {van Zwieten}, J.},
  year = 2013,
  month = may,
  journal = {Astronomy \& Astrophysics},
  volume = {556},
  pages = {53},
  url = {http://arxiv.org/abs/1305.3550}
}

@article{Virtanen2020,
  title = {{{SciPy}} 1.0: Fundamental Algorithms for Scientific Computing in {{Python}}},
  shorttitle = {{{SciPy}} 1.0},
  author = {Virtanen, Pauli and Gommers, Ralf and Oliphant, Travis E. and Haberland, Matt and Reddy, Tyler and Cournapeau, David and Burovski, Evgeni and Peterson, Pearu and Weckesser, Warren and Bright, Jonathan and {van der Walt}, St{\'e}fan J. and Brett, Matthew and Wilson, Joshua and Millman, K. Jarrod and Mayorov, Nikolay and Nelson, Andrew R. J. and Jones, Eric and Kern, Robert and Larson, Eric and Carey, C. J. and Polat, {\.I}lhan and Feng, Yu and Moore, Eric W. and VanderPlas, Jake and Laxalde, Denis and Perktold, Josef and Cimrman, Robert and Henriksen, Ian and Quintero, E. A. and Harris, Charles R. and Archibald, Anne M. and Ribeiro, Ant{\^o}nio H. and Pedregosa, Fabian and {van Mulbregt}, Paul},
  year = 2020,
  month = mar,
  journal = {Nature Methods},
  volume = {17},
  number = {3},
  pages = {261--272},
  publisher = {Nature Publishing Group},
  issn = {1548-7105},
  url = {https://www.nature.com/articles/s41592-019-0686-2},
  urldate = {2022-07-25},
  copyright = {2020 The Author(s)},
  langid = {english}
}

@article{Visbal2012,
  title = {The Signature of the First Stars in Atomic Hydrogen at Redshift 20},
  author = {Visbal, Eli and Barkana, Rennan and Fialkov, Anastasia and Tseliakhovich, Dmitriy and Hirata, Christopher M.},
  year = 2012,
  month = jul,
  journal = {Nature},
  volume = {487},
  pages = {70--73},
  issn = {0028-0836},
  url = {https://ui.adsabs.harvard.edu/abs/2012Natur.487...70V},
  urldate = {2025-04-24}
}

@article{Vruno2023,
  title = {Unintended Electromagnetic Radiation from {{Starlink}} Satellites Detected with {{LOFAR}} between 110 and 188 {{MHz}}},
  author = {Vruno, F. Di and Winkel, B. and Bassa, C. G. and J{\'o}zsa, G. I. G. and Brentjens, M. A. and Jessner, A. and Garrington, S.},
  year = 2023,
  month = aug,
  journal = {Astronomy \& Astrophysics},
  volume = {676},
  eprint = {2307.02316},
  primaryclass = {astro-ph},
  pages = {A75},
  issn = {0004-6361, 1432-0746},
  url = {http://arxiv.org/abs/2307.02316},
  urldate = {2025-05-20},
  archiveprefix = {arXiv}
}

@article{Wilensky2020,
  title = {Quantifying Excess Power from Radio Frequency Interference in {{Epoch}} of {{Reionization}} Measurements},
  author = {Wilensky, Michael J and Barry, Nichole and Morales, Miguel F and Hazelton, Bryna J and Byrne, Ruby},
  year = 2020,
  month = oct,
  journal = {Monthly Notices of the Royal Astronomical Society},
  volume = {498},
  number = {1},
  pages = {265--275},
  issn = {0035-8711, 1365-2966},
  url = {https://academic.oup.com/mnras/article/498/1/265/5893307},
  urldate = {2025-05-12},
  copyright = {https://academic.oup.com/journals/pages/open\_access/funder\_policies/chorus/standard\_publication\_model},
  langid = {english}
}

@article{Wilensky2023,
  title = {Evidence of {{Ultrafaint Radio Frequency Interference}} in {{Deep}} 21 Cm {{Epoch}} of {{Reionization Power Spectra}} with the {{Murchison Wide-field Array}}},
  author = {Wilensky, Michael J. and Morales, Miguel F. and Hazelton, Bryna J. and Star, Pyxie L. and Barry, Nichole and Byrne, Ruby and Jordan, C. H. and Jacobs, Daniel C. and Pober, Jonathan C. and Trott, C. M.},
  year = 2023,
  month = nov,
  journal = {The Astrophysical Journal},
  volume = {957},
  number = {2},
  pages = {78},
  issn = {0004-637X, 1538-4357},
  url = {https://iopscience.iop.org/article/10.3847/1538-4357/acffbd},
  urldate = {2025-05-12}
}

@article{Witstok2025,
  title = {Witnessing the Onset of Reionization through {{Lyman-$\alpha$}} Emission at Redshift 13},
  author = {Witstok, Joris and Jakobsen, Peter and Maiolino, Roberto and Helton, Jakob M. and Johnson, Benjamin D. and Robertson, Brant E. and Tacchella, Sandro and Cameron, Alex J. and Smit, Renske and Bunker, Andrew J. and Saxena, Aayush and Sun, Fengwu and Alberts, Stacey and Arribas, Santiago and Baker, William M. and Bhatawdekar, Rachana and Boyett, Kristan and Cargile, Phillip A. and Carniani, Stefano and Charlot, St{\'e}phane and Chevallard, Jacopo and Curti, Mirko and {Curtis-Lake}, Emma and D'Eugenio, Francesco and Eisenstein, Daniel J. and Hainline, Kevin N. and Jones, Gareth C. and Kumari, Nimisha and Maseda, Michael V. and {P{\'e}rez-Gonz{\'a}lez}, Pablo G. and Rinaldi, Pierluigi and Scholtz, Jan and {\"U}bler, Hannah and Williams, Christina C. and Willmer, Christopher N. A. and Willott, Chris and Zhu, Yongda},
  year = 2025,
  month = mar,
  journal = {Nature},
  volume = {639},
  pages = {897--901},
  issn = {0028-0836},
  url = {https://ui.adsabs.harvard.edu/abs/2025Natur.639..897W},
  urldate = {2025-04-23}
}

@article{Yoshiura2021,
  title = {A New {{MWA}} Limit on the 21 Cm Power Spectrum at Redshifts  13-17},
  author = {Yoshiura, S. and Pindor, B. and Line, J. L. B. and Barry, N. and Trott, C. M. and Beardsley, A. and Bowman, J. and Byrne, R. and Chokshi, A. and Hazelton, B. J. and Hasegawa, K. and Howard, E. and Greig, B. and Jacobs, D. and Jordan, C. H. and Joseph, R. and Kolopanis, M. and Lynch, C. and McKinley, B. and Mitchell, D. A. and Morales, M. F. and Murray, S. G. and Pober, J. C. and Rahimi, M. and Takahashi, K. and Tingay, S. J. and Wayth, R. B. and Webster, R. L. and Wilensky, M. and Wyithe, J. S. B. and Zhang, Z. and Zheng, Q.},
  year = 2021,
  month = aug,
  journal = {Monthly Notices of the Royal Astronomical Society},
  volume = {505},
  pages = {4775--4790},
  issn = {0035-8711},
  url = {https://ui.adsabs.harvard.edu/abs/2021MNRAS.505.4775Y},
  urldate = {2021-11-09}
}

@inproceedings{Zarka2012,
  title = {{{LSS}}/{{NenuFAR}}: {{The LOFAR Super Station}} Project in {{Nan\c cay}}},
  shorttitle = {{{LSS}}/{{NenuFAR}}},
  booktitle = {{{SF2A-2012}}: {{Proceedings}} of the {{Annual}} Meeting of the {{French Society}} of {{Astronomy}} and {{Astrophysics}}},
  author = {Zarka, P. and Girard, J. N. and Tagger, M. and Denis, L.},
  year = 2012,
  month = dec,
  pages = {687--694},
  url = {https://ui.adsabs.harvard.edu/abs/2012sf2a.conf..687Z},
  urldate = {2025-08-05}
}

@article{Zheng2014,
  title = {{{MITEoR}}: A Scalable Interferometer for Precision 21 Cm Cosmology},
  shorttitle = {{{MITEoR}}},
  author = {Zheng, H. and Tegmark, M. and Buza, V. and Dillon, J. S. and Gharibyan, H. and Hickish, J. and Kunz, E. and Liu, A. and Losh, J. and Lutomirski, A. and Morrison, S. and Narayanan, S. and Perko, A. and Rosner, D. and Sanchez, N. and Schutz, K. and Tribiano, S. M. and Valdez, M. and Yang, H. and Adami, K. Zarb and Zelko, I. and Zheng, K. and Armstrong, R. P. and Bradley, R. F. and Dexter, M. R. and {Ewall-Wice}, A. and Magro, A. and Matejek, M. and Morgan, E. and Neben, A. R. and Pan, Q. and Penna, R. F. and Peterson, C. M. and Su, M. and Villasenor, J. and Williams, C. L. and Zhu, Y.},
  year = 2014,
  month = dec,
  journal = {Monthly Notices of the Royal Astronomical Society},
  volume = {445},
  pages = {1084--1103},
  publisher = {OUP},
  issn = {0035-8711},
  url = {https://ui.adsabs.harvard.edu/abs/2014MNRAS.445.1084Z},
  urldate = {2025-04-17}
}

@article{Zhu2022,
  title = {Long {{Dark Gaps}} in the {{Ly$\beta$ Forest}} at z {$<$} 6: {{Evidence}} of {{Ultra-late Reionization}} from {{XQR-30 Spectra}}},
  shorttitle = {Long {{Dark Gaps}} in the {{Ly$\beta$ Forest}} at z {$<$} 6},
  author = {Zhu, Yongda and Becker, George D. and Bosman, Sarah E. I. and Keating, Laura C. and D'Odorico, Valentina and Davies, Rebecca L. and Christenson, Holly M. and Ba{\~n}ados, Eduardo and Bian, Fuyan and Bischetti, Manuela and Chen, Huanqing and Davies, Frederick B. and Eilers, Anna-Christina and Fan, Xiaohui and Gaikwad, Prakash and Greig, Bradley and Haehnelt, Martin G. and Kulkarni, Girish and Lai, Samuel and Pallottini, Andrea and Qin, Yuxiang and {Ryan-Weber}, Emma V. and Walter, Fabian and Wang, Feige and Yang, Jinyi},
  year = 2022,
  month = jun,
  journal = {The Astrophysical Journal},
  volume = {932},
  pages = {76},
  publisher = {IOP},
  issn = {0004-637X},
  url = {https://ui.adsabs.harvard.edu/abs/2022ApJ...932...76Z},
  urldate = {2025-04-17}
}
\bibliographystyle{aasjournalv7}



\end{document}